\documentclass[%
 reprint,
superscriptaddress,
 amsmath,amssymb,
 aps,
]{revtex4-2}
\usepackage{graphicx}
\usepackage{dcolumn}
\usepackage{bm}
\usepackage{wrapfig}
\usepackage{mhchem}
\usepackage{supertabular}
\usepackage[table,xcdraw]{xcolor}
\begin{document}

\title{A first-principles theoretical study on two-dimensional MX and \ce{MX2} metal halides: bandgap engineering, magnetism, and catalytic descriptors}




\author{Yu-Hsiu Lin}
\affiliation{Department of Chemical Engineering \& Materials Science, Michigan State University, East Lansing, Michigan 48824, United States.}
 
\author{Daniel Maldonado-Lopez}%
\affiliation{Department of Chemical Engineering \& Materials Science, Michigan State University, East Lansing, Michigan 48824, United States.}

\author{Jose L. Mendoza-Cortes}
 \email{jmendoza@msu.edu}
\affiliation{Department of Chemical Engineering \& Materials Science, Michigan State University, East Lansing, Michigan 48824, United States.}
\affiliation{Department of Physics \& Astronomy, Michigan State University, East Lansing, Michigan 48824, United States.}

\date{\today}

\begin{abstract}
Metal halides, particularly MX and MX$_2$ compounds (where M represents metal elements and X = F, Cl, Br, I), have attracted significant interest due to their diverse electronic and optoelectronic properties. However, a comprehensive understanding of their structural and electronic behavior, particularly the evolution of these properties from bulk to low-dimensional forms, remains limited. To address this gap, we performed first-principles calculations to develop a database of 60 MX and MX$_2$ metal halides, detailing their structural and electronic properties in both bulk and slab configurations.
Calculations were performed using the advanced \texttt{HSE06-D3} hybrid functional for density functional theory (DFT), ensuring high precision in predicting material properties despite the associated computational cost. The results reveal that these materials are predominantly semiconductors, but their bandgaps range from 0 to 9 eV. A detailed analysis of the transition from bulk to slab structures highlights notable shifts in electronic properties, including bandgap modifications.
Upon dimensional reduction, 9 materials exhibit an indirect-to-direct bandgap transition, enhancing their potential for energy conversion. Beyond structural dimensionality, the influence of chemical composition on bandgap variations was also examined. To further assess their practical applicability, the catalytic and magnetic properties of these metal halides were systematically evaluated. These findings not only illuminate previously underexplored MX and MX$_2$ metal halides but also identify promising candidates for electronic, optoelectronic, catalytic and spintronic applications. This database serves as a valuable resource for guiding future research and technology development in low-dimensional materials.

\end{abstract}

\maketitle

\section{Introduction}

Since the groundbreaking exfoliation of graphene from graphite,\cite{novoselov2004electric} two-dimensional (2D) materials have become a focal point of research due to their unique properties and potential applications.\cite{xu2013graphene,wang2012electronics,chhowalla2013chemistry} These materials often exhibit distinct characteristics compared to their bulk counterparts, making them valuable for various technological applications.\cite{qian2014quantum,kuc2011influence} However, the vast diversity of 2D materials has left many of them underexplored. Among these, pristine MX and MX$_2$ metal halides (MH) represent a significant yet understudied class of 2D materials, with limited comprehensive experimental and computational investigations.\cite{nicolosi2013liquid,lebegue2013two}
Previous studies on MX and MX$_2$ metal halides often focus on specific elements or small subsets of the periodic table, leaving a substantial portion of this material class unexplored.\cite{ahmed2010compton} Despite this, the potential of these materials in applications such as optoelectronics,\cite{huang2021single,yue2021single,kuklin2020two,tao2017computational,mahida2022first,hoat2019assessing,almayyali2021two} spintronics,\cite{zhou2015new,an2022spin,li2018high} thermoelectrics,\cite{tao2022biaxial} energy storage,\cite{wu2021comparative} and magnetism\cite{luo2020electric,xu2020prediction,liu2018screening,liu2018exfoliated,jiang2018screening} has been increasingly recognized.
In addition, in existing theoretical studies, the accuracy of the prediction still has room for refinement. This highlights the need for a comprehensive and precise database that consolidates the structural and electronic properties of MX and MX$_2$ metal halides.

Theoretical studies of metal halides have predominantly employed density functional theory (DFT) with generalized gradient approximation (GGA) functionals, such as \texttt{PBE}, for both geometric optimizations and property calculations.\cite{jiang2018screening,hoat2019tuning,naseri2020theoretical} While these methods offer high computational efficiency, they often yield significant deviations in key properties, such as bandgap underestimation. For example, calculations on monolayer MX$_2$ metal iodides have shown that bandgaps computed using the \texttt{HSE} hybrid functional are 0.33–1.27 eV higher than those obtained with \texttt{PBE}.\cite{rappoport2006approximate}
In some studies, hybrid functionals have been employed solely for property calculations, while structural information is derived from GGA-level optimizations.\cite{tao2017computational,almayyali2021two,tao2022biaxial,liu2018exfoliated} However, inaccuracies in GGA-optimized geometries can propagate to electronic property predictions. Specifically, discrepancies in lattice constants between \texttt{PBE} and hybrid functionals have been reported, with average errors of 7.6 pm for \texttt{PBE} compared to 3.5 pm for hybrid functionals.\cite{rappoport2006approximate} These geometric differences, though seemingly minor, can have a noticeable impact on electronic properties, underscoring the importance of using consistent and accurate methods for both geometry and property calculations.
In this work, we address these limitations by employing DFT with the \texttt{HSE06} hybrid functional, supplemented by Grimme's three-body dispersion correction (\texttt{HSE06-D3}), for both structural optimizations and property calculations. This approach ensures a higher level of accuracy in predicting the characteristics of MX and MX$_2$ metal halides, providing reliable data.

\begin{figure*}
\includegraphics[width=\textwidth]{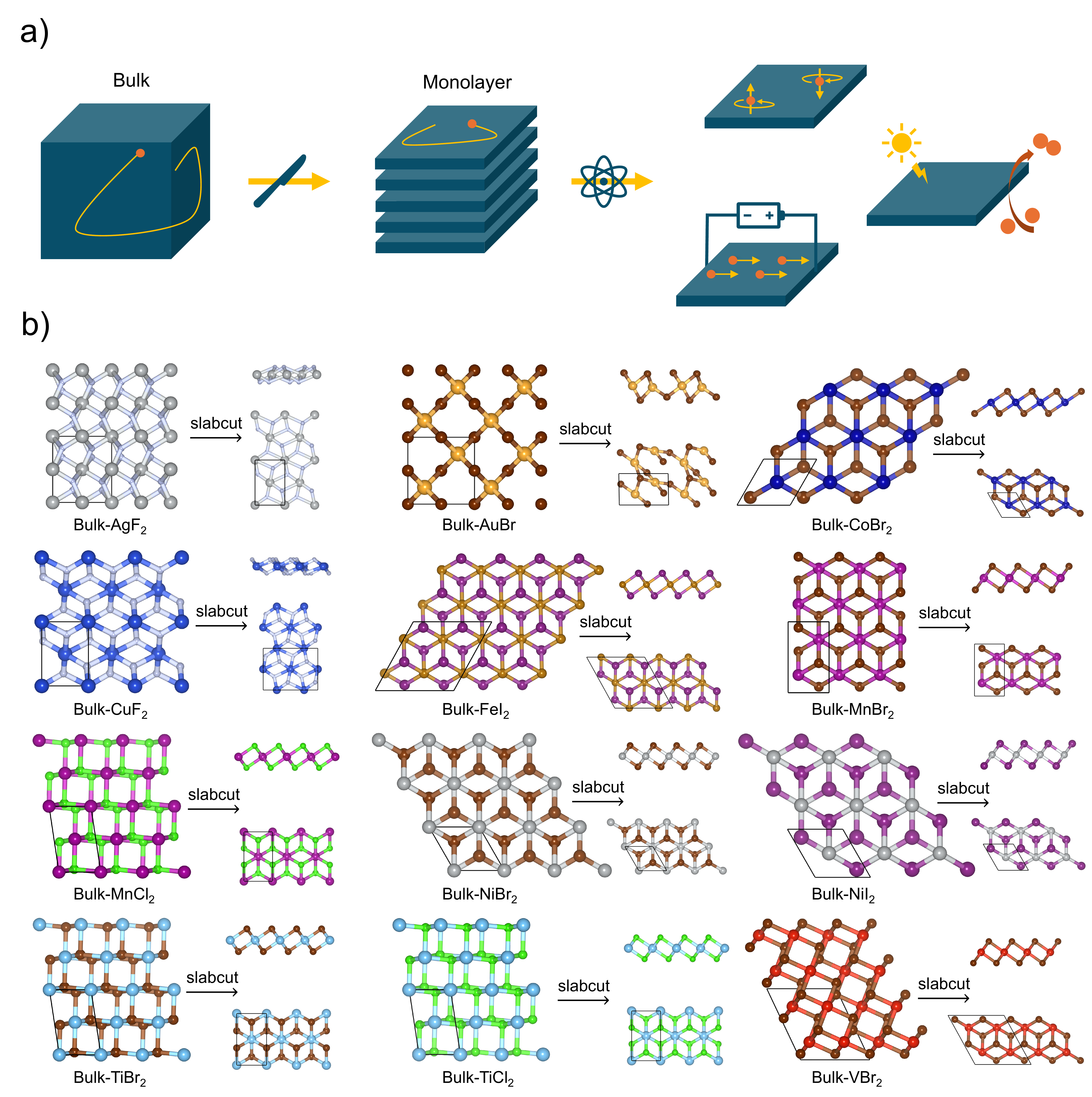}
\centering
\caption{\footnotesize Prediction of isolated 2D monolayers derived from bulk counterparts. (a) Monolayers slabcut from bulk structures and optimized through ab initio calculations, exhibiting distinct electronic, catalytic, and spintronic properties due to electron confinement effects. (b) Examples of optimized bulk and monolayer crystal structures, with spheres representing different elements: Ag (gray), Au (gold), Br (brown), Cl (fluorescent green), Co (dark blue), Cu (blue), F (light blue), Fe (dark yellow), I (dark purple), Mn (light purple), Ni (silver), Ti (aqua blue), and V (red). Bulk structures visualized from a top view while monolayers consisting of a side view and a top view.}
\label{fig:concept}
\end{figure*}

The properties of both 2D and bulk structures of MX and MX$_2$ metal halides were systematically calculated, as significant changes in properties between these structural forms can offer valuable insights. Such differences, particularly in electronic properties, highlight the potential to tune the characteristics of the material by varying the number of layers, making dimensional reduction a viable strategy for achieving desirable properties for specific applications. With the high accuracy provided by the \texttt{HSE06-D3} functional, particular attention was given to identifying shifts in electronic properties during the transition from bulk to slab structures, such as indirect-to-direct bandgap transitions. These findings underscore the potential of MX and MX$_2$ metal halides as candidates for targeted electronic and optoelectronic applications. To further explore other potential applicability, detailed calculations and analyses were conducted to investigate their catalytic and magnetic characteristics from the perspective of dimensionality.
In addition, trends and correlations between characteristics within the database were analyzed. For example, the influence of varying atomic numbers of halogens on the properties of MX and MX$_2$ compounds was explored, providing a deeper understanding of the factors governing their behavior. This comprehensive investigation not only addresses existing ambiguities surrounding MX and MX$_2$ metal halides but also contributes valuable data to the broader field of 2D materials, paving the way for further exploration and utilization.

\section{Methodology}
The calculations were carried out using unrestricted hybrid DFT with the Heyd–Scuseria–Ernzerhof (\texttt{HSE06}) exchange-correlation functional,\cite{heyd2003hybrid,krukau2006influence} as implemented in the CRYSTAL23 ab initio solid-state chemistry modeling software.\cite{erba2022crystal23, dovesi2023crystal23} The \texttt{HSE06} functional was chosen for its proven accuracy in predicting both structural and electronic properties.\cite{rappoport2006approximate} CRYSTAL23 employs Gaussian-type orbitals (GTOs), which facilitate efficient implementation of advanced methods such as hybrid functionals and post-Hartree–Fock approaches. The structural optimizations were performed using triple-zeta with polarization quality (TZVP) Gaussian basis sets for all elements,\cite{vilela2019bsse} ensuring high precision. Additionally, the Grimme-D3 dispersion correction was incorporated to account for van der Waals interactions.\cite{grimme2010consistent}

The initial bulk structures of metal halides were provided by Inorganic Crystal Structure Database
(ICSD)\cite{belsky2002new} and Heine et al.\cite{miro2014atlas} 
Materials were first optimized in their bulk form using the \texttt{HSE06} functional.
Then, preliminary monolayer structures were derived from these optimized bulk counterparts by cleaving the slabs along the main crystallographic plane.
To avoid self-interactions due to periodic boundary conditions, a vacuum of 500 Å was applied in the [001] direction, a default value in CRYSTAL23 for slab calculations. The use of GTOs instead of plane waves mitigates the computational challenges typically associated with adding large vacuum spaces. The convergence thresholds of energies were set at 10$^{-7}$ a.u. for all calculations.
The reciprocal space was sampled using a $\Gamma$–centered Monkhorst–Pack grid with a resolution of approximately $2\pi$×$1/60$ Å$^{-1}$.\cite{monkhorst1976special} All structures were optimized under full relaxation of cell parameters and atom positions with applicable symmetries found from FINDSYM.\cite{stokes2005findsym} The magnetic properties of the materials were analyzed by initially assigning polarized spin configurations corresponding to ferromagnetic and antiferromagnetic alignments during structural optimization. A material was classified as magnetic if the assigned spin configuration persisted after optimization and resulted in a more stable ground state compared to non-magnetic configurations.
The electronic band structures were calculated along k–paths selected based on the crystal symmetry.

Apart from bandgap values, the band alignment relative to vacuum is crucial for determining the practical applications of spin-polarized 2D metal halides. To achieve this, we calculated the work function, defined as the energy difference between an electron at infinity and the Fermi energy level. We adopted a single-point calculation approach, following the CRYSTAL tutorials website\cite{dovesi2023crystal23}, as illustrated in Figure S1, to obtain a better description of the electrostatic potential at vacuum.
Specifically, two ghost layers providing electronic states were placed above and below the optimized monolayer metal halides to induce nonzero electrostatic potentials at positive and negative values, sufficiently far away from the slab. In this case, the electrostatic potential was measured at 100 a.u. ($\approx$~53 Å) from the monolayer. Band alignment was performed by shifting the bands according to the electrostatic potential, with the vacuum energy level set to 0 eV.
The separation distance between the monolayers and ghost layers was adjusted for each compound, ensuring an observable but minimal electron population—typically below 0.01—to achieve an accurate vacuum description without compromising the material’s electronic structure. The electrostatic potential was determined as the average of the absolute potentials at positive and negative infinity. This approach addresses the limitations of Gaussian-based DFT calculations, which primarily focus on localized electron density around the material structure.

\begin{figure}[h]
    \centering
    \includegraphics[width=0.48\textwidth]{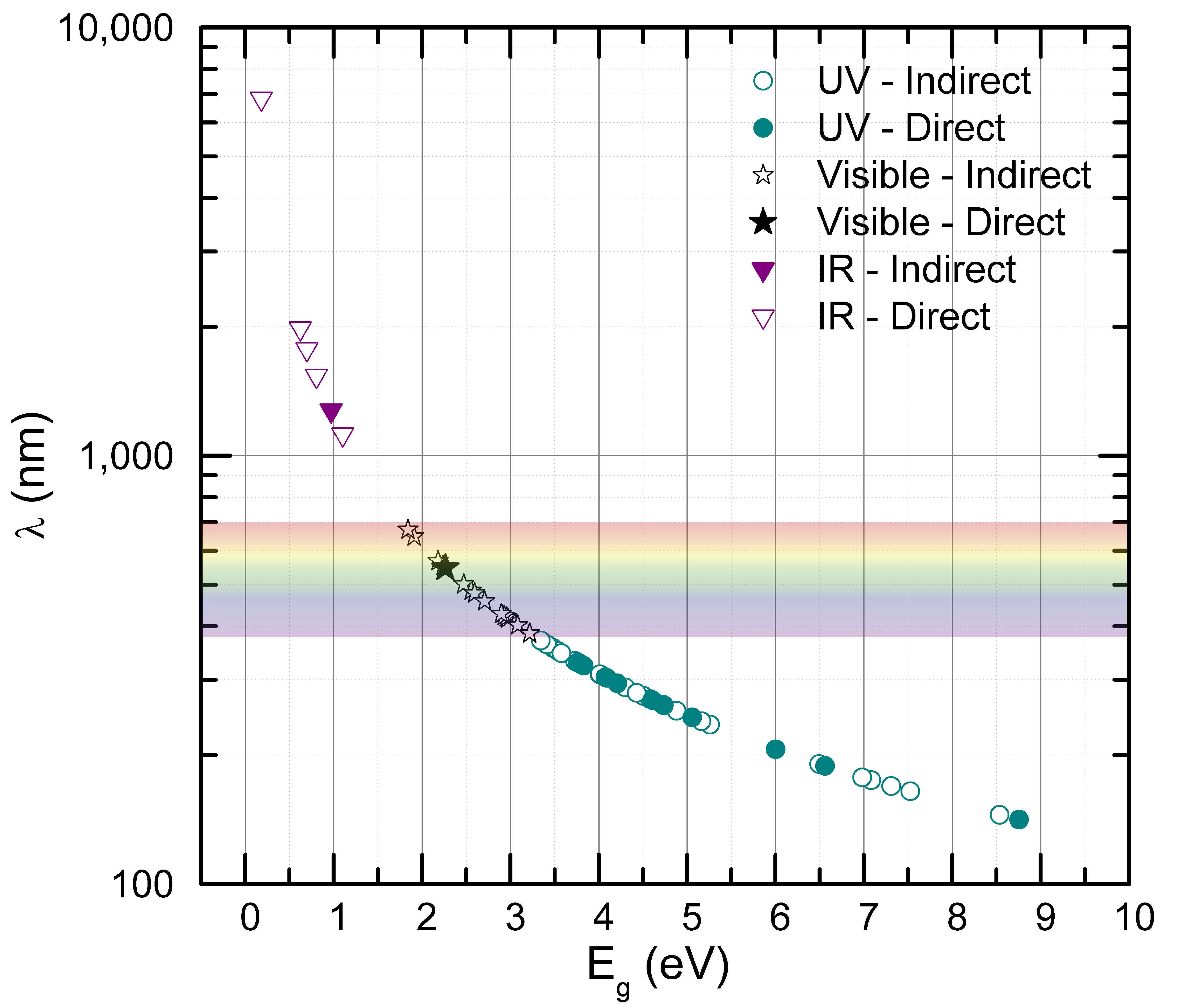}
    \caption{\footnotesize Optoelectronic properties of 2D monolayer semiconductors: Solid markers indicate materials with direct bandgaps, while hollow markers denote those with indirect bandgaps.}
    \label{fig:Eg_wavelength}
\end{figure}

\begin{figure}
    \centering
    \includegraphics[width=0.47\textwidth]{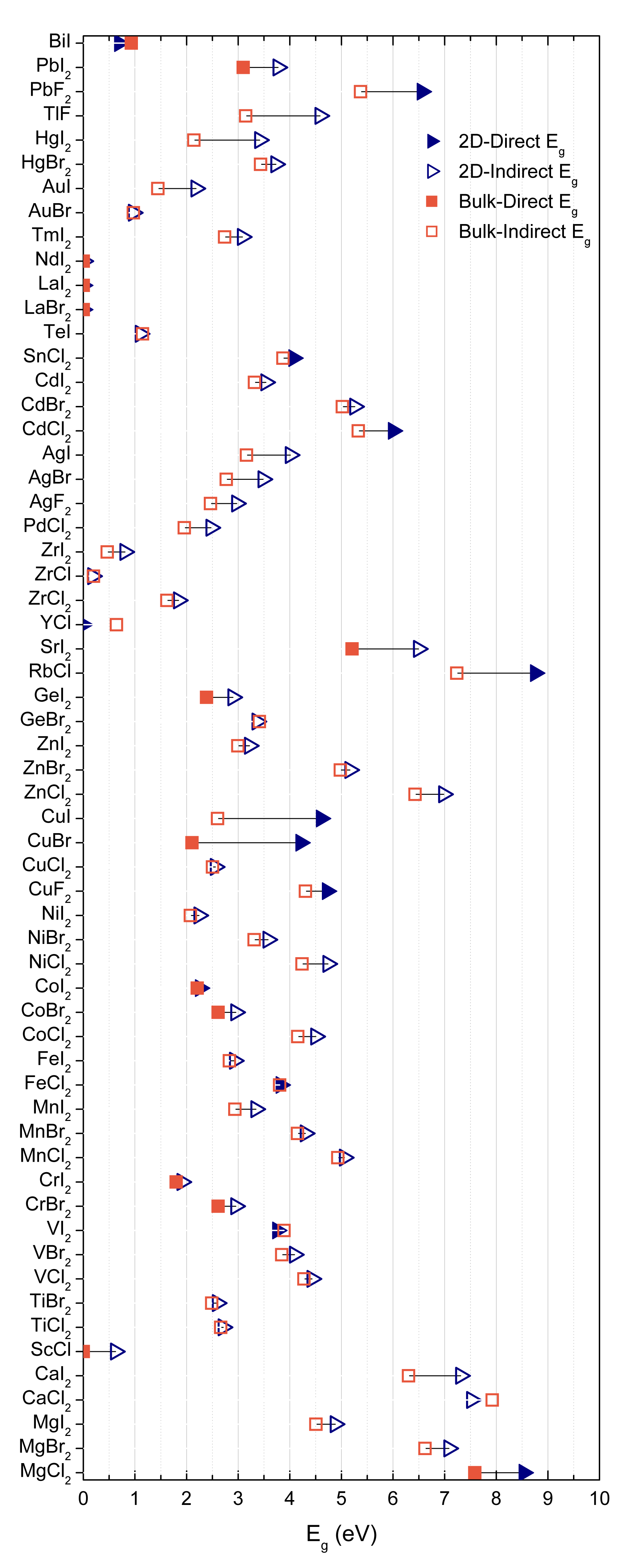}
    \caption{\footnotesize Bandgap transitions between bulk and 2D monolayer materials in the database: Solid markers represent direct bandgaps, hollow markers denote indirect bandgaps, and metallic materials are indicated by solid markers at 0 eV. Dashed lines indicate an increase in bandgap values as materials transition from bulk to monolayer structures.}
    \label{fig:Eg_Transition}
\end{figure}

\section{Discussion}
The transformation of bulk materials into isolated 2D monolayers offers a pathway to explore and exploit unique properties that arise from reduced dimensionality. As shown in Figure~\ref{fig:concept}, the process involves extracting monolayer slabs from bulk structures, followed by ab initio optimization to stabilize their geometry. This dimensional reduction induces significant changes in electronic, catalytic, and spintronic characteristics due to electron confinement effects. The figure also highlights examples of metal halides with optimized bulk and monolayer crystal structures, providing a visual representation of the diversity of materials studied and the atomic compositions involved. These transformations form the foundation for understanding the potential applications of 2D metal halides across various technological domains.
The relationship between electronic and optical properties is crucial in identifying materials suitable for optoelectronic applications. Leveraging this correlation allows for the efficient screening of candidates from the compiled database. A key example is the conversion of electronic bandgaps into their corresponding wavelengths, as shown in Figure~\ref{fig:Eg_wavelength}, which illustrates the distribution of bandgaps for the 2D monolayer semiconductors within the database. 
Among the 60 2D materials analyzed, 57 exhibit nonzero bandgaps, classifying them as semiconductors or insulators, while the remaining materials display metallic behavior. Of the semiconducting materials, 15 possess bandgaps corresponding to wavelengths within the visible (VIS) range, 7 in the infrared (IR), and 35 in the ultraviolet (UV). These findings hold particular significance for applications in energy harvesting, as over 90\% of the solar energy reaching Earth's surface consists of visible and infrared light. Materials with direct bandgaps aligned with these wavelengths have the potential to act as efficient solar energy harvesters. Notably, two materials from the database meet these criteria (VIS: CoI$_2$; IR: BiI).
While 2D materials with bandgaps in the UV range are less suitable for solar energy applications, they offer promise as UV detectors due to their ability to absorb high-energy photons. Specifically, eleven materials (CdCl$_2$, CoBr$_2$, CuBr, CuF$_2$, CuI, FeCl$_2$, MgCl$_2$, PbF$_2$, RbCl, SnCl$_2$, VI$_2$) were identified with direct bandgaps spanning a broad UV range, from 140 nm to 330 nm. Beyond light absorption, Figure~\ref{fig:Eg_wavelength} also provides insights into potential light-emitting applications, offering a preliminary understanding of these materials as emitters at specific wavelengths.

The exfoliation of graphite into graphene has sparked significant interest in exploring the 2D counterparts of various layered materials. While experimental approaches to synthesizing and characterizing these materials are often labor-intensive and time-consuming, advancements in computational techniques have provided an efficient alternative for predicting and analyzing their properties. Figure~\ref{fig:Eg_Transition} offers computational insights into the potential benefits of exfoliating bulk materials into their 2D forms, focusing on changes in electronic properties.
In Figure~\ref{fig:Eg_Transition}, the data points represent the bandgap values of both 2D and bulk materials, along with their bandgap types (direct or indirect). Generally, the bandgap tends to increase slightly as the structure transitions from bulk to monolayer. However, seven materials (BiI, CaCl$_2$, GeBr$_2$, TeI, VI$_2$, YCl, ZrCl) exhibit a notable decrease in bandgap upon this transformation. These findings highlight the potential to fine-tune electronic properties by modulating the number of layers in layered materials. Furthermore, transitions in bandgap types, such as from indirect to direct, are observed, which have significant implications for energy efficiency and optoelectronic applications.
Specifically, nine materials (CdCl$_2$, CoBr$_2$, CuF$_2$, CuI, FeCl$_2$, PbF$_2$, RbCl, SnCl$_2$, VI$_2$) undergo an indirect-to-direct bandgap transition during the bulk-to-slab transformation, enhancing their potential for efficient light emission and absorption. Conversely, some materials experience other types of transitions: BiI, CoI$_2$, CuBr, and MgCl$_2$ remain direct bandgaps, LaBr$_2$ and LaI$_2$ keep zero bandgaps, while six others (CrBr$_2$, CrI$_2$, GeI$_2$, NdI$_2$, PbI$_2$, SrI$_2$) exhibit direct-to-indirect transitions. More extremely, ScCl and YCl can shift between a semiconductor and a conductor with a zero bandgap by undergoing the dimensional transformation. These results demonstrate that both the magnitude and type of bandgap are influenced by the number of layers, underscoring the versatility of these materials for tailored applications in electronic and optoelectronic devices. Apart from the bandgap, the crystal structures are affected fundamentally. The mean bond distance between metal and halide atoms is shown in Figure S2, which quantifies how elemental and dimensional factors impact geometry. 

\begin{figure}[t]
    \centering
    \includegraphics[width=0.48\textwidth]{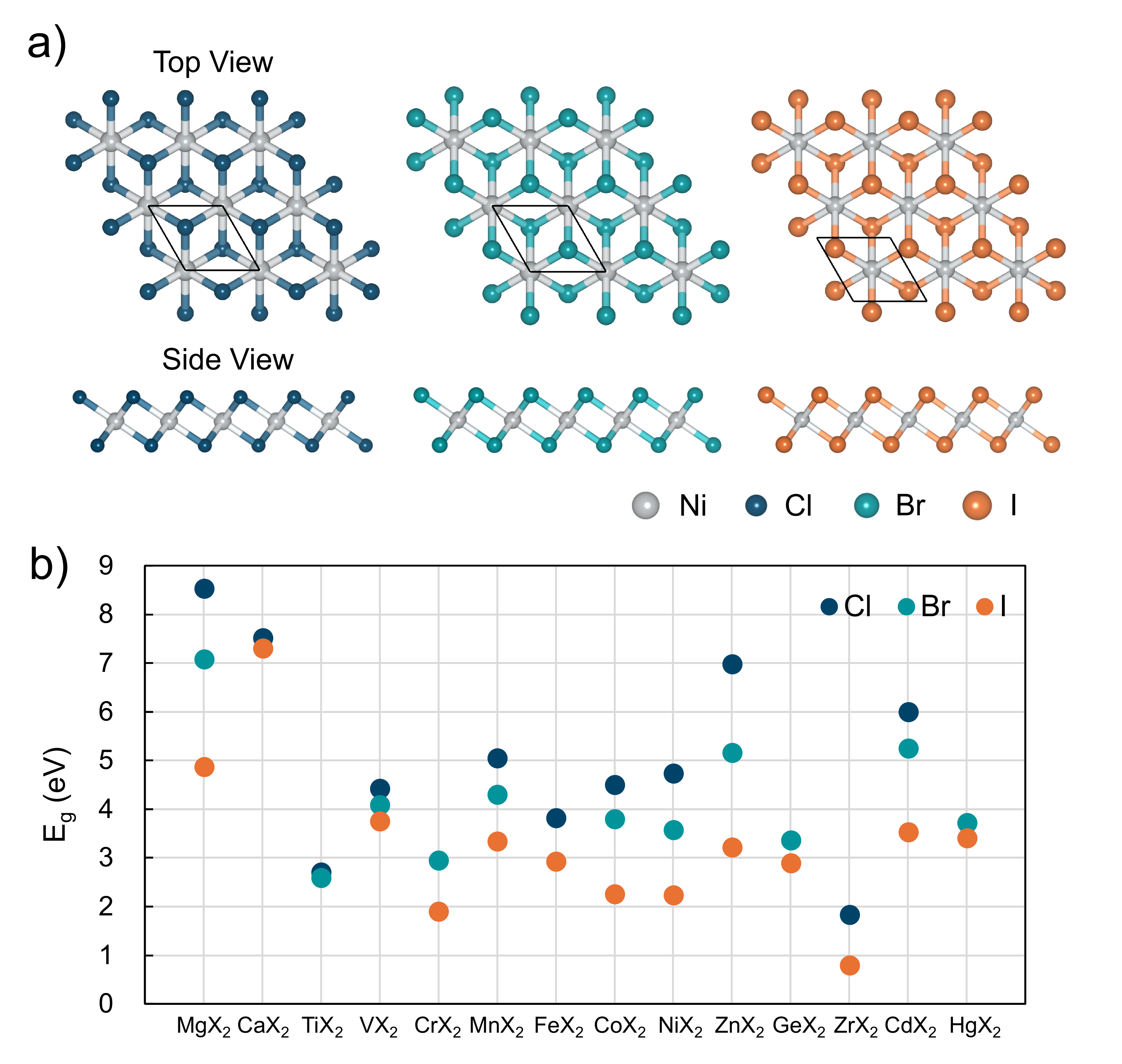}
    \caption{\footnotesize Compositional dependence of bandgap values on halogen elements in MX$_2$ monolayer compounds. (a) An example of optimized NiX$_2$ where X is exchanged among Cl, Br, and I. (b) Dependence of bandgap values.}
    \label{fig:Eg_X}
\end{figure}

\begin{figure}[t]
\includegraphics[width=0.48\textwidth]{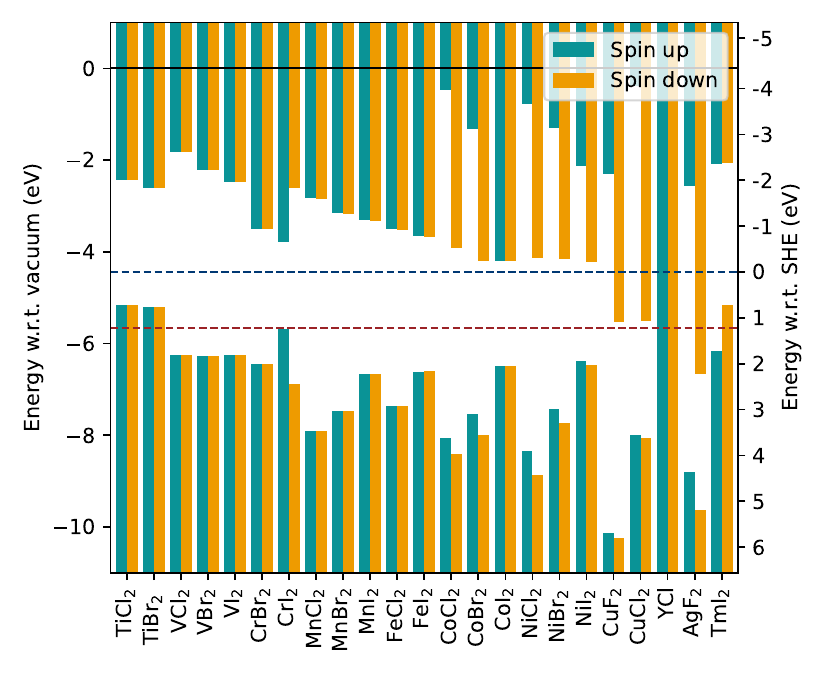}
\centering
\caption{\footnotesize Absolute band edge alignment for spin-polarized monolayer halides. $\alpha$ (spin-up) and $\beta$ (spin-down) electronic bands were plotted in green and yellow, respectively. Energy reference with respect to vacuum is positioned on the left scale and energy reference with respect to a standard hydrogen electrode (-4.44 eV w.r.t. vacuum) is positioned on the right scale. White spaces between the bands indicate the spin-polarized bandgap for each material. The vacuum, standard hydrogen electrode, and standard oxygen electrode reference levels were plotted as horizontal solid black, dashed blue, and red lines, respectively.}
\label{fig:band_alignment}
\end{figure}

Figure~\ref{fig:Eg_X} illustrates the relationship between the bandgaps of MX$_2$ metal halides and the halogen element (X). A general trend emerges in which the bandgap decreases as the atomic number of the halogen increases. This observation is consistent with the differences in electronegativity among halogens (Cl $>$ Br $>$ I). Lower electronegativity weakens the ionic bonding between the metal and halogen atoms, reducing the localization of electronic states and resulting in narrower bandgaps.\cite{lu2016class} These findings highlight the potential for tuning the electronic properties of metal halides by manipulating the strength and nature of their bonding. Such trends are not unique to the \texttt{HSE06-D3} calculations used in this study but have also been observed in prior computational works employing GGA and LDA functionals, such as those investigating monolayer alkaline earth and transition metal halides\cite{lin2014towards}, and organometallic halide perovskites.\cite{castelli2014bandgap} This alignment underscores the robustness of the observed trends and provides a qualitative framework for understanding the electronic behavior of metal halides. 

Figure~\ref{fig:band_alignment} presents the band edge alignments relative to the vacuum level for spin-polarized monolayers in the database. The spin-up and spin-down bands are plotted separately, enabling the identification of materials with potential spintronic applications. In specific, these monolayers exhibit spin-split band structures, where the spin-up and spin-down electrons have distinct bandgap values.
Among these monolayers, the MX$_2$ metal halides, in which M $=$ Co, Cu, and Ni, are more likely to induce high spin-polarization.
Notably, several materials, AgF$_2$, CuCl$_2$, CrI$_2$ and NiI$_2$, demonstrate characteristics of half semiconductors. These materials behave as semiconductors (bandgap $<$ 3 eV) in a spin channel while acting as insulators (bandgap $>$ 3 eV) in the other. This unique behavior arises because of distinct band edges in the two spin channels. Consequently, upon thermal or optical excitation, these materials can generate highly spin-polarized electrons or holes, a property desirable for spintronic devices.
Especially, AgF$_2$ and NiI$_2$ stand out as potential bipolar magnetic semiconductors. In these materials, the VBM is fully polarized in the spin-up channel, while the CBM is fully polarized in the spin-down channel. This configuration makes them possible to achieve fully spin-polarized currents by applying an external gate voltage. Under zero gate voltage, they behave as conventional semiconductors. However, when a negative gate voltage is applied, the Fermi level shifts upward, enabling a conductive spin-down electron channel while maintaining an insulating spin-down channel. Conversely, a positive gate voltage raises the Fermi level, resulting in a conductive spin-up channel and a semiconducting spin-down channel.
This ability to control spin orientation and polarization through gate voltage modulation positions materials as promising candidates for next-generation spintronic applications. The tunability of spin-dependent electronic properties in these materials highlights their potential for developing efficient and versatile spintronic devices.

The potential catalytic activity of the materials can also be assessed using Figure~\ref{fig:band_alignment}, which includes energy references for two key catalytic reactions: hydrogen reduction (H$^+$/H$_2$) and oxygen evolution (O$_2$/H$_2$O). These reactions are represented by the Standard Hydrogen Electrode (SHE) and Standard Oxygen Electrode (SOE) energy levels, plotted as blue and red dashed lines, respectively. A semiconductor with a CBM positioned above the SHE level can supply excited electrons with sufficient energy for hydrogen reduction, making it a viable candidate for this reaction. Conversely, a material with a VBM below the SOE level can facilitate oxygen evolution by providing holes at appropriate energy levels.
Given the growing demand for sustainable energy solutions, identifying materials capable of catalyzing both hydrogen reduction and oxygen evolution is critical for efficiently producing hydrogen from water. For a material to be an excellent photocatalyst, it must meet three criteria: (1) the CBM should be above the SHE level to enable hydrogen reduction, (2) the VBM should be below the SOE level to support oxygen evolution, and (3) the bandgap should fall within the main solar spectrum (1.2–2.8 eV) to harness sunlight effectively for charge carrier generation.\cite{yan2017solar}
Among these spin-polarized monolayers, two materials satisfy these requirements: CrI$_2$ with a bandgap of 1.91 eV in the $\alpha$-electron channel and NiI$_2$ with a bandgap of 2.24 eV in the $\beta$-electron channel, indicating their potential to photo-catalytic applications.

\begin{figure*}
\includegraphics[width=\textwidth]{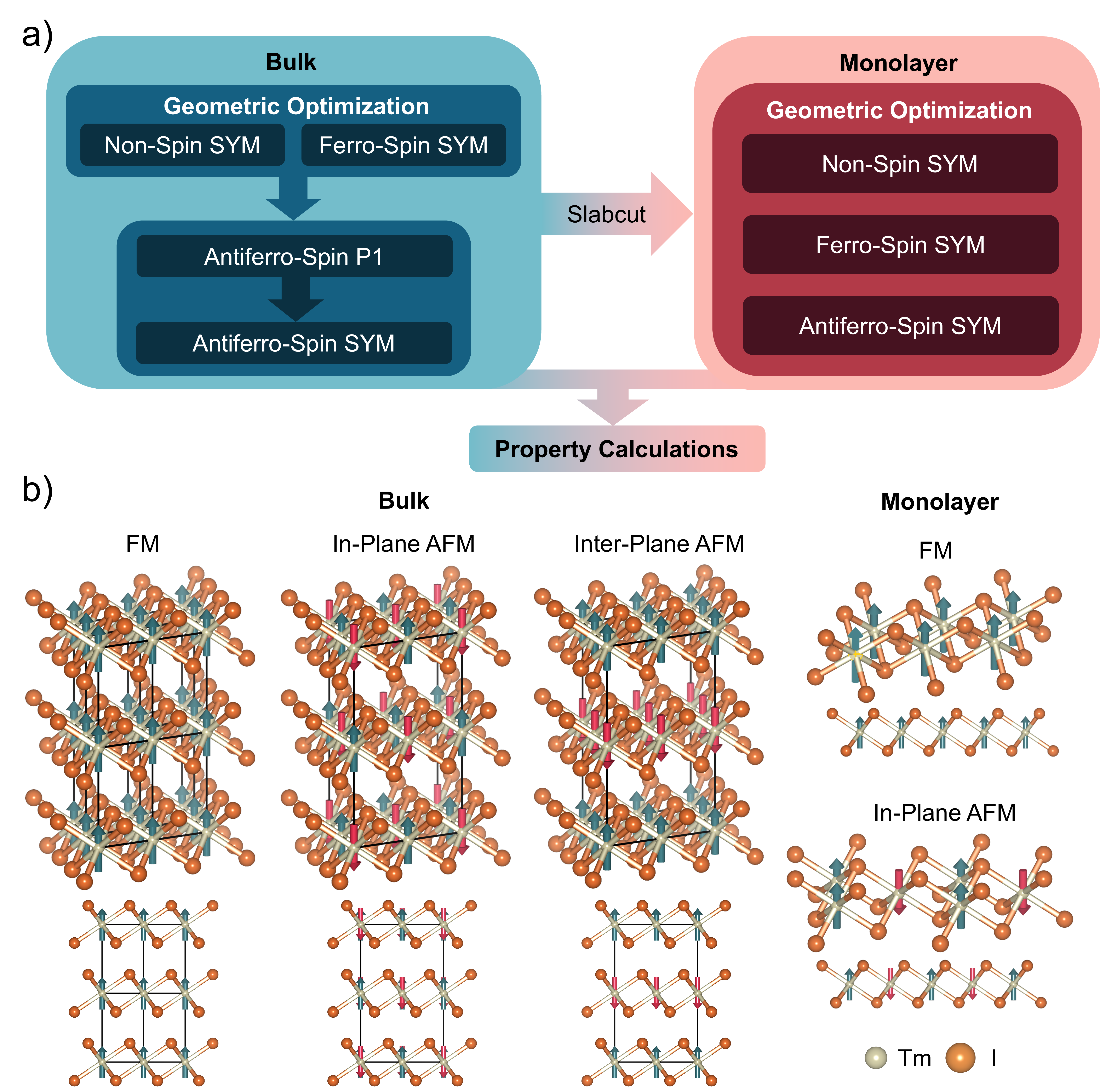}
\centering
\caption{\footnotesize Investigation on magnetic property. (a) Procedure to define the magnetic feature of each bulk and monolayer metal halide. (b) Illustrations of magnetic states in the case of TmI$_2$, including the forms of bulk and monolayer with different spin arrangements. Visualized geometry of each combination consists of a perspective view and a side view. Green and red arrows denote spin-up and spin-down electrons, respectively. Black boxes represent unit cells.}
\label{fig:method}
\end{figure*}

\begin{table}[h]
\caption{\footnotesize Summary of magnetic metal halides, detailing the types of magnetism observed in both 2D monolayer and bulk structures. FM and AFM denote ferromagnetic and antiferromagnetic, respectively. The state of AFM is further specified by inter-plane (Inter) and in-plane (In) spin arrangement.}
\label{table:magnetinism}
\begin{tabular}{lrr} \hline
\textbf{Material}  & \textbf{Magnetism (bulk)} & \textbf{Magnetism (slab)} \\ \hline
\ce{AgF2}    & (Inter) AFM  & FM     \\
\ce{AuBr}    & (Inter) AFM  & -    \\
\ce{CoBr2}   & (Inter) AFM  & FM     \\
\ce{CoCl2}   & (Inter) AFM  & FM     \\
\ce{CoI2}    & (Inter) AFM  & (In) AFM \\
\ce{CrBr2}   & (In) AFM     & (In) AFM \\
\ce{CrI2}    & (In) AFM     & FM     \\
\ce{CuCl2}   & (Inter) AFM  & FM     \\
\ce{CuF2}    & (Inter) AFM  & FM     \\
\ce{FeCl2}   & (In) AFM     & (In) AFM \\
\ce{FeI2}    & (In) AFM     & (In) AFM \\
\ce{MnBr2}   & (In) AFM     & (In) AFM \\
\ce{MnCl2}   & (In) AFM     & (In) AFM \\
\ce{MnI2}    & (Inter) AFM  & (In) AFM \\
\ce{NiBr2}   & (Inter) AFM  & FM     \\
\ce{NiCl2}   & (Inter) AFM  & FM     \\
\ce{NiI2}    & (Inter) AFM  & FM     \\
\ce{ScCl}    &        FM    & -    \\
\ce{TiBr2}   & (In)   AFM   & (In) AFM \\
\ce{TiCl2}   & (In)   AFM   & (In) AFM \\
\ce{TmI2}    &        FM    & FM     \\
\ce{VBr2}    & (In) AFM     & (In) AFM \\
\ce{VCl2}    & (In) AFM     & (In) AFM \\
\ce{VI2}     &        FM    & (In) AFM \\
\ce{YCl}     & (Inter) AFM  & FM     \\\hline   
\end{tabular}
\end{table}

The distribution of spin not only governs the magnetic properties of materials but also plays a critical role in determining their structural stability. To account for this, different initial spin configurations were applied during the design phase prior to geometric optimization. The most stable spin configuration for each material was identified as the one corresponding to the lowest total energy after optimization. To systematically analyze the influence of dimensionality on magnetic properties, the types of magnetism, ferromagnetic (FM) or antiferromagnetic (AFM), were determined for both bulk and 2D monolayer forms.
Figure~\ref{fig:method}(a) illustrates the methodology used to define the magnetic states of 60 metal halides. Initially, bulk structures were optimized both with and without assigned spins corresponding to the FM state. The more stable ground state was then selected between these two configurations. For materials that stabilized in the FM state, an additional optimization was performed with spins assigned to an AFM configuration. Since this reassignment altered atomic differentiation and disrupted the original symmetry, a re-symmetrization process was applied, mapping the structure onto a different space group. A second round of energy comparisons between FM and AFM states was conducted to determine the final magnetic states of the bulk structures. Subsequently, the most stable bulk structures were cleaved to construct the initial monolayer models. Before optimization, each monolayer was assigned spin configurations corresponding to FM, AFM, and a non-magnetic state. The final magnetic states of the monolayers were determined based on the lowest-energy configuration among these possibilities. The energy difference of material structures with varied spin assignments is plotted in Figure S3. 
The variation in spin configurations between bulk and monolayer structures further highlights the effects of electron confinement induced by reduced dimensionality. Figure~\ref{fig:method}(b) provides a visual representation of the assigned spin states in both bulk and monolayer forms. Notably, inter-plane AFM ordering is exclusively observed in bulk structures, as the interlayer interactions necessary to sustain this configuration are absent in monolayers.

Within the database, 25 metal halides were found to exhibit magnetic behavior, as summarized in Table~\ref{table:magnetinism}.
Among the 25 magnetic materials, 12 retain the same type of magnetism when transitioning from bulk to monolayer structures. Interestingly, 10 materials (AgF$_2$, CoBr$_2$, CoCl$_2$, CrI$_2$, CuCl$_2$, CuF$_2$, NiBr$_2$, NiCl$_2$, NiI$_2$, and YCl) undergo a magnetic transition from antiferromagnetic in the bulk to ferromagnetic in their 2D monolayer form, and one material (VI$_2$) undergo an opposite ferromagnetic-to-antiferromagnetic transition. In addition, the magnetic feature disappears in two materials (AuBr and ScCl) under a bulk-to-2D transformation.
This dimensional transition in magnetic behavior aligns with trends observed in experimentally studied metal halides and provides valuable insights into unexplored materials within the database. These findings highlight the potential of 2D metal halides for applications in spintronics and other magnetic technologies, emphasizing the importance of structural dimensions in tailoring magnetic properties.

\section{Conclusion}
This study comprehensively examines the structural, electronic, and magnetic properties of MX and MX$_2$ metal halides in both bulk and 2D forms using high-accuracy hybrid density functional theory (HSE06-D3). Transformations from bulk to monolayer structures were shown to significantly affect material properties, including bandgap values and types, with indirect-to-direct transitions identified in nine compounds. Such findings highlight the potential of these materials for optoelectronic applications, particularly in solar energy harvesting and ultraviolet detection.
The study also identifies promising candidates for spintronic and catalytic applications. Materials like AgF$_2$, NiI$_2$ exhibit spin-polarized electronic properties, while CrI$_2$ and NiI$_2$ show potential for photocatalytic water splitting due to their favorable band alignment and bandgap characteristics. Additionally, magnetic behavior analysis revealed dimensionality-driven transitions between antiferromagnetic and ferromagnetic states in select materials, offering insights for spintronic device design.
In summary, this work provides a detailed database and theoretical framework for MX and MX$_2$ metal halides, paving the way for experimental validation and future material optimization. These findings underscore the versatility of metal halides for diverse technological applications, from energy harvesting to spintronics.

\bibliography{main}

\clearpage
\newpage

\onecolumngrid

\section{Supporting Information}

\renewcommand{\figurename}{Figure S\!\!}
\renewcommand{\tablename}{Table S\!\!}
\setcounter{figure}{0}  
\renewcommand\thesubsection{\arabic{subsection}}


\begin{figure}[h]
    \centering
    \includegraphics[width=0.8\textwidth]{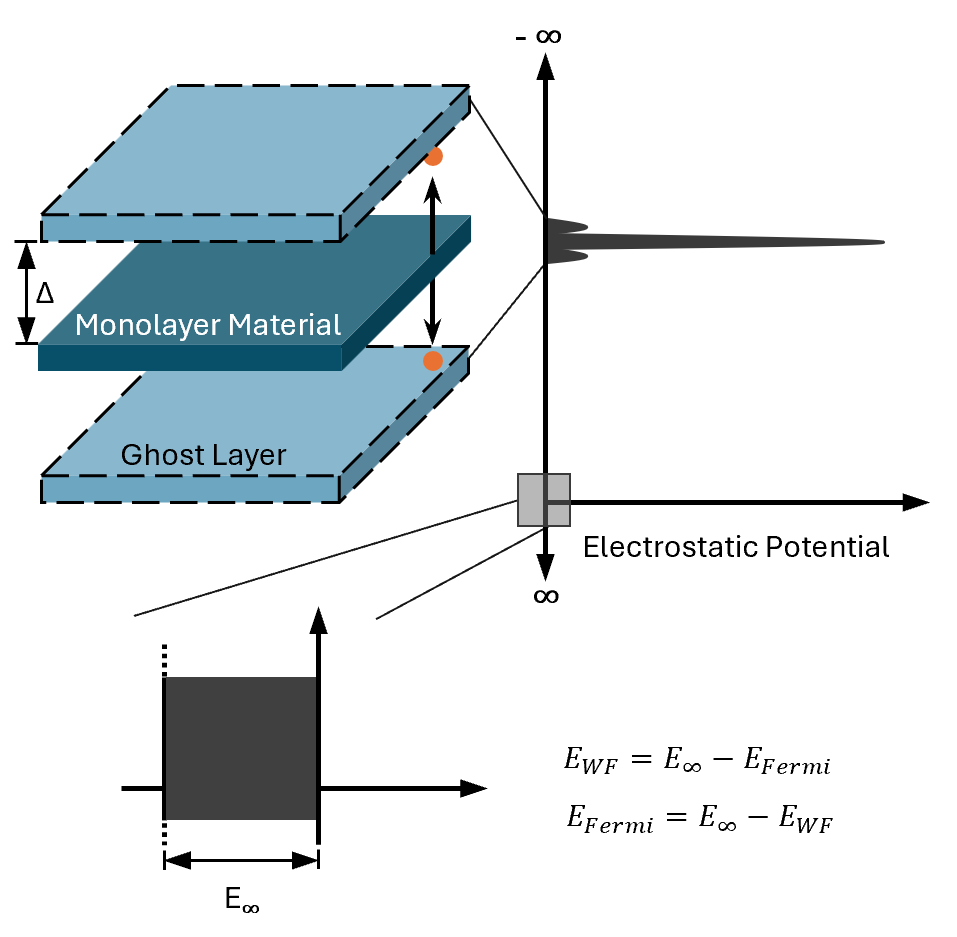}
    \caption{\footnotesize Approach for absolute band-edge alignment. Single-point calculations with ghost layers followed by comparing the energy between the Fermi level and the vacuum. $\Delta$ represents the distance between monolayer material and ghost layer. $E_{WF}$, $E_{\infty}$, and  $E_{Fermi}$ are electrostatic potential of work function, at infinity, and at Fermi energy level, respectively.}
    \label{fig:Ghost}
\end{figure}

\begin{figure}[h]
    \centering
    \includegraphics[width=\textwidth]{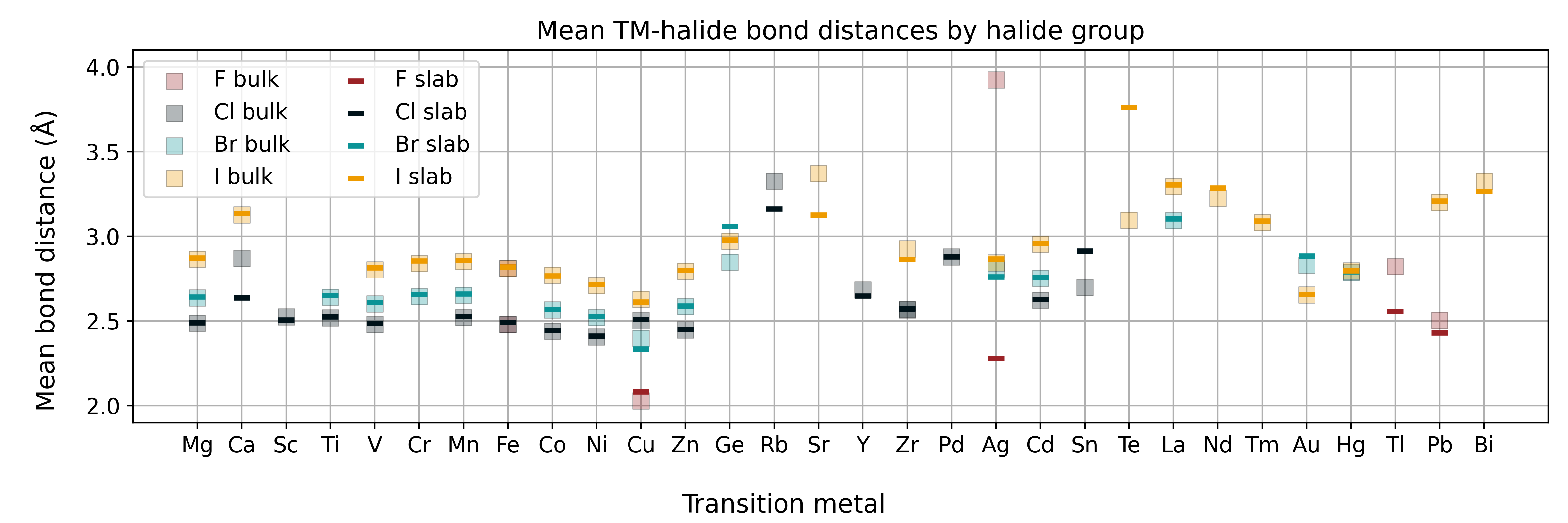}
    \caption{\footnotesize Mean bond distance between transition metal and halide elements for all of our compounds. Data separated by halide group (different colors) and plotted by transition metals in the x-axis. Bulk and slab are shown in the same graph and are represented by a transparent square (bulk) and a solid line (slab) to distinguish the compounds' dimensionality reduction.}
    \label{fig:Ghost}
\end{figure}

\begin{figure}[h]
    \centering
    \includegraphics[width=\textwidth]{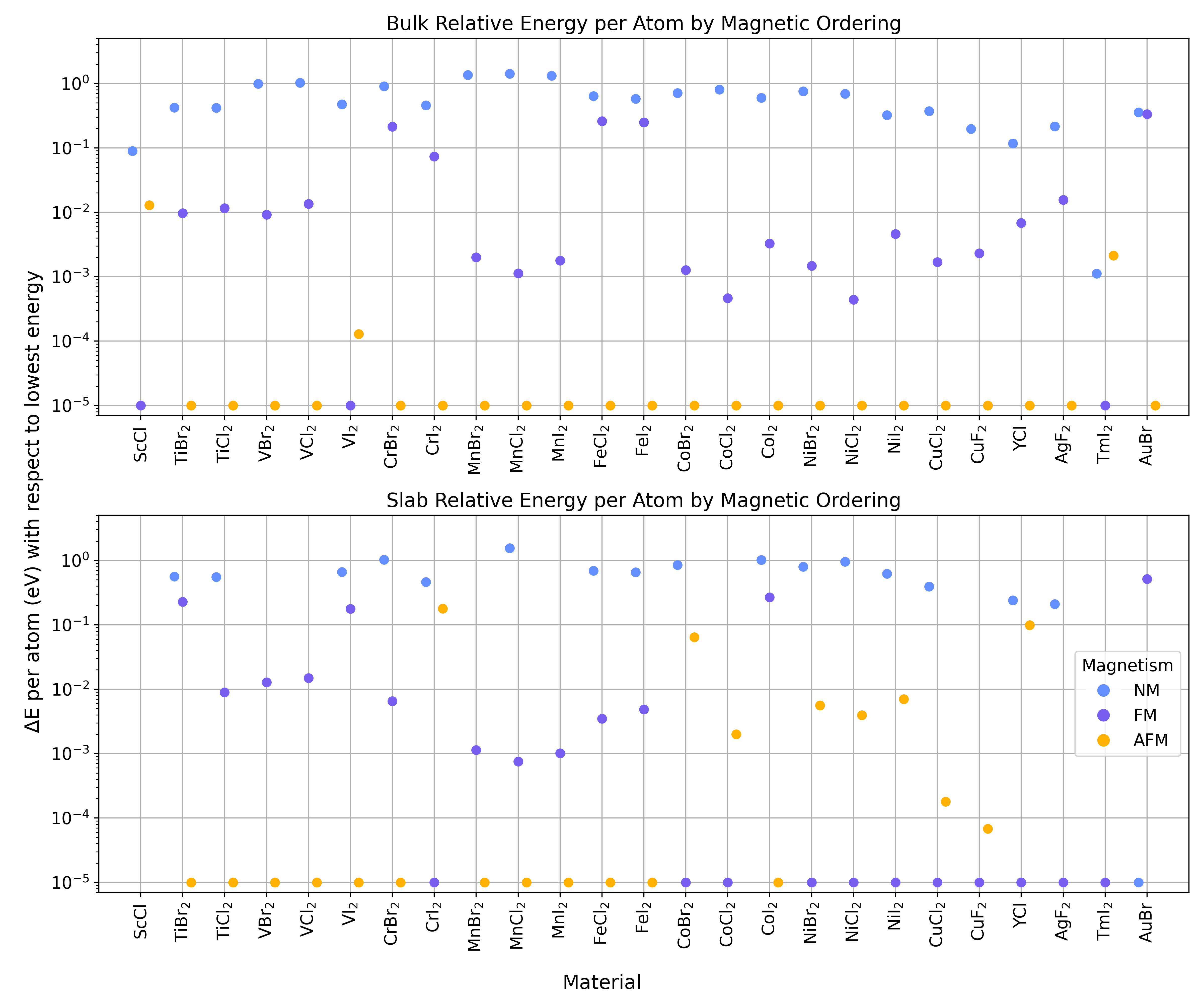}
    \caption{\footnotesize Relative total energies for materials that converged in ferromagnetic (FM) or antiferromagnetic (AFM) configurations. Different magnetic configurations were plotted in different colors for each, allowing to compare their energy differences. The data were plotted on a log scale to enhance differences at smaller energies. The ScCl slab results in non-magnetic configuration even after attempting to converge in other spin states; thus, no data was added for this slab on the spin plot.}
    \label{fig:Ghost}
\end{figure}

\clearpage
\newpage
\section{Optimized Geometries and Band Structures}
\footnotesize

\subsection{AgBr bulk}
\begin{verbatim}
_cell_length_a 3.97211208
_cell_length_b 3.97211208
_cell_length_c 3.97211208
_cell_angle_alpha 60.000000
_cell_angle_beta 60.000000
_cell_angle_gamma 60.000000
_symmetry_space_group_name_H-M         'P 1'
_symmetry_Int_Tables_number            1

loop_
_symmetry_equiv_pos_as_xyz
   'x, y, z'

loop_
_atom_site_label
_atom_site_type_symbol
_atom_site_fract_x
_atom_site_fract_y
_atom_site_fract_z
Ag001 Ag 0.000000000000E+00 0.000000000000E+00 0.000000000000E+00
Br002 Br 5.000000000000E-01 5.000000000000E-01 5.000000000000E-01
\end{verbatim}
\begin{figure}[h]
    \centering
    \includegraphics[width=\textwidth]{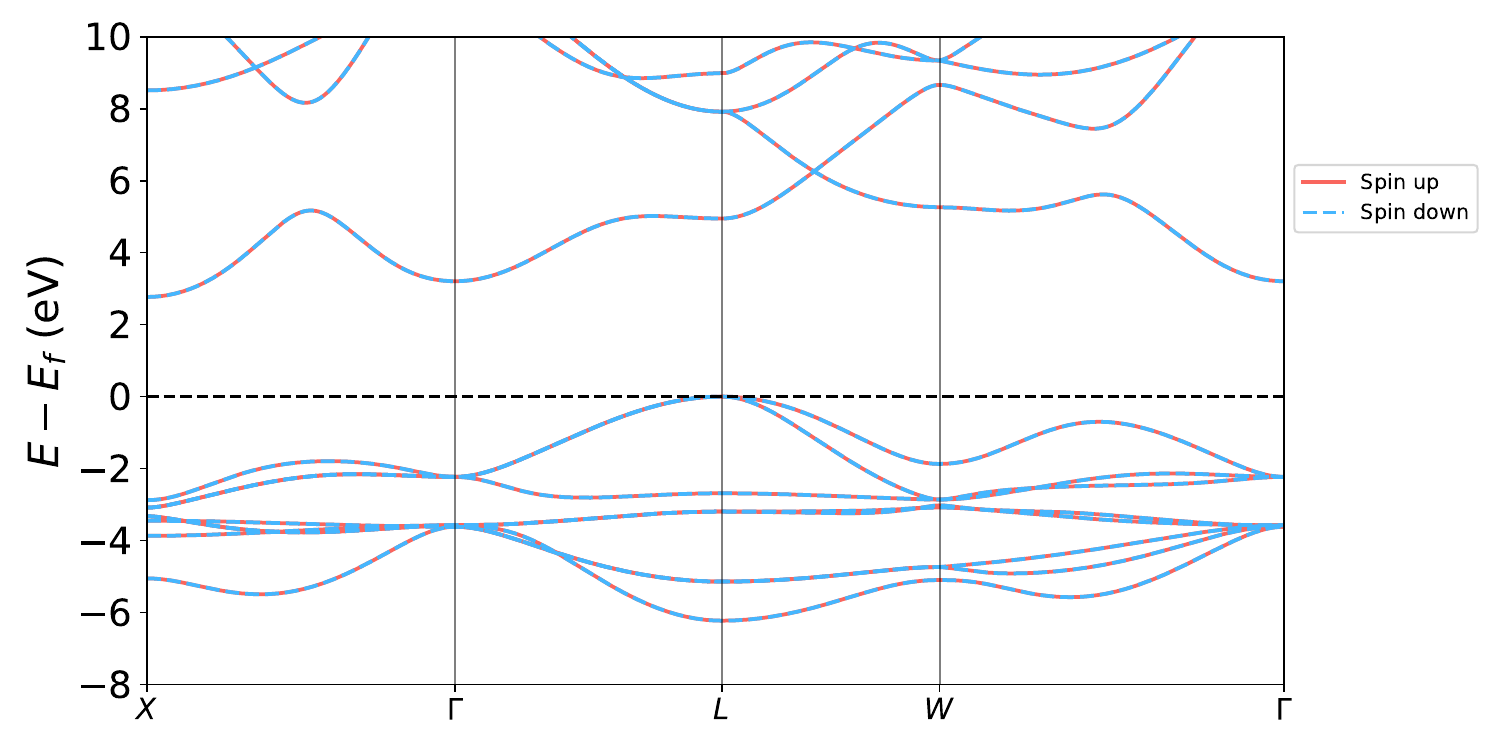}
    \caption{\footnotesize AgBr bulk spin-polarized band structure.}
    \label{fig:AgBr_band_bulk}
\end{figure}
\clearpage
\newpage

\subsection{AgBr slab}
\begin{verbatim}
_cell_length_a                         3.67363881
_cell_length_b                         3.67363881
_cell_length_c                         40.00000000
_cell_angle_alpha                      90.000000
_cell_angle_beta                       90.000000
_cell_angle_gamma                      90.000000
_cell_volume                           'P 1'
_space_group_name_H-M_alt              'P 1'
_space_group_IT_number                 1

loop_
_space_group_symop_operation_xyz
   'x, y, z'

loop_
   _atom_site_label
   _atom_site_occupancy
   _atom_site_fract_x
   _atom_site_fract_y
   _atom_site_fract_z
   _atom_site_adp_type
   _atom_site_B_ios_or_equiv
   _atom_site_type_symbol
Ag001  1.0  0.250000000000  -0.250000000000  0.518543077545  Biso  1.000000  Ag
Br002  1.0  -0.250000000000  0.250000000000  0.548127769537  Biso  1.000000  Br
Ag003  1.0  -0.250000000000  0.250000000000  0.481456922455  Biso  1.000000  Ag
Br004  1.0  0.250000000000  -0.250000000000  0.451872230463  Biso  1.000000  Br

\end{verbatim}

\begin{figure}[h]
    \centering
    \includegraphics[width=\textwidth]{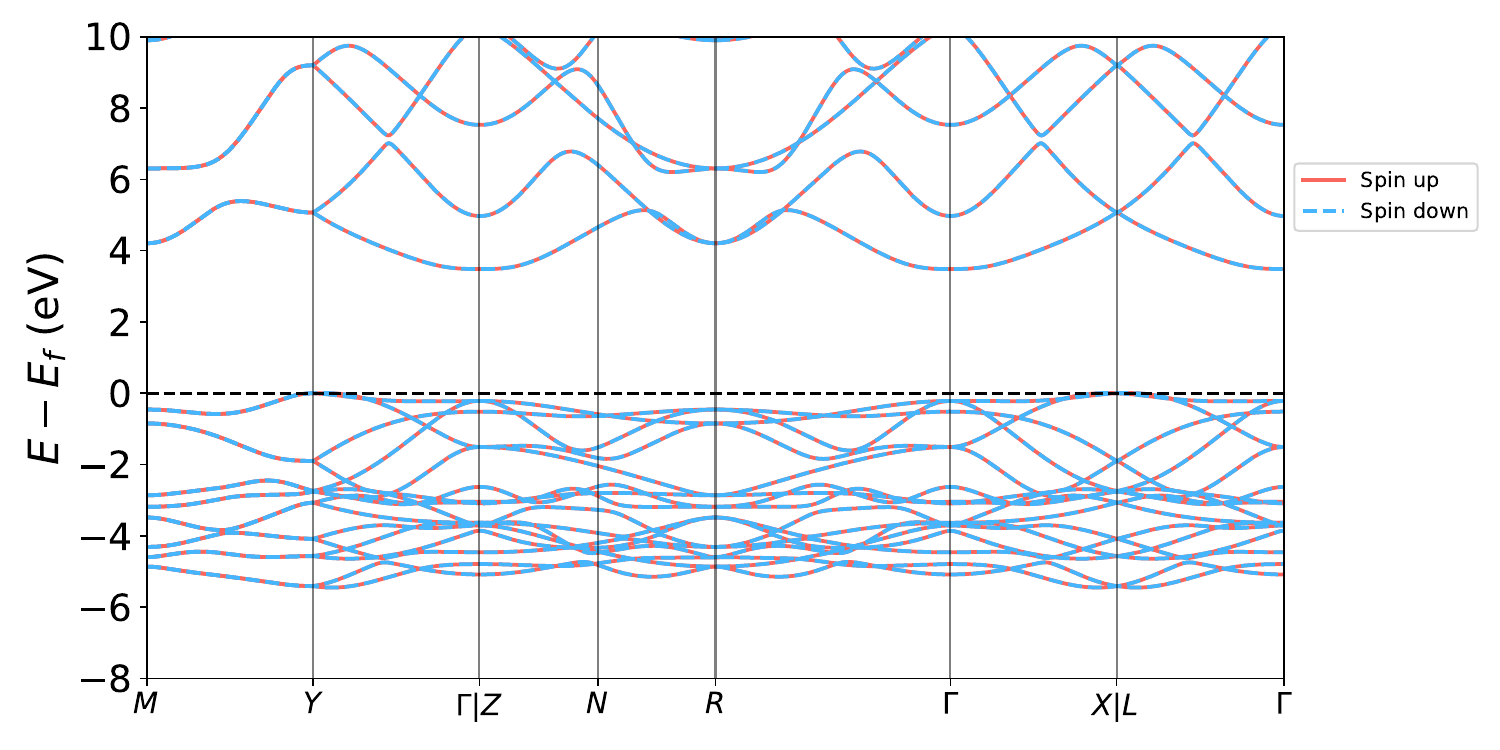}
    \caption{\footnotesize AgBr slab spin-polarized band structure.}
    \label{fig:AgBr_band_slab}
\end{figure}

\clearpage
\newpage

\subsection{\ce{AgF2} bulk}
\begin{verbatim}
_cell_length_a                         5.01699302
_cell_length_b                         5.56305079
_cell_length_c                         5.79168529
_cell_angle_alpha                      90.000316
_cell_angle_beta                       90.000060
_cell_angle_gamma                      90.000393
_cell_volume                           'P 1'
_space_group_name_H-M_alt              'P 1'
_space_group_IT_number                 1

loop_
_space_group_symop_operation_xyz
   'x, y, z'

loop_
   _atom_site_label
   _atom_site_occupancy
   _atom_site_fract_x
   _atom_site_fract_y
   _atom_site_fract_z
   _atom_site_adp_type
   _atom_site_B_ios_or_equiv
   _atom_site_type_symbol
   Ag001   1.0   -0.000000726497  -0.000000992033  0.499998090676  Biso  1.000000  Ag
   Ag002   1.0   0.499998837048  0.000000970354  0.000000191806  Biso  1.000000  Ag
   Ag003   1.0   -0.499998286409  0.499999167759  0.499999144036  Biso  1.000000  Ag
   Ag004   1.0   -0.000001551115  -0.499999050009  -0.000000874189  Biso  1.000000  Ag
    F005   1.0   -0.323994096899  -0.190290918188  -0.370865089091  Biso  1.000000   F
    F006   1.0   -0.176005905134  0.190292178232  0.129135189083  Biso  1.000000   F
    F007   1.0   0.176010051186  -0.309723203847  0.370872552069  Biso  1.000000   F
    F008   1.0   0.323990942634  0.309722142161  -0.129127128229  Biso  1.000000   F
    F009   1.0   0.323996524681  0.190288049649  0.370867426287  Biso  1.000000   F
    F010   1.0   0.176003992465  -0.190289949355  -0.129133915953  Biso  1.000000   F
    F011   1.0   -0.176006618278  0.309722346432  -0.370873090305  Biso  1.000000   F
    F012   1.0   -0.323993163683  -0.309720741155  0.129127503809  Biso  1.000000   F

\end{verbatim}

\begin{figure}[h]
    \centering
    \includegraphics[width=\textwidth]{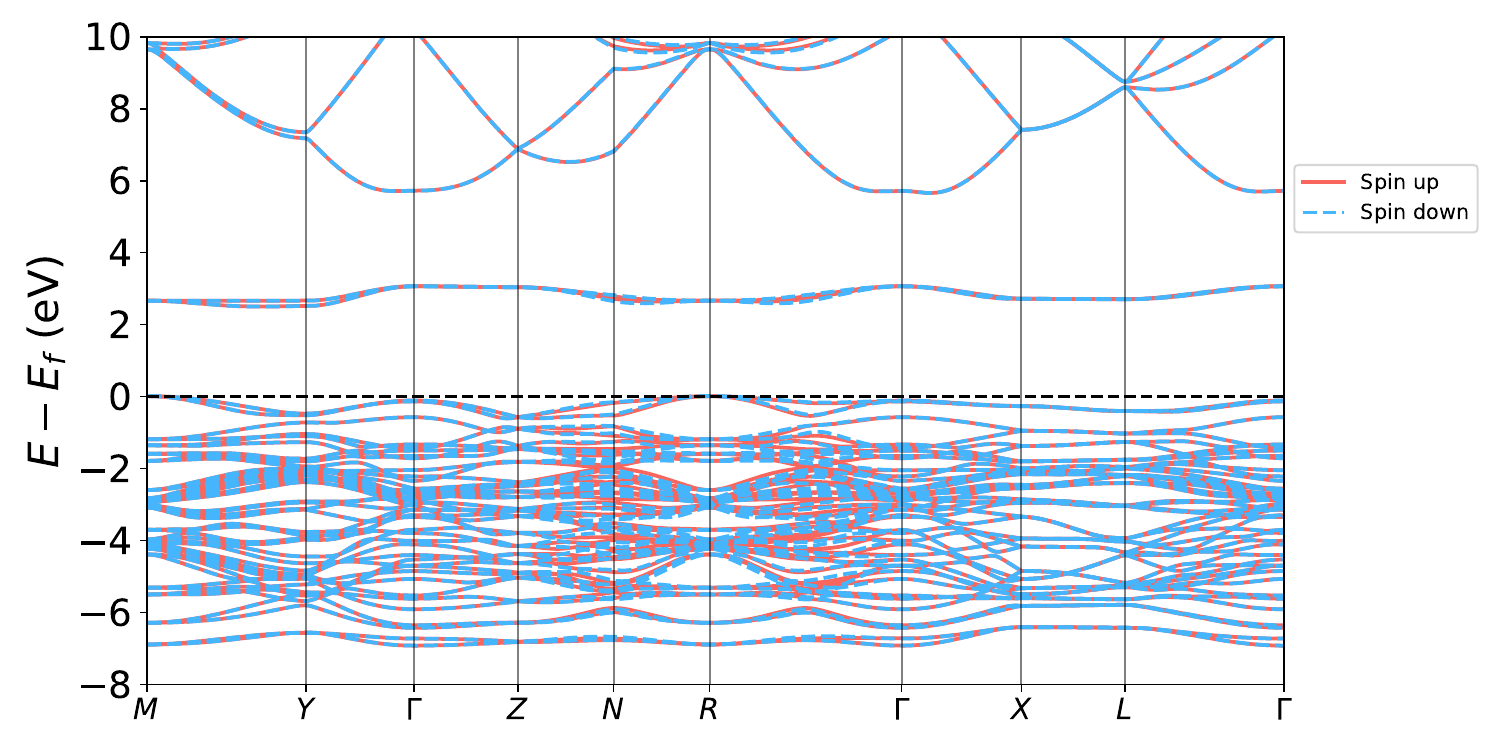}
    \caption{\footnotesize \ce{AgF2} bulk spin-polarized band structure.}
    \label{fig:AgF2_band_bulk}
\end{figure}
\clearpage
\newpage

\subsection{\ce{AgF2} slab}
\begin{verbatim}
_cell_length_a                         3.79249244
_cell_length_b                         5.73484784
_cell_length_c                         40.00000000
_cell_angle_alpha                      90.000000
_cell_angle_beta                       90.000000
_cell_angle_gamma                      90.000000
_cell_volume                           'P 1'
_space_group_name_H-M_alt              'P 1'
_space_group_IT_number                 1

loop_
_space_group_symop_operation_xyz
   'x, y, z'

loop_
   _atom_site_label
   _atom_site_occupancy
   _atom_site_fract_x
   _atom_site_fract_y
   _atom_site_fract_z
   _atom_site_adp_type
   _atom_site_B_ios_or_equiv
   _atom_site_type_symbol
F001  1.0  -0.401836947935  -0.182054639002  0.523811463869  Biso  1.000000  F
F002  1.0  -0.098163052065  0.317945360998  0.523811463869  Biso  1.000000  F
Ag003  1.0  0.000000000000  -0.000000000000  0.5  Biso  1.000000  Ag
Ag004  1.0  -0.500000000000  -0.500000000000  0.5  Biso  1.000000  Ag
F005  1.0  0.098163052065  -0.317945360998  0.476188536131  Biso  1.000000  F
F006  1.0  0.401836947935  0.182054639002  0.476188536131  Biso  1.000000  F

\end{verbatim}
\begin{figure}[h]
    \centering
    \includegraphics[width=\textwidth]{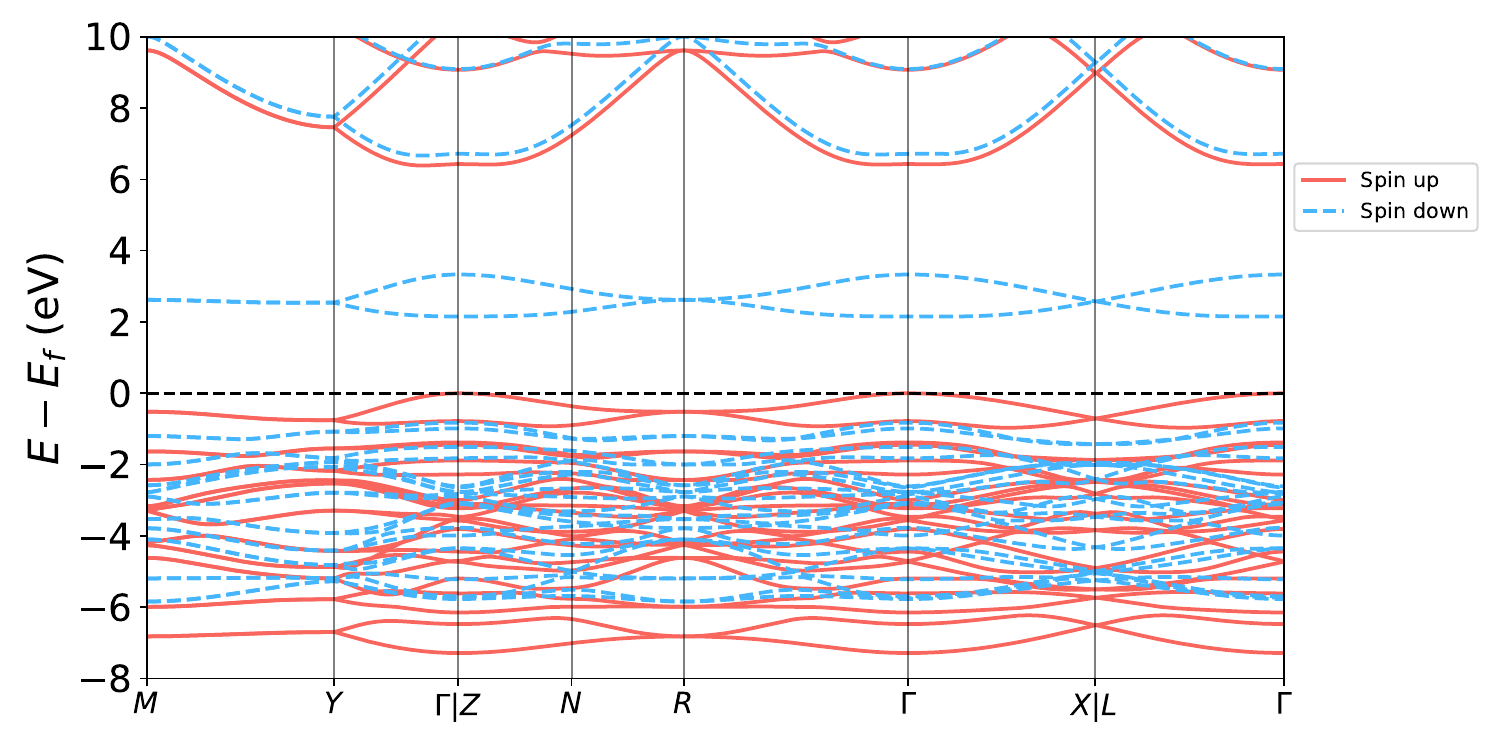}
    \caption{\footnotesize \ce{AgF2} bulk spin-polarized band structure.}
    \label{fig:AgF2_band_slab}
\end{figure}
\clearpage
\newpage

\subsection{AgI bulk}
\begin{verbatim}
_cell_length_a 4.23091474
_cell_length_b 4.23091474
_cell_length_c 6.15360030
_cell_angle_alpha 90.000000
_cell_angle_beta 90.000000
_cell_angle_gamma 90.000000
_symmetry_space_group_name_H-M         'P 1'
_symmetry_Int_Tables_number            1

loop_
_symmetry_equiv_pos_as_xyz
   'x, y, z'

loop_
_atom_site_label
_atom_site_type_symbol
_atom_site_fract_x
_atom_site_fract_y
_atom_site_fract_z
I001 I 2.500000000000E-01 2.500000000000E-01 3.088479906138E-01
I002 I -2.500000000000E-01 -2.500000000000E-01 -3.088479906138E-01
Ag003 Ag -2.500000000000E-01 2.500000000000E-01 0.000000000000E+00
Ag004 Ag 2.500000000000E-01 -2.500000000000E-01 0.000000000000E+00

\end{verbatim}
\begin{figure}[h]
    \centering
    \includegraphics[width=\textwidth]{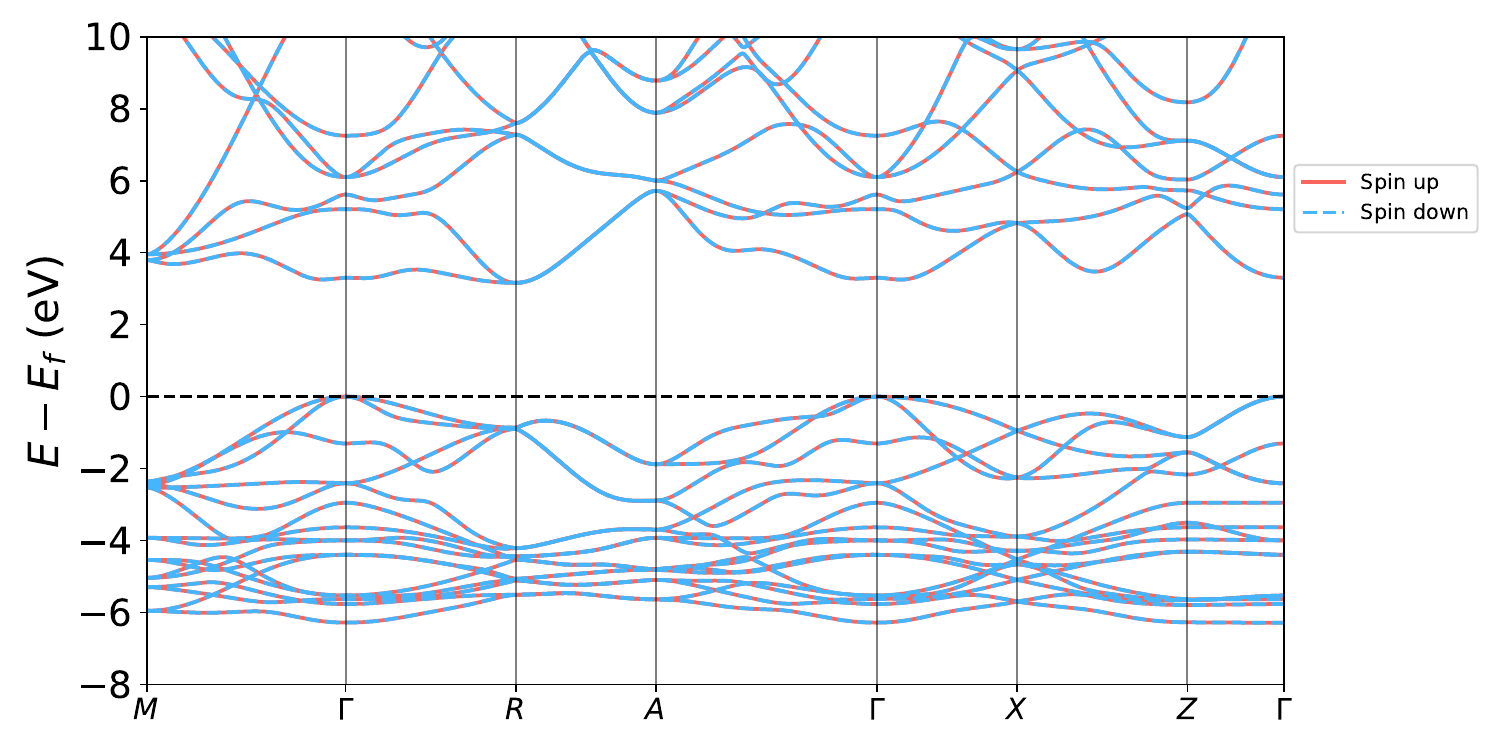}
    \caption{\footnotesize AgI bulk spin-polarized band structure.}
    \label{fig:AgI_band_bulk}
\end{figure}
\clearpage
\newpage

\subsection{AgI slab}
\begin{verbatim}
_cell_length_a                         4.15210956
_cell_length_b                         4.15210956
_cell_length_c                         40.00000000
_cell_angle_alpha                      90.000000
_cell_angle_beta                       90.000000
_cell_angle_gamma                      90.000000
_cell_volume                           'P 1'
_space_group_name_H-M_alt              'P 1'
_space_group_IT_number                 1

loop_
_space_group_symop_operation_xyz
   'x, y, z'

loop_
   _atom_site_label
   _atom_site_occupancy
   _atom_site_fract_x
   _atom_site_fract_y
   _atom_site_fract_z
   _atom_site_adp_type
   _atom_site_B_ios_or_equiv
   _atom_site_type_symbol
I001  1.0  0.250000000000  0.250000000000  0.549448788288  Biso  1.000000  I
Ag002  1.0  -0.250000000000  0.250000000000  0.5  Biso  1.000000  Ag
Ag003  1.0  0.250000000000  -0.250000000000  0.5  Biso  1.000000  Ag
I004  1.0  -0.250000000000  -0.250000000000  0.450551211712  Biso  1.000000  I
\end{verbatim}
\begin{figure}[h]
    \centering
    \includegraphics[width=\textwidth]{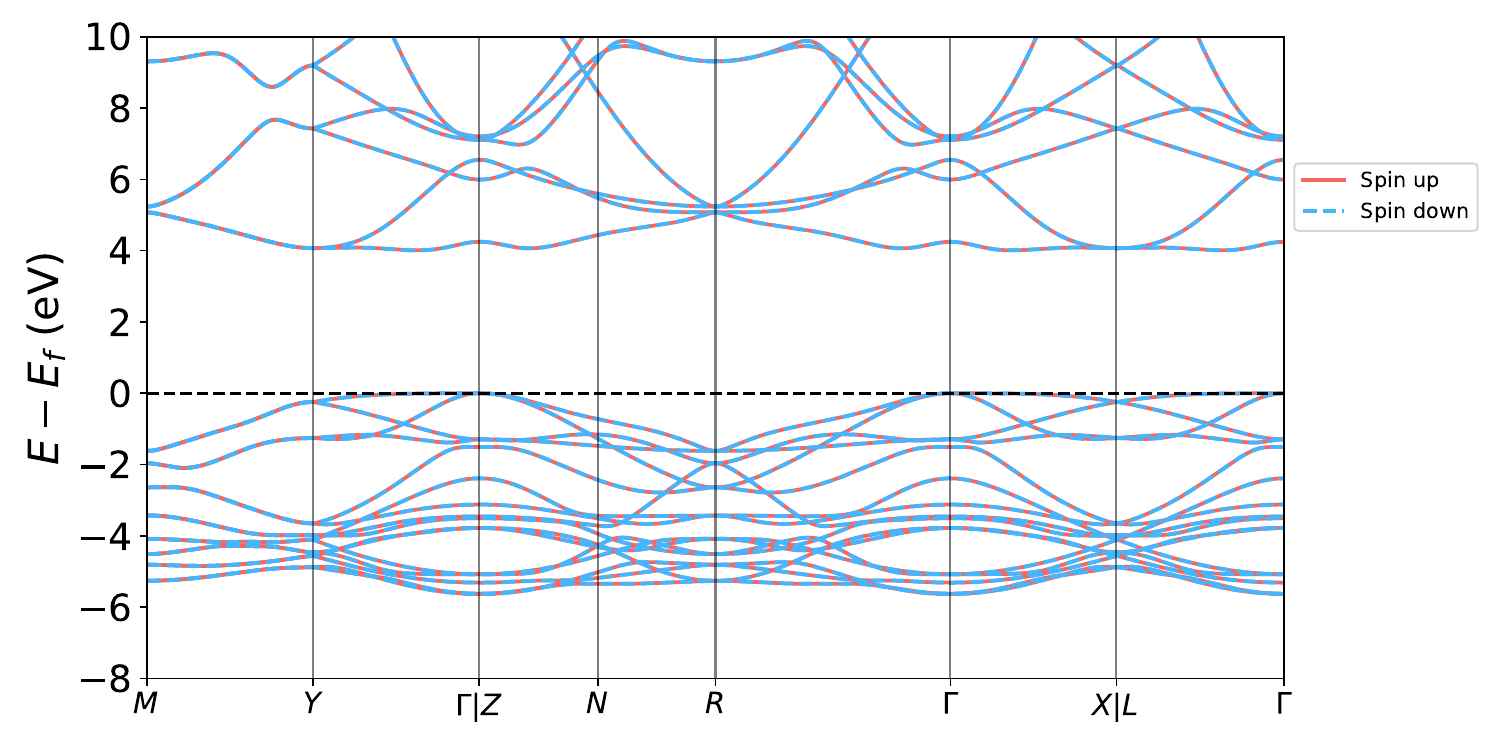}
    \caption{\footnotesize AgI slab spin-polarized band structure.}
    \label{fig:AgI_band_slab}
\end{figure}
\clearpage
\newpage

\subsection{AuBr bulk}
\begin{verbatim}
_cell_length_a                         4.05890484
_cell_length_b                         6.38255430
_cell_length_c                         11.69247193
_cell_angle_alpha                      95.976593
_cell_angle_beta                       91.209758
_cell_angle_gamma                      89.916857
_cell_volume                           'P 1'
_space_group_name_H-M_alt              'P 1'
_space_group_IT_number                 1

loop_
_space_group_symop_operation_xyz
   'x, y, z'

loop_
   _atom_site_label
   _atom_site_occupancy
   _atom_site_fract_x
   _atom_site_fract_y
   _atom_site_fract_z
   _atom_site_adp_type
   _atom_site_B_ios_or_equiv
   _atom_site_type_symbol
   Au001   1.0   0.209030141782  0.125162166921  0.250075309697  Biso  1.000000  Au
   Au002   1.0   -0.290932448346  -0.374762973116  0.250052010647  Biso  1.000000  Au
   Au003   1.0   0.457027374117  -0.124912385373  -0.250233115135  Biso  1.000000  Au
   Au004   1.0   -0.042218260880  0.374851446646  -0.249986919105  Biso  1.000000  Au
   Br005   1.0   0.111643761405  0.413214365855  0.116283766836  Biso  1.000000  Br
   Br006   1.0   -0.388662189535  -0.086690626157  0.116294676776  Biso  1.000000  Br
   Br007   1.0   -0.193325688315  0.337208911949  0.383811236863  Biso  1.000000  Br
   Br008   1.0   0.305893925367  -0.163139987897  0.383758409931  Biso  1.000000  Br
   Br009   1.0   -0.445280938355  0.213377173180  -0.116503466643  Biso  1.000000  Br
   Br010   1.0   0.054951168689  -0.286403857875  -0.116336750007  Biso  1.000000  Br
   Br011   1.0   0.360833513284  -0.463852917024  -0.383551389702  Biso  1.000000  Br
   Br012   1.0   -0.138960359213  0.035948682893  -0.383663770158  Biso  1.000000  Br

\end{verbatim}
\begin{figure}[h]
    \centering
    \includegraphics[width=\textwidth]{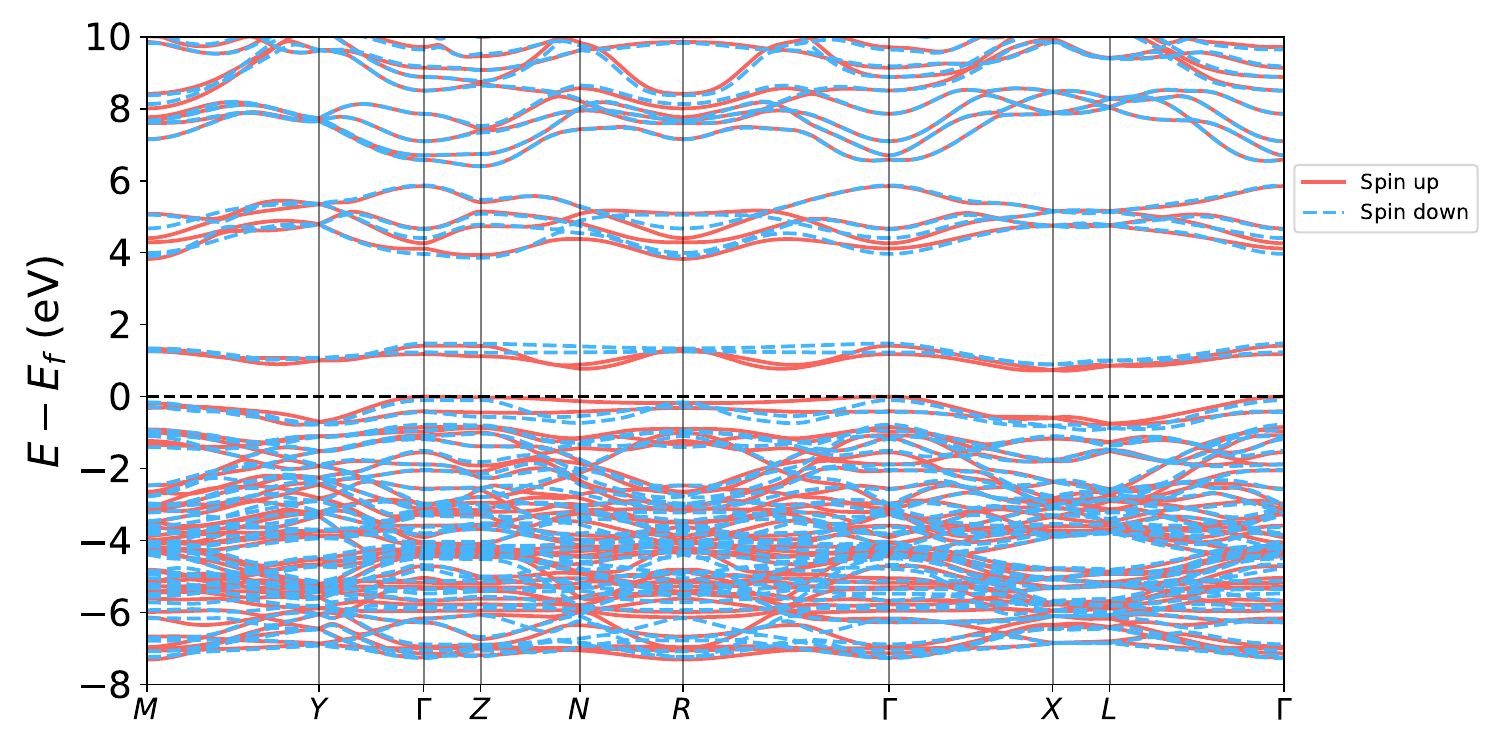}
    \caption{\footnotesize AuBr bulk spin-polarized band structure.}
    \label{fig:AuBr_band_bulk}
\end{figure}
\clearpage
\newpage

\subsection{AuBr slab}
\begin{verbatim}
_cell_length_a                         6.33823343
_cell_length_b                         4.02827898
_cell_length_c                         40.00000000
_cell_angle_alpha                      90.000000
_cell_angle_beta                       90.000000
_cell_angle_gamma                      90.903758
_cell_volume                           'P 1'
_space_group_name_H-M_alt              'P 1'
_space_group_IT_number                 1

loop_
_space_group_symop_operation_xyz
   'x, y, z'

loop_
   _atom_site_label
   _atom_site_occupancy
   _atom_site_fract_x
   _atom_site_fract_y
   _atom_site_fract_z
   _atom_site_adp_type
   _atom_site_B_ios_or_equiv
   _atom_site_type_symbol
Br001  1.0  0.432630891411  -0.150640989165  0.539570203402  Biso  1.000000  Br
Br002  1.0  -0.096033863558  0.380528587957  0.54255542394  Biso  1.000000  Br
Au003  1.0  -0.269183969722  -0.274702642300  0.501429360296  Biso  1.000000  Au
Au004  1.0  0.325442085367  0.348378080129  0.496125540867  Biso  1.000000  Au
Br005  1.0  0.037556628166  -0.334892905395  0.465300193154  Biso  1.000000  Br
Br006  1.0  -0.430439767114  0.031284870429  0.455045949349  Biso  1.000000  Br

\end{verbatim}
\begin{figure}[h]
    \centering
    \includegraphics[width=\textwidth]{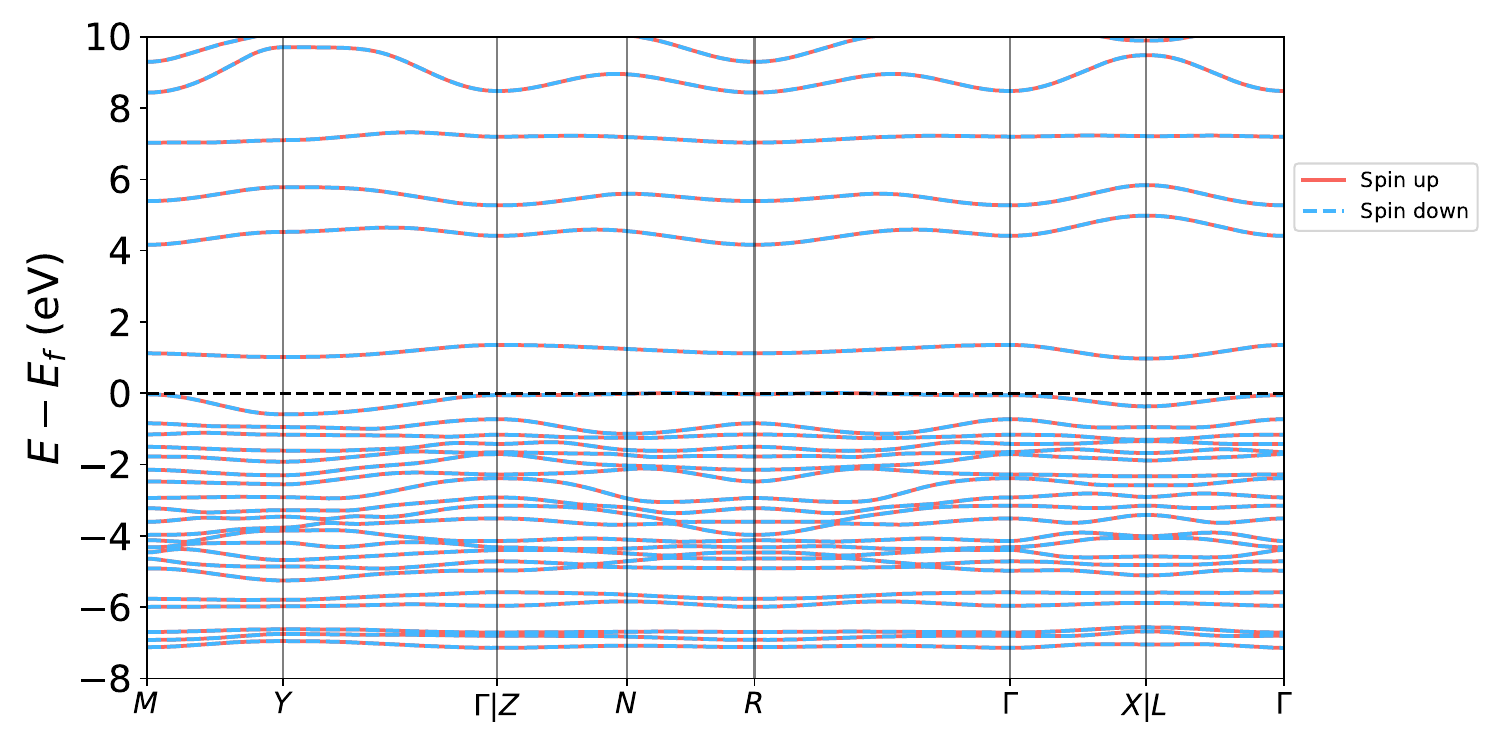}
    \caption{\footnotesize AuBr slab spin-polarized band structure.}
    \label{fig:AuBr_band_slab}
\end{figure}
\clearpage
\newpage

\subsection{AuI bulk}
\begin{verbatim}
_cell_length_a 3.91507393
_cell_length_b 3.91507393
_cell_length_c 14.08162589
_cell_angle_alpha 90.000000
_cell_angle_beta 90.000000
_cell_angle_gamma 90.000000
_symmetry_space_group_name_H-M         'P 1'
_symmetry_Int_Tables_number            1

loop_
_symmetry_equiv_pos_as_xyz
   'x, y, z'

loop_
_atom_site_label
_atom_site_type_symbol
_atom_site_fract_x
_atom_site_fract_y
_atom_site_fract_z
Au001 Au 0.000000000000E+00 0.000000000000E+00 -5.000000000000E-01
Au002 Au -5.000000000000E-01 -5.000000000000E-01 -5.000000000000E-01
Au003 Au -5.000000000000E-01 0.000000000000E+00 0.000000000000E+00
Au004 Au 0.000000000000E+00 -5.000000000000E-01 0.000000000000E+00
I005 I 2.500000000000E-01 2.500000000000E-01 1.609486818214E-01
I006 I -2.500000000000E-01 -2.500000000000E-01 3.390513181786E-01
I007 I -2.500000000000E-01 -2.500000000000E-01 -1.609486818214E-01
I008 I 2.500000000000E-01 2.500000000000E-01 -3.390513181786E-01

\end{verbatim}
\begin{figure}[h]
    \centering
    \includegraphics[width=\textwidth]{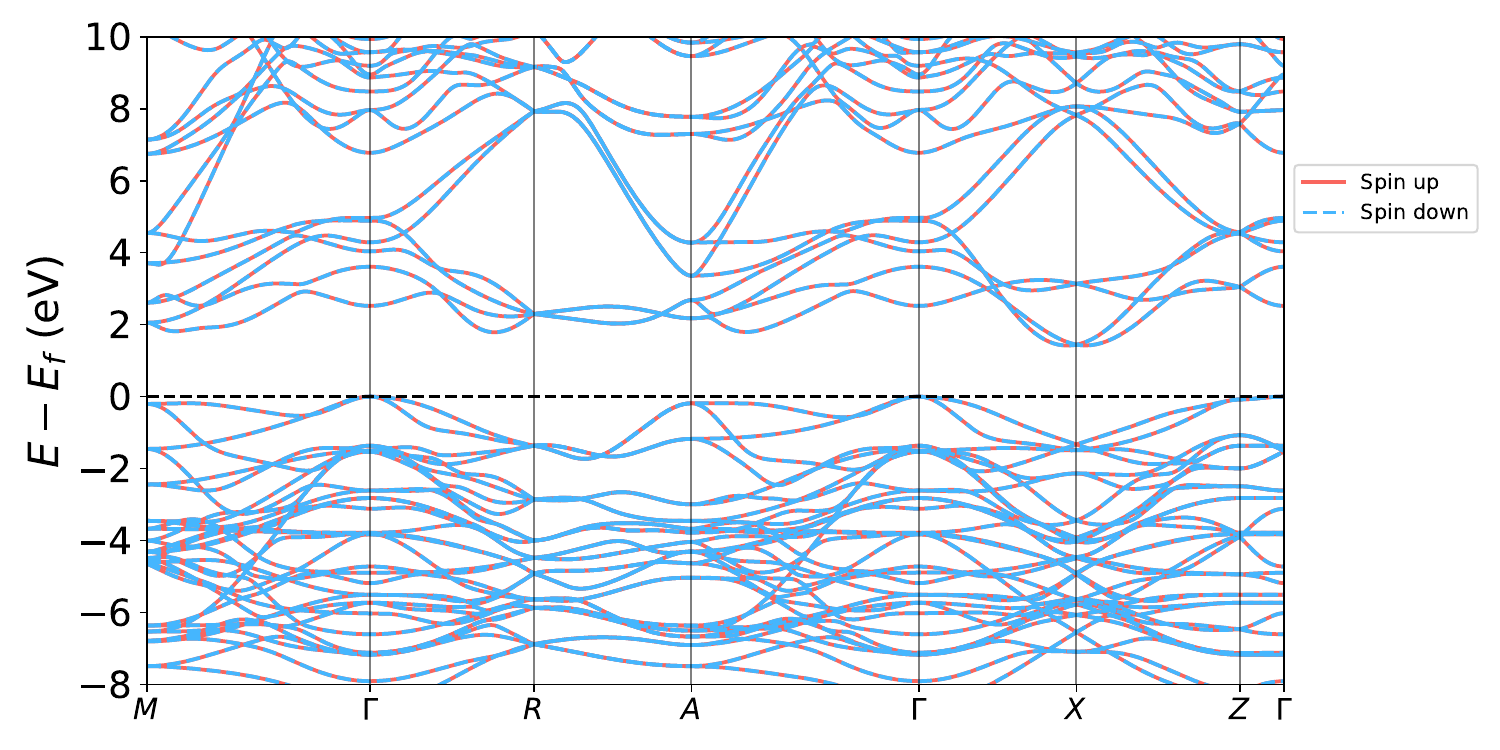}
    \caption{\footnotesize AuI bulk spin-polarized band structure.}
    \label{fig:AuI_band_bulk}
\end{figure}
\clearpage
\newpage

\subsection{AuI slab}
\begin{verbatim}
_cell_length_a                         3.92169193
_cell_length_b                         3.92169193
_cell_length_c                         40.00000000
_cell_angle_alpha                      90.000000
_cell_angle_beta                       90.000000
_cell_angle_gamma                      90.131863
_cell_volume                           'P 1'
_space_group_name_H-M_alt              'P 1'
_space_group_IT_number                 1

loop_
_space_group_symop_operation_xyz
   'x, y, z'

loop_
   _atom_site_label
   _atom_site_occupancy
   _atom_site_fract_x
   _atom_site_fract_y
   _atom_site_fract_z
   _atom_site_adp_type
   _atom_site_B_ios_or_equiv
   _atom_site_type_symbol
I001  1.0  0.250000000000  0.250000000000  0.556705940809  Biso  1.000000  I
Au002  1.0  0.000000000000  0.000000000000  0.5  Biso  1.000000  Au
Au003  1.0  -0.500000000000  -0.500000000000  0.5  Biso  1.000000  Au
I004  1.0  -0.250000000000  -0.250000000000  0.443294059191  Biso  1.000000  I

\end{verbatim}
\begin{figure}[h]
    \centering
    \includegraphics[width=\textwidth]{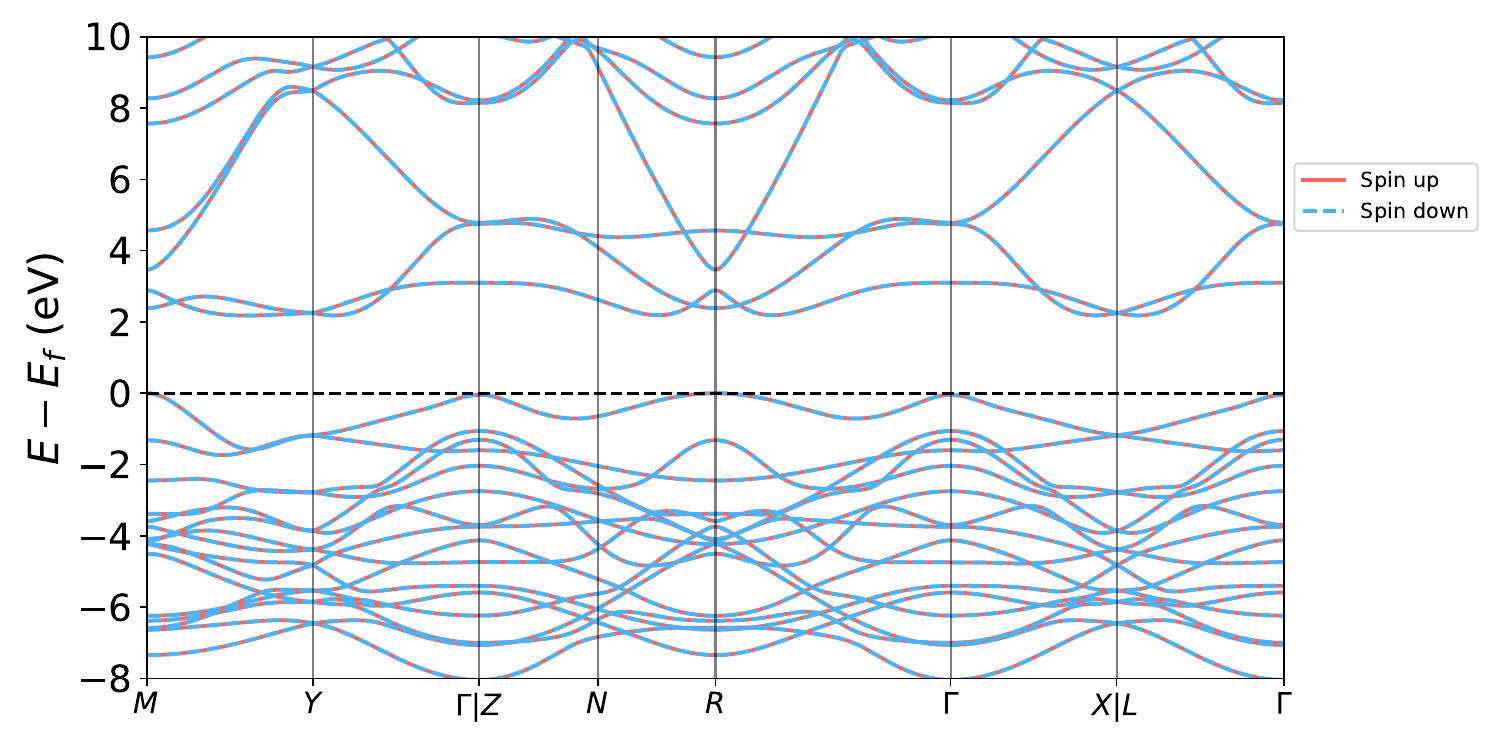}
    \caption{\footnotesize AuI slab spin-polarized band structure.}
    \label{fig:AuI_band_slab}
\end{figure}
\clearpage
\newpage

\subsection{BiI bulk}
\begin{verbatim}
_cell_length_a 7.13826972
_cell_length_b 7.13826972
_cell_length_c 19.07564814
_cell_angle_alpha 86.788811
_cell_angle_beta 93.211189
_cell_angle_gamma 145.072786
_symmetry_space_group_name_H-M         'P 1'
_symmetry_Int_Tables_number            1

loop_
_symmetry_equiv_pos_as_xyz
   'x, y, z'

loop_
_atom_site_label
_atom_site_type_symbol
_atom_site_fract_x
_atom_site_fract_y
_atom_site_fract_z
I001 I -3.510995494853E-01 3.510995494853E-01 5.171084833644E-02
I002 I 3.510995494853E-01 -3.510995494853E-01 -5.171084833644E-02
I003 I 3.602699353061E-01 -3.602699353061E-01 1.493664329074E-01
I004 I -3.602699353061E-01 3.602699353061E-01 -1.493664329074E-01
I005 I 1.036421440156E-01 -1.036421440156E-01 4.413711017711E-01
I006 I -1.036421440156E-01 1.036421440156E-01 -4.413711017711E-01
I007 I 3.981131668922E-01 -3.981131668922E-01 3.598697976183E-01
I008 I -3.981131668922E-01 3.981131668922E-01 -3.598697976183E-01
Bi009 Bi 5.222960199984E-02 -5.222960199984E-02 2.524138642022E-01
Bi010 Bi -5.222960199984E-02 5.222960199984E-02 -2.524138642022E-01
Bi011 Bi -3.007310893291E-01 3.007310893291E-01 2.391221613847E-01
Bi012 Bi 3.007310893291E-01 -3.007310893291E-01 -2.391221613847E-01
Bi013 Bi 1.523977674472E-03 -1.523977674472E-03 9.761890941700E-02
Bi014 Bi -1.523977674472E-03 1.523977674472E-03 -9.761890941700E-02
Bi015 Bi 2.522124292982E-01 -2.522124292982E-01 -3.931984573113E-01
Bi016 Bi -2.522124292982E-01 2.522124292982E-01 3.931984573113E-01

\end{verbatim}
\begin{figure}[h]
    \centering
    \includegraphics[width=\textwidth]{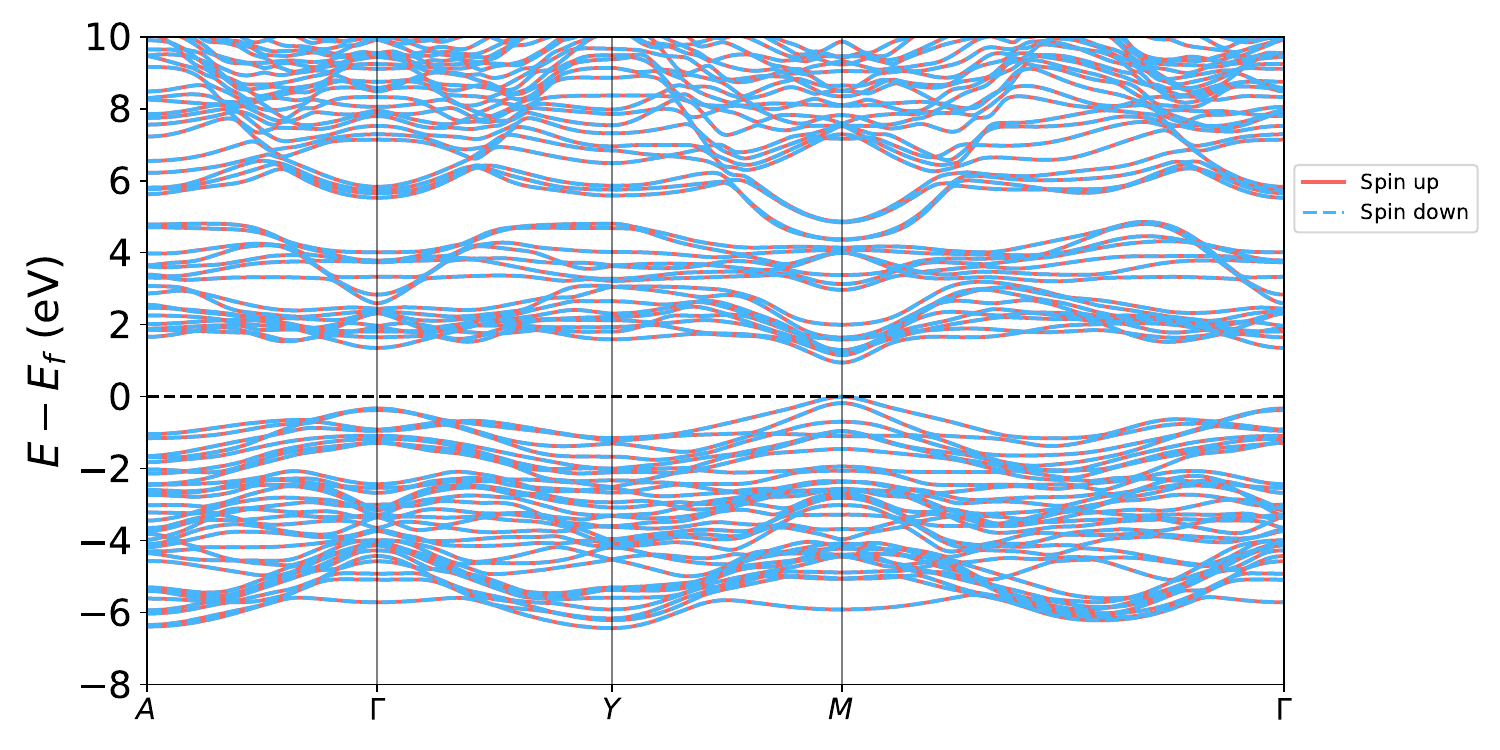}
    \caption{\footnotesize BiI bulk spin-polarized band structure.}
    \label{fig:BiI_band_bulk}
\end{figure}
\clearpage
\newpage

\subsection{BiI slab}
\begin{verbatim}
_cell_length_a                         4.18825172
_cell_length_b                         5.92296673
_cell_length_c                         40.00000000
_cell_angle_alpha                      90.000000
_cell_angle_beta                       90.000000
_cell_angle_gamma                      110.705234
_cell_volume                           'P 1'
_space_group_name_H-M_alt              'P 1'
_space_group_IT_number                 1

loop_
_space_group_symop_operation_xyz
   'x, y, z'

loop_
   _atom_site_label
   _atom_site_occupancy
   _atom_site_fract_x
   _atom_site_fract_y
   _atom_site_fract_z
   _atom_site_adp_type
   _atom_site_B_ios_or_equiv
   _atom_site_type_symbol
Bi001  1.0  0.421479118106  -0.157041763787  0.60874423956  Biso  1.000000  Bi
I002  1.0  -0.319974375551  0.360051248898  0.622605889584  Biso  1.000000  I
Bi003  1.0  -0.099816316546  -0.199632633093  0.555962283959  Biso  1.000000  Bi
I004  1.0  0.149793030517  0.299586061035  0.533635996841  Biso  1.000000  I
I005  1.0  -0.149793030517  -0.299586061035  0.466364003159  Biso  1.000000  I
Bi006  1.0  0.099816316546  0.199632633093  0.444037716041  Biso  1.000000  Bi
I007  1.0  0.319974375551  -0.360051248898  0.377394110416  Biso  1.000000  I
Bi008  1.0  -0.421479118106  0.157041763787  0.39125576044  Biso  1.000000  Bi

\end{verbatim}

\begin{figure}[h]
    \centering
    \includegraphics[width=\textwidth]{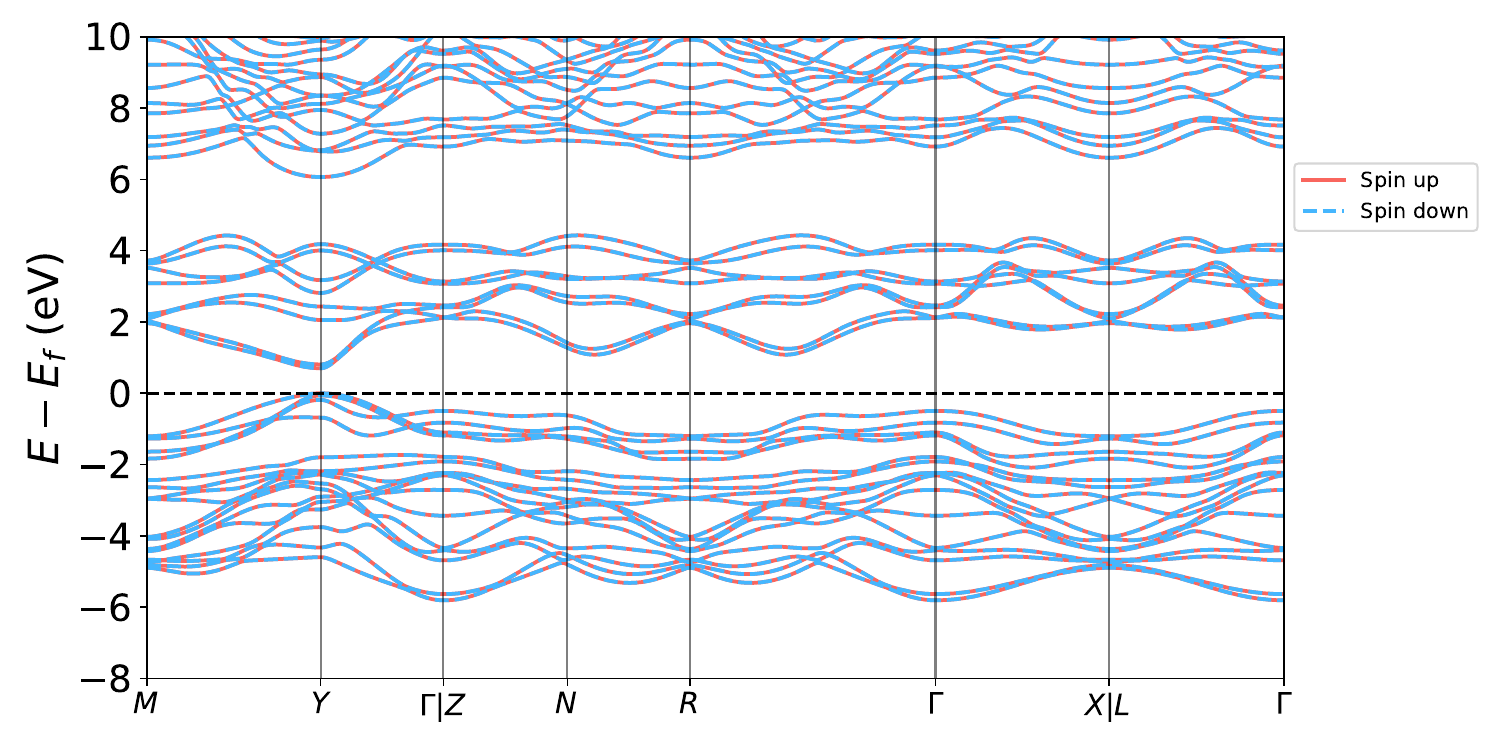}
    \caption{\footnotesize BiI slab spin-polarized band structure.}
    \label{fig:BiI_band_slab}
\end{figure}
\clearpage
\newpage

\subsection{\ce{CaCl2} bulk}
\begin{verbatim}
_cell_length_a 4.68582413
_cell_length_b 4.68582413
_cell_length_c 4.68582413
_cell_angle_alpha 60.000000
_cell_angle_beta 60.000000
_cell_angle_gamma 60.000000
_symmetry_space_group_name_H-M         'P 1'
_symmetry_Int_Tables_number            1

loop_
_symmetry_equiv_pos_as_xyz
   'x, y, z'

loop_
_atom_site_label
_atom_site_type_symbol
_atom_site_fract_x
_atom_site_fract_y
_atom_site_fract_z
Cl001 Cl 2.500000000000E-01 2.500000000000E-01 2.500000000000E-01
Cl002 Cl -2.500000000000E-01 -2.500000000000E-01 -2.500000000000E-01
Ca003 Ca 0.000000000000E+00 0.000000000000E+00 0.000000000000E+00

\end{verbatim}
\begin{figure}[h]
    \centering
    \includegraphics[width=\textwidth]{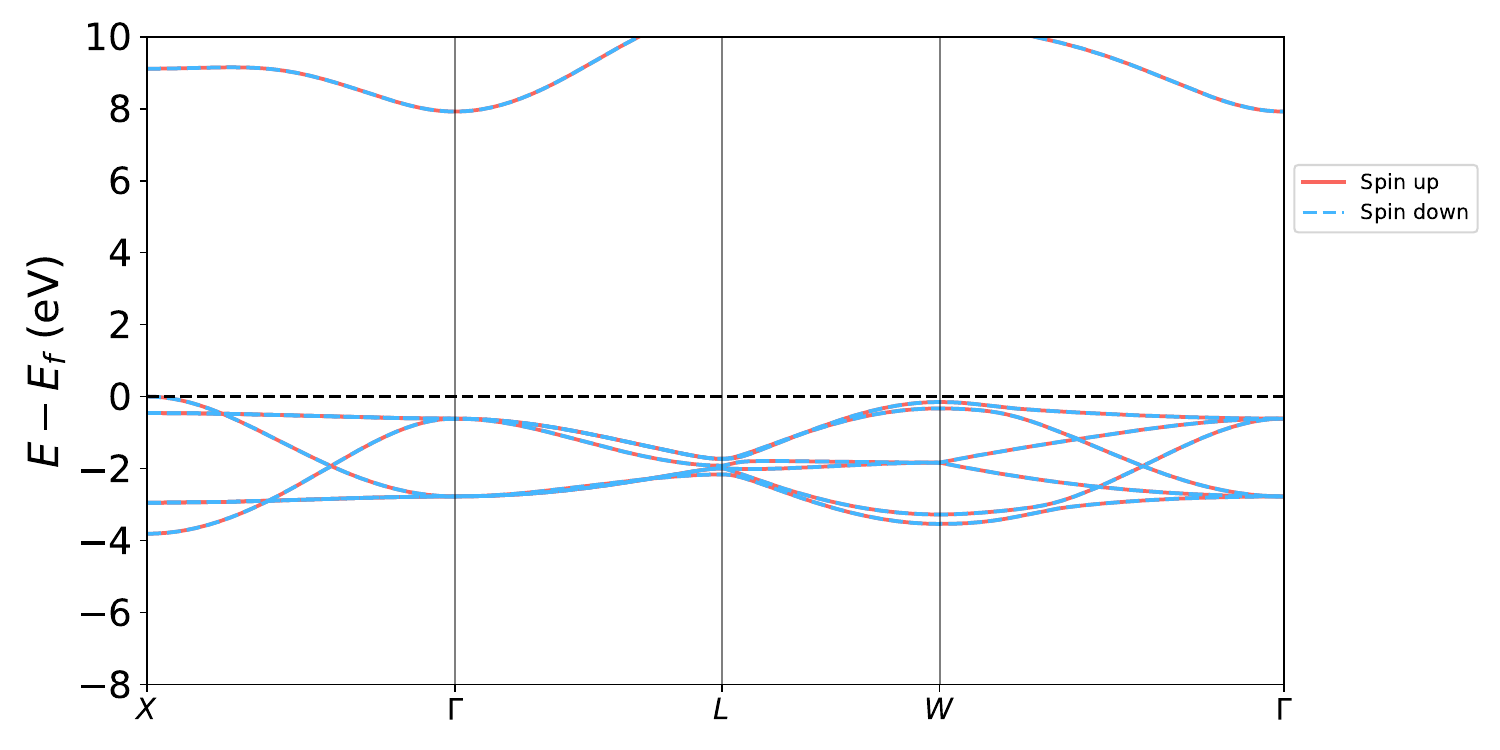}
    \caption{\footnotesize \ce{CaCl2} bulk spin-polarized band structure.}
    \label{fig:CaCl2_band_bulk}
\end{figure}
\clearpage
\newpage

\subsection{\ce{CaCl2} slab}
\begin{verbatim}
_cell_length_a                         5.27551575
_cell_length_b                         5.27551575
_cell_length_c                         40.00000000
_cell_angle_alpha                      90.000000
_cell_angle_beta                       90.000000
_cell_angle_gamma                      90.000000
_cell_volume                           'P 1'
_space_group_name_H-M_alt              'P 1'
_space_group_IT_number                 1

loop_
_space_group_symop_operation_xyz
   'x, y, z'

loop_
   _atom_site_label
   _atom_site_occupancy
   _atom_site_fract_x
   _atom_site_fract_y
   _atom_site_fract_z
   _atom_site_adp_type
   _atom_site_B_ios_or_equiv
   _atom_site_type_symbol
Ca001  1.0  0.125000000000  -0.125000000000  0.493100895428  Biso  1.000000  Ca
Cl002  1.0  -0.375000000000  -0.125000000000  0.493095245989  Biso  1.000000  Cl
Cl003  1.0  0.125000000000  0.375000000000  0.493095245989  Biso  1.000000  Cl

\end{verbatim}
\begin{figure}[h]
    \centering
    \includegraphics[width=\textwidth]{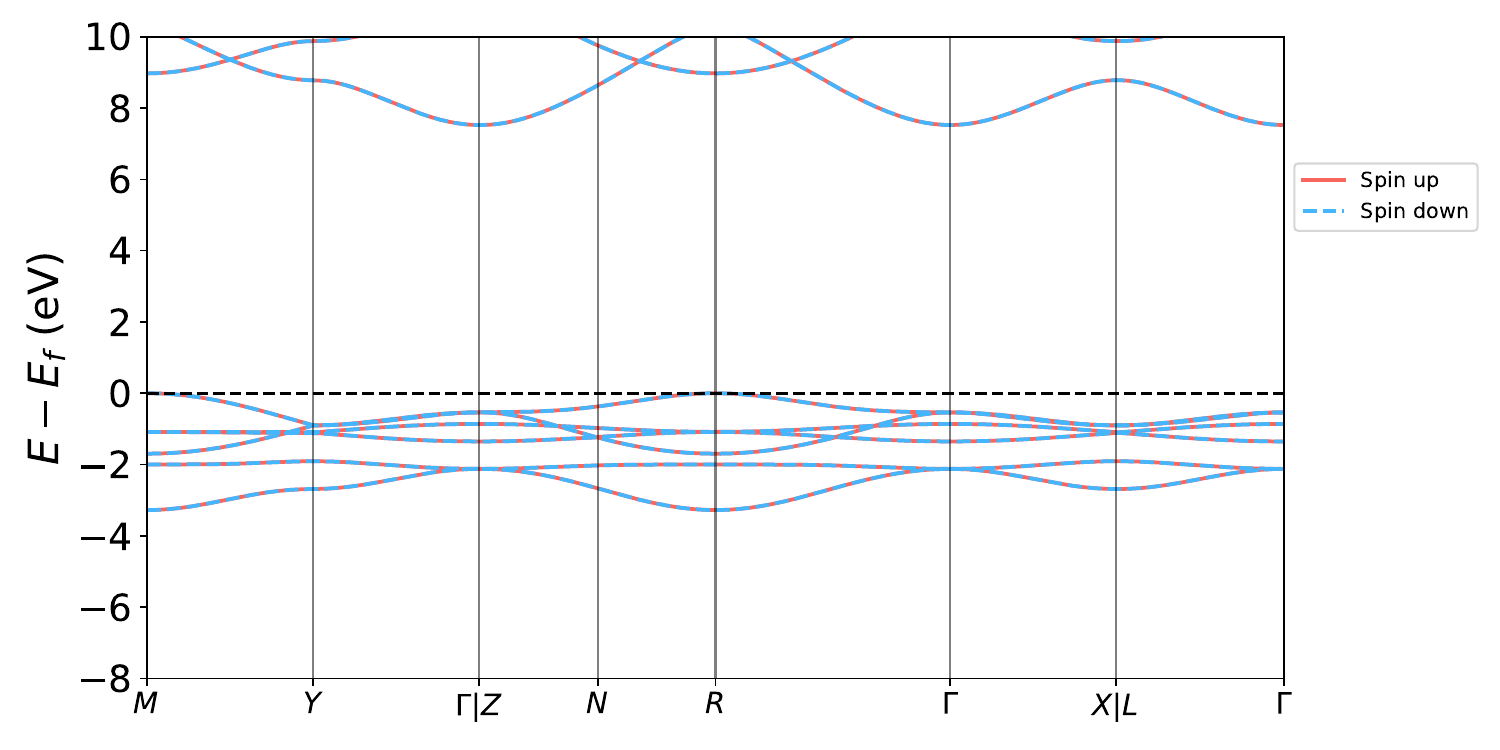}
    \caption{\footnotesize \ce{CaCl2} slab spin-polarized band structure.}
    \label{fig:CaCl2_band_slab}
\end{figure}
\clearpage
\newpage

\subsection{\ce{CaI2} bulk}
\begin{verbatim}
_cell_length_a 4.50334183
_cell_length_b 4.50334183
_cell_length_c 6.39510396
_cell_angle_alpha 90.034900
_cell_angle_beta 89.965100
_cell_angle_gamma 119.979586
_symmetry_space_group_name_H-M         'P 1'
_symmetry_Int_Tables_number            1

loop_
_symmetry_equiv_pos_as_xyz
   'x, y, z'

loop_
_atom_site_label
_atom_site_type_symbol
_atom_site_fract_x
_atom_site_fract_y
_atom_site_fract_z
I001 I 3.332176826578E-01 -3.332176826578E-01 2.719773325923E-01
I002 I -3.332176826578E-01 3.332176826578E-01 -2.719773325923E-01
Ca003 Ca 0.000000000000E+00 0.000000000000E+00 0.000000000000E+00

\end{verbatim}
\begin{figure}[h]
    \centering
    \includegraphics[width=\textwidth]{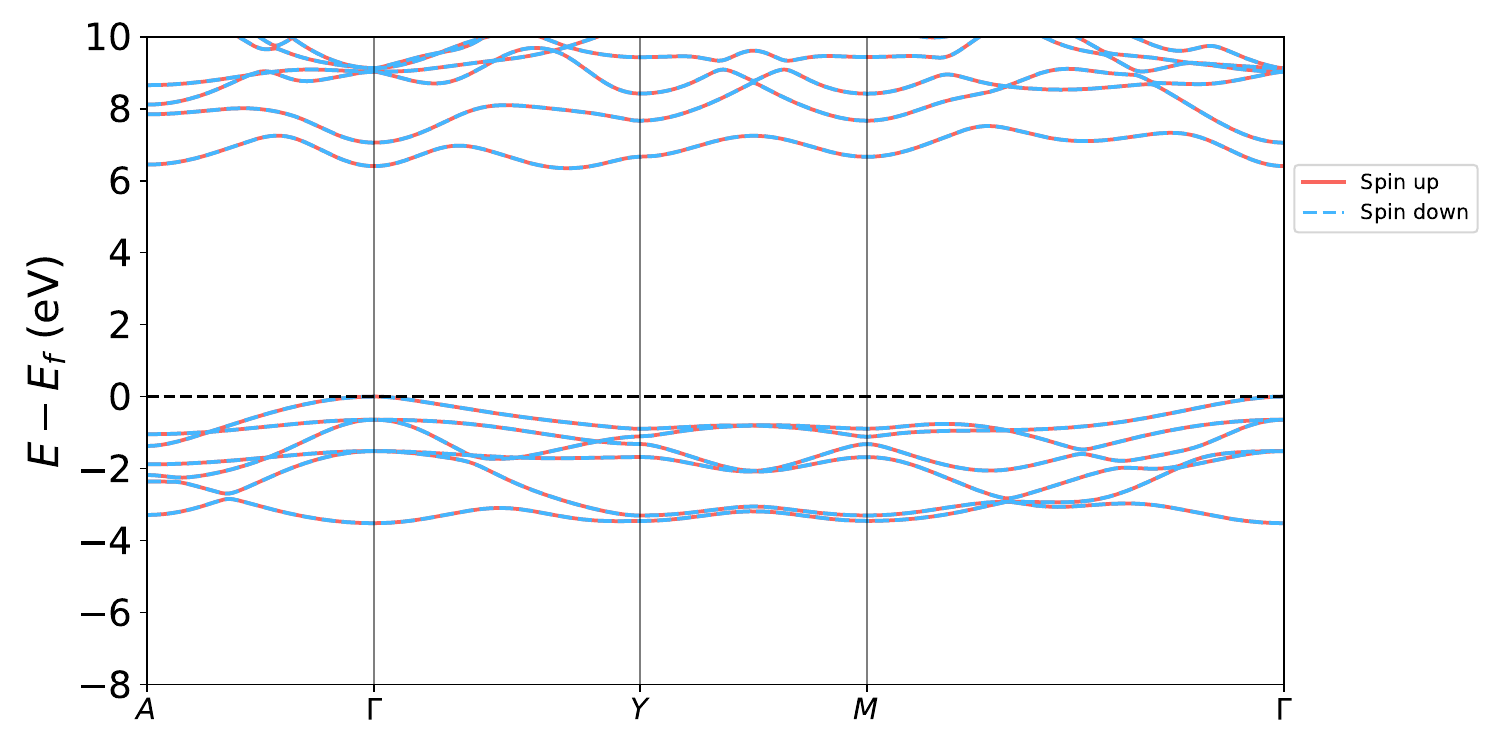}
    \caption{\footnotesize \ce{CaI2} bulk spin-polarized band structure.}
    \label{fig:CaI2_band_bulk}
\end{figure}
\clearpage
\newpage

\subsection{\ce{CaI2} slab}
\begin{verbatim}
_cell_length_a                         4.50292439
_cell_length_b                         4.50292439
_cell_length_c                         40.00000000
_cell_angle_alpha                      90.000000
_cell_angle_beta                       90.000000
_cell_angle_gamma                      119.998256
_cell_volume                           'P 1'
_space_group_name_H-M_alt              'P 1'
_space_group_IT_number                 1

loop_
_space_group_symop_operation_xyz
   'x, y, z'

loop_
   _atom_site_label
   _atom_site_occupancy
   _atom_site_fract_x
   _atom_site_fract_y
   _atom_site_fract_z
   _atom_site_adp_type
   _atom_site_B_ios_or_equiv
   _atom_site_type_symbol
I001  1.0  0.333310201515  -0.333310201515  0.54386724338  Biso  1.000000  I
Ca002  1.0  -0.000000000000  -0.000000000000  0.5  Biso  1.000000  Ca
I003  1.0  -0.333310201515  0.333310201515  0.45613275662  Biso  1.000000  I

\end{verbatim}

\begin{figure}[h]
    \centering
    \includegraphics[width=\textwidth]{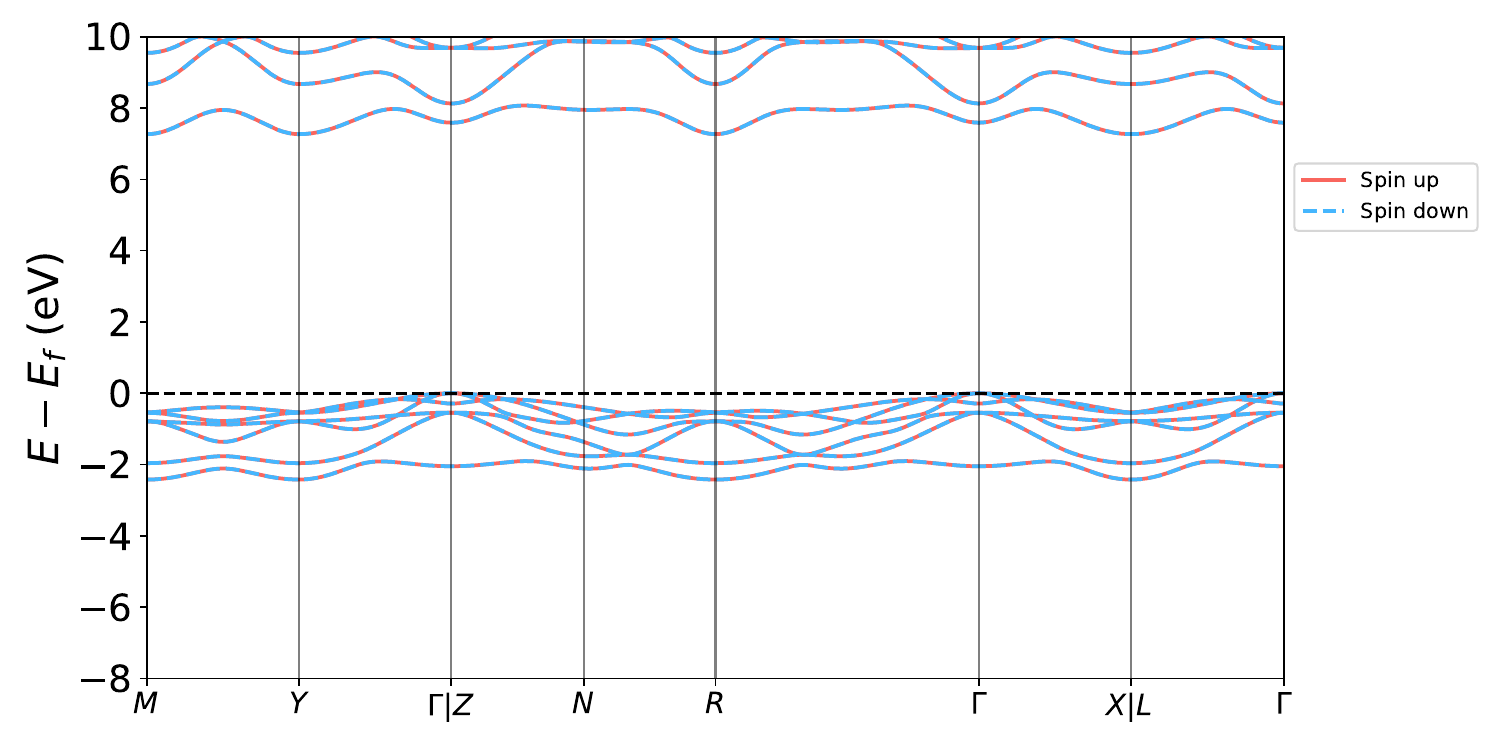}
    \caption{\footnotesize \ce{CaI2} slab spin-polarized band structure.}
    \label{fig:CaI2_band_slab}
\end{figure}

\clearpage
\newpage

\subsection{\ce{CdBr2} bulk}
\begin{verbatim}
_cell_length_a 3.93349289
_cell_length_b 3.93349289
_cell_length_c 6.04064620
_cell_angle_alpha 90.000000
_cell_angle_beta 90.000000
_cell_angle_gamma 120.000000
_symmetry_space_group_name_H-M         'P 1'
_symmetry_Int_Tables_number            1

loop_
_symmetry_equiv_pos_as_xyz
   'x, y, z'

loop_
_atom_site_label
_atom_site_type_symbol
_atom_site_fract_x
_atom_site_fract_y
_atom_site_fract_z
Br001 Br 3.333333333333E-01 -3.333333333333E-01 -2.575187047630E-01
Br002 Br -3.333333333333E-01 3.333333333333E-01 2.575187047630E-01
Cd003 Cd 0.000000000000E+00 0.000000000000E+00 0.000000000000E+00

\end{verbatim}
\begin{figure}[h]
    \centering
    \includegraphics[width=\textwidth]{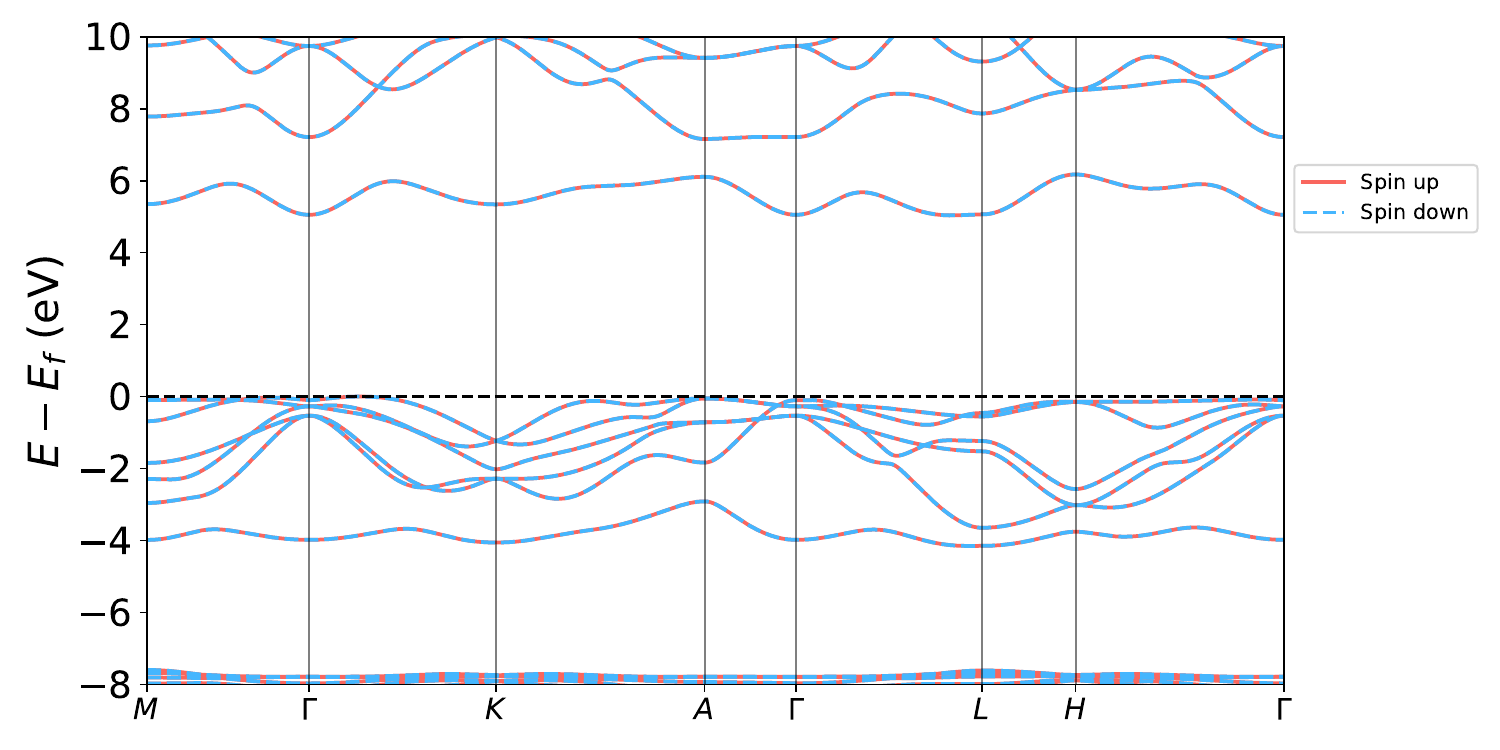}
    \caption{\footnotesize \ce{CdBr2} bulk spin-polarized band structure.}
    \label{fig:CdBr2_band_bulk}
\end{figure}
\clearpage
\newpage

\subsection{\ce{CdBr2} slab}
\begin{verbatim}
_cell_length_a                         3.92678010
_cell_length_b                         3.92678010
_cell_length_c                         40.00000000
_cell_angle_alpha                      90.000000
_cell_angle_beta                       90.000000
_cell_angle_gamma                      120.000000
_cell_volume                           'P 1'
_space_group_name_H-M_alt              'P 1'
_space_group_IT_number                 1

loop_
_space_group_symop_operation_xyz
   'x, y, z'

loop_
   _atom_site_label
   _atom_site_occupancy
   _atom_site_fract_x
   _atom_site_fract_y
   _atom_site_fract_z
   _atom_site_adp_type
   _atom_site_B_ios_or_equiv
   _atom_site_type_symbol
Br001  1.0  -0.333333333333  0.333333333333  0.539275159531  Biso  1.000000  Br
Cd002  1.0  0.000000000000  0.000000000000  0.5  Biso  1.000000  Cd
Br003  1.0  0.333333333333  -0.333333333333  0.460724840469  Biso  1.000000  Br

\end{verbatim}
\begin{figure}[h]
    \centering
    \includegraphics[width=\textwidth]{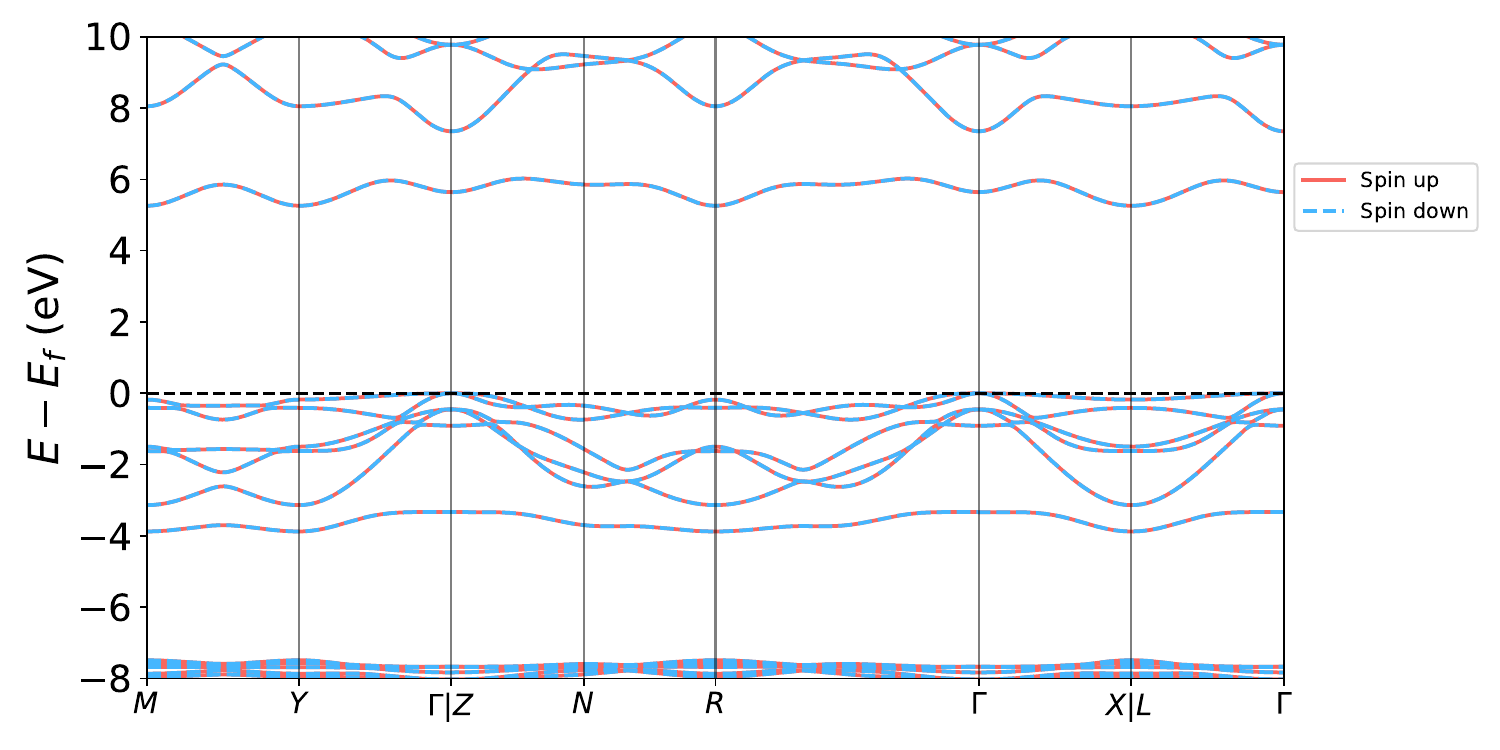}
    \caption{\footnotesize \ce{CdBr2} slab spin-polarized band structure.}
    \label{fig:CdBr2_band_slab}
\end{figure}

\clearpage
\newpage

\subsection{\ce{CdCl2} bulk}
\begin{verbatim}
_cell_length_a 3.80716055
_cell_length_b 3.80716055
_cell_length_c 5.60377014
_cell_angle_alpha 90.000000
_cell_angle_beta 90.000000
_cell_angle_gamma 120.000000
_symmetry_space_group_name_H-M         'P 1'
_symmetry_Int_Tables_number            1

loop_
_symmetry_equiv_pos_as_xyz
   'x, y, z'

loop_
_atom_site_label
_atom_site_type_symbol
_atom_site_fract_x
_atom_site_fract_y
_atom_site_fract_z
Cl001 Cl 3.333333333333E-01 -3.333333333333E-01 -2.556027803311E-01
Cl002 Cl -3.333333333333E-01 3.333333333333E-01 2.556027803311E-01
Cd003 Cd 0.000000000000E+00 0.000000000000E+00 0.000000000000E+00

\end{verbatim}
\begin{figure}[h]
    \centering
    \includegraphics[width=\textwidth]{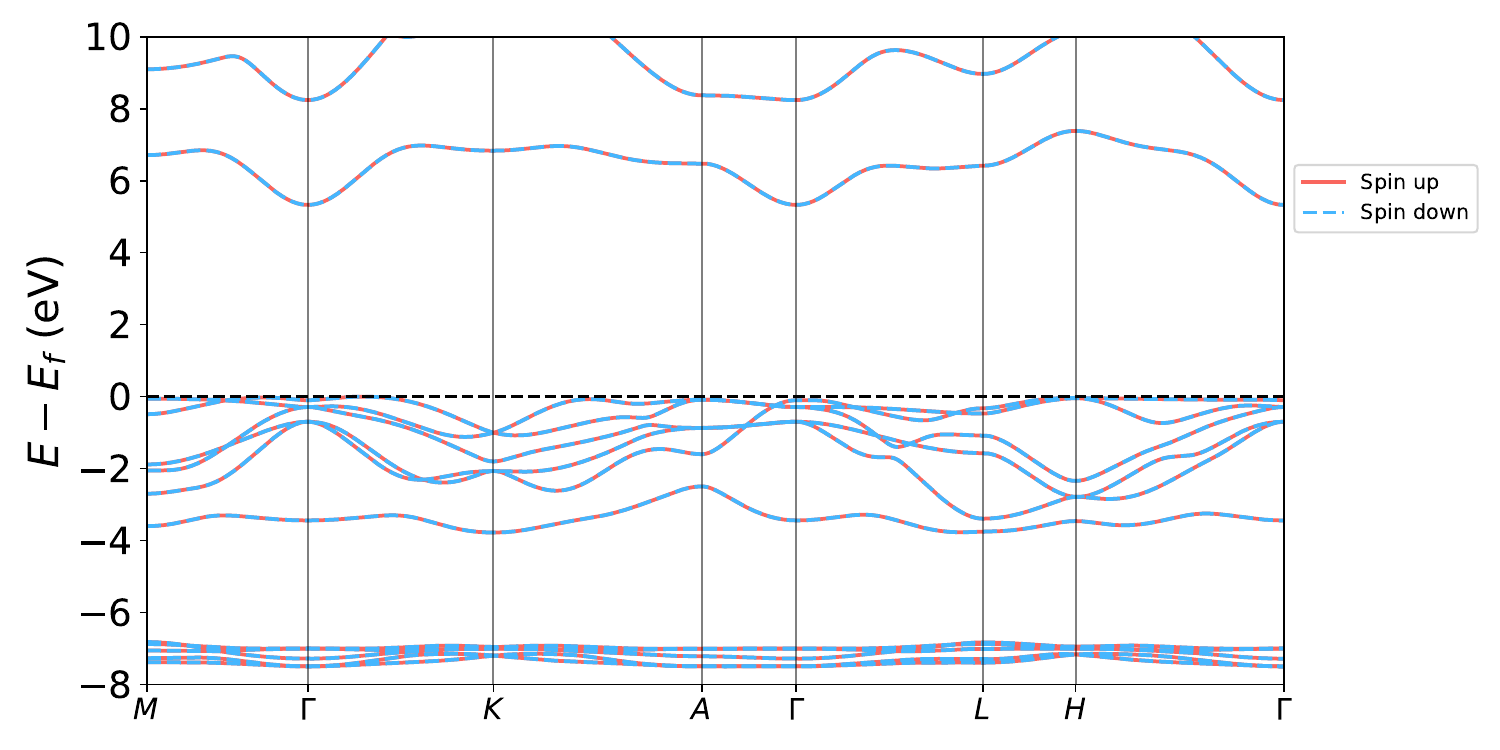}
    \caption{\footnotesize \ce{CdCl2} bulk spin-polarized band structure.}
    \label{fig:CdCl2_band_bulk}
\end{figure}
\clearpage
\newpage

\subsection{\ce{CdCl2} slab}
\begin{verbatim}
_cell_length_a                         3.78224223
_cell_length_b                         3.78224223
_cell_length_c                         40.00000000
_cell_angle_alpha                      90.000000
_cell_angle_beta                       90.000000
_cell_angle_gamma                      120.000000
_cell_volume                           'P 1'
_space_group_name_H-M_alt              'P 1'
_space_group_IT_number                 1

loop_
_space_group_symop_operation_xyz
   'x, y, z'

loop_
   _atom_site_label
   _atom_site_occupancy
   _atom_site_fract_x
   _atom_site_fract_y
   _atom_site_fract_z
   _atom_site_adp_type
   _atom_site_B_ios_or_equiv
   _atom_site_type_symbol
Cl001  1.0  -0.333333333333  0.333333333333  0.536550038773  Biso  1.000000  Cl
Cd002  1.0  0.000000000000  0.000000000000  0.5  Biso  1.000000  Cd
Cl003  1.0  0.333333333333  -0.333333333333  0.463449961227  Biso  1.000000  Cl

\end{verbatim}
\begin{figure}[h]
    \centering
    \includegraphics[width=\textwidth]{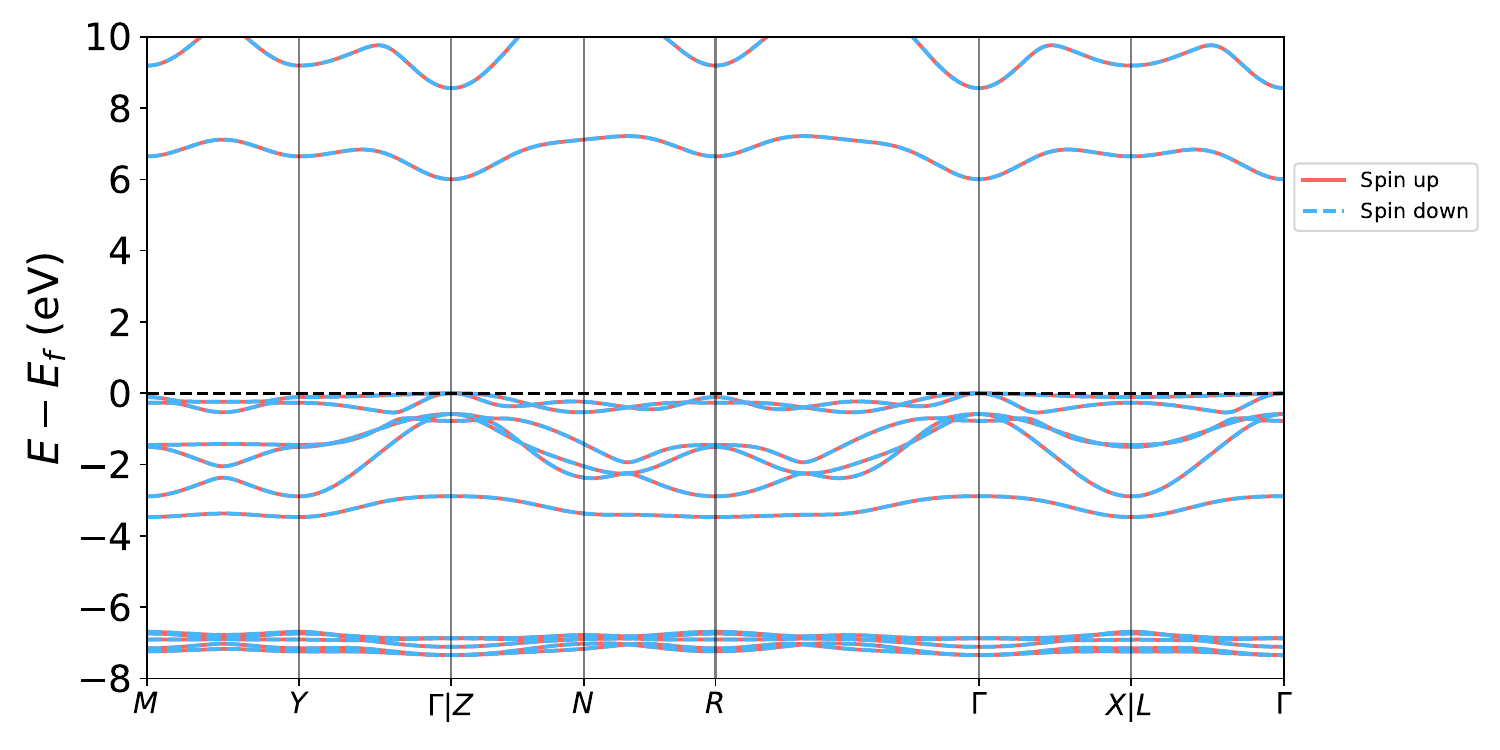}
    \caption{\footnotesize \ce{CdCl2} slab spin-polarized band structure.}
    \label{fig:CdCl2_band_slab}
\end{figure}
\clearpage
\newpage

\subsection{\ce{CdI2} bulk}
\begin{verbatim}
_cell_length_a 4.15520331
_cell_length_b 4.15520331
_cell_length_c 6.55565856
_cell_angle_alpha 90.000000
_cell_angle_beta 90.000000
_cell_angle_gamma 120.000000
_symmetry_space_group_name_H-M         'P 1'
_symmetry_Int_Tables_number            1

loop_
_symmetry_equiv_pos_as_xyz
   'x, y, z'

loop_
_atom_site_label
_atom_site_type_symbol
_atom_site_fract_x
_atom_site_fract_y
_atom_site_fract_z
I001 I 3.333333333333E-01 -3.333333333333E-01 2.626996827134E-01
I002 I -3.333333333333E-01 3.333333333333E-01 -2.626996827134E-01
Cd003 Cd 0.000000000000E+00 0.000000000000E+00 0.000000000000E+00

\end{verbatim}
\begin{figure}[h]
    \centering
    \includegraphics[width=\textwidth]{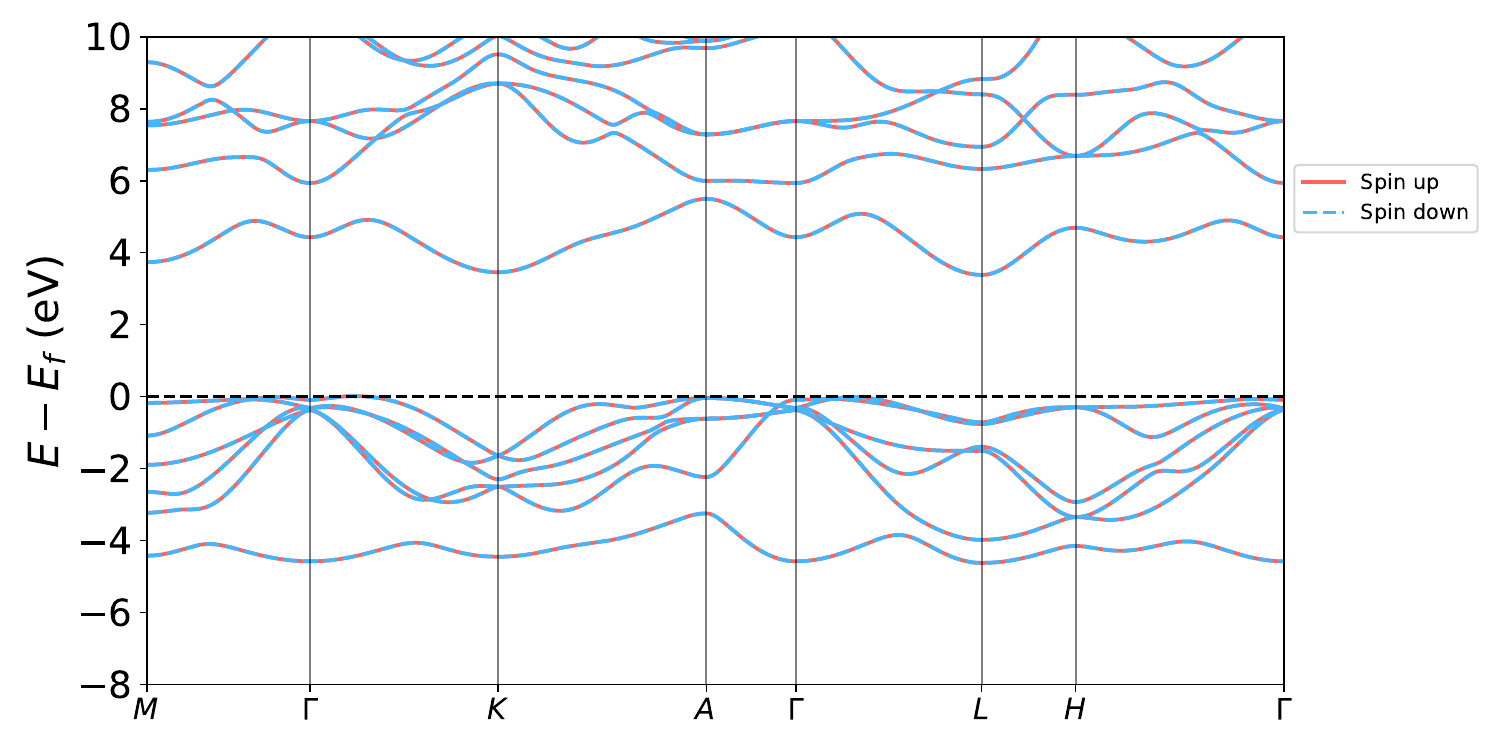}
    \caption{\footnotesize \ce{CdI2} bulk spin-polarized band structure.}
    \label{fig:CdI2_band_bulk}
\end{figure}
\clearpage
\newpage

\subsection{\ce{CdI2} slab}
\begin{verbatim}
_cell_length_a                         4.15718107
_cell_length_b                         4.15718107
_cell_length_c                         40.00000000
_cell_angle_alpha                      90.000000
_cell_angle_beta                       90.000000
_cell_angle_gamma                      120.000000
_cell_volume                           'P 1'
_space_group_name_H-M_alt              'P 1'
_space_group_IT_number                 1

loop_
_space_group_symop_operation_xyz
   'x, y, z'

loop_
   _atom_site_label
   _atom_site_occupancy
   _atom_site_fract_x
   _atom_site_fract_y
   _atom_site_fract_z
   _atom_site_adp_type
   _atom_site_B_ios_or_equiv
   _atom_site_type_symbol
I001  1.0  0.333333333333  -0.333333333333  0.543303483974  Biso  1.000000  I
Cd002  1.0  0.000000000000  0.000000000000  0.5  Biso  1.000000  Cd
I003  1.0  -0.333333333333  0.333333333333  0.456696516026  Biso  1.000000  I

\end{verbatim}
\begin{figure}[h]
    \centering
    \includegraphics[width=\textwidth]{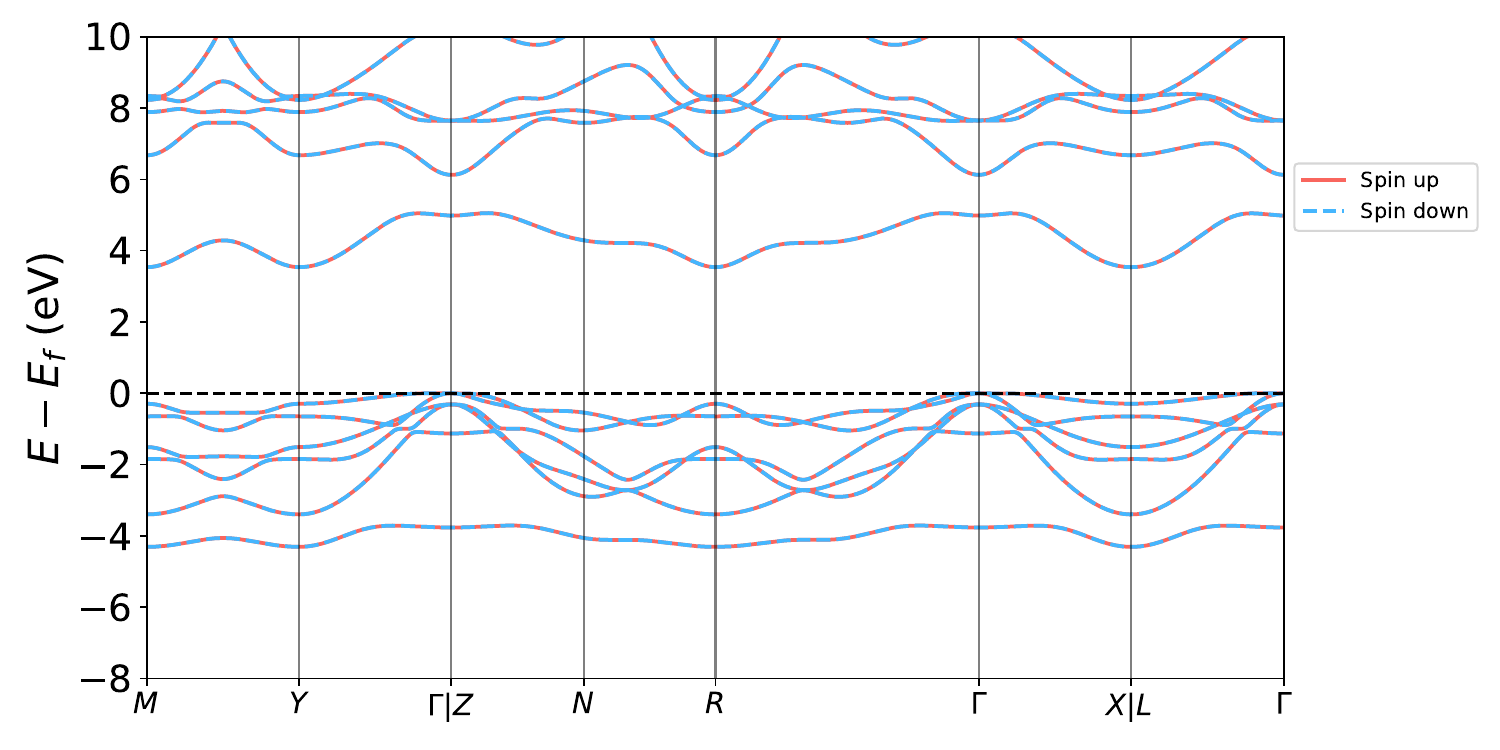}
    \caption{\footnotesize \ce{CdI2} slab spin-polarized band structure.}
    \label{fig:CdI2_band_slab}
\end{figure}
\clearpage
\newpage

\subsection{\ce{CoBr2} bulk}
\begin{verbatim}
_cell_length_a                         3.690534
_cell_length_b                         3.690668
_cell_length_c                         11.681402
_cell_angle_alpha                      89.985153
_cell_angle_beta                       89.983513
_cell_angle_gamma                      60.003063
_cell_volume                           137.794929
_space_group_name_H-M_alt              'P -1'
_space_group_IT_number                 2

loop_
_space_group_symop_operation_xyz
   'x, y, z'
   '-x, -y, -z'

loop_
   _atom_site_label
   _atom_site_occupancy
   _atom_site_fract_x
   _atom_site_fract_y
   _atom_site_fract_z
   _atom_site_adp_type
   _atom_site_B_iso_or_equiv
   _atom_site_type_symbol
   Br1         1.0     0.166540     0.666560     0.122500    Biso  1.000000 Br
   Br2         1.0     0.166540     0.666560     0.622500    Biso  1.000000 Br
   Co1         1.0     0.500000     0.000000     0.000000    Biso  1.000000 Co
   Co2         1.0     0.500000     0.000000     0.500000    Biso  1.000000 Co

\end{verbatim}

\begin{figure}[h]
    \centering
    \includegraphics[width=\textwidth]{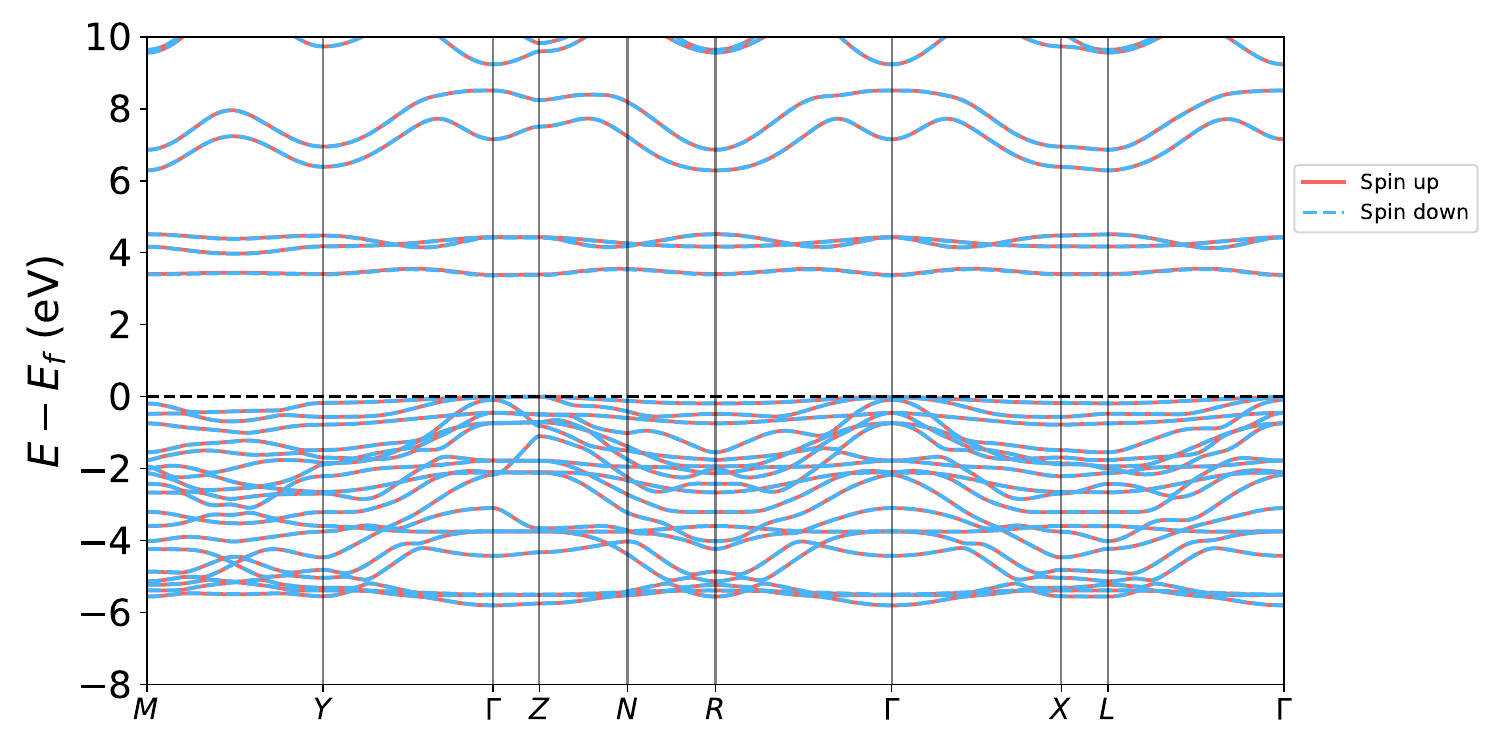}
    \caption{\footnotesize \ce{CoBr2} bulk spin-polarized band structure.}
    \label{fig:CoBr2_band_bulk}
\end{figure}
\clearpage
\newpage

\subsection{\ce{CoBr2} slab}
\begin{verbatim}
_cell_length_a                         3.69096289
_cell_length_b                         3.69108340
_cell_length_c                         40.00000000
_cell_angle_alpha                      90.000000
_cell_angle_beta                       90.000000
_cell_angle_gamma                      120.003882
_cell_volume                           'P 1'
_space_group_name_H-M_alt              'P 1'
_space_group_IT_number                 1

loop_
_space_group_symop_operation_xyz
   'x, y, z'

loop_
   _atom_site_label
   _atom_site_occupancy
   _atom_site_fract_x
   _atom_site_fract_y
   _atom_site_fract_z
   _atom_site_adp_type
   _atom_site_B_ios_or_equiv
   _atom_site_type_symbol
Br001  1.0  -0.166733699758  0.166648947011  0.535841009677  Biso  1.000000  Br
Co002  1.0  0.500000000000  0.500000000000  0.5  Biso  1.000000  Co
Br003  1.0  0.166733699758  -0.166648947011  0.464158990323  Biso  1.000000  Br

\end{verbatim}
\begin{figure}[h]
    \centering
    \includegraphics[width=\textwidth]{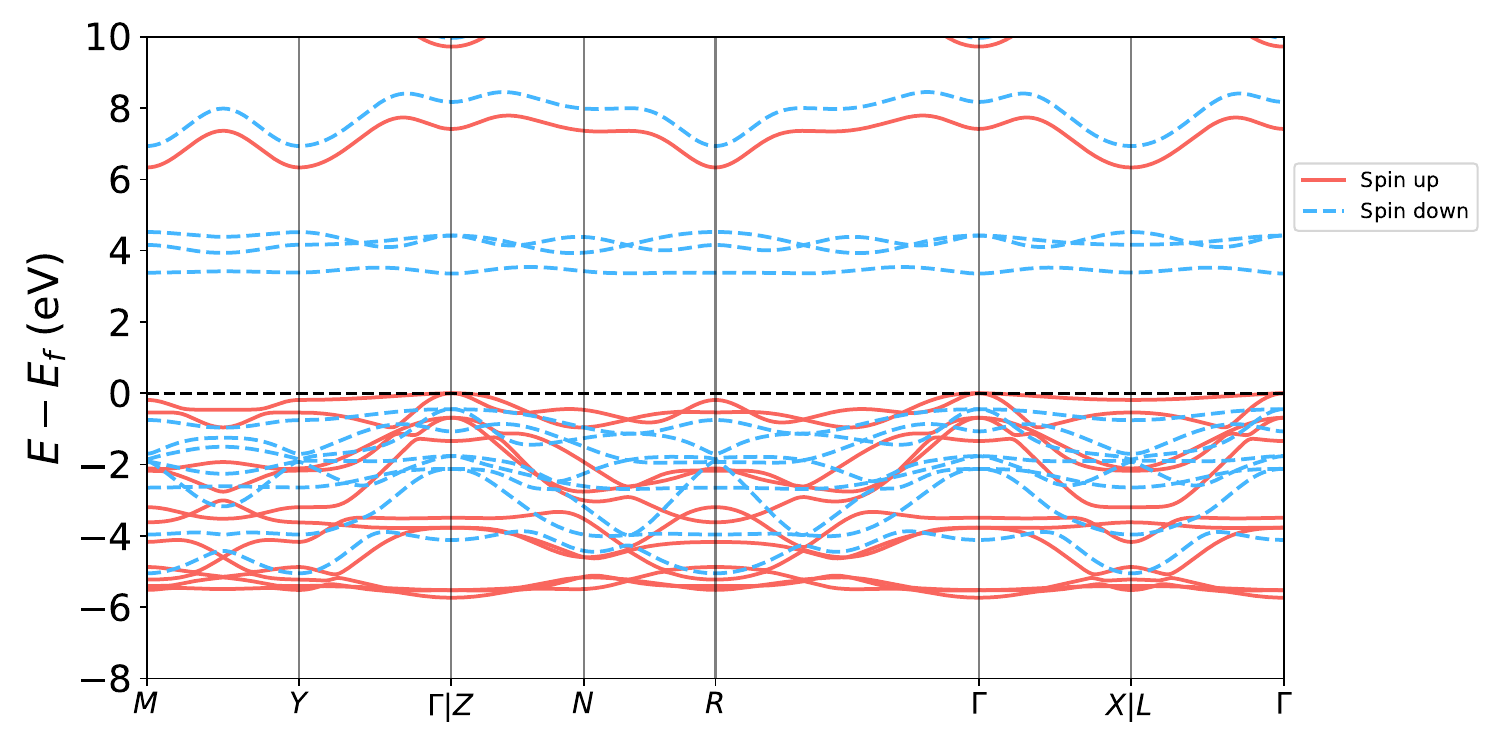}
    \caption{\footnotesize \ce{CoBr2} slab spin-polarized band structure.}
    \label{fig:CoBr2_band_slab}
\end{figure}
\clearpage
\newpage

\subsection{\ce{CoCl2} bulk}
\begin{verbatim}
_cell_length_a                         6.109289
_cell_length_b                         3.527209
_cell_length_c                         11.195854
_cell_angle_alpha                      90.000000
_cell_angle_beta                       90.049957
_cell_angle_gamma                      90.000000
_cell_volume                           241.256427
_space_group_name_H-M_alt              'C 2/m'
_space_group_IT_number                 12

loop_
_space_group_symop_operation_xyz
   'x, y, z'
   '-x, -y, -z'
   '-x, y, -z'
   'x, -y, z'
   'x+1/2, y+1/2, z'
   '-x+1/2, -y+1/2, -z'
   '-x+1/2, y+1/2, -z'
   'x+1/2, -y+1/2, z'

loop_
   _atom_site_label
   _atom_site_occupancy
   _atom_site_fract_x
   _atom_site_fract_y
   _atom_site_fract_z
   _atom_site_adp_type
   _atom_site_B_iso_or_equiv
   _atom_site_type_symbol
   Cl1         1.0     0.333560     0.000000     0.120210    Biso  1.000000 Cl
   Cl2         1.0     0.333560     0.000000     0.620210    Biso  1.000000 Cl
   Co1         1.0     0.000000     0.000000     0.000000    Biso  1.000000 Co
   Co2         1.0     0.000000     0.000000     0.500000    Biso  1.000000 Co

\end{verbatim}

\begin{figure}[h]
    \centering
    \includegraphics[width=\textwidth]{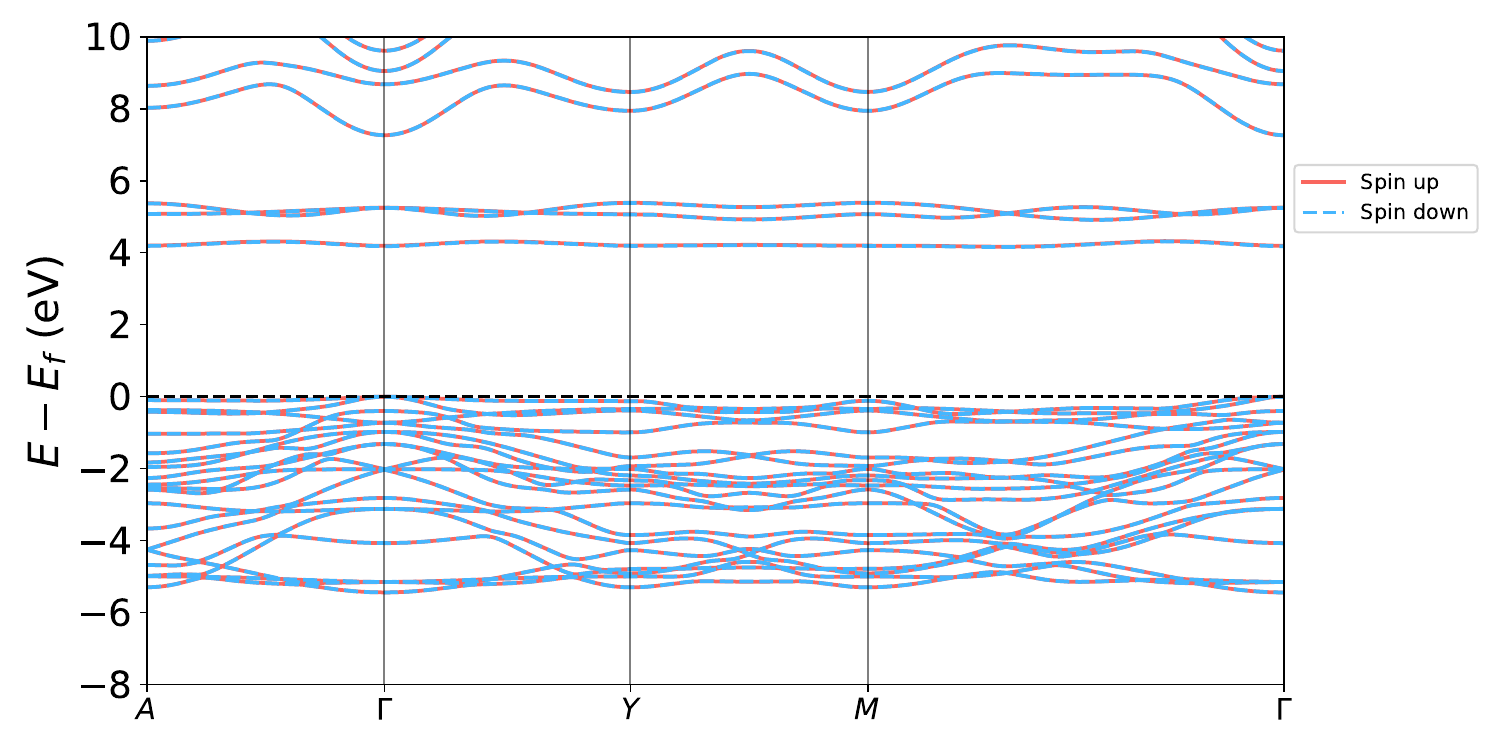}
    \caption{\footnotesize \ce{CoCl2} bulk spin-polarized band structure.}
    \label{fig:CoCl2_band_bulk}
\end{figure}
\clearpage
\newpage

\subsection{\ce{CoCl2} slab}
\begin{verbatim}
_cell_length_a                         3.52663281
_cell_length_b                         3.52663281
_cell_length_c                         40.00000000
_cell_angle_alpha                      90.000000
_cell_angle_beta                       90.000000
_cell_angle_gamma                      120.005303
_cell_volume                           'P 1'
_space_group_name_H-M_alt              'P 1'
_space_group_IT_number                 1

loop_
_space_group_symop_operation_xyz
   'x, y, z'

loop_
   _atom_site_label
   _atom_site_occupancy
   _atom_site_fract_x
   _atom_site_fract_y
   _atom_site_fract_z
   _atom_site_adp_type
   _atom_site_B_ios_or_equiv
   _atom_site_type_symbol
Cl001  1.0  -0.333319696886  0.333319696886  0.533872264726  Biso  1.000000  Cl
Co002  1.0  0.000000000000  -0.000000000000  0.5  Biso  1.000000  Co
Cl003  1.0  0.333319696886  -0.333319696886  0.466127735274  Biso  1.000000  Cl

\end{verbatim}
\begin{figure}[h]
    \centering
    \includegraphics[width=\textwidth]{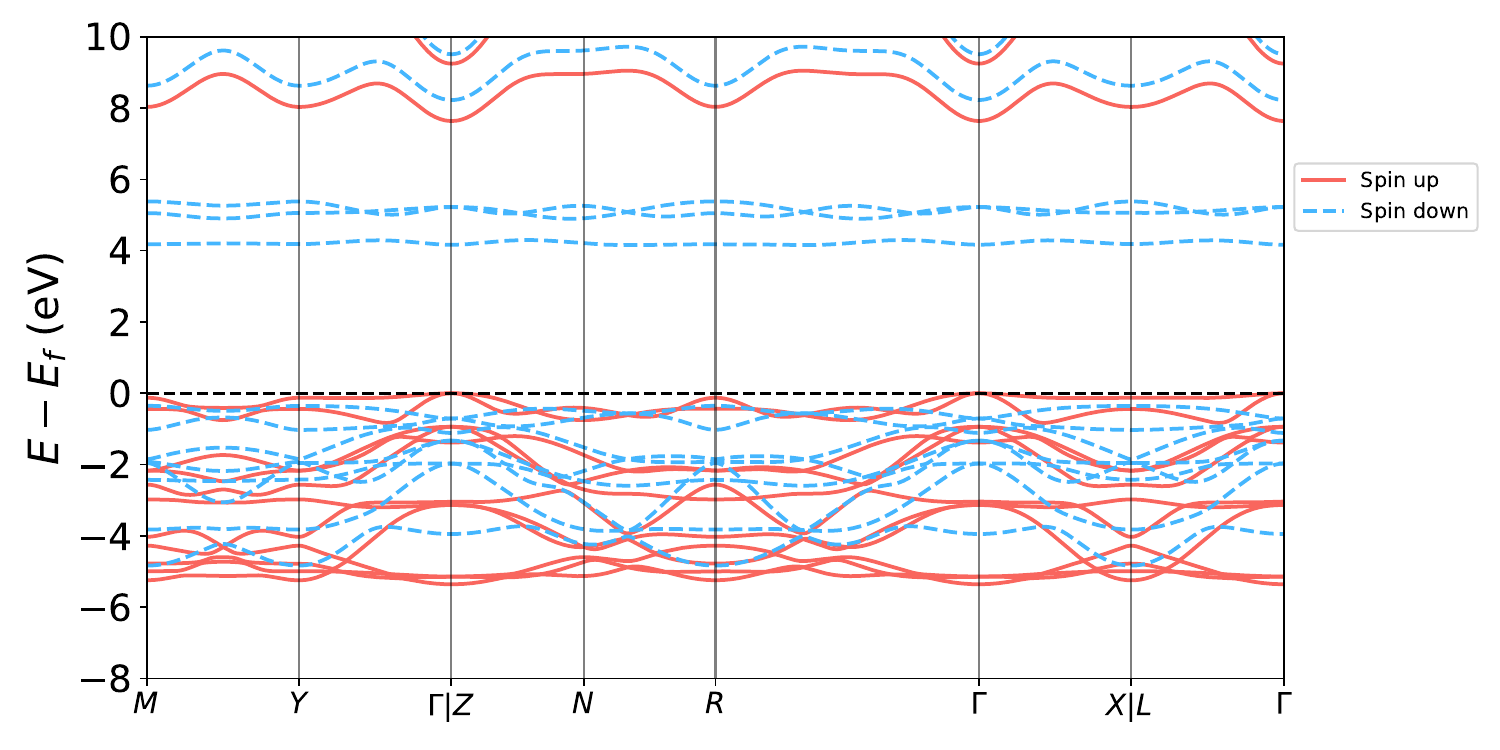}
    \caption{\footnotesize \ce{CoCl2} slab spin-polarized band structure.}
    \label{fig:CoCl2_band_slab}
\end{figure}

\clearpage
\newpage

\subsection{\ce{CoI2} bulk}
\begin{verbatim}
_cell_length_a                         3.95832100
_cell_length_b                         3.95996500
_cell_length_c                         12.40816200
_cell_angle_alpha                      89.813545
_cell_angle_beta                       89.996956
_cell_angle_gamma                      60.038494
_cell_volume                           'P 1'
_space_group_name_H-M_alt              'P 1'
_space_group_IT_number                 1

loop_
_space_group_symop_operation_xyz
   'x, y, z'

loop_
   _atom_site_label
   _atom_site_occupancy
   _atom_site_fract_x
   _atom_site_fract_y
   _atom_site_fract_z
   _atom_site_adp_type
   _atom_site_B_ios_or_equiv
   _atom_site_type_symbol
   Co001   1.0   0.000000000000  0.000000000000  0.000000000000  Biso  1.000000  Co
   Co002   1.0   0.000000000000  0.000000000000  0.500000000000  Biso  1.000000  Co
    I003   1.0   -0.334120000000  -0.331900000000  -0.126010000000  Biso  1.000000   I
    I004   1.0   0.334120000000  0.331900000000  0.126010000000  Biso  1.000000   I
    I005   1.0   -0.334020000000  -0.332100000000  0.373990000000  Biso  1.000000   I
    I006   1.0   0.334020000000  0.332100000000  -0.373990000000  Biso  1.000000   I

\end{verbatim}
\begin{figure}[h]
    \centering
    \includegraphics[width=\textwidth]{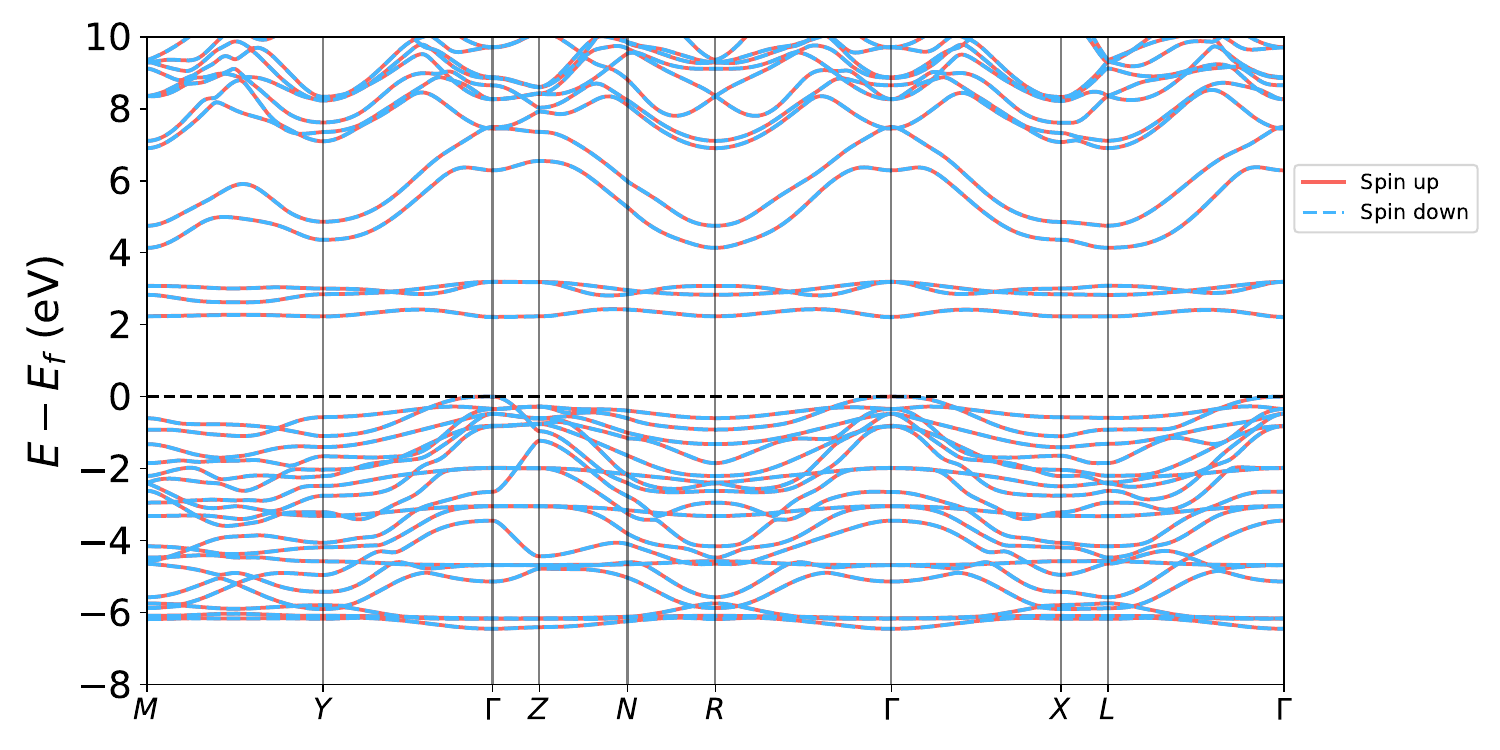}
    \caption{\footnotesize \ce{CoI2} bulk spin-polarized band structure.}
    \label{fig:CoI2_band_bulk}
\end{figure}
\clearpage
\newpage

\subsection{\ce{CoI2} slab}
\begin{verbatim}
_cell_length_a                         7.89871364
_cell_length_b                         7.89566590
_cell_length_c                         40.00000000
_cell_angle_alpha                      90.000000
_cell_angle_beta                       90.000000
_cell_angle_gamma                      121.277215
_cell_volume                           'P 1'
_space_group_name_H-M_alt              'P 1'
_space_group_IT_number                 1

loop_
_space_group_symop_operation_xyz
   'x, y, z'

loop_
   _atom_site_label
   _atom_site_occupancy
   _atom_site_fract_x
   _atom_site_fract_y
   _atom_site_fract_z
   _atom_site_adp_type
   _atom_site_B_ios_or_equiv
   _atom_site_type_symbol
I001  1.0  -0.418124699406  0.167844090362  0.539750713185  Biso  1.000000  I
I002  1.0  0.081927609021  -0.332124240573  0.539745519549  Biso  1.000000  I
I003  1.0  -0.418139416251  -0.332154415832  0.539749174259  Biso  1.000000  I
I004  1.0  0.081906926635  0.167880218887  0.539745608991  Biso  1.000000  I
Co005  1.0  -0.250020623256  0.499979619215  0.50001043127  Biso  1.000000  Co
Co006  1.0  0.249975376498  0.000011508732  0.499993747811  Biso  1.000000  Co
Co007  1.0  -0.249974951451  -0.000010395616  0.499991219279  Biso  1.000000  Co
Co008  1.0  0.250018426074  -0.499981580950  0.499973506726  Biso  1.000000  Co
I009  1.0  0.418116462015  -0.167813099541  0.460238650022  Biso  1.000000  I
I010  1.0  -0.081937747020  0.332131800786  0.460241223683  Biso  1.000000  I
I011  1.0  0.418150492337  0.332146821845  0.460232600552  Biso  1.000000  I
I012  1.0  -0.081894075705  -0.167902891202  0.460235384411  Biso  1.000000  I

\end{verbatim}
\begin{figure}[h]
    \centering
    \includegraphics[width=\textwidth]{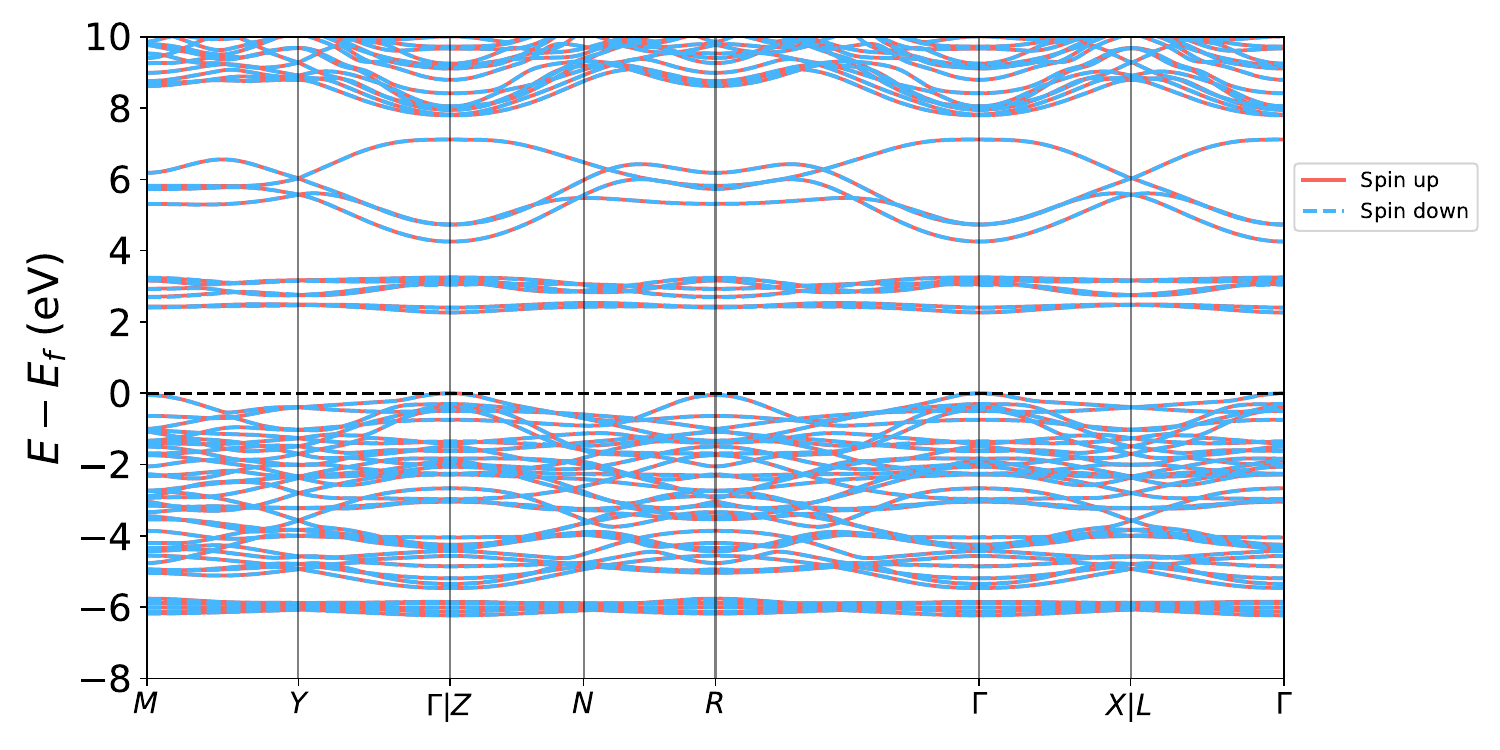}
    \caption{\footnotesize \ce{CoI2} slab spin-polarized band structure.}
    \label{fig:CoI2_band_slab}
\end{figure}
\clearpage
\newpage

\subsection{\ce{CrBr2} bulk}
\begin{verbatim}
_cell_length_a                         7.190969
_cell_length_b                         7.679319
_cell_length_c                         11.907789
_cell_angle_alpha                      92.733368
_cell_angle_beta                       90.299667
_cell_angle_gamma                      117.350349
_cell_volume                           583.124268
_space_group_name_H-M_alt              'P -1'
_space_group_IT_number                 2

loop_
_space_group_symop_operation_xyz
   'x, y, z'
   '-x, -y, -z'

loop_
   _atom_site_label
   _atom_site_occupancy
   _atom_site_fract_x
   _atom_site_fract_y
   _atom_site_fract_z
   _atom_site_adp_type
   _atom_site_B_iso_or_equiv
   _atom_site_type_symbol
   Cr1         1.0    -0.000080     0.250050     0.249960    Biso  1.000000 Cr
   Cr2         1.0    -0.000150     0.249970     0.749980    Biso  1.000000 Cr
   Cr3         1.0     0.499870     0.250050     0.250010    Biso  1.000000 Cr
   Cr4         1.0     0.499760     0.249950     0.749990    Biso  1.000000 Cr
   Br1         1.0     0.826940    -0.097130     0.376250    Biso  1.000000 Br
   Br2         1.0     0.826930    -0.097290     0.876250    Biso  1.000000 Br
   Br3         1.0     0.826780     0.402730     0.376210    Biso  1.000000 Br
   Br4         1.0     0.826590     0.402750     0.876200    Biso  1.000000 Br
   Br5         1.0     0.327020    -0.097230     0.376230    Biso  1.000000 Br
   Br6         1.0     0.326870    -0.097230     0.876210    Biso  1.000000 Br
   Br7         1.0     0.326660     0.402840     0.376210    Biso  1.000000 Br
   Br8         1.0     0.326680     0.402670     0.876240    Biso  1.000000 Br

\end{verbatim}
\begin{figure}[h]
    \centering
    \includegraphics[width=\textwidth]{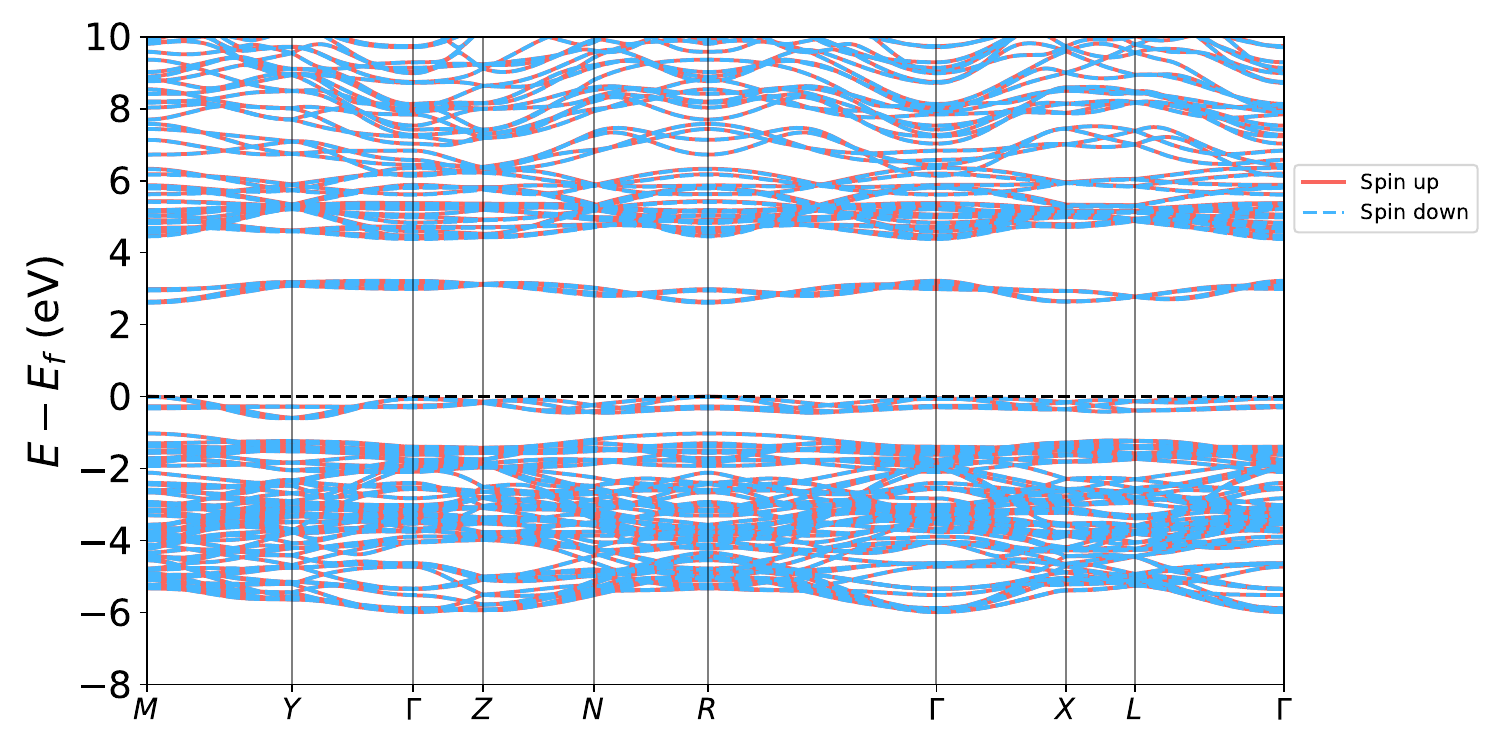}
    \caption{\footnotesize \ce{CrBr2} bulk spin-polarized band structure.}
    \label{fig:CrBr2_band_bulk}
\end{figure}
\clearpage
\newpage

\subsection{\ce{CrBr2} slab}
\begin{verbatim}
_cell_length_a                         7.19206184
_cell_length_b                         7.68914318
_cell_length_c                         40.00000000
_cell_angle_alpha                      90.000000
_cell_angle_beta                       90.000000
_cell_angle_gamma                      117.363269
_cell_volume                           'P 1'
_space_group_name_H-M_alt              'P 1'
_space_group_IT_number                 1

loop_
_space_group_symop_operation_xyz
   'x, y, z'

loop_
   _atom_site_label
   _atom_site_occupancy
   _atom_site_fract_x
   _atom_site_fract_y
   _atom_site_fract_z
   _atom_site_adp_type
   _atom_site_B_ios_or_equiv
   _atom_site_type_symbol
Br001  1.0  -0.181728797129  -0.112405156694  0.537989254085  Biso  1.000000  Br
Br002  1.0  -0.181679874514  0.388060195436  0.537933097344  Biso  1.000000  Br
Br003  1.0  0.318474607249  -0.111990614314  0.537917478287  Biso  1.000000  Br
Br004  1.0  0.318223869839  0.387728870443  0.537994548242  Biso  1.000000  Br
Cr005  1.0  -0.000047162261  -0.250201215057  0.500016019449  Biso  1.000000  Cr
Cr006  1.0  -0.000108650636  0.249878934102  0.500002283623  Biso  1.000000  Cr
Cr007  1.0  0.499985517276  -0.250259446801  0.499996581482  Biso  1.000000  Cr
Cr008  1.0  0.499837859138  0.249931895028  0.500016591039  Biso  1.000000  Cr
Br009  1.0  0.181651798377  0.112231544923  0.462017642814  Biso  1.000000  Br
Br010  1.0  0.181643669047  -0.387810660153  0.462013222233  Biso  1.000000  Br
Br011  1.0  -0.318449348352  0.112292702093  0.46201984597  Biso  1.000000  Br
Br012  1.0  -0.318294354402  -0.387920078291  0.462020128238  Biso  1.000000  Br

\end{verbatim}
\begin{figure}[h]
    \centering
    \includegraphics[width=\textwidth]{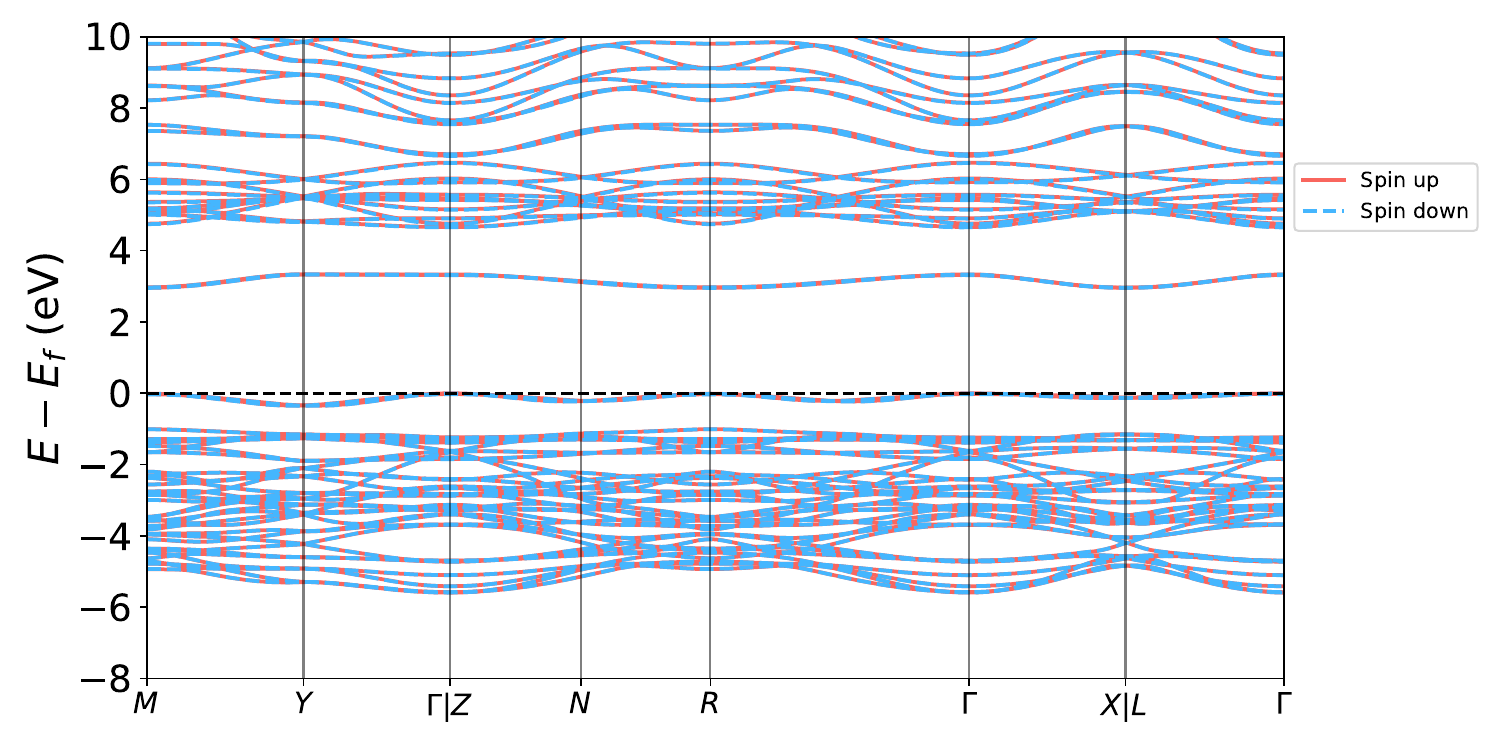}
    \caption{\footnotesize \ce{CrBr2} slab spin-polarized band structure.}
    \label{fig:CrBr2_band_slab}
\end{figure}
\clearpage
\newpage

\subsection{\ce{CrI2} bulk}
\begin{verbatim}
_cell_length_a                         7.74623678
_cell_length_b                         8.15488264
_cell_length_c                         12.75113852
_cell_angle_alpha                      87.096997
_cell_angle_beta                       89.819997
_cell_angle_gamma                      117.901605
_cell_volume                           'P 1'
_space_group_name_H-M_alt              'P 1'
_space_group_IT_number                 1

loop_
_space_group_symop_operation_xyz
   'x, y, z'

loop_
   _atom_site_label
   _atom_site_occupancy
   _atom_site_fract_x
   _atom_site_fract_y
   _atom_site_fract_z
   _atom_site_adp_type
   _atom_site_B_ios_or_equiv
   _atom_site_type_symbol
   Cr001   1.0   -0.249566196239  -0.249912957784  -0.250018294179  Biso  1.000000  Cr
   Cr002   1.0   -0.250594946432  -0.250204052794  0.249584602775  Biso  1.000000  Cr
   Cr003   1.0   -0.249545614579  0.250116759978  -0.249711779860  Biso  1.000000  Cr
   Cr004   1.0   -0.250268818770  0.249840425144  0.250123210110  Biso  1.000000  Cr
   Cr005   1.0   0.250587928175  -0.249426744031  -0.249835997772  Biso  1.000000  Cr
   Cr006   1.0   0.249487103473  -0.250579447164  0.249781113910  Biso  1.000000  Cr
   Cr007   1.0   0.250295282168  0.250527993117  -0.249843808817  Biso  1.000000  Cr
   Cr008   1.0   0.249650115612  0.249658936776  0.249917879029  Biso  1.000000  Cr
    I009   1.0   0.077996914283  -0.094440674984  0.121431572799  Biso  1.000000   I
    I010   1.0   0.078772959579  -0.093526107036  -0.378193358436  Biso  1.000000   I
    I011   1.0   0.078155366476  0.405730268773  0.121610692310  Biso  1.000000   I
    I012   1.0   0.078851423786  0.406208344013  -0.378274553501  Biso  1.000000   I
    I013   1.0   -0.422164437687  -0.094167421560  0.121389161728  Biso  1.000000   I
    I014   1.0   -0.421216650991  -0.093886440001  -0.378230769870  Biso  1.000000   I
    I015   1.0   -0.421886210180  0.404991451010  0.121464774999  Biso  1.000000   I
    I016   1.0   -0.421309357075  0.405337819699  -0.378337331041  Biso  1.000000   I
    I017   1.0   -0.078330105064  0.093926567108  -0.121440815509  Biso  1.000000   I
    I018   1.0   -0.078933945291  0.093899818960  0.378349042784  Biso  1.000000   I
    I019   1.0   -0.077866558939  -0.405663542407  -0.121573649788  Biso  1.000000   I
    I020   1.0   -0.078509621632  -0.405181182479  0.378314510387  Biso  1.000000   I
    I021   1.0   0.421887587050  0.095075308044  -0.121350341015  Biso  1.000000   I
    I022   1.0   0.421082177497  0.093577445451  0.378211370331  Biso  1.000000   I
    I023   1.0   0.422181226462  -0.405676735211  -0.121572748899  Biso  1.000000   I
    I024   1.0   0.421244378317  -0.406225832622  0.378205517524  Biso  1.000000   I

\end{verbatim}

\begin{figure}[h]
    \centering
    \includegraphics[width=\textwidth]{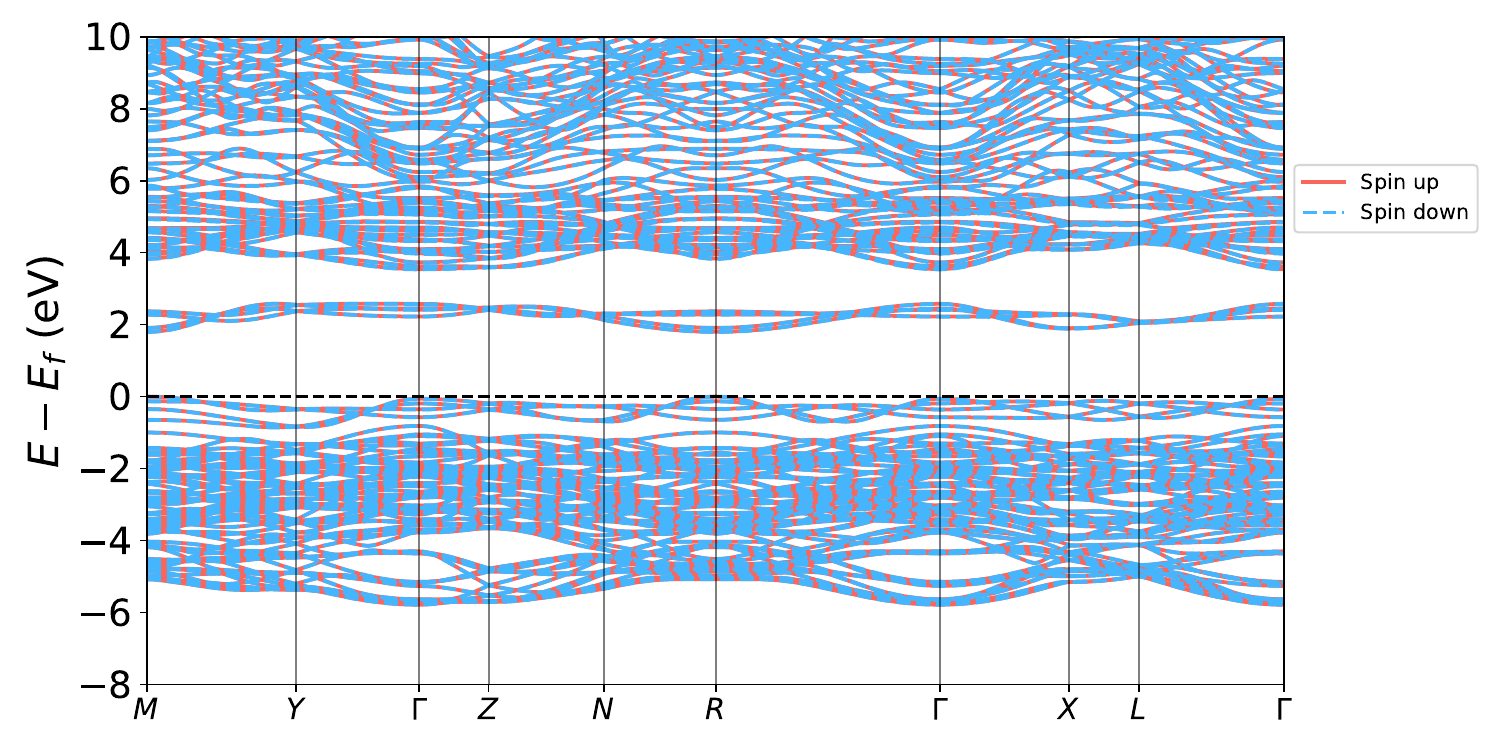}
    \caption{\footnotesize \ce{CrI2} bulk spin-polarized band structure.}
    \label{fig:CrI2_band_bulk}
\end{figure}
\clearpage
\newpage

\subsection{\ce{CrI2} slab}
\begin{verbatim}
_cell_length_a                         7.77390423
_cell_length_b                         8.22036191
_cell_length_c                         40.00000000
_cell_angle_alpha                      90.000000
_cell_angle_beta                       90.000000
_cell_angle_gamma                      118.226146
_cell_volume                           'P 1'
_space_group_name_H-M_alt              'P 1'
_space_group_IT_number                 1

loop_
_space_group_symop_operation_xyz
   'x, y, z'

loop_
   _atom_site_label
   _atom_site_occupancy
   _atom_site_fract_x
   _atom_site_fract_y
   _atom_site_fract_z
   _atom_site_adp_type
   _atom_site_B_ios_or_equiv
   _atom_site_type_symbol
I001  1.0  0.430724584697  0.110967036540  0.541177693755  Biso  1.000000  I
I002  1.0  -0.069261039049  0.110921362906  0.541175090876  Biso  1.000000  I
I003  1.0  -0.069289649163  -0.389063836654  0.541177659344  Biso  1.000000  I
I004  1.0  0.430752496323  -0.389019367978  0.541171427417  Biso  1.000000  I
Cr005  1.0  -0.249615659145  0.250233030666  0.499999027569  Biso  1.000000  Cr
Cr006  1.0  0.250371806334  -0.249727159894  0.499995552284  Biso  1.000000  Cr
Cr007  1.0  0.250433132777  0.250359434388  0.500001277108  Biso  1.000000  Cr
Cr008  1.0  -0.249592945246  -0.249786269617  0.499980928425  Biso  1.000000  Cr
I009  1.0  0.070085847243  -0.110419375608  0.45880631613  Biso  1.000000  I
I010  1.0  -0.429969018001  -0.110464819026  0.458813450233  Biso  1.000000  I
I011  1.0  0.070040759552  0.389567011254  0.458811039358  Biso  1.000000  I
I012  1.0  -0.429954138122  0.389505616900  0.458807163301  Biso  1.000000  I

\end{verbatim}
\begin{figure}[h]
    \centering
    \includegraphics[width=\textwidth]{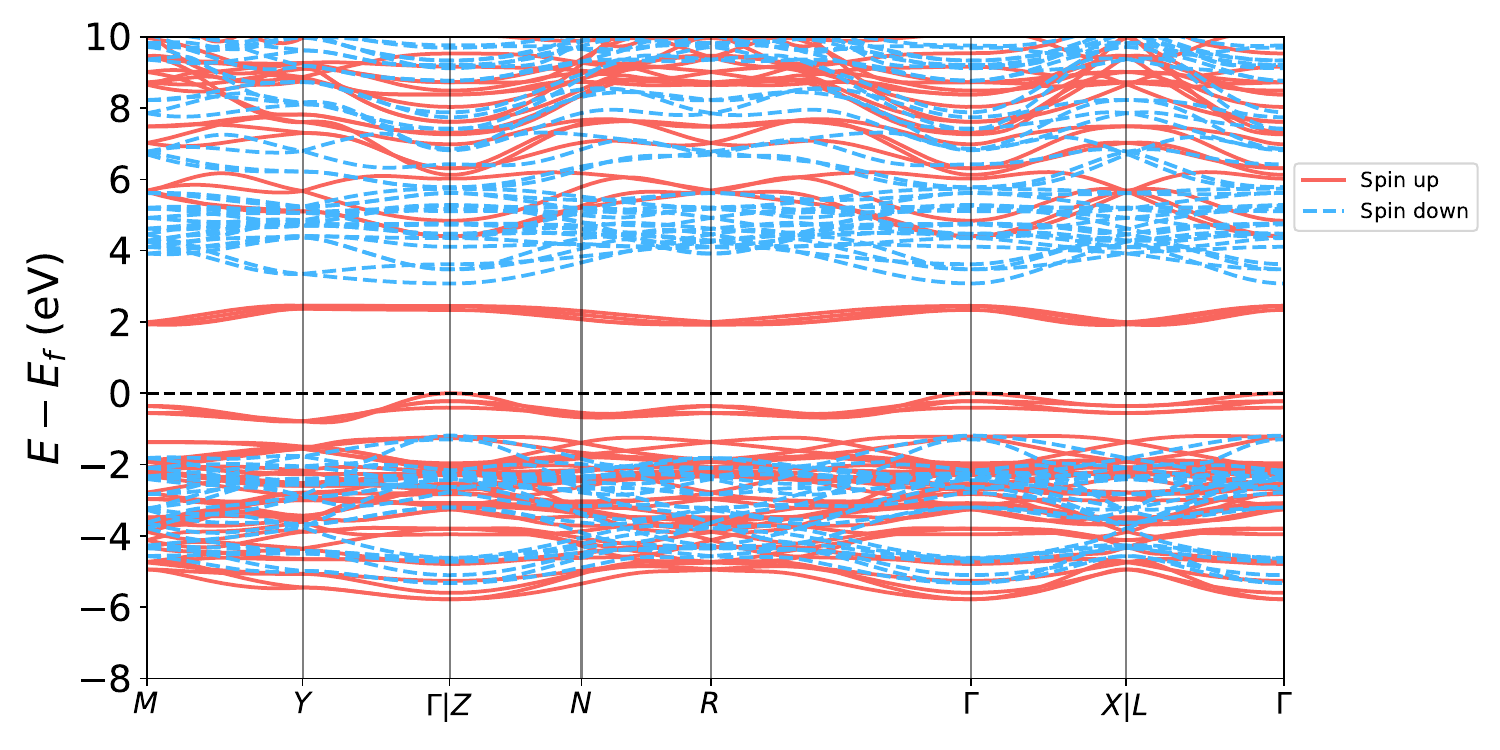}
    \caption{\footnotesize \ce{CrI2} slab spin-polarized band structure.}
    \label{fig:CrI2_band_slab}
\end{figure}

\clearpage
\newpage

\subsection{CuBr bulk}
\begin{verbatim}
_cell_length_a 6.49786661
_cell_length_b 6.49786661
_cell_length_c 6.49786661
_cell_angle_alpha 33.660943
_cell_angle_beta 33.660943
_cell_angle_gamma 33.660943
_symmetry_space_group_name_H-M         'P 1'
_symmetry_Int_Tables_number            1

loop_
_symmetry_equiv_pos_as_xyz
   'x, y, z'

loop_
_atom_site_label
_atom_site_type_symbol
_atom_site_fract_x
_atom_site_fract_y
_atom_site_fract_z
Br001 Br 9.291150962897E-02 9.291150962897E-02 9.291150962897E-02
Br002 Br -9.291150962897E-02 -9.291150962897E-02 -9.291150962897E-02
Cu003 Cu 3.709649477915E-01 3.709649477915E-01 3.709649477915E-01
Cu004 Cu -3.709649477915E-01 -3.709649477915E-01 -3.709649477915E-01

\end{verbatim}
\begin{figure}[h]
    \centering
    \includegraphics[width=\textwidth]{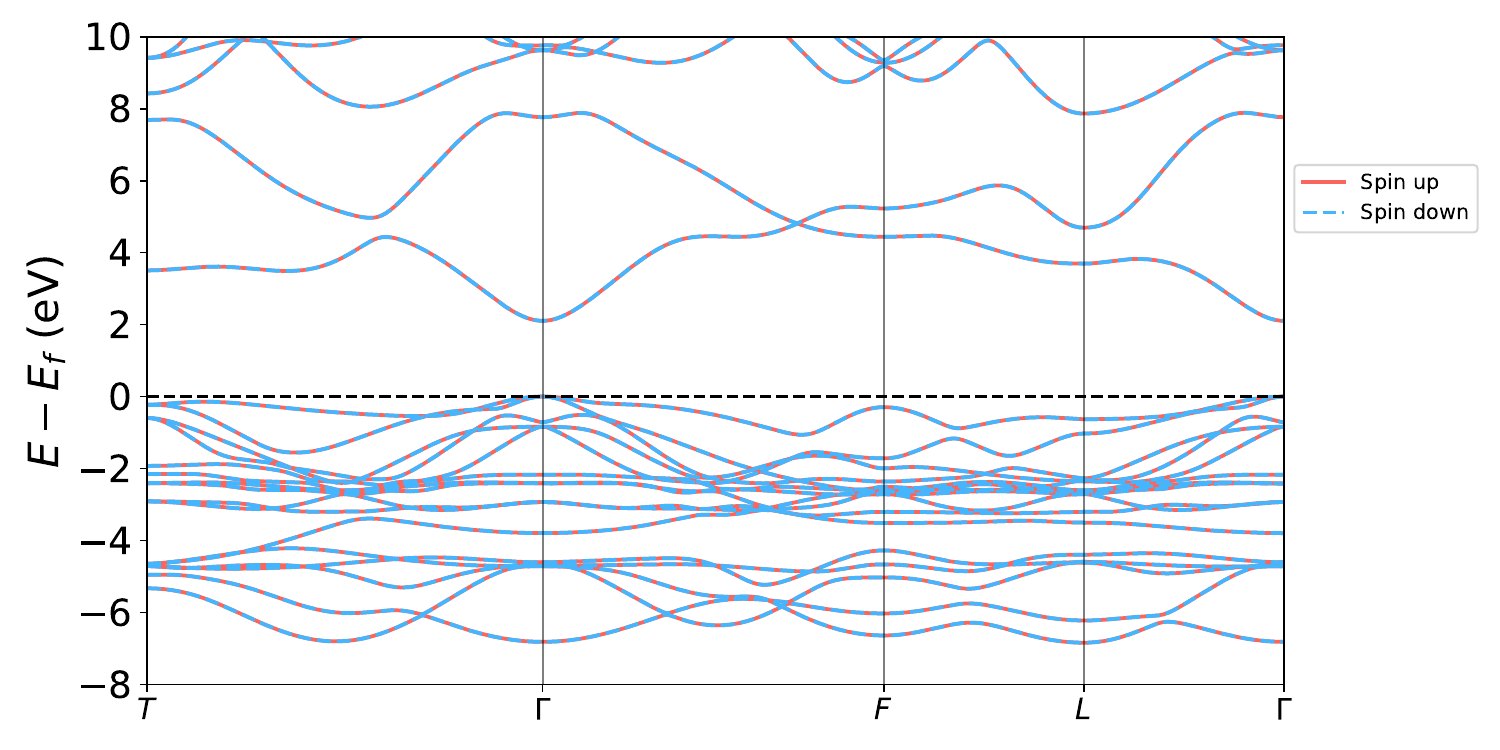}
    \caption{\footnotesize \ce{CuBr} bulk spin-polarized band structure.}
    \label{fig:CuBr_band_bulk}
\end{figure}
\clearpage
\newpage

\subsection{CuBr slab}
\begin{verbatim}
_cell_length_a                         4.04326928
_cell_length_b                         4.04326928
_cell_length_c                         40.00000000
_cell_angle_alpha                      90.000000
_cell_angle_beta                       90.000000
_cell_angle_gamma                      120.000000
_cell_volume                           'P 1'
_space_group_name_H-M_alt              'P 1'
_space_group_IT_number                 1

loop_
_space_group_symop_operation_xyz
   'x, y, z'

loop_
   _atom_site_label
   _atom_site_occupancy
   _atom_site_fract_x
   _atom_site_fract_y
   _atom_site_fract_z
   _atom_site_adp_type
   _atom_site_B_ios_or_equiv
   _atom_site_type_symbol
Cu001  1.0  0.202790209246  -0.398604895377  0.500001765258  Biso  1.000000  Cu
Br002  1.0  -0.130543124087  -0.065271562044  0.499998234742  Biso  1.000000  Br

\end{verbatim}
\begin{figure}[h]
    \centering
    \includegraphics[width=\textwidth]{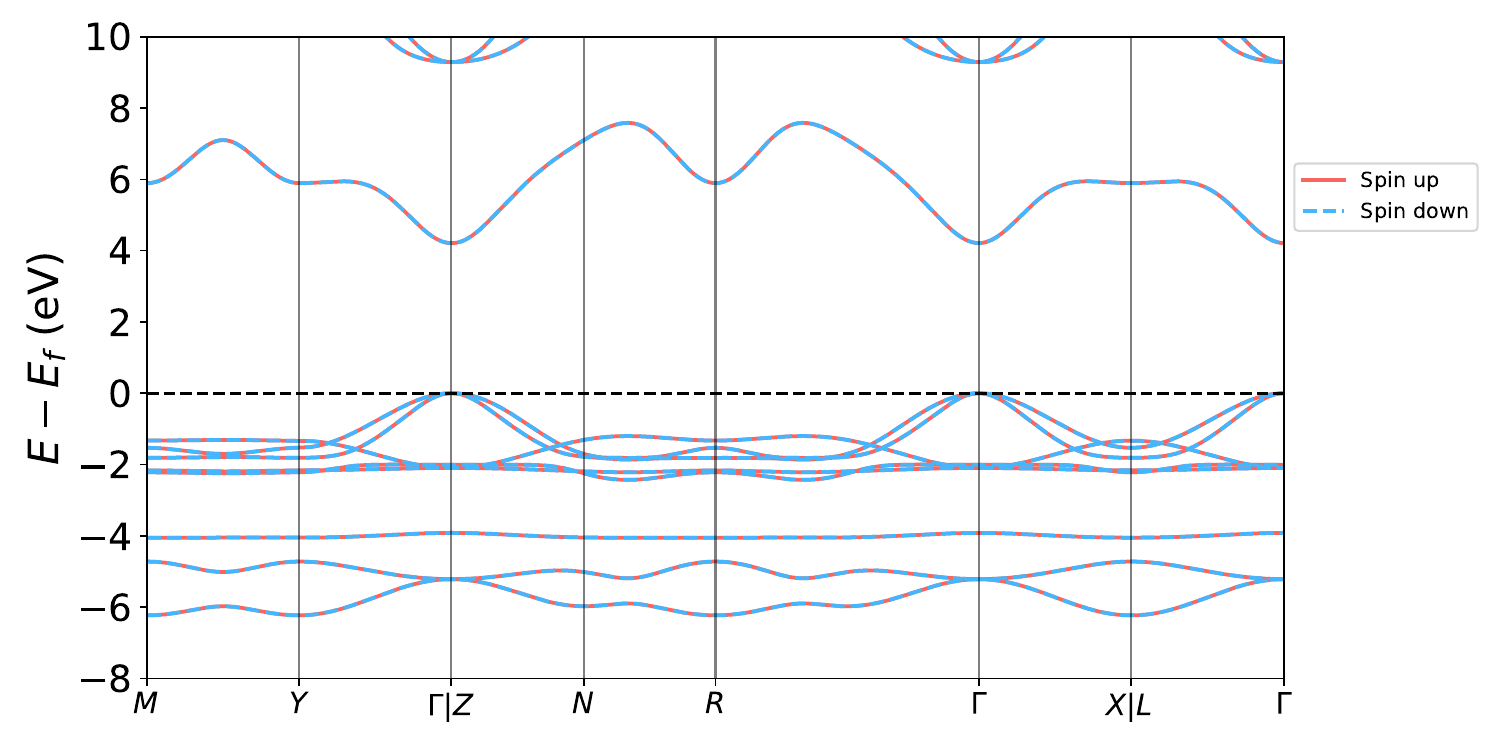}
    \caption{\footnotesize CuBr slab spin-polarized band structure.}
    \label{fig:CuBr_band_slab}
\end{figure}
\clearpage
\newpage

\subsection{\ce{CuCl2} bulk}
\begin{verbatim}
_cell_length_a                         6.391266
_cell_length_b                         3.333986
_cell_length_c                         11.175098
_cell_angle_alpha                      90.000000
_cell_angle_beta                       94.253868
_cell_angle_gamma                      90.000000
_cell_volume                           237.467348
_space_group_name_H-M_alt              'C 2/m'
_space_group_IT_number                 12

loop_
_space_group_symop_operation_xyz
   'x, y, z'
   '-x, -y, -z'
   '-x, y, -z'
   'x, -y, z'
   'x+1/2, y+1/2, z'
   '-x+1/2, -y+1/2, -z'
   '-x+1/2, y+1/2, -z'
   'x+1/2, -y+1/2, z'

loop_
   _atom_site_label
   _atom_site_occupancy
   _atom_site_fract_x
   _atom_site_fract_y
   _atom_site_fract_z
   _atom_site_adp_type
   _atom_site_B_iso_or_equiv
   _atom_site_type_symbol
   Cl1         1.0     0.651070     0.000000     0.120990    Biso  1.000000 Cl
   Cl2         1.0     0.651070     0.000000     0.620990    Biso  1.000000 Cl
   Cu1         1.0     0.000000     0.000000     0.000000    Biso  1.000000 Cu
   Cu2         1.0     0.000000     0.000000     0.500000    Biso  1.000000 Cu

\end{verbatim}
\begin{figure}[h]
    \centering
    \includegraphics[width=\textwidth]{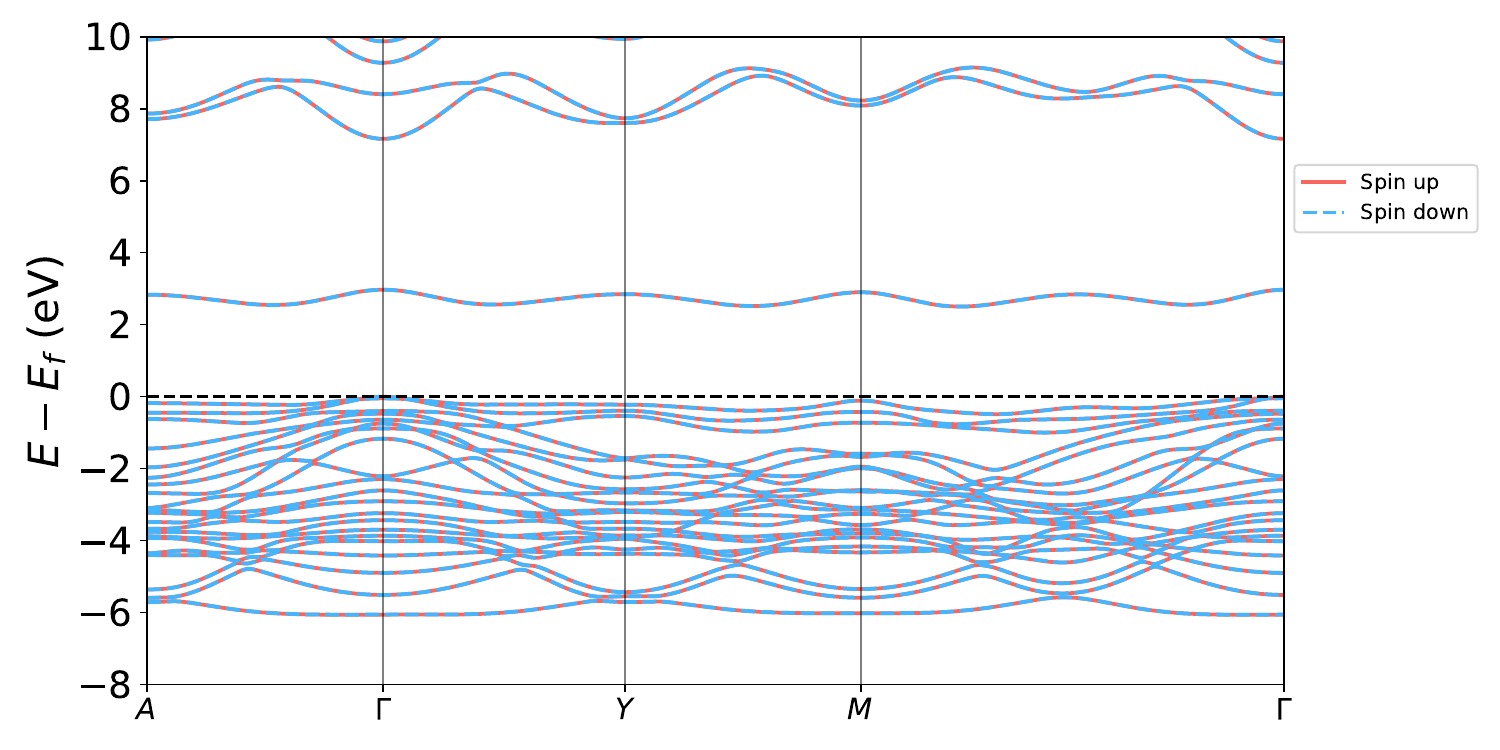}
    \caption{\footnotesize \ce{CuCl2} bulk spin-polarized band structure.}
    \label{fig:CuCl2_band_bulk}
\end{figure}
\clearpage
\newpage

\subsection{\ce{CuCl2} slab}
\begin{verbatim}
_cell_length_a                         3.32892897
_cell_length_b                         3.60314919
_cell_length_c                         40.00000000
_cell_angle_alpha                      90.000000
_cell_angle_beta                       90.000000
_cell_angle_gamma                      117.512825
_cell_volume                           'P 1'
_space_group_name_H-M_alt              'P 1'
_space_group_IT_number                 1

loop_
_space_group_symop_operation_xyz
   'x, y, z'

loop_
   _atom_site_label
   _atom_site_occupancy
   _atom_site_fract_x
   _atom_site_fract_y
   _atom_site_fract_z
   _atom_site_adp_type
   _atom_site_B_ios_or_equiv
   _atom_site_type_symbol
Cl001  1.0  0.365991213041  -0.268017573919  0.534019661878  Biso  1.000000  Cl
Cu002  1.0  -0.000000000000  -0.000000000000  0.5  Biso  1.000000  Cu
Cl003  1.0  -0.365991213041  0.268017573919  0.465980338122  Biso  1.000000  Cl

\end{verbatim}
\begin{figure}[h]
    \centering
    \includegraphics[width=\textwidth]{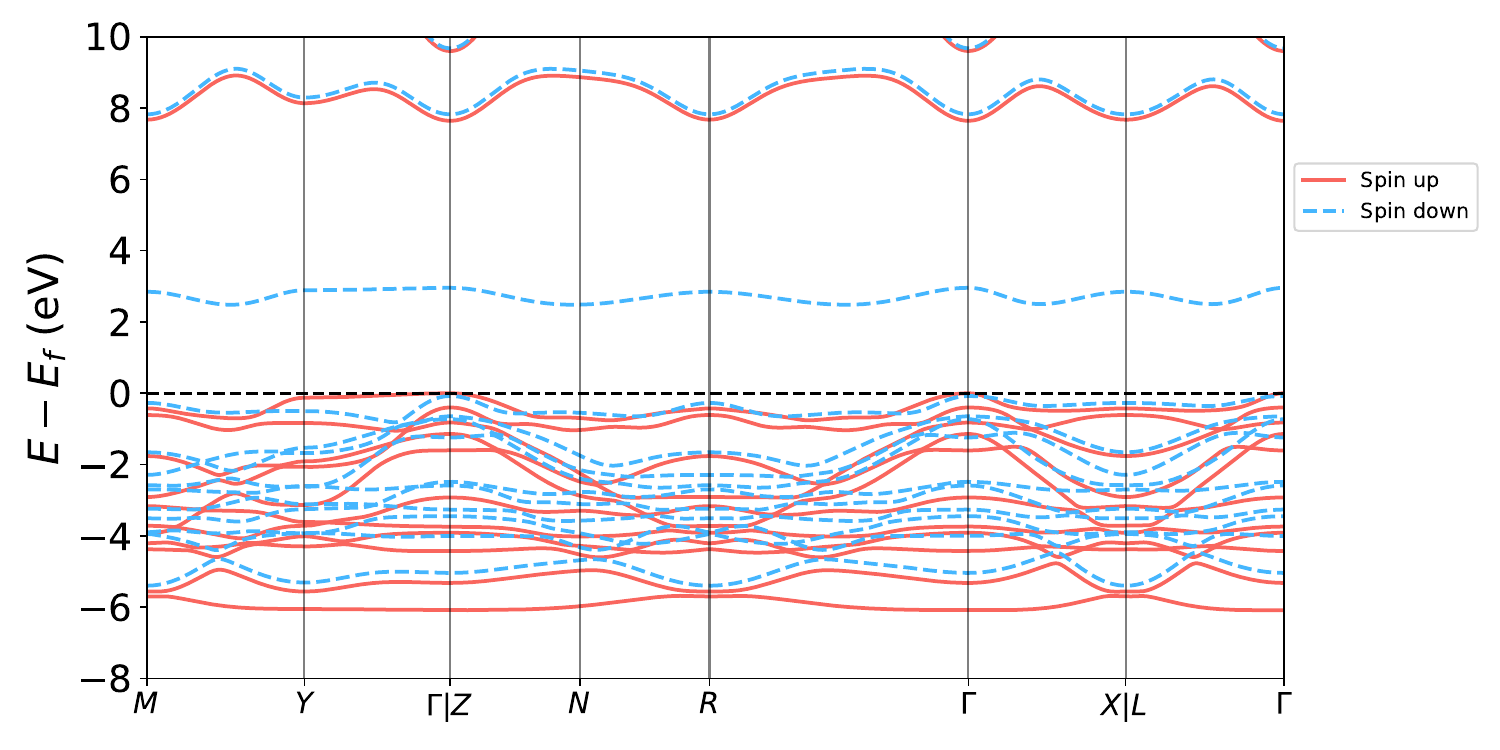}
    \caption{\footnotesize \ce{CuCl2} slab spin-polarized band structure.}
    \label{fig:CuCl2_band_slab}
\end{figure}
\clearpage
\newpage

\subsection{\ce{CuF2} bulk}
\begin{verbatim}
_cell_length_a                         3.296694
_cell_length_b                         4.504817
_cell_length_c                         4.530087
_cell_angle_alpha                      89.999954
_cell_angle_beta                       95.815804
_cell_angle_gamma                      90.000801
_cell_volume                           66.930059
_space_group_name_H-M_alt              'P -1'
_space_group_IT_number                 2

loop_
_space_group_symop_operation_xyz
   'x, y, z'
   '-x, -y, -z'

loop_
   _atom_site_label
   _atom_site_occupancy
   _atom_site_fract_x
   _atom_site_fract_y
   _atom_site_fract_z
   _atom_site_adp_type
   _atom_site_B_iso_or_equiv
   _atom_site_type_symbol
   F1          1.0     0.464780     0.197140     0.206670    Biso  1.000000 F
   F2          1.0     0.035240     0.697140     0.293320    Biso  1.000000 F
   Cu1         1.0     0.000000     0.000000     0.000000    Biso  1.000000 Cu
   Cu2         1.0     0.500000     0.500000     0.500000    Biso  1.000000 Cu

\end{verbatim}
\begin{figure}[h]
    \centering
    \includegraphics[width=\textwidth]{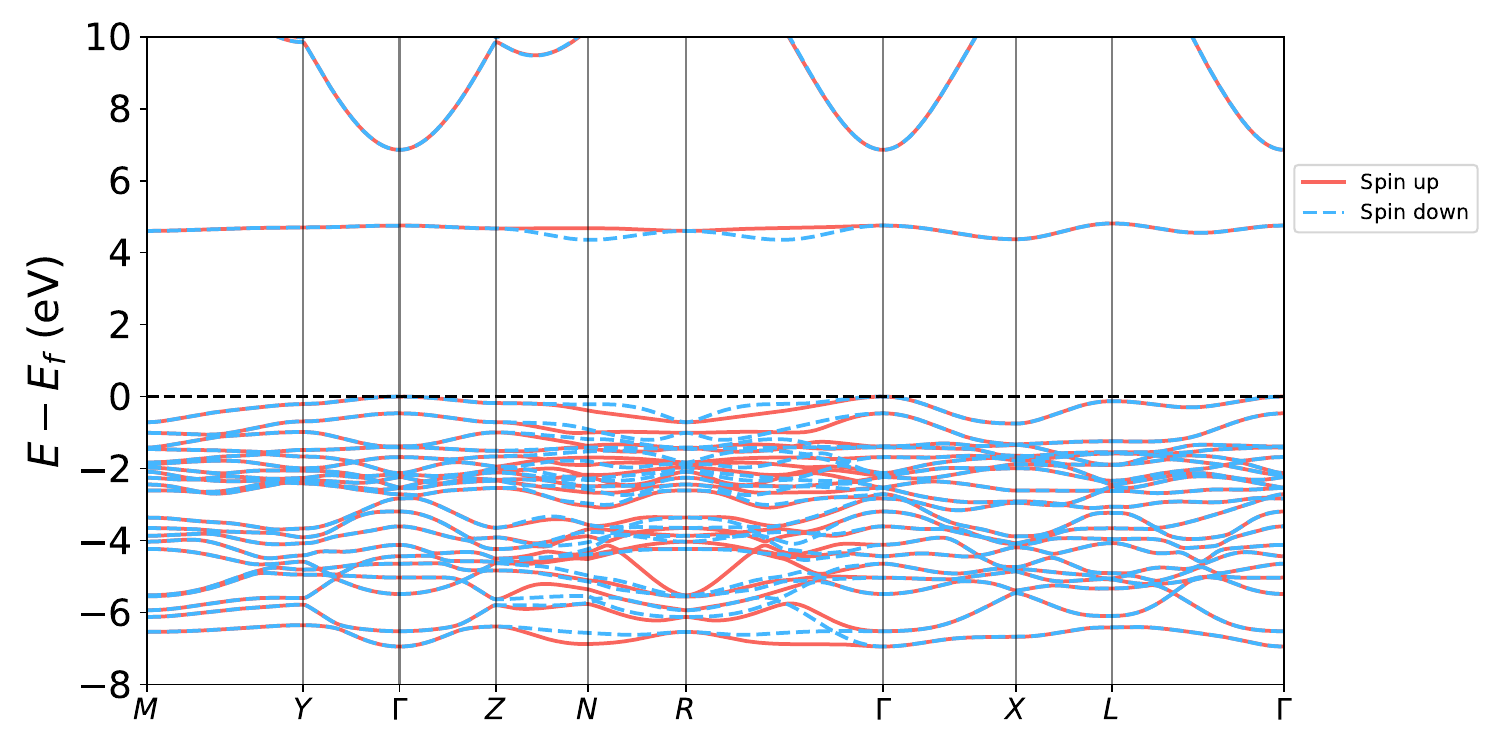}
    \caption{\footnotesize \ce{CuF2} bulk spin-polarized band structure.}
    \label{fig:CuF2_band_bulk}
\end{figure}

\clearpage
\newpage

\subsection{\ce{CuF2} slab}
\begin{verbatim}
_cell_length_a                         5.27000643
_cell_length_b                         6.92244774
_cell_length_c                         40.00000000
_cell_angle_alpha                      90.000000
_cell_angle_beta                       90.000000
_cell_angle_gamma                      90.000242
_cell_volume                           'P 1'
_space_group_name_H-M_alt              'P 1'
_space_group_IT_number                 1

loop_
_space_group_symop_operation_xyz
   'x, y, z'

loop_
   _atom_site_label
   _atom_site_occupancy
   _atom_site_fract_x
   _atom_site_fract_y
   _atom_site_fract_z
   _atom_site_adp_type
   _atom_site_B_ios_or_equiv
   _atom_site_type_symbol
F001  1.0  0.183087622390  -0.451786200847  0.521613314635  Biso  1.000000  F
F002  1.0  -0.316826963385  -0.298134628954  0.521597877803  Biso  1.000000  F
F003  1.0  0.183106410524  0.048104410619  0.521607570597  Biso  1.000000  F
F004  1.0  -0.316967738238  0.201904427221  0.521622384581  Biso  1.000000  F
Cu005  1.0  -0.000014393290  0.250064286146  0.500034049432  Biso  1.000000  Cu
Cu006  1.0  -0.499999277719  0.000018408607  0.500001439404  Biso  1.000000  Cu
Cu007  1.0  0.499974490368  -0.499971484743  0.50000207919  Biso  1.000000  Cu
Cu008  1.0  -0.000004692980  -0.250022807405  0.499967155171  Biso  1.000000  Cu
F009  1.0  0.316948213071  -0.201864260595  0.478381073495  Biso  1.000000  F
F010  1.0  -0.183116875651  -0.048057624106  0.478393550167  Biso  1.000000  F
F011  1.0  0.316826845057  0.298174939158  0.47840411387  Biso  1.000000  F
F012  1.0  -0.183118928168  0.451834543296  0.47839202192  Biso  1.000000  F

\end{verbatim}
\begin{figure}[h]
    \centering
    \includegraphics[width=\textwidth]{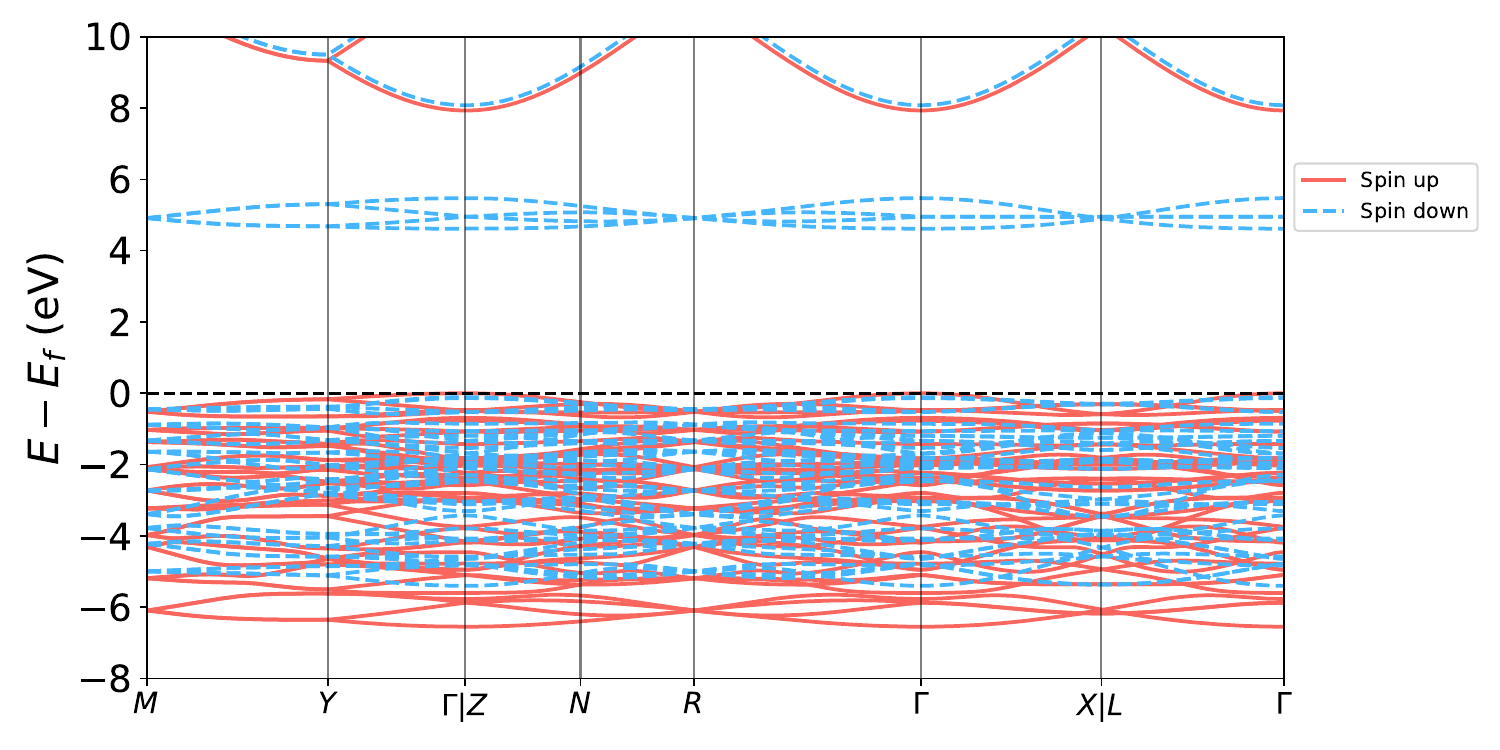}
    \caption{\footnotesize \ce{CuF2} slab spin-polarized band structure.}
    \label{fig:CuF2_band_slab}
\end{figure}

\clearpage
\newpage

\subsection{CuI bulk}
\begin{verbatim}
_cell_length_a 4.51554660
_cell_length_b 4.51554660
_cell_length_c 4.43300549
_cell_angle_alpha 87.015885
_cell_angle_beta 92.984115
_cell_angle_gamma 91.780686
_symmetry_space_group_name_H-M         'P 1'
_symmetry_Int_Tables_number            1

loop_
_symmetry_equiv_pos_as_xyz
   'x, y, z'

loop_
_atom_site_label
_atom_site_type_symbol
_atom_site_fract_x
_atom_site_fract_y
_atom_site_fract_z
I001 I 2.330923436875E-01 -2.330923436875E-01 2.957304159502E-01
I002 I -2.330923436875E-01 2.330923436875E-01 -2.957304159502E-01
Cu003 Cu 2.924197382011E-01 2.924197382011E-01 -7.562927798820E-35
Cu004 Cu -2.924197382011E-01 -2.924197382011E-01 7.562927798820E-35

\end{verbatim}
\begin{figure}[h]
    \centering
    \includegraphics[width=\textwidth]{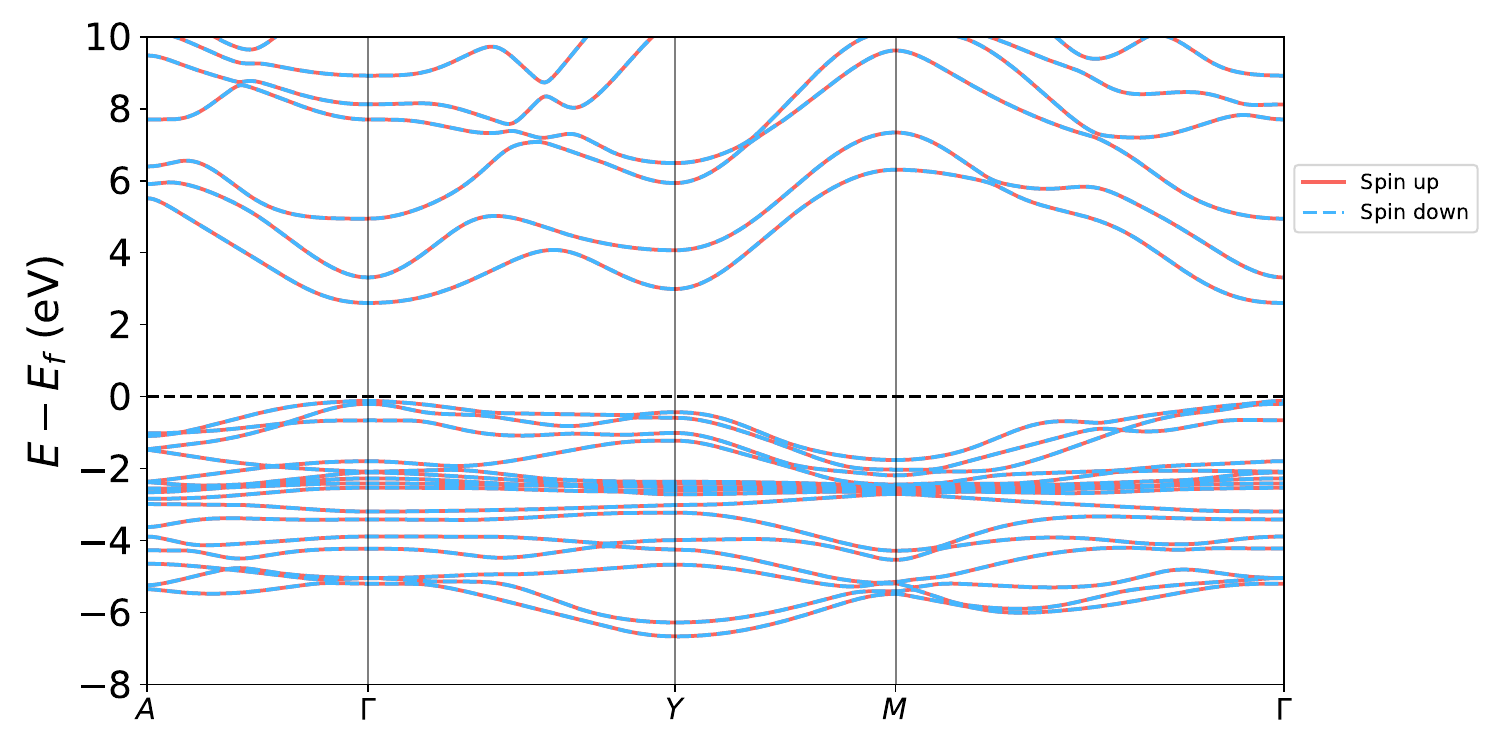}
    \caption{\footnotesize CuI bulk spin-polarized band structure.}
    \label{fig:CuI_band_bulk}
\end{figure}
\clearpage
\newpage

\subsection{CuI slab}
\begin{verbatim}
_cell_length_a                         4.15374454
_cell_length_b                         4.15374454
_cell_length_c                         40.00000000
_cell_angle_alpha                      90.000000
_cell_angle_beta                       90.000000
_cell_angle_gamma                      92.976388
_cell_volume                           'P 1'
_space_group_name_H-M_alt              'P 1'
_space_group_IT_number                 1

loop_
_space_group_symop_operation_xyz
   'x, y, z'

loop_
   _atom_site_label
   _atom_site_occupancy
   _atom_site_fract_x
   _atom_site_fract_y
   _atom_site_fract_z
   _atom_site_adp_type
   _atom_site_B_ios_or_equiv
   _atom_site_type_symbol
I001  1.0  0.228911827406  -0.228911827406  0.539308267261  Biso  1.000000  I
Cu002  1.0  0.276851841850  0.276851841850  0.5  Biso  1.000000  Cu
Cu003  1.0  -0.276851841850  -0.276851841850  0.5  Biso  1.000000  Cu
I004  1.0  -0.228911827406  0.228911827406  0.460691732739  Biso  1.000000  I

\end{verbatim}
\begin{figure}[h]
    \centering
    \includegraphics[width=\textwidth]{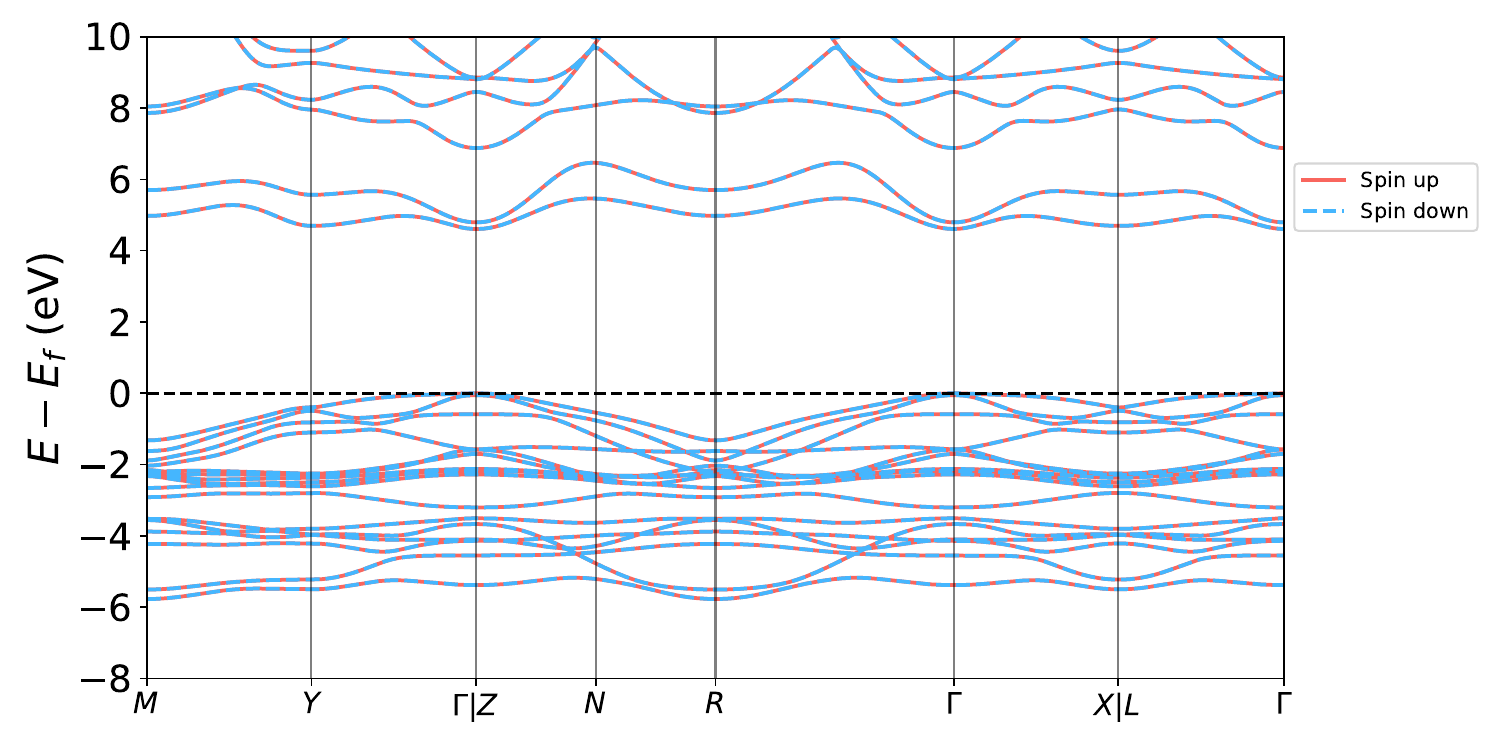}
    \caption{\footnotesize \ce{CuI} slab spin-polarized band structure.}
    \label{fig:CuI_band_slab}
\end{figure}
\clearpage
\newpage

\subsection{\ce{FeCl2} bulk}
\begin{verbatim}
_cell_length_a                         7.09410179
_cell_length_b                         7.09403382
_cell_length_c                         11.33324720
_cell_angle_alpha                      89.953593
_cell_angle_beta                       89.954529
_cell_angle_gamma                      61.267982
_cell_volume                           'P -1'
_space_group_name_H-M_alt              'P -1'
_space_group_IT_number                 2

loop_
_space_group_symop_operation_xyz
   'x, y, z'
   '-x, -y, -z'

loop_
   _atom_site_label
   _atom_site_occupancy
   _atom_site_fract_x
   _atom_site_fract_y
   _atom_site_fract_z
   _atom_site_adp_type
   _atom_site_B_ios_or_equiv
   _atom_site_type_symbol
   Cl001   1.0   0.166088142567  0.416137401754  -0.372869406985  Biso  1.000000  Cl
   Cl002   1.0   0.166023044338  0.416075023045  0.127145783393  Biso  1.000000  Cl
   Cl003   1.0   -0.333982394232  0.416055016861  -0.372856333192  Biso  1.000000  Cl
   Cl004   1.0   -0.333914679214  0.416119310820  0.127128209403  Biso  1.000000  Cl
   Cl005   1.0   0.166033200226  -0.083986631458  -0.372857395710  Biso  1.000000  Cl
   Cl006   1.0   0.166106288757  -0.083924894181  0.127127413826  Biso  1.000000  Cl
   Cl007   1.0   -0.333891579912  -0.083910007968  -0.372872688926  Biso  1.000000  Cl
   Cl008   1.0   -0.333961535923  -0.083967588390  0.127146493241  Biso  1.000000  Cl
   Fe009   1.0   -0.000010009468  -0.249986373827  -0.249963218723  Biso  1.000000  Fe
   Fe010   1.0   -0.000008203139  -0.249986258099  0.250034606126  Biso  1.000000  Fe
   Fe011   1.0   -0.499989856181  -0.250001398204  -0.249994996329  Biso  1.000000  Fe
   Fe012   1.0   -0.499988192858  -0.249995736209  0.250009776596  Biso  1.000000  Fe

\end{verbatim}
\begin{figure}[h]
    \centering
    \includegraphics[width=\textwidth]{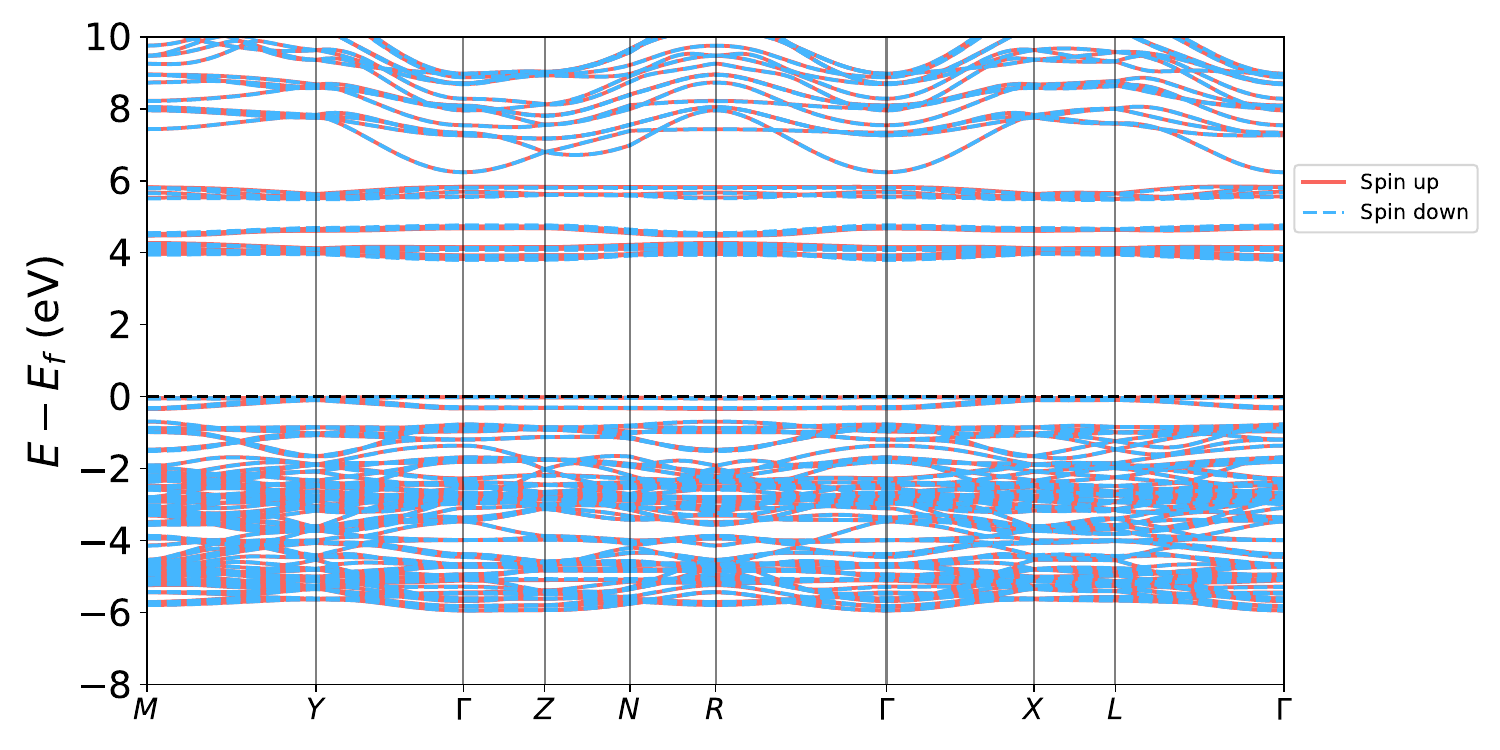}
    \caption{\footnotesize \ce{FeCl2} bulk spin-polarized band structure.}
    \label{fig:FeCl2_band_bulk}
\end{figure}
\clearpage
\newpage

\subsection{\ce{FeCl2} slab}
\begin{verbatim}
_cell_length_a                         7.08630918
_cell_length_b                         7.08659884
_cell_length_c                         40.00000000
_cell_angle_alpha                      90.000000
_cell_angle_beta                       90.000000
_cell_angle_gamma                      118.809784
_cell_volume                           'P 1'
_space_group_name_H-M_alt              'P 1'
_space_group_IT_number                 1

loop_
_space_group_symop_operation_xyz
   'x, y, z'

loop_
   _atom_site_label
   _atom_site_occupancy
   _atom_site_fract_x
   _atom_site_fract_y
   _atom_site_fract_z
   _atom_site_adp_type
   _atom_site_B_ios_or_equiv
   _atom_site_type_symbol
Cl001  1.0  -0.415745194890  0.165705048923  0.53520169043  Biso  1.000000  Cl
Cl002  1.0  -0.415804404466  -0.334230874743  0.535200876556  Biso  1.000000  Cl
Cl003  1.0  0.084242811430  0.165794245391  0.5351958554239999  Biso  1.000000  Cl
Cl004  1.0  0.084299465563  -0.334266732004  0.535196975466  Biso  1.000000  Cl
Fe005  1.0  -0.250002394196  0.000004310768  0.500009185252  Biso  1.000000  Fe
Fe006  1.0  -0.250006513467  0.499993884144  0.500003275907  Biso  1.000000  Fe
Fe007  1.0  0.249999333397  -0.499992638172  0.49999512925  Biso  1.000000  Fe
Fe008  1.0  0.249998978490  -0.000004018600  0.499988668989  Biso  1.000000  Fe
Cl009  1.0  0.415812314154  -0.165770815748  0.464796626303  Biso  1.000000  Cl
Cl010  1.0  0.415727212588  0.334297835459  0.46479666647  Biso  1.000000  Cl
Cl011  1.0  -0.084317643105  -0.165729898365  0.464801567401  Biso  1.000000  Cl
Cl012  1.0  -0.084234971855  0.334204662473  0.464801580469  Biso  1.000000  Cl

\end{verbatim}
\begin{figure}[h]
    \centering
    \includegraphics[width=\textwidth]{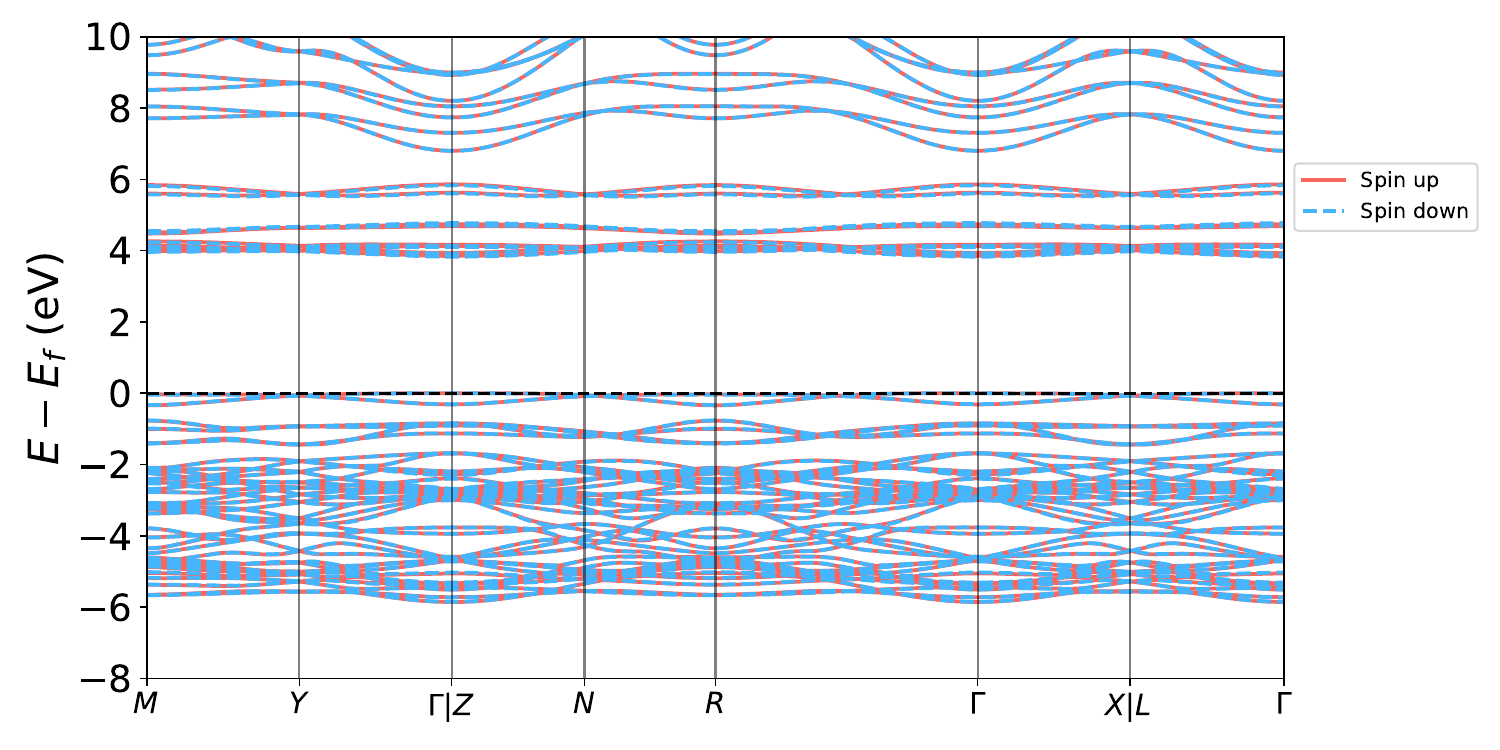}
    \caption{\footnotesize \ce{FeCl2} slab spin-polarized band structure.}
    \label{fig:FeCl2_band_slab}
\end{figure}
\clearpage
\newpage

\subsection{\ce{FeI2} bulk}
\begin{verbatim}
_cell_length_a                         7.920233
_cell_length_b                         7.920624
_cell_length_c                         12.774837
_cell_angle_alpha                      89.966568
_cell_angle_beta                       89.965584
_cell_angle_gamma                      61.051086
_cell_volume                           701.271610
_space_group_name_H-M_alt              'P -1'
_space_group_IT_number                 2

loop_
_space_group_symop_operation_xyz
   'x, y, z'
   '-x, -y, -z'

loop_
   _atom_site_label
   _atom_site_occupancy
   _atom_site_fract_x
   _atom_site_fract_y
   _atom_site_fract_z
   _atom_site_adp_type
   _atom_site_B_iso_or_equiv
   _atom_site_type_symbol
   I1          1.0     0.333240     0.583230     0.122790    Biso  1.000000 I
   I2          1.0     0.333240     0.583270     0.622800    Biso  1.000000 I
   I3          1.0     0.333260     0.083250     0.122810    Biso  1.000000 I
   I4          1.0     0.333200     0.083220     0.622810    Biso  1.000000 I
   I5          1.0     0.833290     0.583140     0.122800    Biso  1.000000 I
   I6          1.0     0.833250     0.583100     0.622820    Biso  1.000000 I
   I7          1.0     0.833200     0.083180     0.122790    Biso  1.000000 I
   I8          1.0     0.833210     0.083230     0.622810    Biso  1.000000 I
   Fe1         1.0     0.000030     0.750020     0.249950    Biso  1.000000 Fe
   Fe2         1.0     0.000010     0.750040     0.749970    Biso  1.000000 Fe
   Fe3         1.0     0.500050     0.749990     0.250000    Biso  1.000000 Fe
   Fe4         1.0     0.500030     0.749990     0.750000    Biso  1.000000 Fe

\end{verbatim}
\begin{figure}[h]
    \centering
    \includegraphics[width=\textwidth]{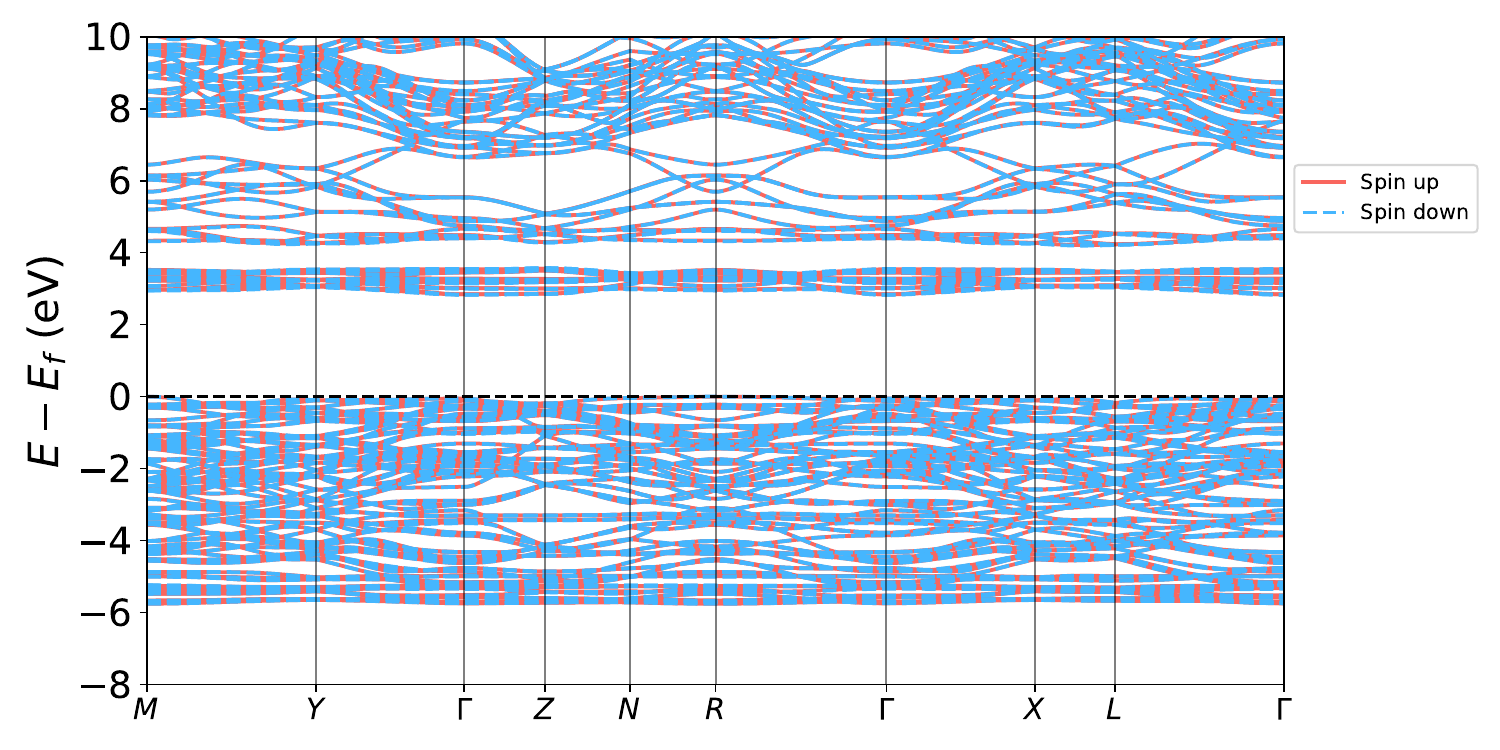}
    \caption{\footnotesize \ce{FeI2} bulk spin-polarized band structure.}
    \label{fig:FeI2_band_bulk}
\end{figure}
\clearpage
\newpage

\subsection{\ce{FeI2} slab}
\begin{verbatim}
_cell_length_a                         7.92262246
_cell_length_b                         7.92243795
_cell_length_c                         40.00000000
_cell_angle_alpha                      90.000000
_cell_angle_beta                       90.000000
_cell_angle_gamma                      119.031865
_cell_volume                           'P 1'
_space_group_name_H-M_alt              'P 1'
_space_group_IT_number                 1

loop_
_space_group_symop_operation_xyz
   'x, y, z'

loop_
   _atom_site_label
   _atom_site_occupancy
   _atom_site_fract_x
   _atom_site_fract_y
   _atom_site_fract_z
   _atom_site_adp_type
   _atom_site_B_ios_or_equiv
   _atom_site_type_symbol
I001  1.0  0.416835620565  0.333181128250  0.540823071669  Biso  1.000000  I
I002  1.0  -0.083193821108  0.333204842562  0.540823024905  Biso  1.000000  I
I003  1.0  0.416846461373  -0.166767429704  0.540822416845  Biso  1.000000  I
I004  1.0  -0.083163639428  -0.166850653892  0.540823322331  Biso  1.000000  I
Fe005  1.0  0.249982675123  0.000018821641  0.500008527103  Biso  1.000000  Fe
Fe006  1.0  -0.249968271750  0.000000746315  0.499991126891  Biso  1.000000  Fe
Fe007  1.0  0.250008447922  -0.499963462222  0.499997007648  Biso  1.000000  Fe
Fe008  1.0  -0.250001997045  0.499983148607  0.499997853461  Biso  1.000000  Fe
I009  1.0  -0.416791377268  -0.333205012413  0.459169936185  Biso  1.000000  I
I010  1.0  0.083165894489  -0.333126249752  0.459177408129  Biso  1.000000  I
I011  1.0  -0.416880430162  0.166828504924  0.459179123844  Biso  1.000000  I
I012  1.0  0.083211406456  0.166820649203  0.459171212446  Biso  1.000000  I

\end{verbatim}
\begin{figure}[h]
    \centering
    \includegraphics[width=\textwidth]{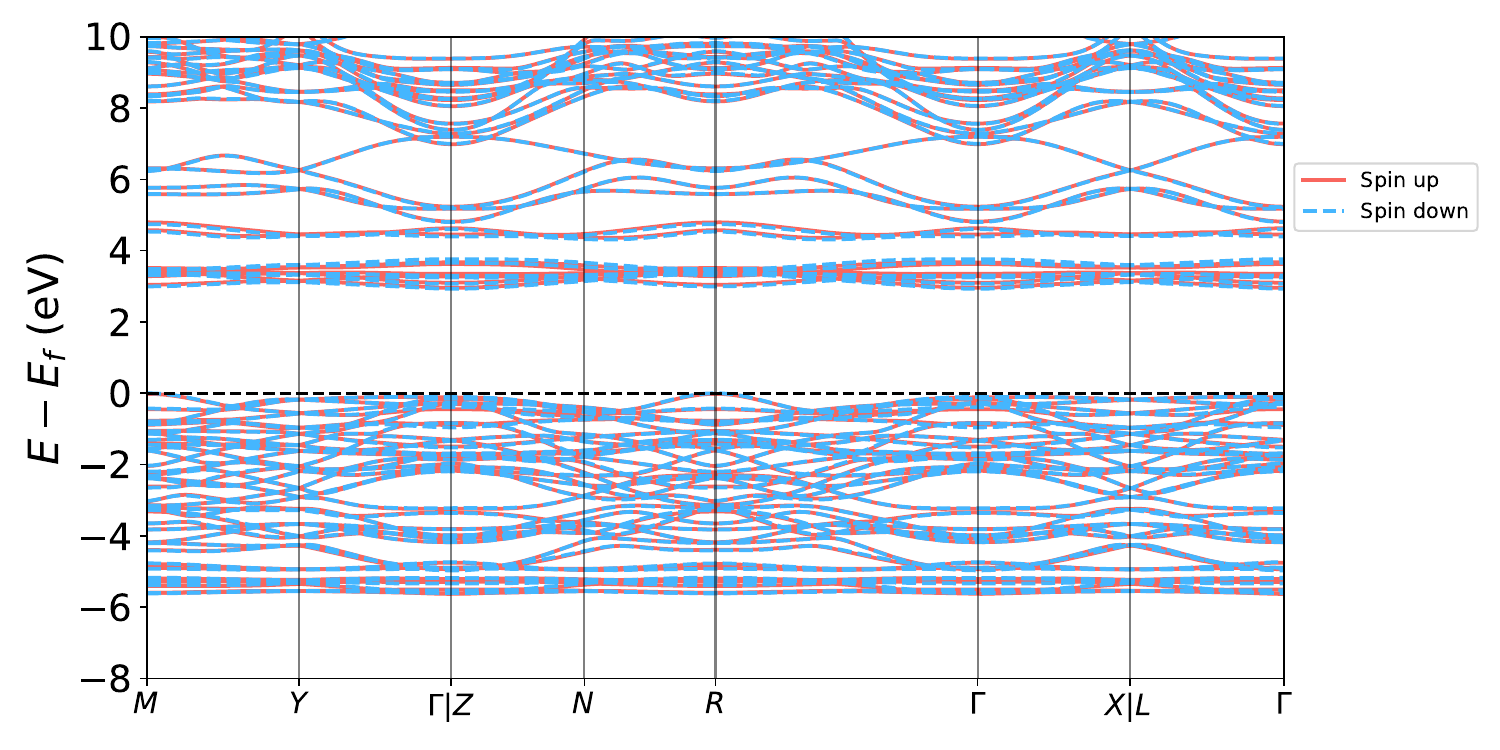}
    \caption{\footnotesize \ce{FeI2} slab spin-polarized band structure.}
    \label{fig:FeI2_band_slab}
\end{figure}
\clearpage
\newpage

\subsection{\ce{GeBr2} bulk}
\begin{verbatim}
_cell_length_a 11.35170337
_cell_length_b 8.81149285
_cell_length_c 6.87946680
_cell_angle_alpha 90.000000
_cell_angle_beta 103.028018
_cell_angle_gamma 90.000000
_symmetry_space_group_name_H-M         'P 1'
_symmetry_Int_Tables_number            1

loop_
_symmetry_equiv_pos_as_xyz
   'x, y, z'

loop_
_atom_site_label
_atom_site_type_symbol
_atom_site_fract_x
_atom_site_fract_y
_atom_site_fract_z
Ge001 Ge 2.248837022803E-02 -3.730556400933E-01 -2.263252505495E-01
Ge002 Ge -2.248837022803E-02 1.269443599067E-01 -2.736747494505E-01
Ge003 Ge -2.248837022803E-02 3.730556400933E-01 2.263252505495E-01
Ge004 Ge 2.248837022803E-02 -1.269443599067E-01 2.736747494505E-01
Ge005 Ge 4.490393587368E-01 -1.454038220557E-01 -2.845016192401E-01
Ge006 Ge -4.490393587368E-01 3.545961779443E-01 -2.154983807599E-01
Ge007 Ge -4.490393587368E-01 1.454038220557E-01 2.845016192401E-01
Ge008 Ge 4.490393587368E-01 -3.545961779443E-01 2.154983807599E-01
Br009 Br 3.321320979258E-01 -3.834582510541E-01 -2.508041887243E-01
Br010 Br -3.321320979258E-01 1.165417489459E-01 -2.491958112757E-01
Br011 Br -3.321320979258E-01 3.834582510541E-01 2.508041887243E-01
Br012 Br 3.321320979258E-01 -1.165417489459E-01 2.491958112757E-01
Br013 Br -1.128158070286E-01 -1.686555741706E-01 -1.042028736089E-01
Br014 Br 1.128158070286E-01 3.313444258294E-01 -3.957971263911E-01
Br015 Br 1.128158070286E-01 1.686555741706E-01 1.042028736089E-01
Br016 Br -1.128158070286E-01 -3.313444258294E-01 3.957971263911E-01
Br017 Br 1.415983128079E-01 -6.340208525663E-02 -3.175927067343E-01
Br018 Br -1.415983128079E-01 4.365979147434E-01 -1.824072932657E-01
Br019 Br -1.415983128079E-01 6.340208525663E-02 3.175927067343E-01
Br020 Br 1.415983128079E-01 -4.365979147434E-01 1.824072932657E-01
Br021 Br -3.985150423586E-01 -2.933744566123E-01 -4.394712888811E-01
Br022 Br 3.985150423586E-01 2.066255433877E-01 -6.052871111890E-02
Br023 Br 3.985150423586E-01 2.933744566123E-01 4.394712888811E-01
Br024 Br -3.985150423586E-01 -2.066255433877E-01 6.052871111890E-02

\end{verbatim}
\begin{figure}[h]
    \centering
    \includegraphics[width=\textwidth]{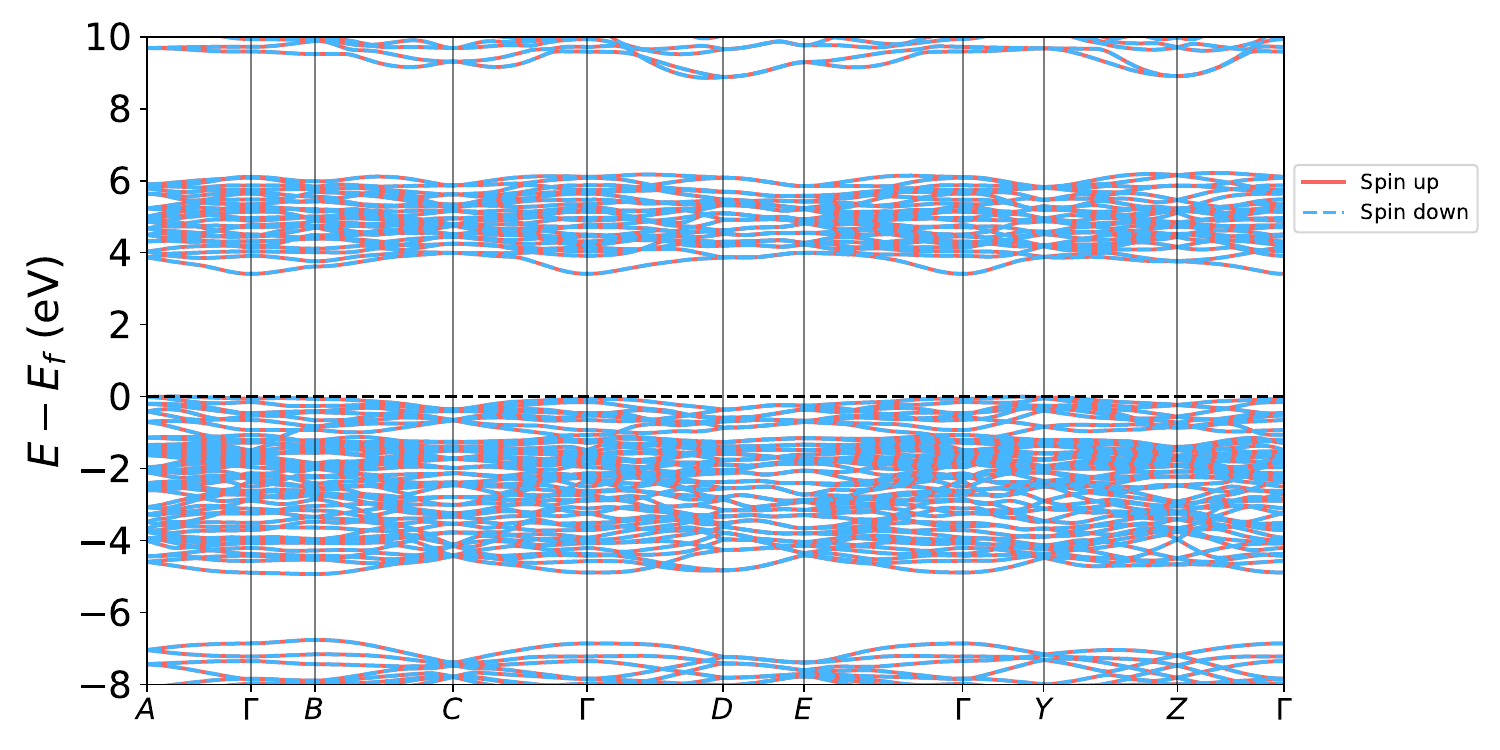}
    \caption{\footnotesize \ce{GeBr2} bulk spin-polarized band structure.}
    \label{fig:GeBr2_band_bulk}
\end{figure}
\clearpage
\newpage

\subsection{\ce{GeBr2} slab}
\begin{verbatim}
_cell_length_a                         7.55342853
_cell_length_b                         7.84825930
_cell_length_c                         40.00000000
_cell_angle_alpha                      90.000000
_cell_angle_beta                       90.000000
_cell_angle_gamma                      90.000000
_cell_volume                           'P 1'
_space_group_name_H-M_alt              'P 1'
_space_group_IT_number                 1

loop_
_space_group_symop_operation_xyz
   'x, y, z'

loop_
   _atom_site_label
   _atom_site_occupancy
   _atom_site_fract_x
   _atom_site_fract_y
   _atom_site_fract_z
   _atom_site_adp_type
   _atom_site_B_ios_or_equiv
   _atom_site_type_symbol
Br001  1.0  0.267766061826  -0.399992119142  0.54304195796  Biso  1.000000  Br
Br002  1.0  -0.243941651369  0.099763682380  0.542945542312  Biso  1.000000  Br
Br003  1.0  0.248687835305  0.130876748295  0.540636066885  Biso  1.000000  Br
Ge004  1.0  0.453407648005  0.380746796008  0.516954323101  Biso  1.000000  Ge
Ge005  1.0  -0.429245190890  -0.119727875662  0.516831056638  Biso  1.000000  Ge
Br006  1.0  0.275975467880  0.369115464377  0.459436187387  Biso  1.000000  Br
Br007  1.0  -0.224024532120  -0.369115464377  0.540563812613  Biso  1.000000  Br
Ge008  1.0  0.070754809110  0.119727875662  0.483168943362  Biso  1.000000  Ge
Ge009  1.0  -0.046592351994  -0.380746796008  0.483045676899  Biso  1.000000  Ge
Br010  1.0  -0.251312164695  -0.130876748295  0.459363933115  Biso  1.000000  Br
Br011  1.0  0.256058348631  -0.099763682380  0.457054457688  Biso  1.000000  Br
Br012  1.0  -0.232233938174  0.399992119142  0.45695804204  Biso  1.000000  Br

\end{verbatim}
\begin{figure}[h]
    \centering
    \includegraphics[width=\textwidth]{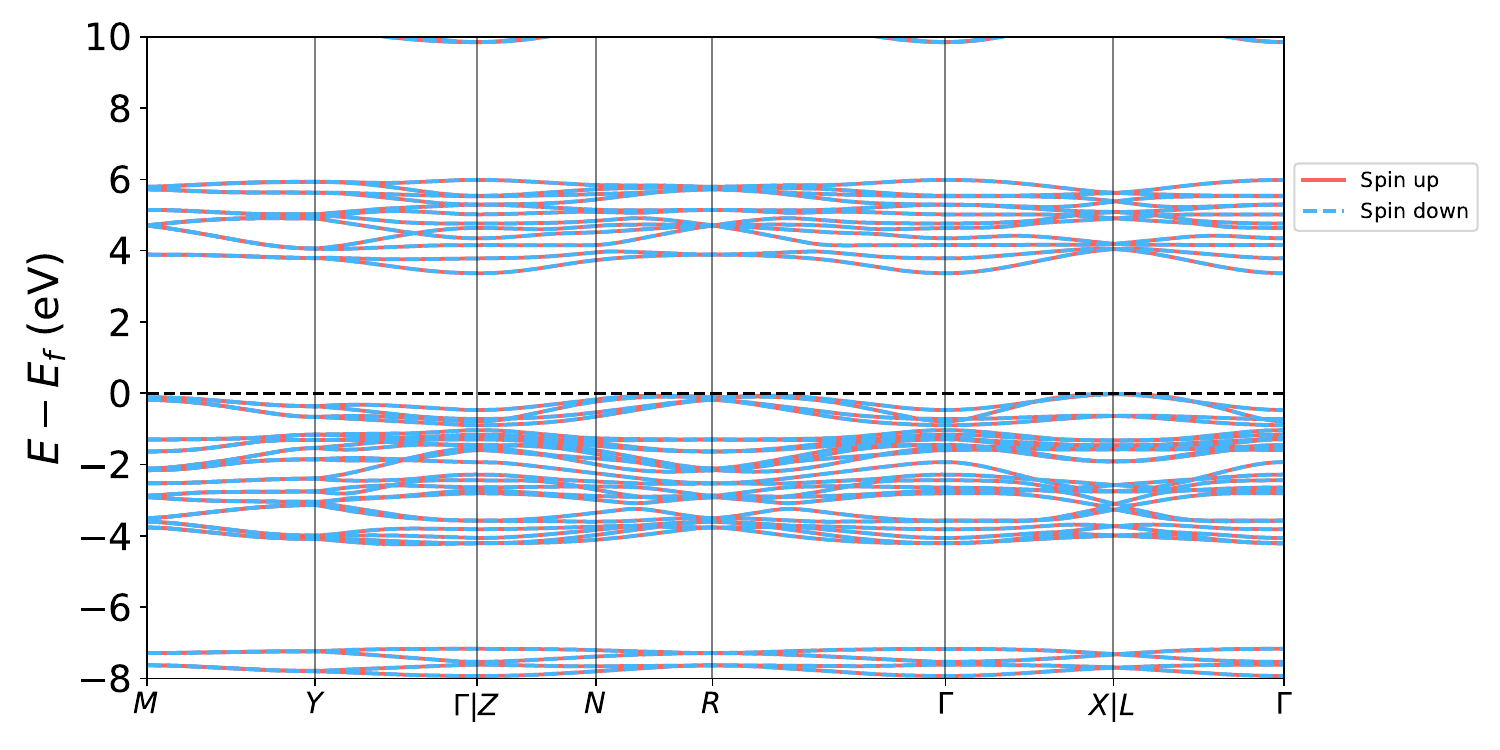}
    \caption{\footnotesize \ce{GeBr2} slab spin-polarized band structure.}
    \label{fig:GeBr2_band_slab}
\end{figure}
\clearpage
\newpage

\subsection{\ce{GeI2} bulk}
\begin{verbatim}
_cell_length_a 4.14016947
_cell_length_b 4.14016947
_cell_length_c 6.60428214
_cell_angle_alpha 90.000000
_cell_angle_beta 90.000000
_cell_angle_gamma 120.000000
_symmetry_space_group_name_H-M         'P 1'
_symmetry_Int_Tables_number            1

loop_
_symmetry_equiv_pos_as_xyz
   'x, y, z'

loop_
_atom_site_label
_atom_site_type_symbol
_atom_site_fract_x
_atom_site_fract_y
_atom_site_fract_z
I001 I 3.333333333333E-01 -3.333333333333E-01 -2.673608194881E-01
I002 I -3.333333333333E-01 3.333333333333E-01 2.673608194881E-01
Ge003 Ge 0.000000000000E+00 0.000000000000E+00 6.570502606307E-35

\end{verbatim}
\begin{figure}[h]
    \centering
    \includegraphics[width=\textwidth]{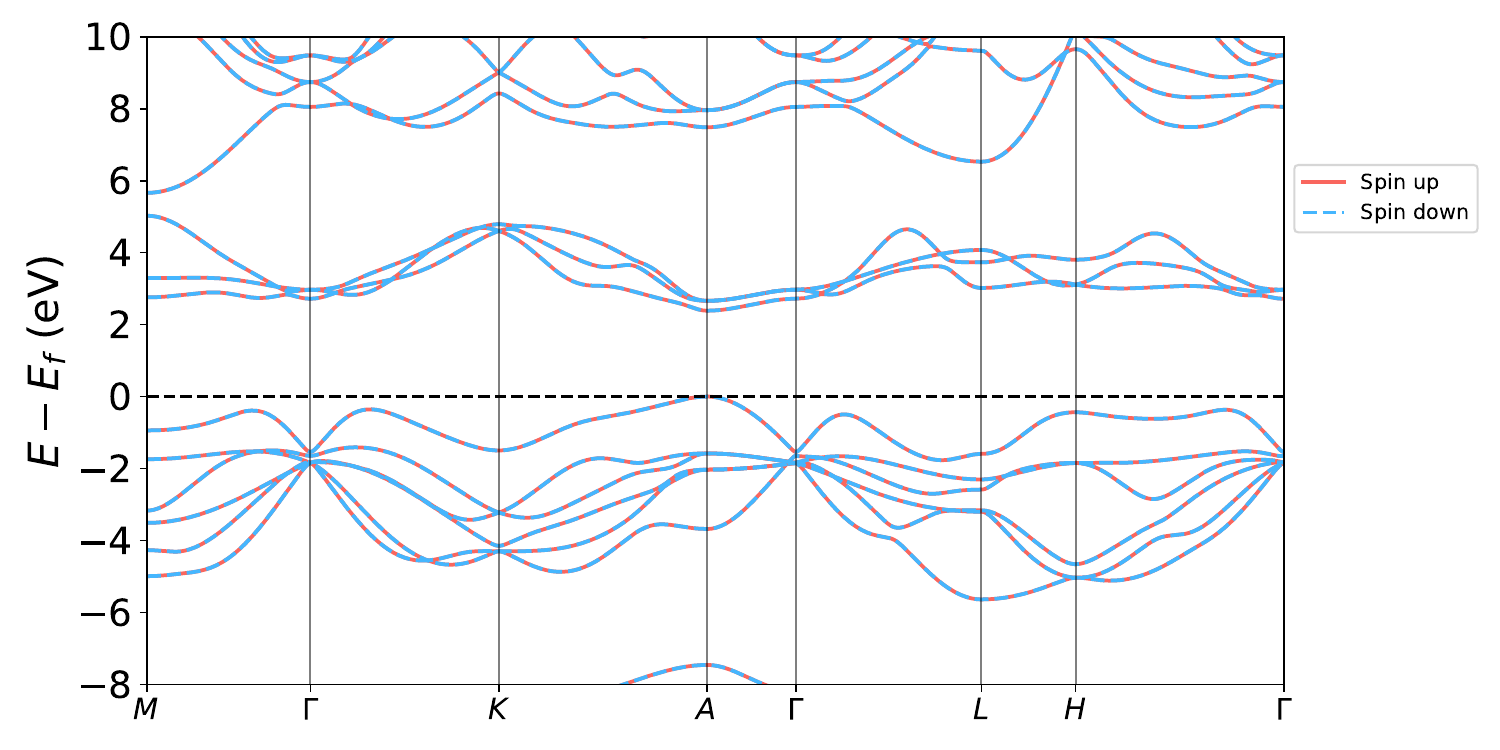}
    \caption{\footnotesize \ce{GeI2} bulk spin-polarized band structure.}
    \label{fig:GeI2_band_bulk}
\end{figure}
\clearpage
\newpage

\subsection{\ce{GeI2} slab}
\begin{verbatim}
_cell_length_a                         4.14639388
_cell_length_b                         4.14639388
_cell_length_c                         40.00000000
_cell_angle_alpha                      90.000000
_cell_angle_beta                       90.000000
_cell_angle_gamma                      120.000000
_cell_volume                           'P 1'
_space_group_name_H-M_alt              'P 1'
_space_group_IT_number                 1

loop_
_space_group_symop_operation_xyz
   'x, y, z'

loop_
   _atom_site_label
   _atom_site_occupancy
   _atom_site_fract_x
   _atom_site_fract_y
   _atom_site_fract_z
   _atom_site_adp_type
   _atom_site_B_ios_or_equiv
   _atom_site_type_symbol
I001  1.0  -0.333333333333  0.333333333333  0.544360452583  Biso  1.000000  I
Ge002  1.0  0.000000000000  0.000000000000  0.5  Biso  1.000000  Ge
I003  1.0  0.333333333333  -0.333333333333  0.455639547417  Biso  1.000000  I

\end{verbatim}

\begin{figure}[h]
    \centering
    \includegraphics[width=\textwidth]{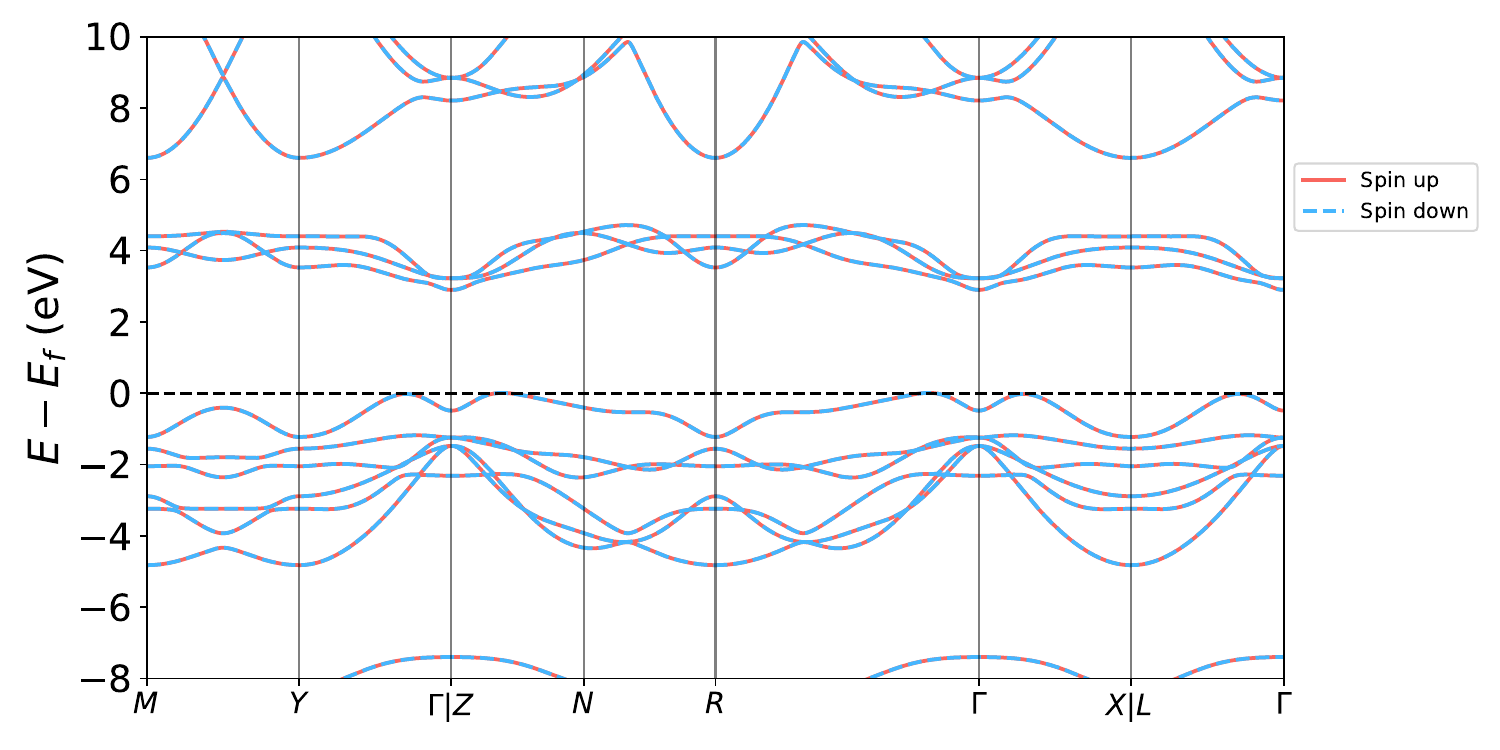}
    \caption{\footnotesize \ce{GeI2} slab spin-polarized band structure.}
    \label{fig:GeI2_band_slab}
\end{figure}
\clearpage
\newpage

\subsection{\ce{HgBr2} bulk}
\begin{verbatim}
_cell_length_a 3.98247525
_cell_length_b 3.98247525
_cell_length_c 5.95291313
_cell_angle_alpha 90.000000
_cell_angle_beta 90.000000
_cell_angle_gamma 120.000000
_symmetry_space_group_name_H-M         'P 1'
_symmetry_Int_Tables_number            1

loop_
_symmetry_equiv_pos_as_xyz
   'x, y, z'

loop_
_atom_site_label
_atom_site_type_symbol
_atom_site_fract_x
_atom_site_fract_y
_atom_site_fract_z
Br001 Br 3.333333333333E-01 -3.333333333333E-01 2.645495966198E-01
Br002 Br -3.333333333333E-01 3.333333333333E-01 -2.645495966198E-01
Hg003 Hg 0.000000000000E+00 0.000000000000E+00 0.000000000000E+00

\end{verbatim}

\begin{figure}[h]
    \centering
    \includegraphics[width=\textwidth]{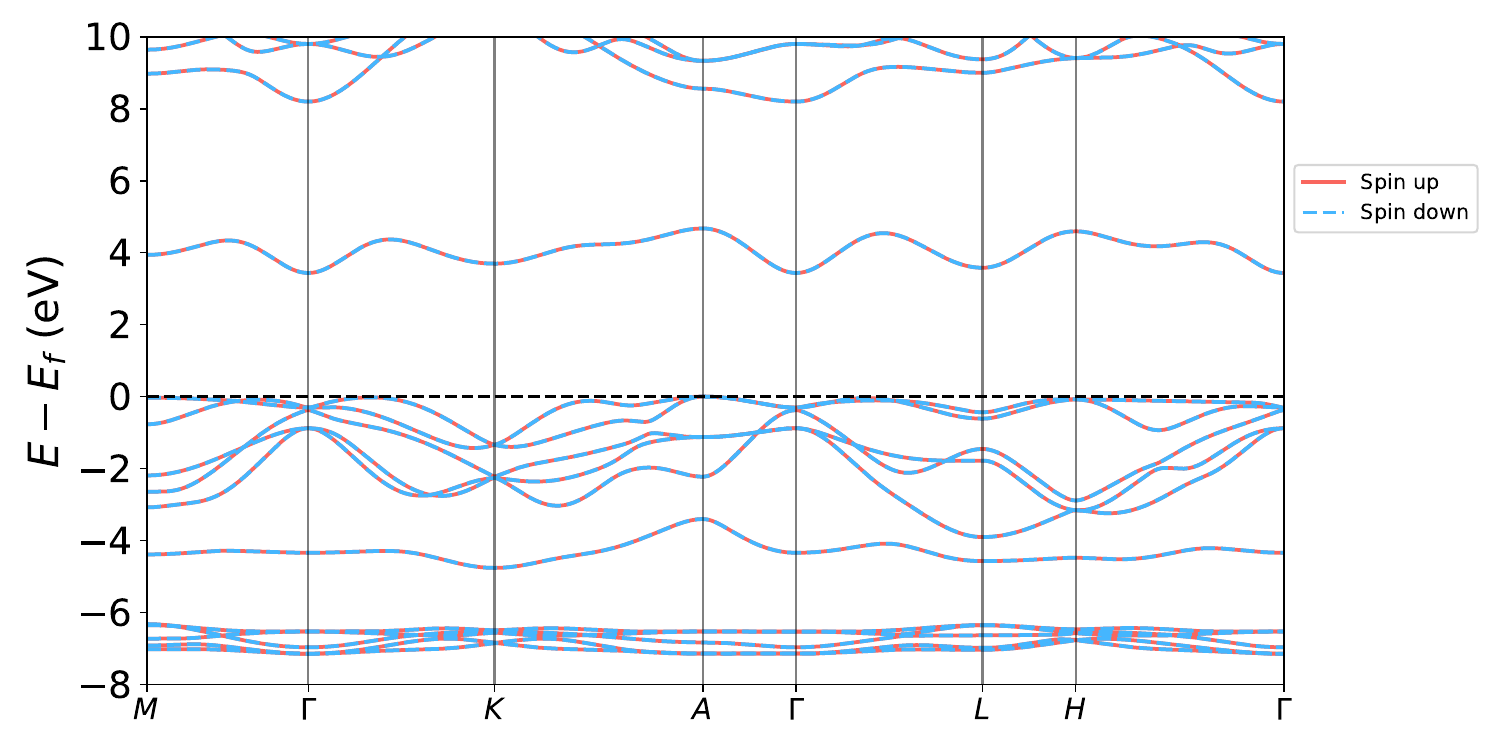}
    \caption{\footnotesize \ce{HgBr2} bulk spin-polarized band structure.}
    \label{fig:HgBr2_band_bulk}
\end{figure}
\clearpage
\newpage

\subsection{\ce{HgBr2} slab}
\begin{verbatim}
_cell_length_a                         3.97619061
_cell_length_b                         3.97619061
_cell_length_c                         40.00000000
_cell_angle_alpha                      90.000000
_cell_angle_beta                       90.000000
_cell_angle_gamma                      120.000000
_cell_volume                           'P 1'
_space_group_name_H-M_alt              'P 1'
_space_group_IT_number                 1

loop_
_space_group_symop_operation_xyz
   'x, y, z'

loop_
   _atom_site_label
   _atom_site_occupancy
   _atom_site_fract_x
   _atom_site_fract_y
   _atom_site_fract_z
   _atom_site_adp_type
   _atom_site_B_ios_or_equiv
   _atom_site_type_symbol
Br001  1.0  0.333333333333  -0.333333333333  0.539750073825  Biso  1.000000  Br
Hg002  1.0  0.000000000000  0.000000000000  0.5  Biso  1.000000  Hg
Br003  1.0  -0.333333333333  0.333333333333  0.460249926175  Biso  1.000000  Br

\end{verbatim}
\begin{figure}[h]
    \centering
    \includegraphics[width=\textwidth]{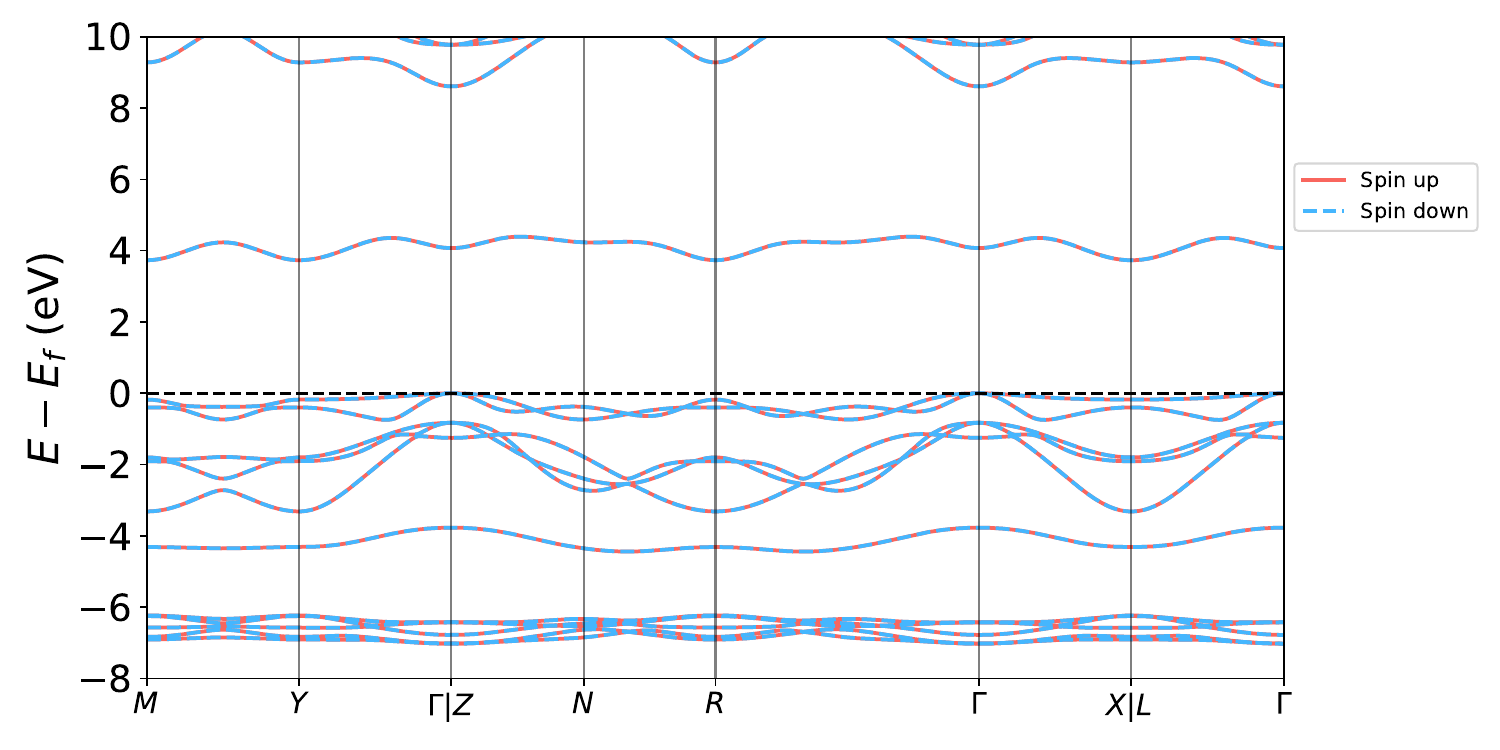}
    \caption{\footnotesize \ce{HgBr2} slab spin-polarized band structure.}
    \label{fig:HgBr2_band_slab}
\end{figure}
\clearpage
\newpage

\subsection{\ce{HgI2} bulk}
\begin{verbatim}
_cell_length_a 4.29175440
_cell_length_b 4.29175440
_cell_length_c 6.01602875
_cell_angle_alpha 90.000000
_cell_angle_beta 90.000000
_cell_angle_gamma 90.000000
_symmetry_space_group_name_H-M         'P 1'
_symmetry_Int_Tables_number            1

loop_
_symmetry_equiv_pos_as_xyz
   'x, y, z'

loop_
_atom_site_label
_atom_site_type_symbol
_atom_site_fract_x
_atom_site_fract_y
_atom_site_fract_z
I001 I 0.000000000000E+00 -5.000000000000E-01 -2.977997304831E-01
I002 I -5.000000000000E-01 0.000000000000E+00 2.977997304831E-01
Hg003 Hg -5.000000000000E-01 -5.000000000000E-01 0.000000000000E+00

\end{verbatim}
\begin{figure}[h]
    \centering
    \includegraphics[width=\textwidth]{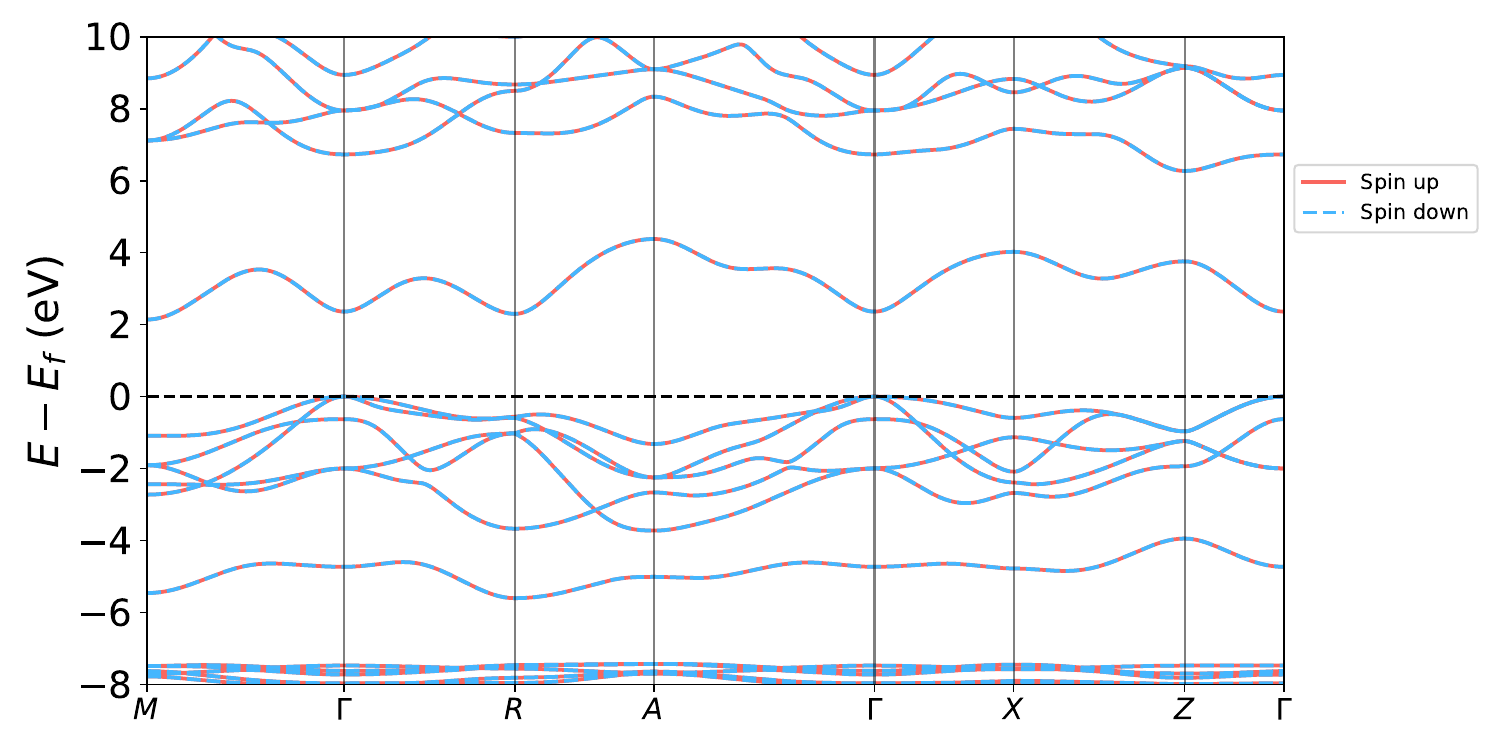}
    \caption{\footnotesize \ce{HgI2} bulk spin-polarized band structure.}
    \label{fig:HgI2_band_bulk}
\end{figure}
\clearpage
\newpage

\subsection{\ce{HgI2} slab}
\begin{verbatim}
_cell_length_a                         4.24416749
_cell_length_b                         4.24416749
_cell_length_c                         40.00000000
_cell_angle_alpha                      90.000000
_cell_angle_beta                       90.000000
_cell_angle_gamma                      90.000000
_cell_volume                           'P 1'
_space_group_name_H-M_alt              'P 1'
_space_group_IT_number                 1

loop_
_space_group_symop_operation_xyz
   'x, y, z'

loop_
   _atom_site_label
   _atom_site_occupancy
   _atom_site_fract_x
   _atom_site_fract_y
   _atom_site_fract_z
   _atom_site_adp_type
   _atom_site_B_ios_or_equiv
   _atom_site_type_symbol
I001  1.0  -0.500000000000  0.000000000000  0.545610255268  Biso  1.000000  I
Hg002  1.0  -0.500000000000  -0.500000000000  0.5  Biso  1.000000  Hg
I003  1.0  0.000000000000  -0.500000000000  0.454389744732  Biso  1.000000  I

\end{verbatim}
\begin{figure}[h]
    \centering
    \includegraphics[width=\textwidth]{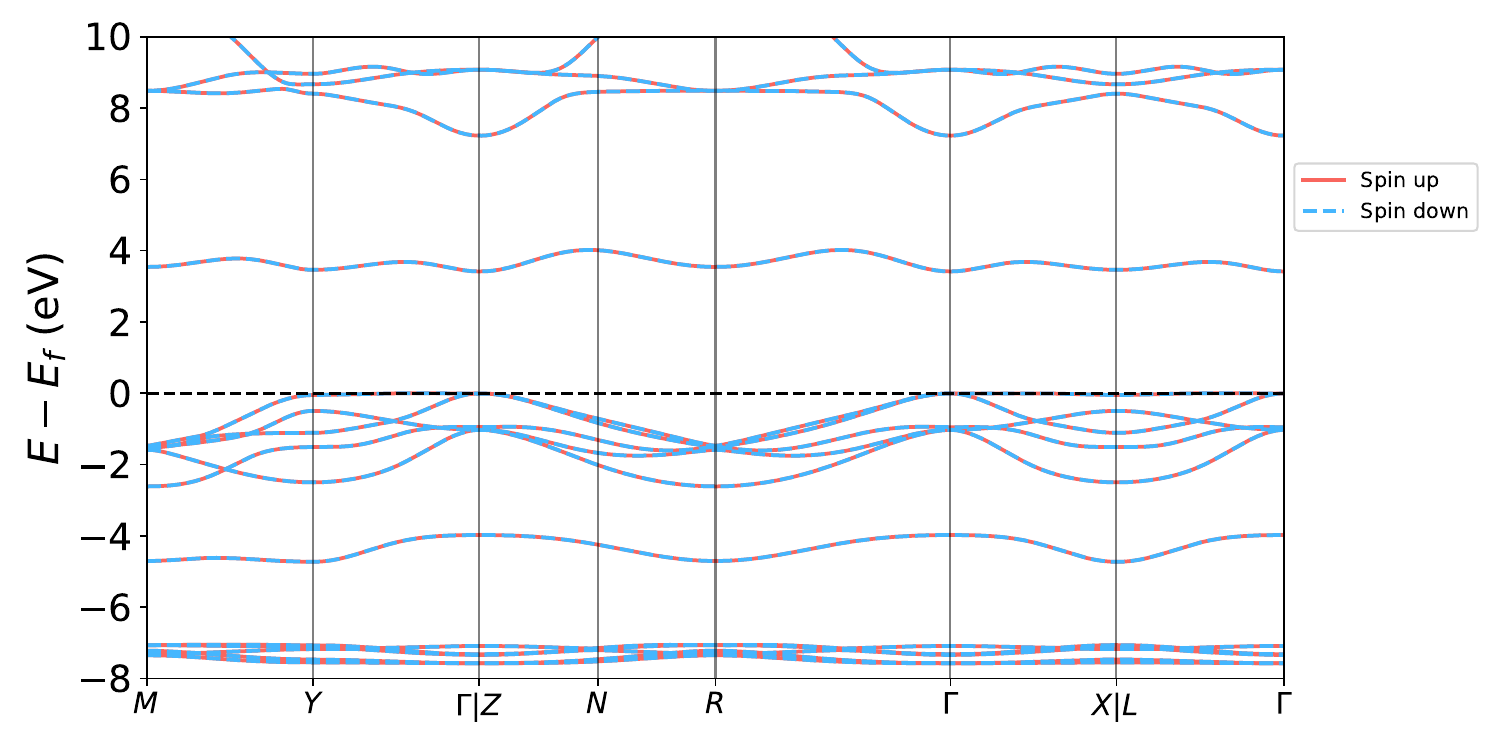}
    \caption{\footnotesize \ce{HgI2} slab spin-polarized band structure.}
    \label{fig:HgI2_band_slab}
\end{figure}
\clearpage
\newpage

\subsection{\ce{LaBr2} bulk}
\begin{verbatim}
_cell_length_a 3.65907877
_cell_length_b 6.48823721
_cell_length_c 6.48826763
_cell_angle_alpha 32.756494
_cell_angle_beta 73.621296
_cell_angle_gamma 73.622210
_symmetry_space_group_name_H-M         'P 1'
_symmetry_Int_Tables_number            1

loop_
_symmetry_equiv_pos_as_xyz
   'x, y, z'

loop_
_atom_site_label
_atom_site_type_symbol
_atom_site_fract_x
_atom_site_fract_y
_atom_site_fract_z
La001 La -2.857715687403E-22 -2.857715687403E-22 2.857715687403E-22
Br002 Br 3.574639852923E-01 -3.574639852923E-01 -3.574639852923E-01
Br003 Br -3.574639852923E-01 3.574639852923E-01 3.574639852923E-01

\end{verbatim}
\begin{figure}[h]
    \centering
    \includegraphics[width=\textwidth]{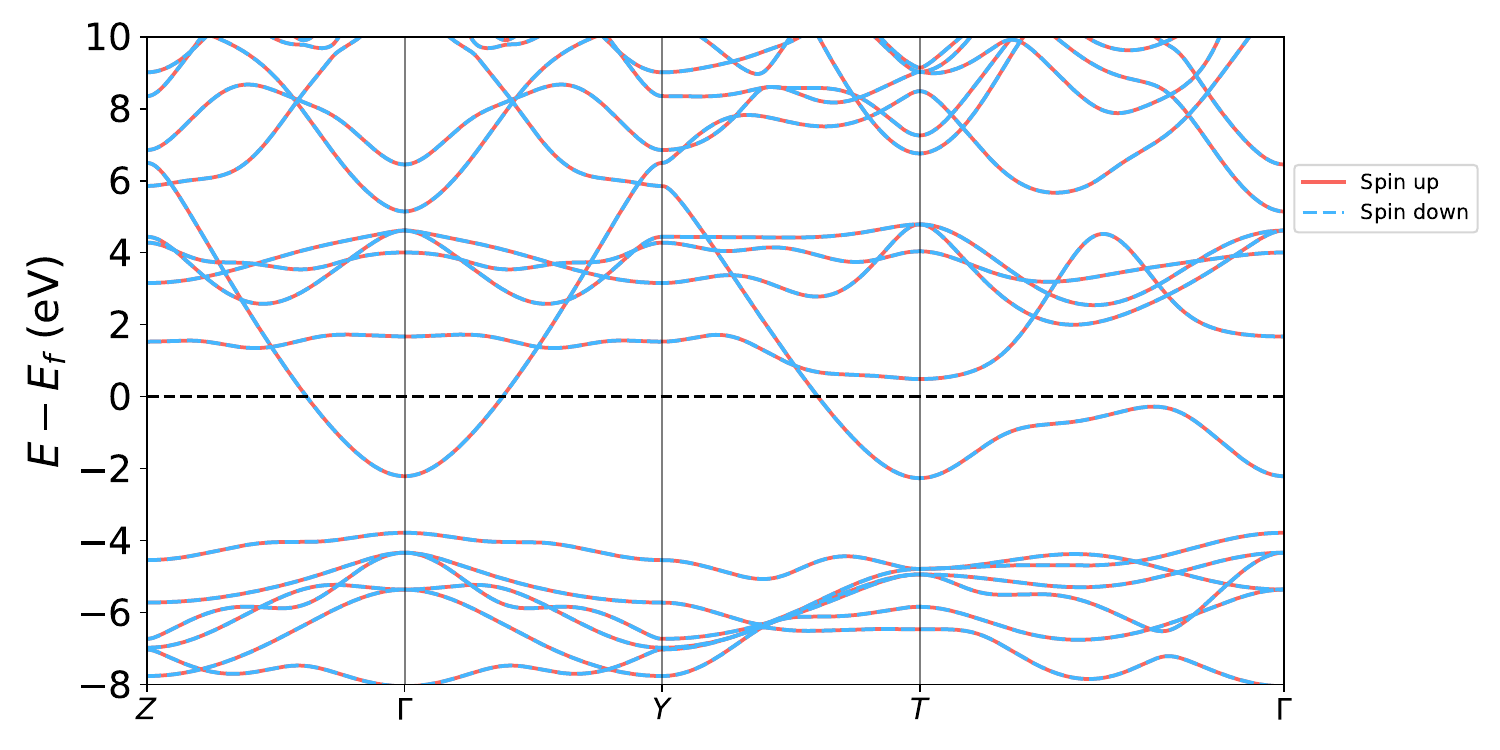}
    \caption{\footnotesize \ce{LaBr2} bulk spin-polarized band structure.}
    \label{fig:LaBr2_band_bulk}
\end{figure}
\clearpage
\newpage

\subsection{\ce{LaBr2} slab}
\begin{verbatim}
_cell_length_a                         3.63987510
_cell_length_b                         3.63987510
_cell_length_c                         40.00000000
_cell_angle_alpha                      90.000000
_cell_angle_beta                       90.000000
_cell_angle_gamma                      89.996116
_cell_volume                           'P 1'
_space_group_name_H-M_alt              'P 1'
_space_group_IT_number                 1

loop_
_space_group_symop_operation_xyz
   'x, y, z'

loop_
   _atom_site_label
   _atom_site_occupancy
   _atom_site_fract_x
   _atom_site_fract_y
   _atom_site_fract_z
   _atom_site_adp_type
   _atom_site_B_ios_or_equiv
   _atom_site_type_symbol
Br001  1.0  -0.500000000000  -0.500000000000  0.543422600341  Biso  1.000000  Br
La002  1.0  0.000000000000  -0.000000000000  0.5  Biso  1.000000  La
Br003  1.0  -0.500000000000  -0.500000000000  0.456577399659  Biso  1.000000  Br

\end{verbatim}
\begin{figure}[h]
    \centering
    \includegraphics[width=\textwidth]{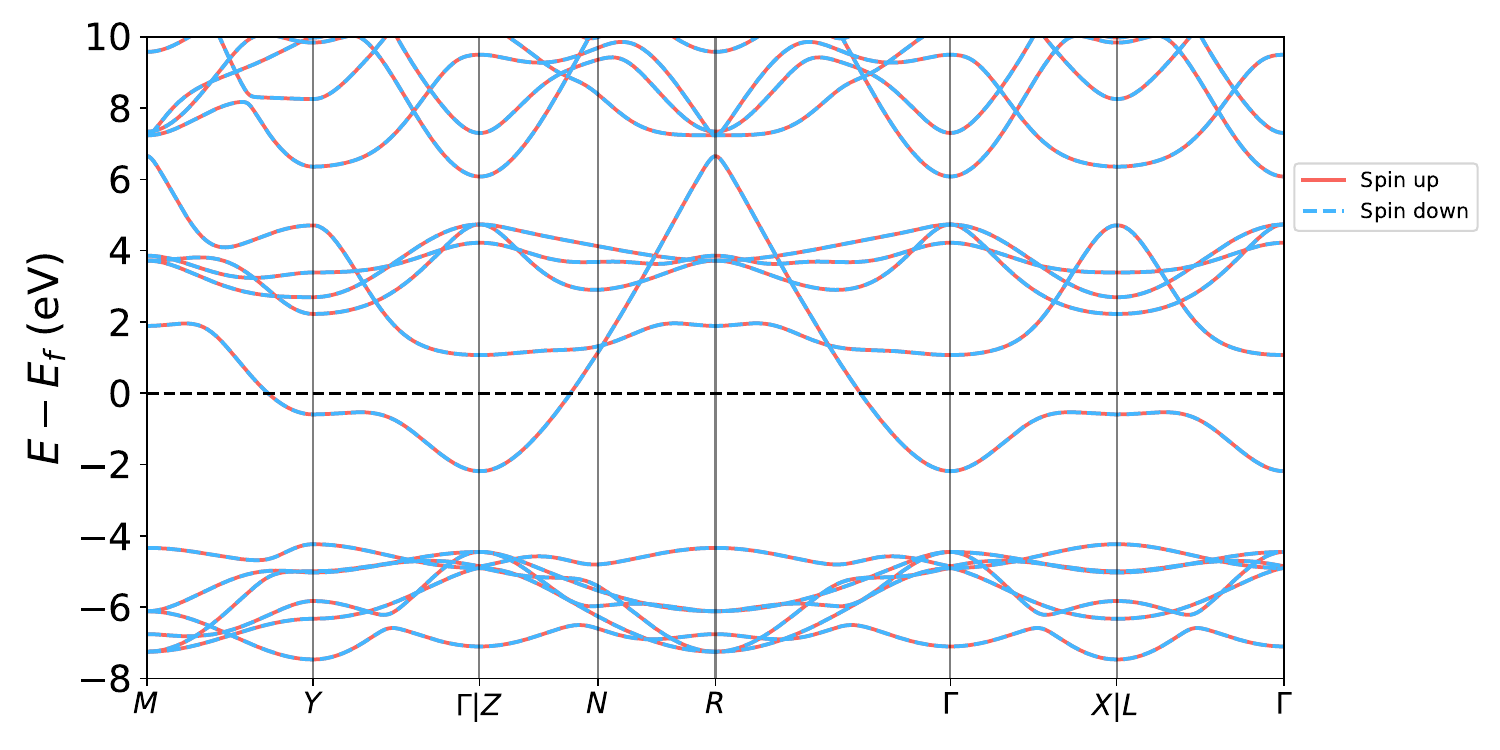}
    \caption{\footnotesize \ce{LaBr2} slab spin-polarized band structure.}
    \label{fig:LaBr2_band_slab}
\end{figure}
\clearpage
\newpage

\subsection{\ce{LaI2} bulk}
\begin{verbatim}
_cell_length_a 3.82435952
_cell_length_b 3.82435952
_cell_length_c 7.50493738
_cell_angle_alpha 89.852452
_cell_angle_beta 90.147548
_cell_angle_gamma 89.993362
_symmetry_space_group_name_H-M         'P 1'
_symmetry_Int_Tables_number            1

loop_
_symmetry_equiv_pos_as_xyz
   'x, y, z'

loop_
_atom_site_label
_atom_site_type_symbol
_atom_site_fract_x
_atom_site_fract_y
_atom_site_fract_z
I001 I -4.987891519395E-01 4.987891519395E-01 2.505564923248E-01
I002 I 4.987891519395E-01 -4.987891519395E-01 -2.505564923248E-01
La003 La 2.847386314662E-36 1.091864073107E-36 3.371346240334E-55

\end{verbatim}
\begin{figure}[h]
    \centering
    \includegraphics[width=\textwidth]{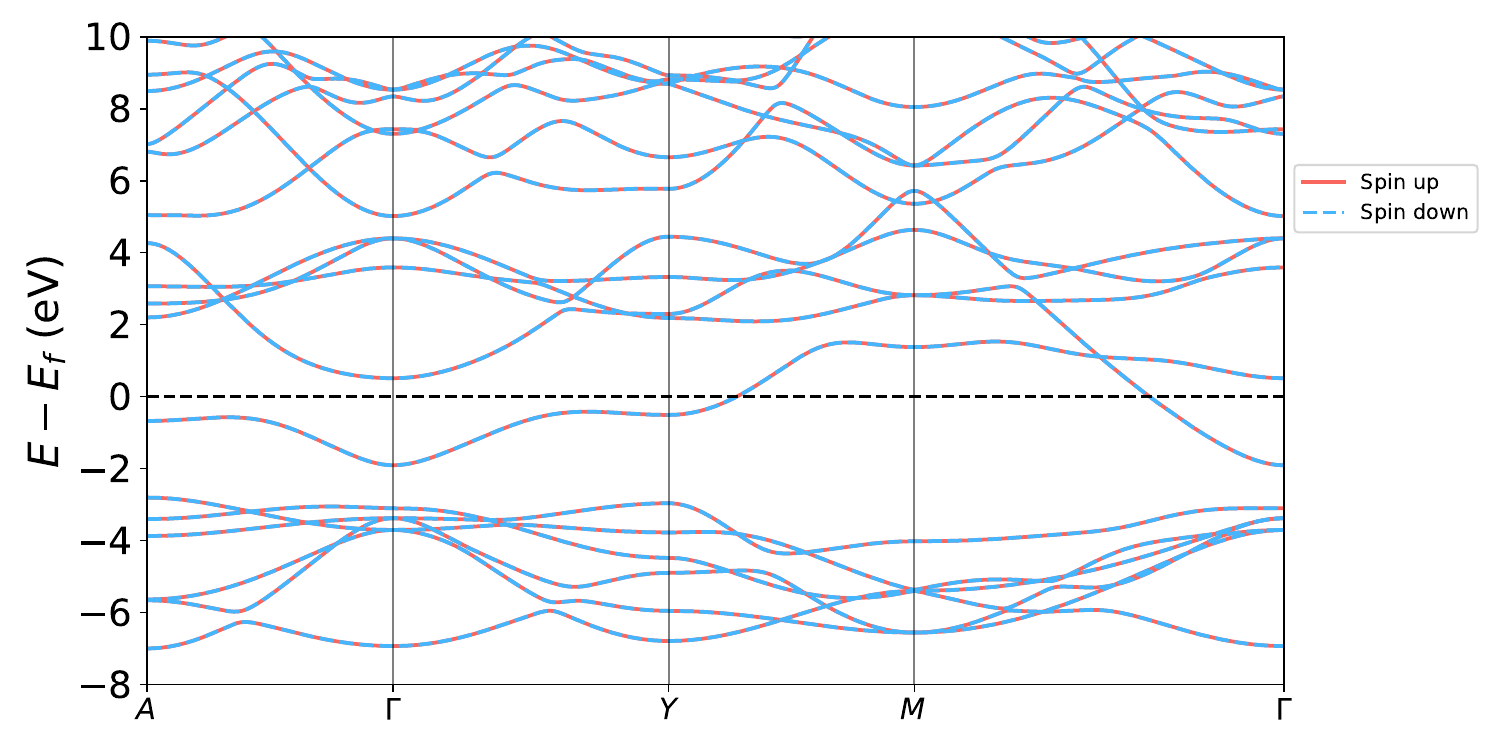}
    \caption{\footnotesize \ce{LaI2} bulk spin-polarized band structure.}
    \label{fig:LaI2_band_bulk}
\end{figure}
\clearpage
\newpage

\subsection{\ce{LaI2} slab}
\begin{verbatim}
_cell_length_a                         3.83912808
_cell_length_b                         3.83912808
_cell_length_c                         40.00000000
_cell_angle_alpha                      90.000000
_cell_angle_beta                       90.000000
_cell_angle_gamma                      89.997168
_cell_volume                           'P 1'
_space_group_name_H-M_alt              'P 1'
_space_group_IT_number                 1

loop_
_space_group_symop_operation_xyz
   'x, y, z'

loop_
   _atom_site_label
   _atom_site_occupancy
   _atom_site_fract_x
   _atom_site_fract_y
   _atom_site_fract_z
   _atom_site_adp_type
   _atom_site_B_ios_or_equiv
   _atom_site_type_symbol
I001  1.0  0.499980198136  -0.499980198136  0.547173276352  Biso  1.000000  I
La002  1.0  -0.000000000000  0.000000000000  0.5  Biso  1.000000  La
I003  1.0  -0.499980198136  0.499980198136  0.45282672364800003  Biso  1.000000  I

\end{verbatim}
\begin{figure}[h]
    \centering
    \includegraphics[width=\textwidth]{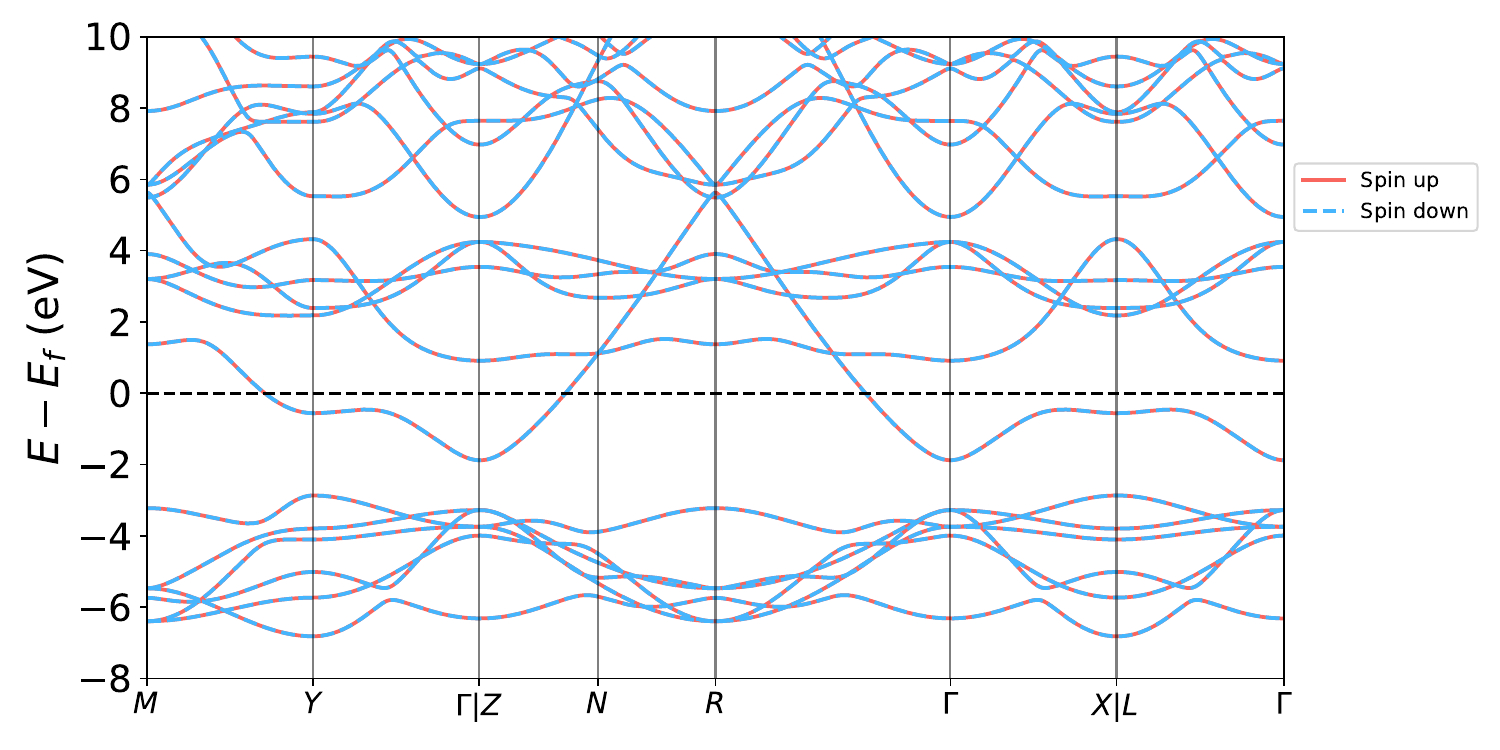}
    \caption{\footnotesize \ce{LaI2} slab spin-polarized band structure.}
    \label{fig:LaI2_band_slab}
\end{figure}
\clearpage
\newpage

\subsection{\ce{MgBr2} bulk}
\begin{verbatim}
_cell_length_a 3.75983713
_cell_length_b 3.75983713
_cell_length_c 6.00878689
_cell_angle_alpha 90.000000
_cell_angle_beta 90.000000
_cell_angle_gamma 120.000000
_symmetry_space_group_name_H-M         'P 1'
_symmetry_Int_Tables_number            1

loop_
_symmetry_equiv_pos_as_xyz
   'x, y, z'

loop_
_atom_site_label
_atom_site_type_symbol
_atom_site_fract_x
_atom_site_fract_y
_atom_site_fract_z
Br001 Br 3.333333333333E-01 -3.333333333333E-01 2.491408425935E-01
Br002 Br -3.333333333333E-01 3.333333333333E-01 -2.491408425935E-01
Mg003 Mg 0.000000000000E+00 0.000000000000E+00 0.000000000000E+00

\end{verbatim}
\begin{figure}[h]
    \centering
    \includegraphics[width=\textwidth]{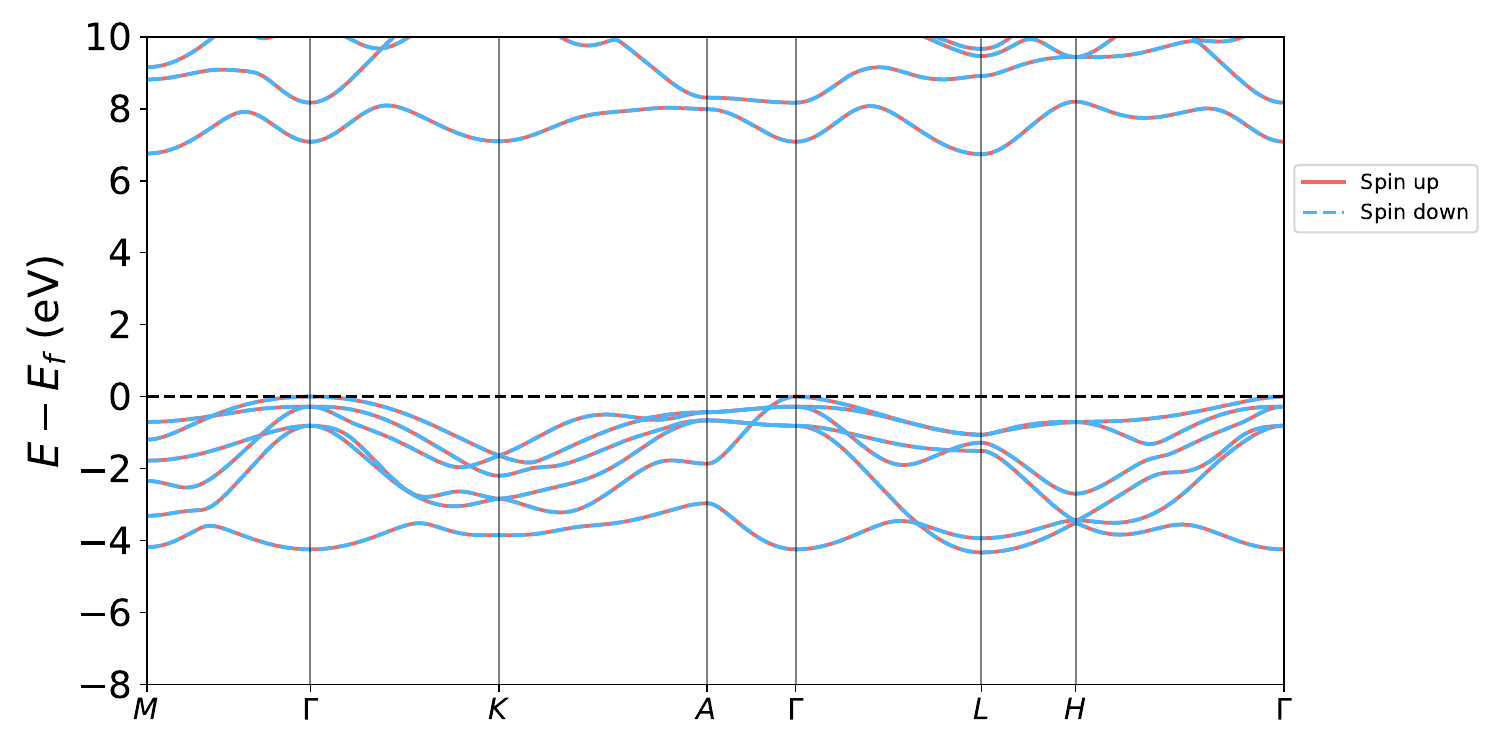}
    \caption{\footnotesize \ce{MgBr2} bulk spin-polarized band structure.}
    \label{fig:MgBr2_band_bulk}
\end{figure}
\clearpage
\newpage

\subsection{\ce{MgBr2} slab}
\begin{verbatim}
_cell_length_a                         3.76400188
_cell_length_b                         3.76400188
_cell_length_c                         40.00000000
_cell_angle_alpha                      90.000000
_cell_angle_beta                       90.000000
_cell_angle_gamma                      120.000000
_cell_volume                           'P 1'
_space_group_name_H-M_alt              'P 1'
_space_group_IT_number                 1

loop_
_space_group_symop_operation_xyz
   'x, y, z'

loop_
   _atom_site_label
   _atom_site_occupancy
   _atom_site_fract_x
   _atom_site_fract_y
   _atom_site_fract_z
   _atom_site_adp_type
   _atom_site_B_ios_or_equiv
   _atom_site_type_symbol
Br001  1.0  0.333333333333  -0.333333333333  0.537642649549  Biso  1.000000  Br
Mg002  1.0  0.000000000000  0.000000000000  0.5  Biso  1.000000  Mg
Br003  1.0  -0.333333333333  0.333333333333  0.462357350451  Biso  1.000000  Br

\end{verbatim}
\begin{figure}[h]
    \centering
    \includegraphics[width=\textwidth]{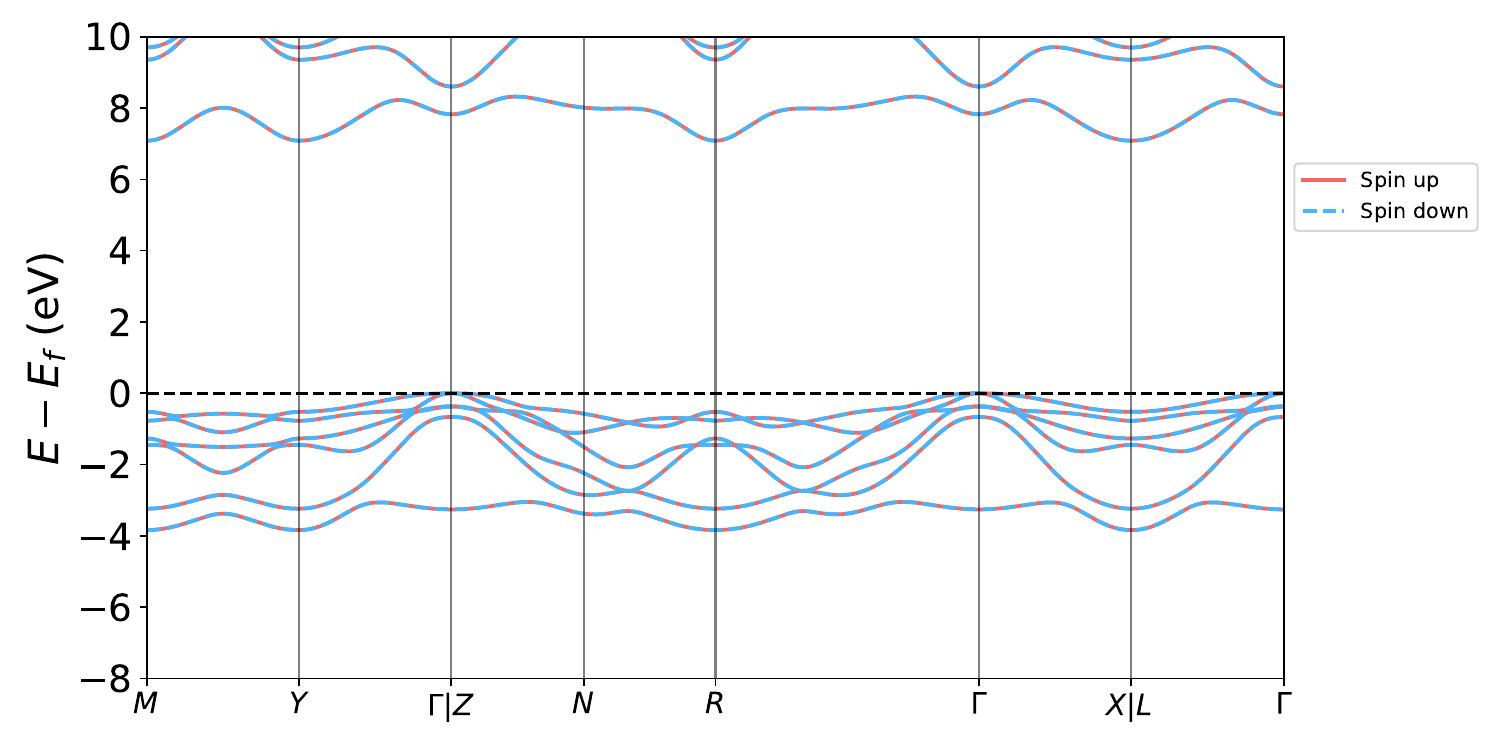}
    \caption{\footnotesize \ce{MgBr2} slab spin-polarized band structure.}
    \label{fig:MgBr2_band_slab}
\end{figure}
\clearpage
\newpage

\subsection{\ce{MgCl2} bulk}
\begin{verbatim}
_cell_length_a 3.58823868
_cell_length_b 3.58823868
_cell_length_c 5.65463336
_cell_angle_alpha 90.000000
_cell_angle_beta 90.000000
_cell_angle_gamma 120.000000
_symmetry_space_group_name_H-M         'P 1'
_symmetry_Int_Tables_number            1

loop_
_symmetry_equiv_pos_as_xyz
   'x, y, z'

loop_
_atom_site_label
_atom_site_type_symbol
_atom_site_fract_x
_atom_site_fract_y
_atom_site_fract_z
Mg001 Mg 0.000000000000E+00 0.000000000000E+00 0.000000000000E+00
Cl002 Cl 3.333333333333E-01 -3.333333333333E-01 2.431792030733E-01
Cl003 Cl -3.333333333333E-01 3.333333333333E-01 -2.431792030733E-01

\end{verbatim}
\begin{figure}[h]
    \centering
    \includegraphics[width=\textwidth]{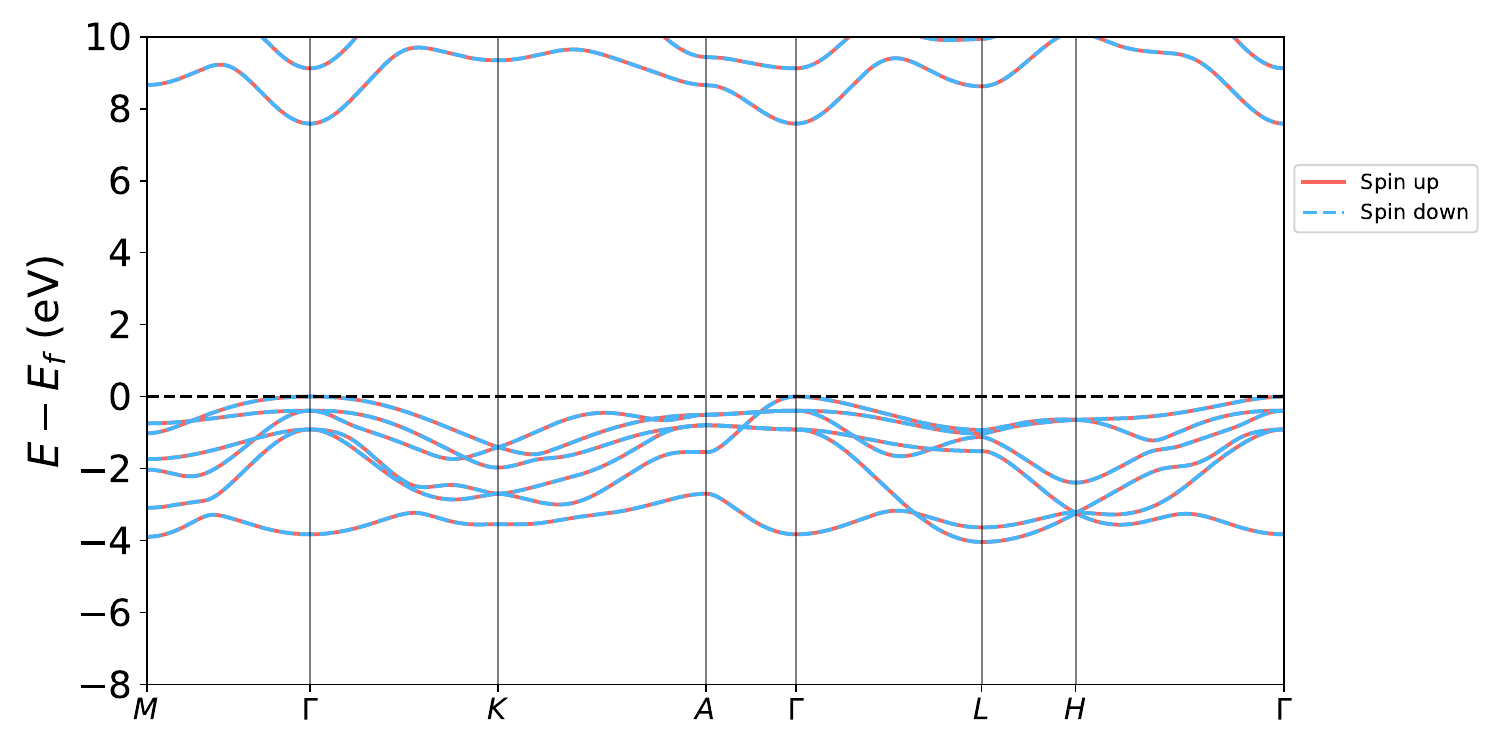}
    \caption{\footnotesize \ce{MgCl2} bulk spin-polarized band structure.}
    \label{fig:MgCl2_band_bulk}
\end{figure}
\clearpage
\newpage

\subsection{\ce{MgCl2} slab}
\begin{verbatim}
_cell_length_a                         3.58340225
_cell_length_b                         3.58340225
_cell_length_c                         40.00000000
_cell_angle_alpha                      90.000000
_cell_angle_beta                       90.000000
_cell_angle_gamma                      120.000000
_cell_volume                           'P 1'
_space_group_name_H-M_alt              'P 1'
_space_group_IT_number                 1

loop_
_space_group_symop_operation_xyz
   'x, y, z'

loop_
   _atom_site_label
   _atom_site_occupancy
   _atom_site_fract_x
   _atom_site_fract_y
   _atom_site_fract_z
   _atom_site_adp_type
   _atom_site_B_ios_or_equiv
   _atom_site_type_symbol
Cl001  1.0  0.333333333333  -0.333333333333  0.534689464635  Biso  1.000000  Cl
Mg002  1.0  0.000000000000  0.000000000000  0.5  Biso  1.000000  Mg
Cl003  1.0  -0.333333333333  0.333333333333  0.465310535365  Biso  1.000000  Cl

\end{verbatim}
\begin{figure}[h]
    \centering
    \includegraphics[width=\textwidth]{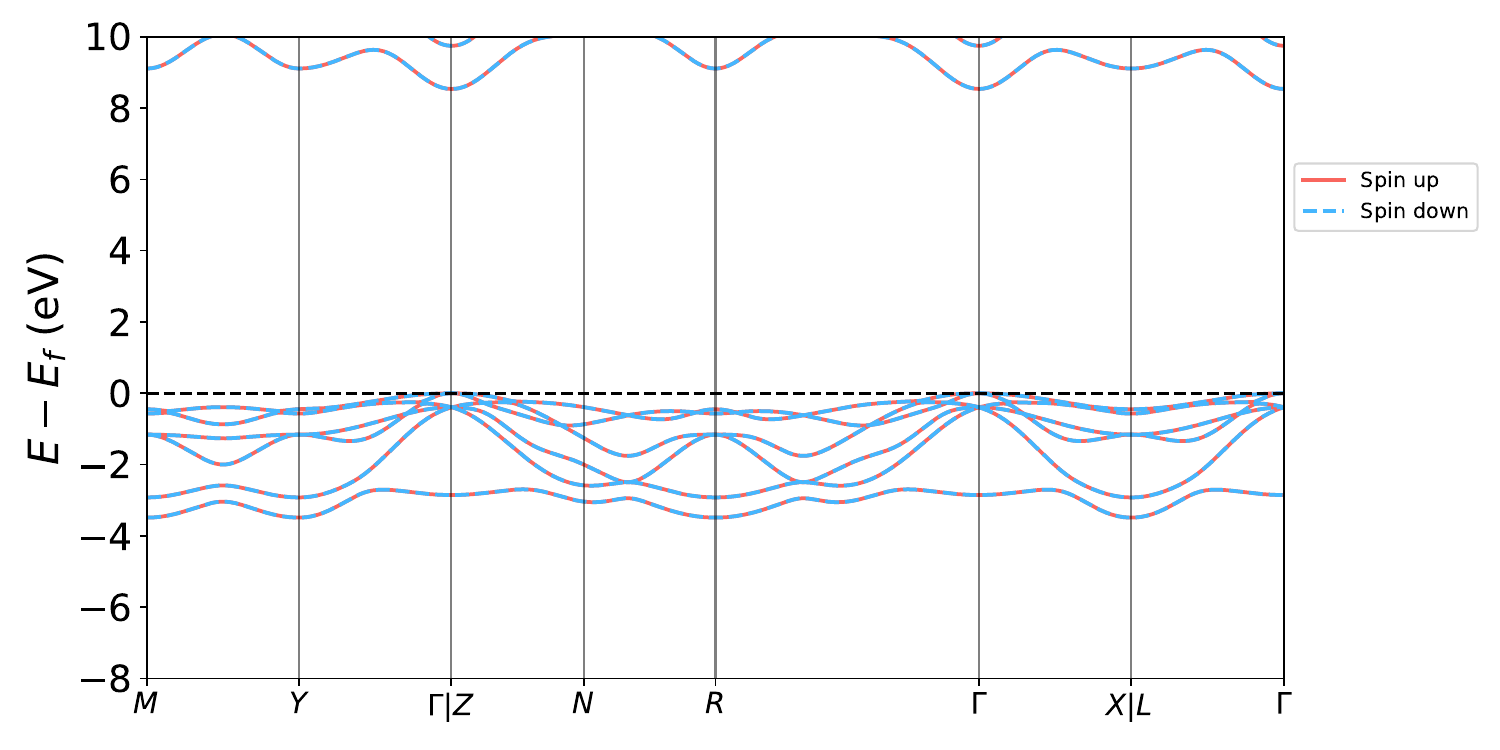}
    \caption{\footnotesize \ce{MgCl2} slab spin-polarized band structure.}
    \label{fig:MgCl2_band_slab}
\end{figure}
\clearpage
\newpage

\subsection{\ce{MgI2} bulk}
\begin{verbatim}
_cell_length_a 4.04651915
_cell_length_b 4.04651915
_cell_length_c 6.47666530
_cell_angle_alpha 90.000000
_cell_angle_beta 90.000000
_cell_angle_gamma 120.000000
_symmetry_space_group_name_H-M         'P 1'
_symmetry_Int_Tables_number            1

loop_
_symmetry_equiv_pos_as_xyz
   'x, y, z'

loop_
_atom_site_label
_atom_site_type_symbol
_atom_site_fract_x
_atom_site_fract_y
_atom_site_fract_z
I001 I 3.333333333333E-01 -3.333333333333E-01 -2.560847339330E-01
I002 I -3.333333333333E-01 3.333333333333E-01 2.560847339330E-01
Mg003 Mg 0.000000000000E+00 0.000000000000E+00 0.000000000000E+00

\end{verbatim}
\begin{figure}[h]
    \centering
    \includegraphics[width=\textwidth]{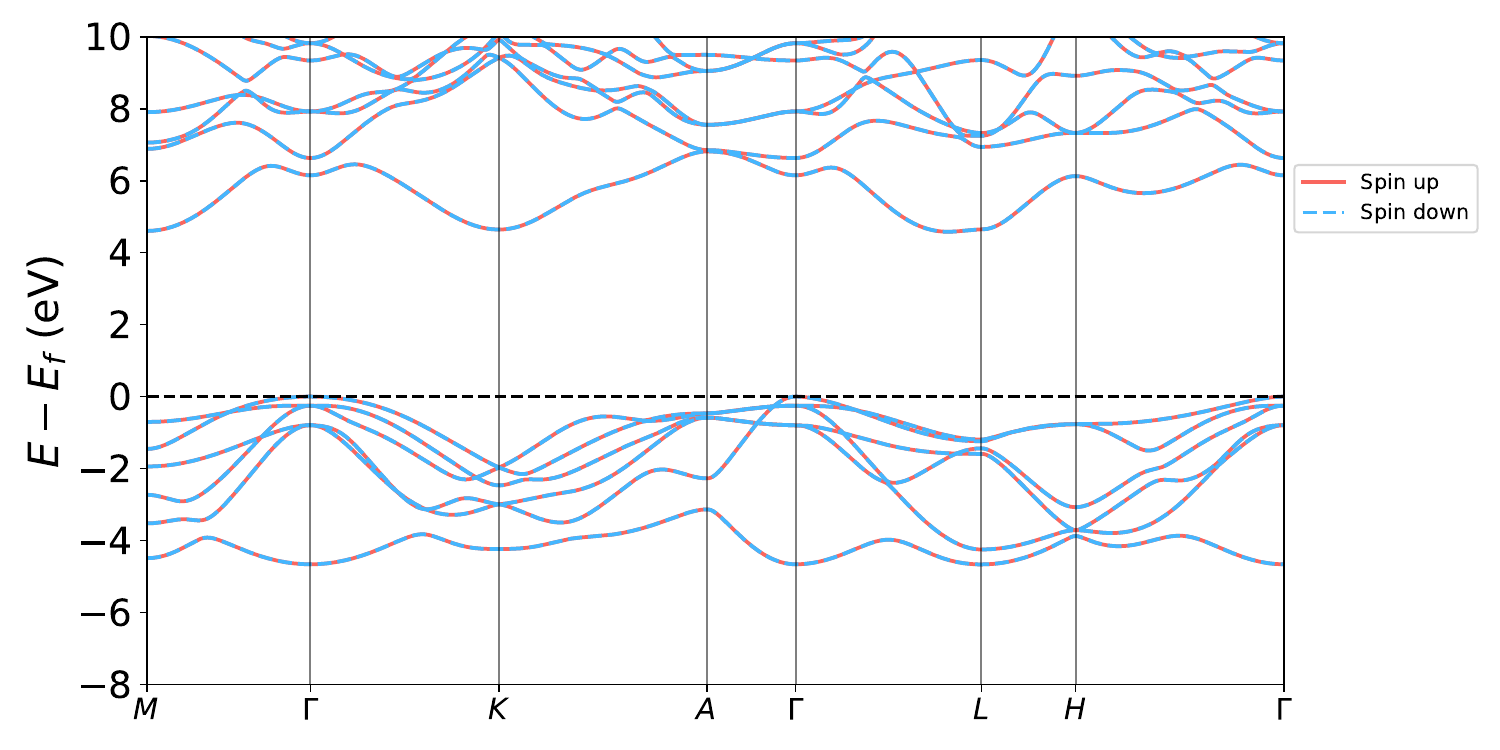}
    \caption{\footnotesize \ce{MgI2} bulk spin-polarized band structure.}
    \label{fig:MgI2_band_bulk}
\end{figure}
\clearpage
\newpage

\subsection{\ce{MgI2} slab}
\begin{verbatim}
_cell_length_a                         4.05510609
_cell_length_b                         4.05510609
_cell_length_c                         40.00000000
_cell_angle_alpha                      90.000000
_cell_angle_beta                       90.000000
_cell_angle_gamma                      120.000000
_cell_volume                           'P 1'
_space_group_name_H-M_alt              'P 1'
_space_group_IT_number                 1

loop_
_space_group_symop_operation_xyz
   'x, y, z'

loop_
   _atom_site_label
   _atom_site_occupancy
   _atom_site_fract_x
   _atom_site_fract_y
   _atom_site_fract_z
   _atom_site_adp_type
   _atom_site_B_ios_or_equiv
   _atom_site_type_symbol
I001  1.0  -0.333333333333  0.333333333333  0.541627425338  Biso  1.000000  I
Mg002  1.0  0.000000000000  0.000000000000  0.5  Biso  1.000000  Mg
I003  1.0  0.333333333333  -0.333333333333  0.458372574662  Biso  1.000000  I

\end{verbatim}
\begin{figure}[h]
    \centering
    \includegraphics[width=\textwidth]{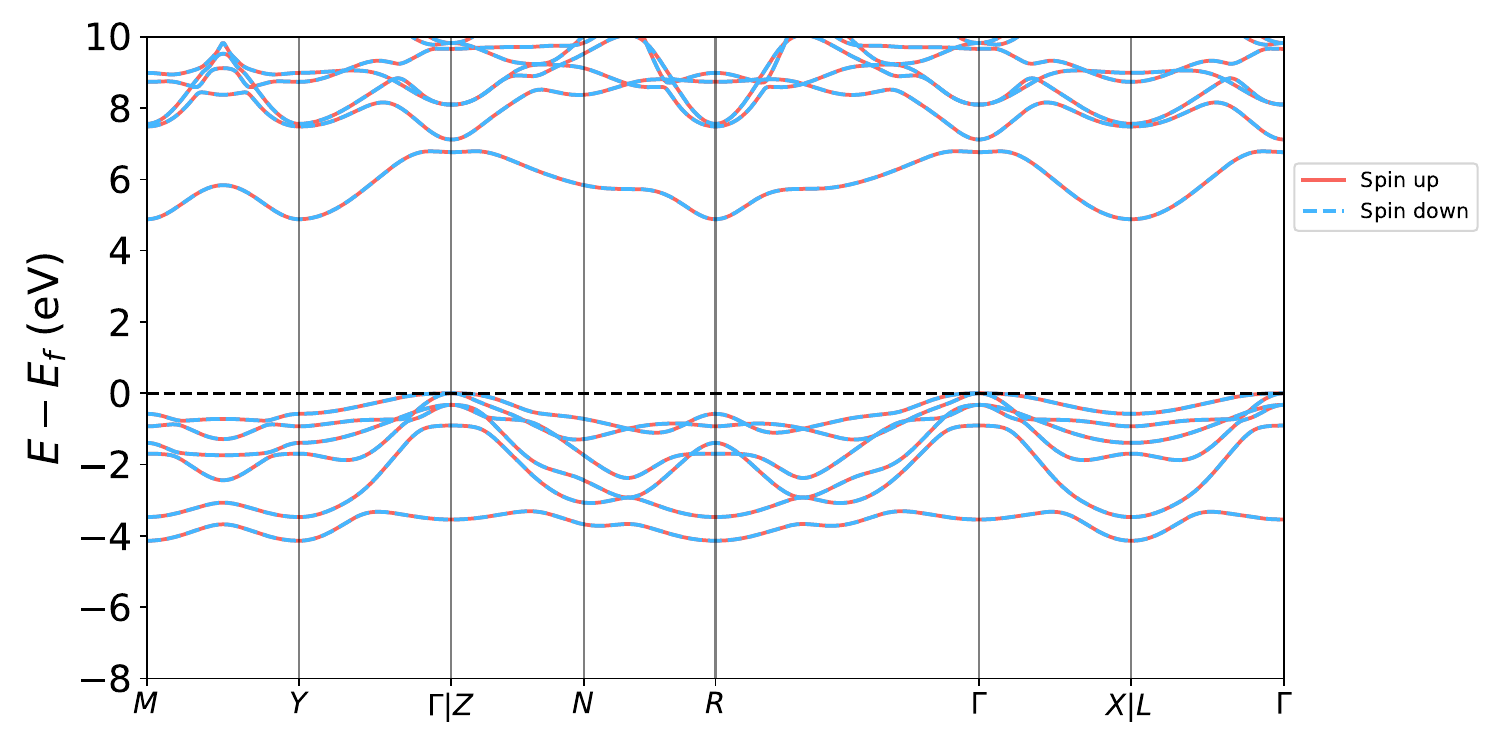}
    \caption{\footnotesize \ce{MgI2} slab spin-polarized band structure.}
    \label{fig:MgI2_band_slab}
\end{figure}
\clearpage
\newpage

\subsection{\ce{MnBr2} bulk}
\begin{verbatim}
_cell_length_a                         3.786839
_cell_length_b                         6.560785
_cell_length_c                         11.866511
_cell_angle_alpha                      90.067055
_cell_angle_beta                       90.005417
_cell_angle_gamma                      90.001045
_cell_volume                           294.818934
_space_group_name_H-M_alt              'P -1'
_space_group_IT_number                 2

loop_
_space_group_symop_operation_xyz
   'x, y, z'
   '-x, -y, -z'

loop_
   _atom_site_label
   _atom_site_occupancy
   _atom_site_fract_x
   _atom_site_fract_y
   _atom_site_fract_z
   _atom_site_adp_type
   _atom_site_B_iso_or_equiv
   _atom_site_type_symbol
   Mn1         1.0     0.749990     0.749940     0.750000    Biso  1.000000 Mn
   Mn2         1.0     0.750010     0.750070     0.250000    Biso  1.000000 Mn
   Br1         1.0     0.250040    -0.083510     0.876480    Biso  1.000000 Br
   Br2         1.0     0.250050    -0.083220     0.376490    Biso  1.000000 Br
   Br3         1.0     0.750050     0.416620     0.876490    Biso  1.000000 Br
   Br4         1.0     0.750070     0.416610     0.376480    Biso  1.000000 Br

\end{verbatim}
\begin{figure}[h]
    \centering
    \includegraphics[width=\textwidth]{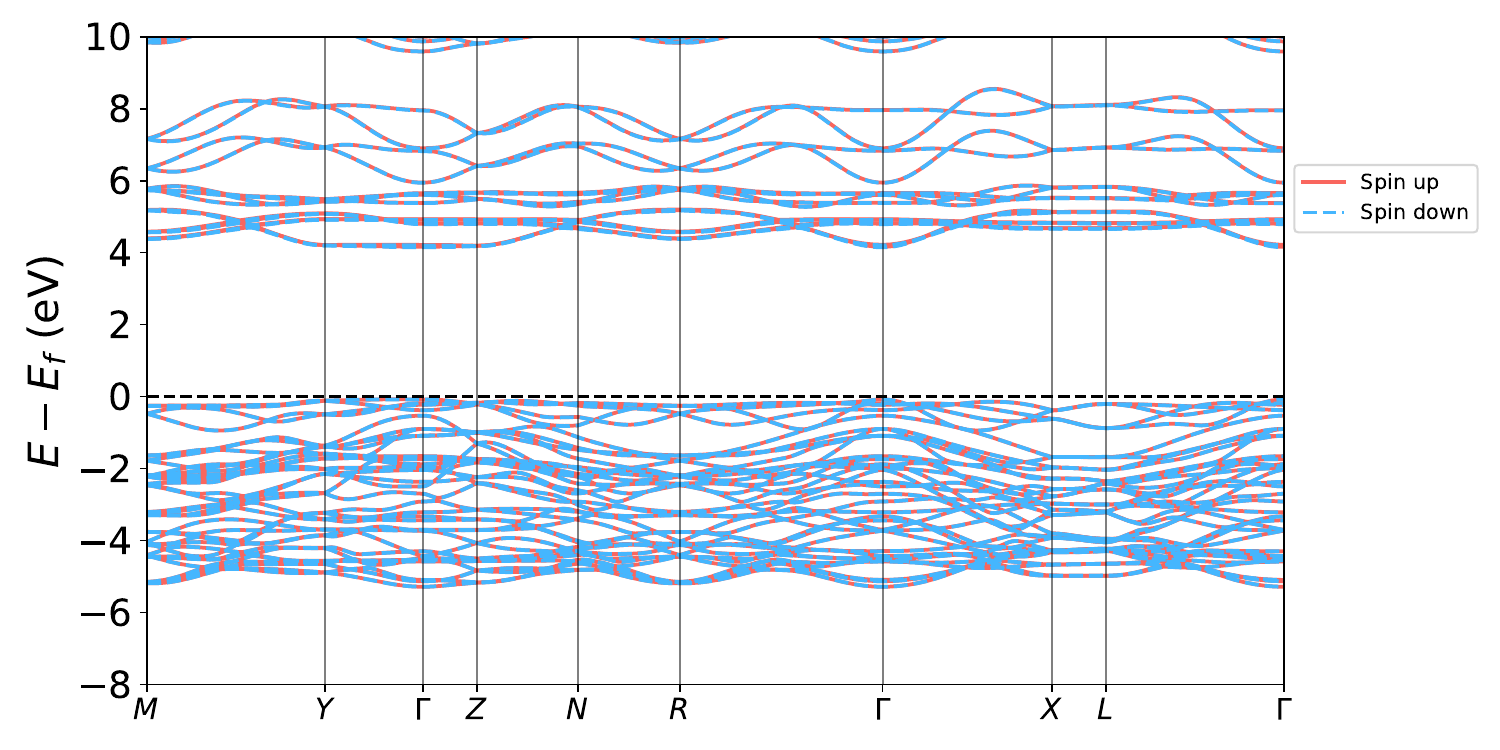}
    \caption{\footnotesize \ce{MnBr2} bulk spin-polarized band structure.}
    \label{fig:MnBr2_band_bulk}
\end{figure}
\clearpage
\newpage

\subsection{\ce{MnBr2} slab}
\begin{verbatim}
_cell_length_a                         3.79011976
_cell_length_b                         6.57107857
_cell_length_c                         40.00000000
_cell_angle_alpha                      90.000000
_cell_angle_beta                       90.000000
_cell_angle_gamma                      90.000717
_cell_volume                           'P 1'
_space_group_name_H-M_alt              'P 1'
_space_group_IT_number                 1

loop_
_space_group_symop_operation_xyz
   'x, y, z'

loop_
   _atom_site_label
   _atom_site_occupancy
   _atom_site_fract_x
   _atom_site_fract_y
   _atom_site_fract_z
   _atom_site_adp_type
   _atom_site_B_ios_or_equiv
   _atom_site_type_symbol
Br001  1.0  0.249994758423  -0.083962894553  0.537734447945  Biso  1.000000  Br
Br002  1.0  -0.249995480975  0.416220274022  0.537744327421  Biso  1.000000  Br
Mn003  1.0  -0.250010489671  -0.250069465889  0.500000355579  Biso  1.000000  Mn
Mn004  1.0  0.249996339858  0.249928894694  0.499999514697  Biso  1.000000  Mn
Br005  1.0  -0.250014790590  0.083631997113  0.462255134019  Biso  1.000000  Br
Br006  1.0  0.249977662955  -0.416176805388  0.462266220339  Biso  1.000000  Br

\end{verbatim}
\begin{figure}[h]
    \centering
    \includegraphics[width=\textwidth]{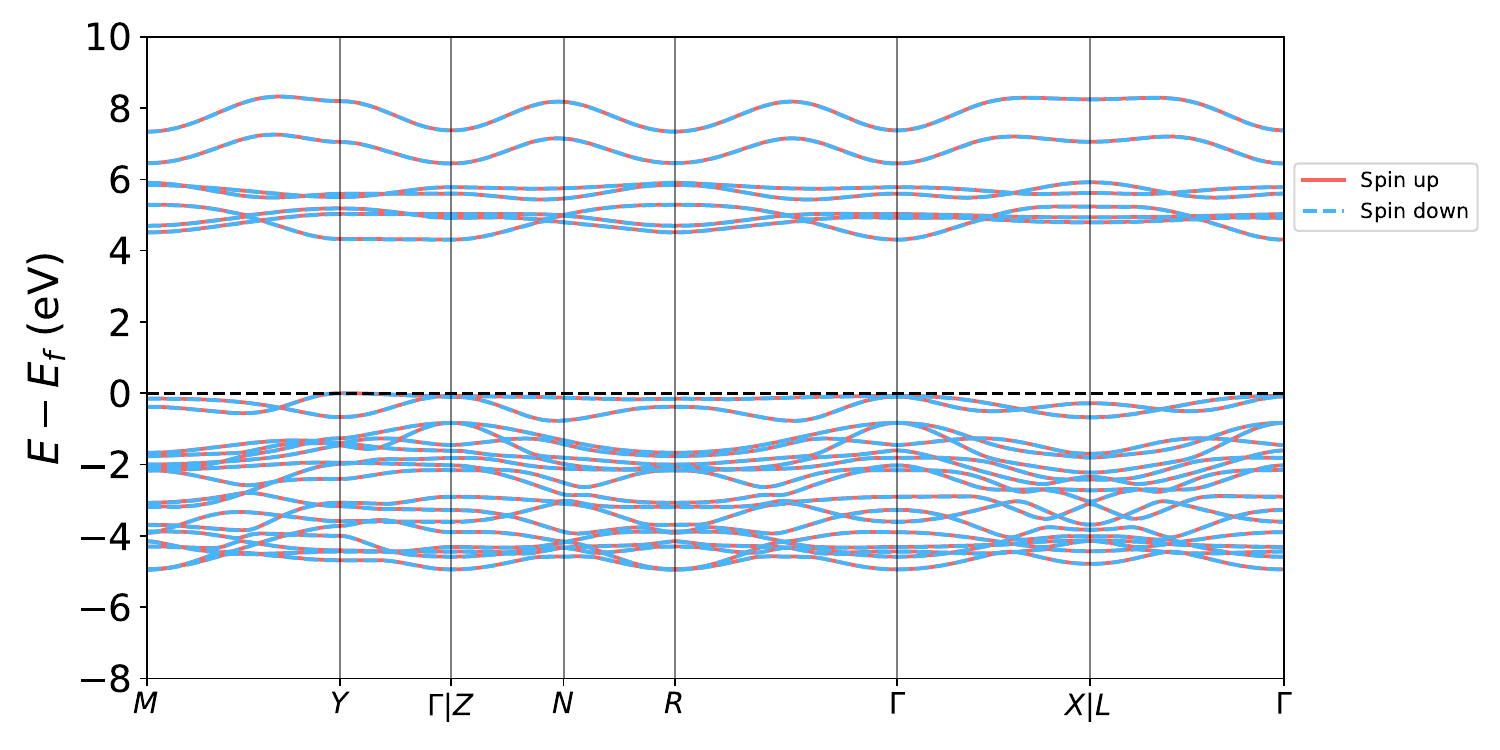}
    \caption{\footnotesize \ce{MnBr2} slab spin-polarized band structure.}
    \label{fig:MnBr2_band_slab}
\end{figure}
\clearpage
\newpage

\subsection{\ce{MnCl2} bulk}
\begin{verbatim}
_cell_length_a                         3.633635
_cell_length_b                         6.291469
_cell_length_c                         6.705541
_cell_angle_alpha                      117.916595
_cell_angle_beta                       105.716438
_cell_angle_gamma                      90.001068
_cell_volume                           128.933811
_space_group_name_H-M_alt              'P -1'
_space_group_IT_number                 2

loop_
_space_group_symop_operation_xyz
   'x, y, z'
   '-x, -y, -z'

loop_
   _atom_site_label
   _atom_site_occupancy
   _atom_site_fract_x
   _atom_site_fract_y
   _atom_site_fract_z
   _atom_site_adp_type
   _atom_site_B_iso_or_equiv
   _atom_site_type_symbol
   Mn1         1.0     0.000000     0.000000     0.000000    Biso  1.000000 Mn
   Mn2         1.0     0.500000     0.500000     0.000000    Biso  1.000000 Mn
   Cl1         1.0     0.375870     0.042390     0.751690    Biso  1.000000 Cl
   Cl2         1.0     0.875860     0.542550     0.751680    Biso  1.000000 Cl

\end{verbatim}
\begin{figure}[h]
    \centering
    \includegraphics[width=\textwidth]{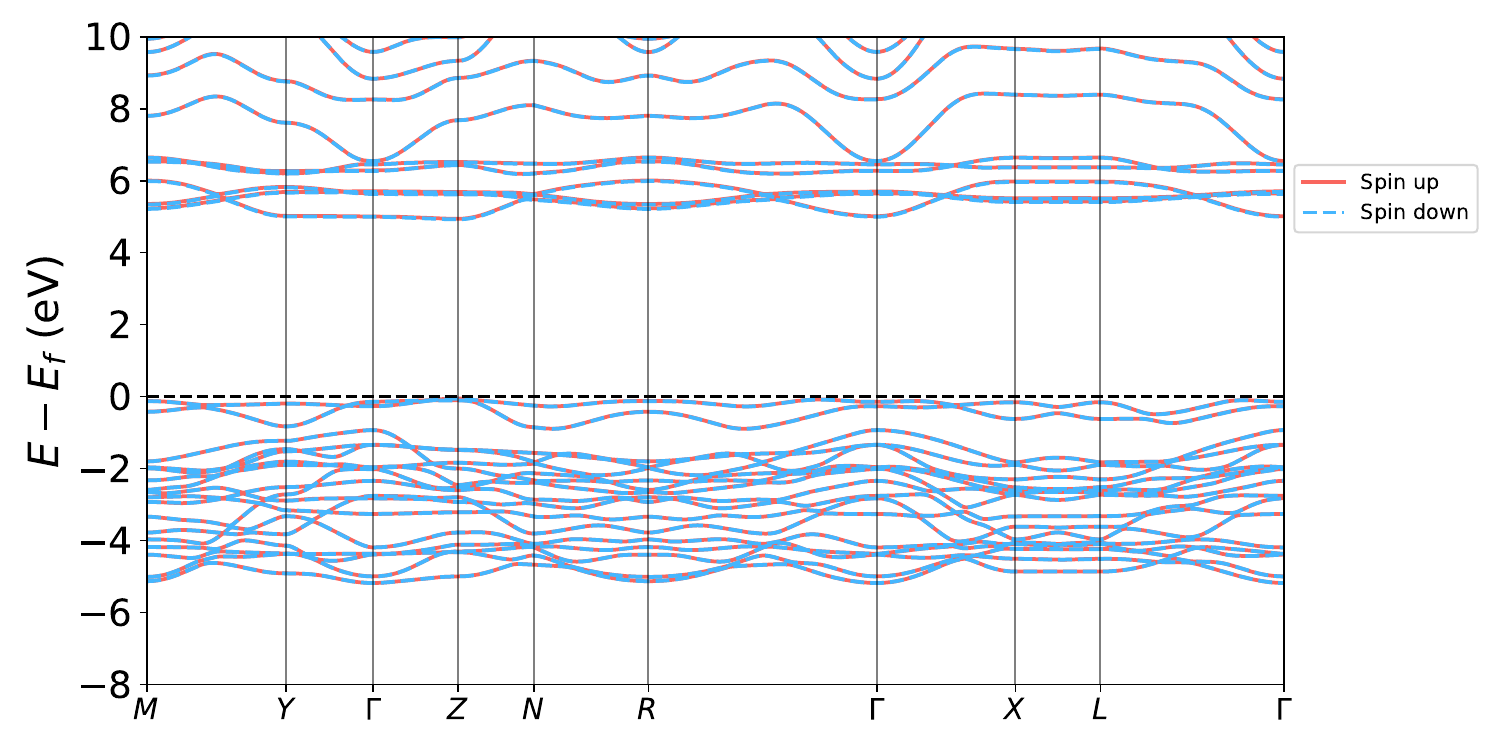}
    \caption{\footnotesize \ce{MnCl2} bulk spin-polarized band structure.}
    \label{fig:MnCl2_band_bulk}
\end{figure}
\clearpage
\newpage

\subsection{\ce{MnCl2} slab}
\begin{verbatim}
_cell_length_a                         3.62842963
_cell_length_b                         6.27588133
_cell_length_c                         40.00000000
_cell_angle_alpha                      90.000000
_cell_angle_beta                       90.000000
_cell_angle_gamma                      89.999808
_cell_volume                           'P 1'
_space_group_name_H-M_alt              'P 1'
_space_group_IT_number                 1

loop_
_space_group_symop_operation_xyz
   'x, y, z'

loop_
   _atom_site_label
   _atom_site_occupancy
   _atom_site_fract_x
   _atom_site_fract_y
   _atom_site_fract_z
   _atom_site_adp_type
   _atom_site_B_ios_or_equiv
   _atom_site_type_symbol
Cl001  1.0  0.499994763750  -0.166130919351  0.535379464246  Biso  1.000000  Cl
Cl002  1.0  -0.000001096973  0.333706846094  0.535384333126  Biso  1.000000  Cl
Mn003  1.0  -0.000003735619  -0.000002614528  0.500000718973  Biso  1.000000  Mn
Mn004  1.0  0.499998066142  0.499997135052  0.50000068447  Biso  1.000000  Mn
Cl005  1.0  0.499999345897  0.166126054537  0.464621929494  Biso  1.000000  Cl
Cl006  1.0  -0.000002340023  -0.333711471945  0.464617099637  Biso  1.000000  Cl

\end{verbatim}
\begin{figure}[h]
    \centering
    \includegraphics[width=\textwidth]{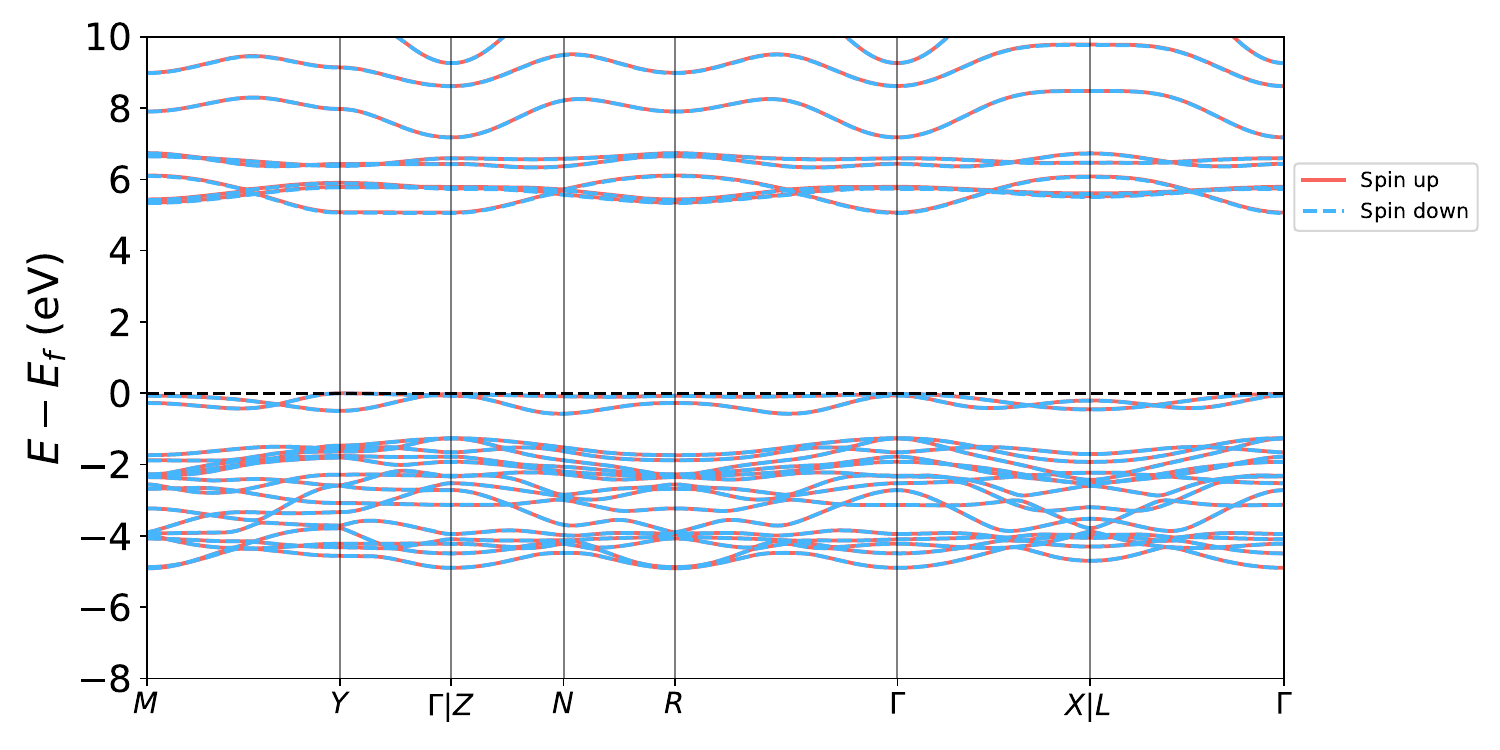}
    \caption{\footnotesize \ce{MnCl2} slab spin-polarized band structure.}
    \label{fig:MnCl2_band_slab}
\end{figure}
\clearpage
\newpage

\subsection{\ce{MnI2} bulk}
\begin{verbatim}
_cell_length_a                         6.969405
_cell_length_b                         4.025836
_cell_length_c                         12.752758
_cell_angle_alpha                      90.000000
_cell_angle_beta                       90.014229
_cell_angle_gamma                      90.000000
_cell_volume                           357.812821
_space_group_name_H-M_alt              'C 2/m'
_space_group_IT_number                 12

loop_
_space_group_symop_operation_xyz
   'x, y, z'
   '-x, -y, -z'
   '-x, y, -z'
   'x, -y, z'
   'x+1/2, y+1/2, z'
   '-x+1/2, -y+1/2, -z'
   '-x+1/2, y+1/2, -z'
   'x+1/2, -y+1/2, z'

loop_
   _atom_site_label
   _atom_site_occupancy
   _atom_site_fract_x
   _atom_site_fract_y
   _atom_site_fract_z
   _atom_site_adp_type
   _atom_site_B_iso_or_equiv
   _atom_site_type_symbol
   Mn1         1.0     0.000000     0.000000     0.000000    Biso  1.000000 Mn
   Mn2         1.0     0.000000     0.000000     0.500000    Biso  1.000000 Mn
   I1          1.0     0.333240     0.000000     0.870350    Biso  1.000000 I
   I2          1.0     0.333240     0.000000     0.370350    Biso  1.000000 I

\end{verbatim}

\begin{figure}[h]
    \centering
    \includegraphics[width=\textwidth]{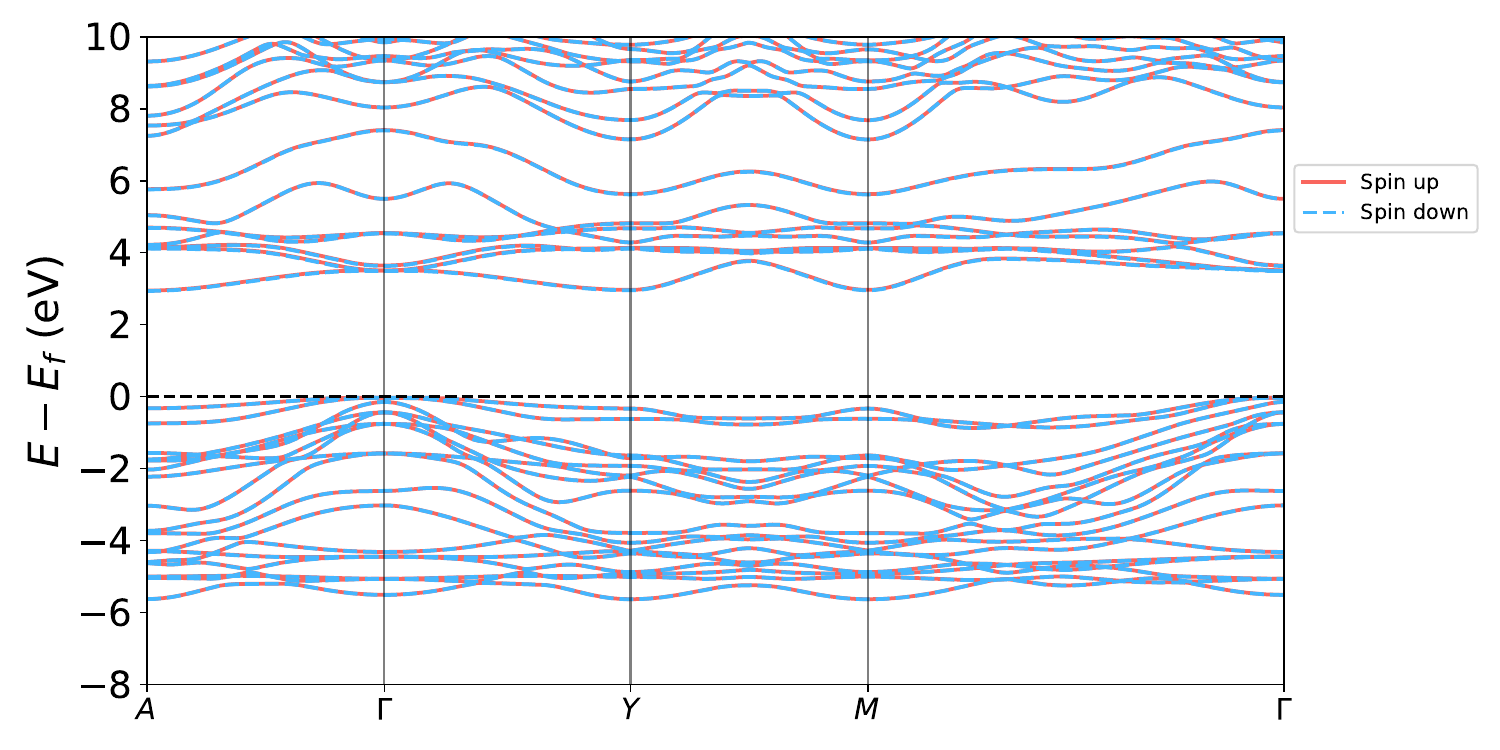}
    \caption{\footnotesize \ce{MnI2} bulk spin-polarized band structure.}
    \label{fig:MnI2_band_bulk}
\end{figure}

\clearpage
\newpage

\subsection{\ce{MnI2} slab}
\begin{verbatim}
_cell_length_a                         4.02897935
_cell_length_b                         6.99329522
_cell_length_c                         40.00000000
_cell_angle_alpha                      90.000000
_cell_angle_beta                       90.000000
_cell_angle_gamma                      90.000584
_cell_volume                           'P 1'
_space_group_name_H-M_alt              'P 1'
_space_group_IT_number                 1

loop_
_space_group_symop_operation_xyz
   'x, y, z'

loop_
   _atom_site_label
   _atom_site_occupancy
   _atom_site_fract_x
   _atom_site_fract_y
   _atom_site_fract_z
   _atom_site_adp_type
   _atom_site_B_ios_or_equiv
   _atom_site_type_symbol
I001  1.0  0.000000499869  0.333532970973  0.541457858115  Biso  1.000000  I
I002  1.0  -0.499999219270  -0.166311680133  0.541450540593  Biso  1.000000  I
Mn003  1.0  0.000001898322  0.000001660766  0.49999946025  Biso  1.000000  Mn
Mn004  1.0  -0.499998243004  -0.499998340389  0.499999452144  Biso  1.000000  Mn
I005  1.0  0.000002784367  -0.333529330414  0.458541004646  Biso  1.000000  I
I006  1.0  -0.499997219978  0.166315209285  0.458548339382  Biso  1.000000  I

\end{verbatim}
\begin{figure}[h]
    \centering
    \includegraphics[width=\textwidth]{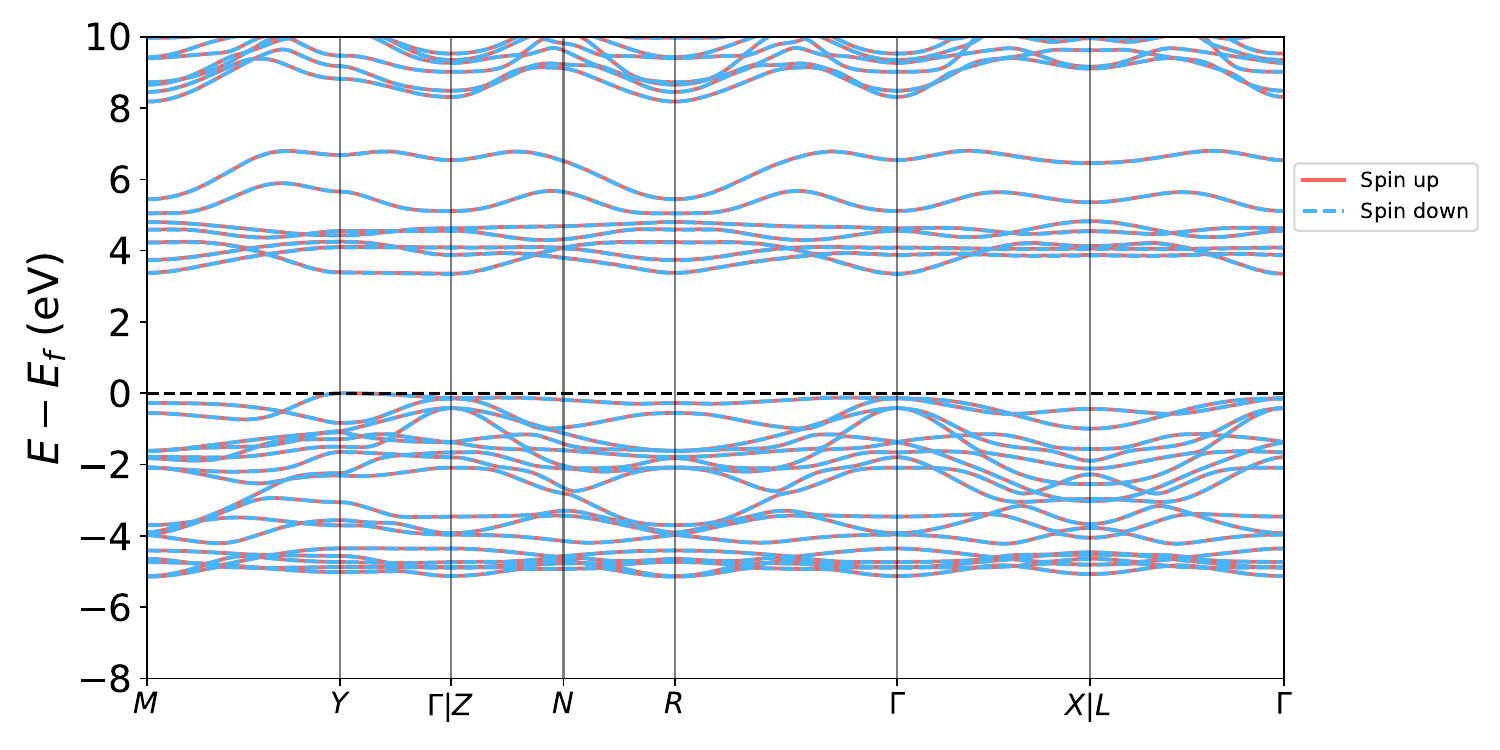}
    \caption{\footnotesize \ce{MnI2} slab spin-polarized band structure.}
    \label{fig:MnI2_band_slab}
\end{figure}
\clearpage
\newpage

\subsection{\ce{NdI2} bulk}
\begin{verbatim}
_cell_length_a 3.74647003
_cell_length_b 3.74404835
_cell_length_c 7.43501297
_cell_angle_alpha 90.000000
_cell_angle_beta 89.999908
_cell_angle_gamma 90.000000
_symmetry_space_group_name_H-M         'P 1'
_symmetry_Int_Tables_number            1

loop_
_symmetry_equiv_pos_as_xyz
   'x, y, z'

loop_
_atom_site_label
_atom_site_type_symbol
_atom_site_fract_x
_atom_site_fract_y
_atom_site_fract_z
I001 I -4.996052486877E-01 5.000000000000E-01 2.478458966789E-01
I002 I 4.996052486877E-01 5.000000000000E-01 -2.478458966789E-01
Nd003 Nd -3.021161315235E-37 -5.213831158029E-51 8.367616329763E-57

\end{verbatim}
\begin{figure}[h]
    \centering
    \includegraphics[width=\textwidth]{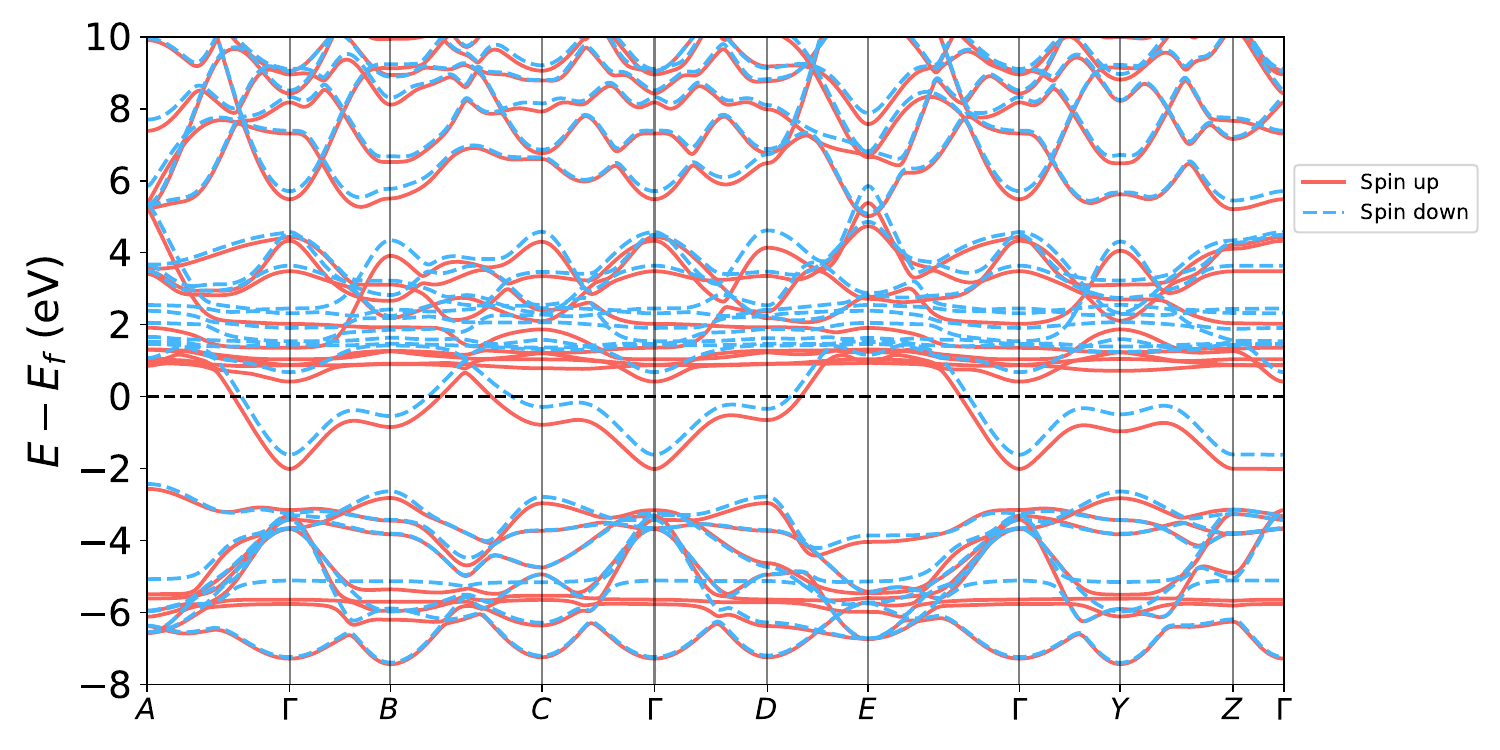}
    \caption{\footnotesize \ce{NdI2} bulk spin-polarized band structure.}
    \label{fig:NdI2_band_bulk}
\end{figure}

\clearpage
\newpage

\subsection{\ce{NdI2} slab}
\begin{verbatim}
_cell_length_a                         3.82641854
_cell_length_b                         3.83732518
_cell_length_c                         40.00000000
_cell_angle_alpha                      90.000000
_cell_angle_beta                       90.000000
_cell_angle_gamma                      90.000000
_cell_volume                           'P 1'
_space_group_name_H-M_alt              'P 1'
_space_group_IT_number                 1

loop_
_space_group_symop_operation_xyz
   'x, y, z'

loop_
   _atom_site_label
   _atom_site_occupancy
   _atom_site_fract_x
   _atom_site_fract_y
   _atom_site_fract_z
   _atom_site_adp_type
   _atom_site_B_ios_or_equiv
   _atom_site_type_symbol
I001  1.0  0.500000000000  0.499973509441  0.546512045712  Biso  1.000000  I
Nd002  1.0  -0.000000000000  0.000000000000  0.5  Biso  1.000000  Nd
I003  1.0  0.500000000000  -0.499973509441  0.453487954288  Biso  1.000000  I

\end{verbatim}
\begin{figure}[h]
    \centering
    \includegraphics[width=\textwidth]{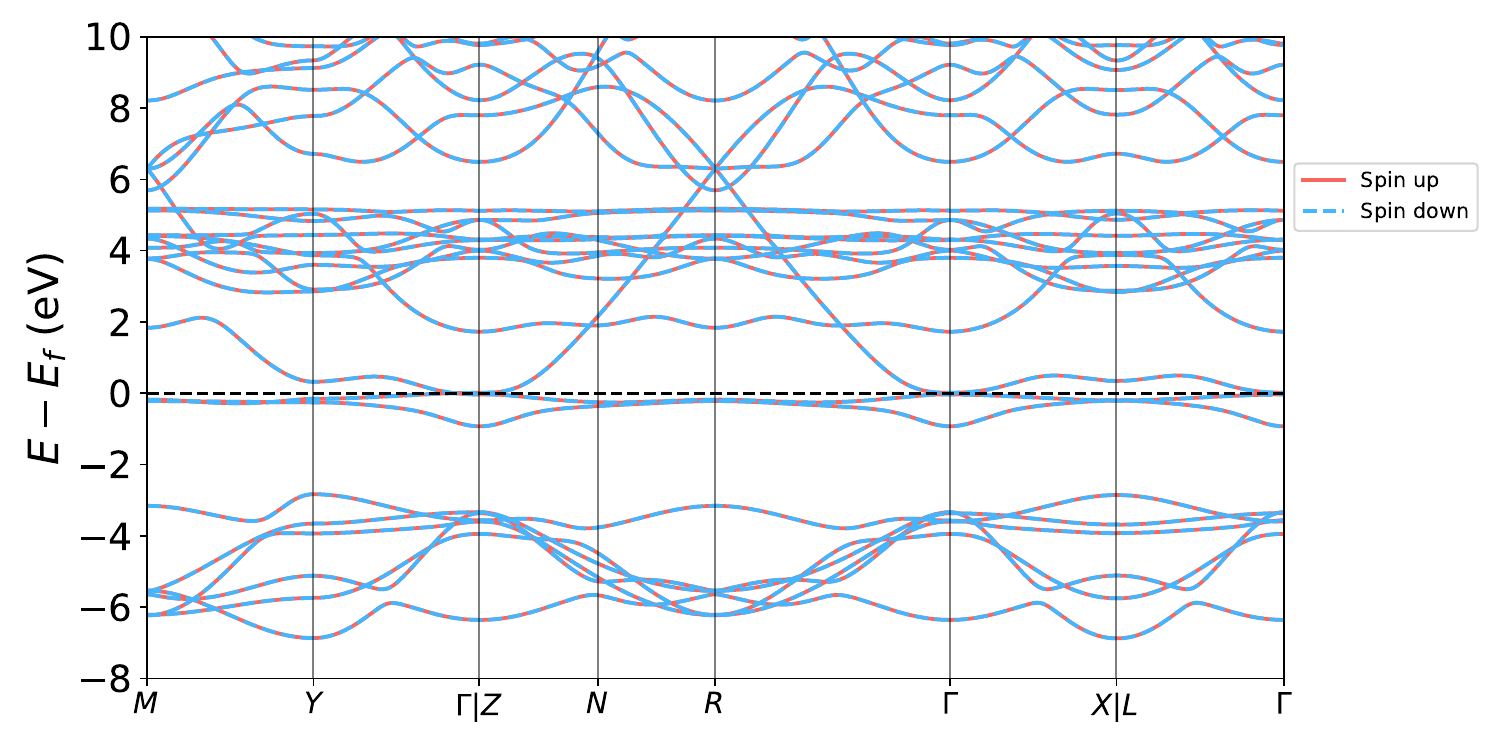}
    \caption{\footnotesize \ce{NdI2} slab spin-polarized band structure.}
    \label{fig:NdI2_band_slab}
\end{figure}
\clearpage
\newpage

\subsection{\ce{NiBr2} bulk}
\begin{verbatim}
_cell_length_a                         3.611711
_cell_length_b                         3.611711
_cell_length_c                         11.678998
_cell_angle_alpha                      90.000000
_cell_angle_beta                       90.000000
_cell_angle_gamma                      120.000000
_cell_volume                           131.935663
_space_group_name_H-M_alt              'P -3 m 1'
_space_group_IT_number                 164

loop_
_space_group_symop_operation_xyz
   'x, y, z'
   '-x, -y, -z'
   '-y, x-y, z'
   'y, -x+y, -z'
   '-x+y, -x, z'
   'x-y, x, -z'
   'y, x, -z'
   '-y, -x, z'
   'x-y, -y, -z'
   '-x+y, y, z'
   '-x, -x+y, -z'
   'x, x-y, z'

loop_
   _atom_site_label
   _atom_site_occupancy
   _atom_site_fract_x
   _atom_site_fract_y
   _atom_site_fract_z
   _atom_site_adp_type
   _atom_site_B_iso_or_equiv
   _atom_site_type_symbol
   Br1         1.0     0.333333     0.666667     0.121530    Biso  1.000000 Br
   Br2         1.0     0.333333     0.666667     0.621530    Biso  1.000000 Br
   Ni1         1.0     0.000000     0.000000     0.000000    Biso  1.000000 Ni
   Ni2         1.0     0.000000     0.000000     0.500000    Biso  1.000000 Ni

\end{verbatim}
\begin{figure}[h]
    \centering
    \includegraphics[width=\textwidth]{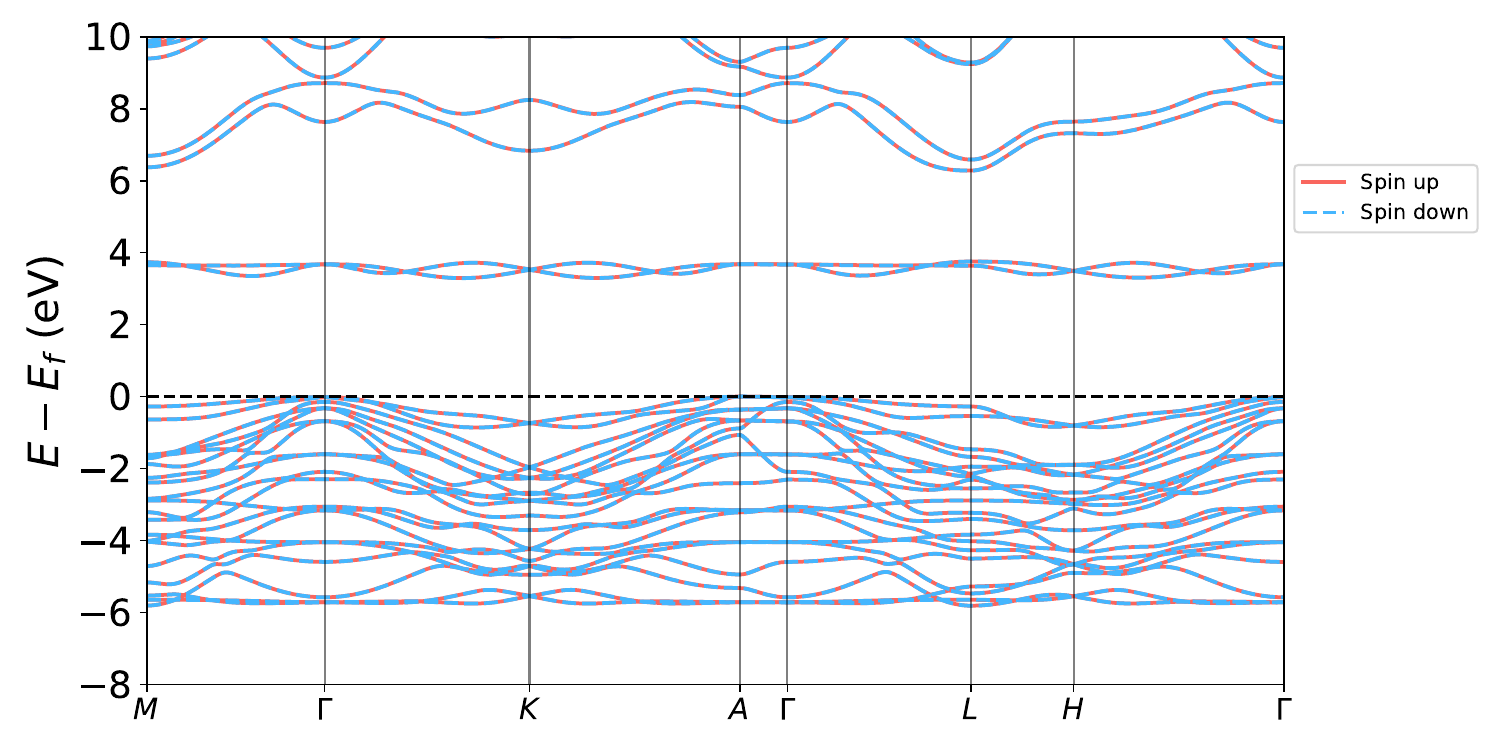}
    \caption{\footnotesize \ce{NiBr2} bulk spin-polarized band structure.}
    \label{fig:NiBr2_band_bulk}
\end{figure}
\clearpage
\newpage

\subsection{\ce{NiBr2} slab}
\begin{verbatim}
_cell_length_a                         3.61445988
_cell_length_b                         3.61445988
_cell_length_c                         40.00000000
_cell_angle_alpha                      90.000000
_cell_angle_beta                       90.000000
_cell_angle_gamma                      120.000000
_cell_volume                           'P 1'
_space_group_name_H-M_alt              'P 1'
_space_group_IT_number                 1

loop_
_space_group_symop_operation_xyz
   'x, y, z'

loop_
   _atom_site_label
   _atom_site_occupancy
   _atom_site_fract_x
   _atom_site_fract_y
   _atom_site_fract_z
   _atom_site_adp_type
   _atom_site_B_ios_or_equiv
   _atom_site_type_symbol
Br001  1.0  -0.333333333333  0.333333333333  0.535671653283  Biso  1.000000  Br
Ni002  1.0  0.000000000000  0.000000000000  0.5  Biso  1.000000  Ni
Br003  1.0  0.333333333333  -0.333333333333  0.464328346717  Biso  1.000000  Br

\end{verbatim}
\begin{figure}[h]
    \centering
    \includegraphics[width=\textwidth]{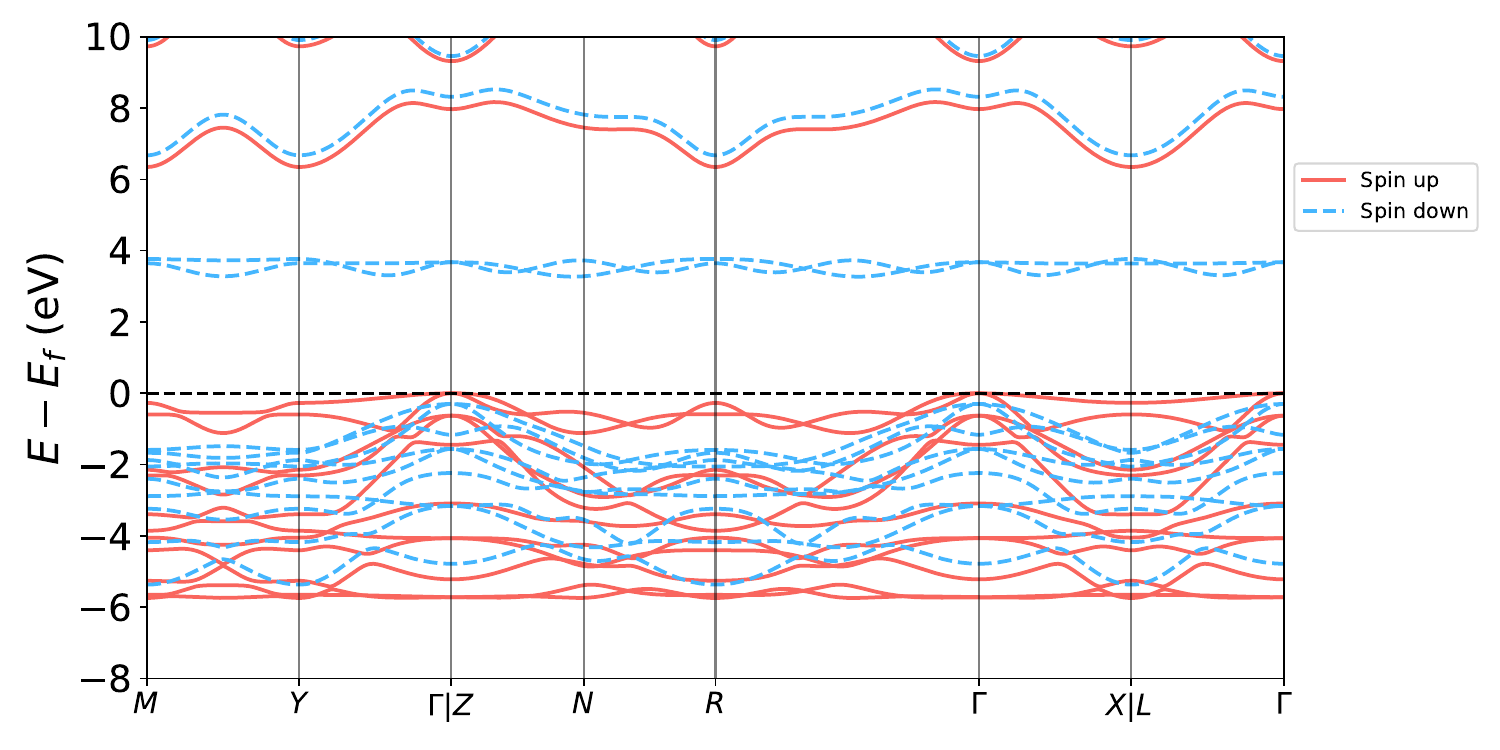}
    \caption{\footnotesize \ce{NiBr2} slab spin-polarized band structure.}
    \label{fig:NiBr2_band_slab}
\end{figure}
\clearpage
\newpage

\subsection{\ce{NiCl2} bulk}
\begin{verbatim}
_cell_length_a                         5.993321
_cell_length_b                         3.461847
_cell_length_c                         11.228026
_cell_angle_alpha                      90.000000
_cell_angle_beta                       90.156700
_cell_angle_gamma                      90.000000
_cell_volume                           232.957777
_space_group_name_H-M_alt              'C 2/m'
_space_group_IT_number                 12

loop_
_space_group_symop_operation_xyz
   'x, y, z'
   '-x, -y, -z'
   '-x, y, -z'
   'x, -y, z'
   'x+1/2, y+1/2, z'
   '-x+1/2, -y+1/2, -z'
   '-x+1/2, y+1/2, -z'
   'x+1/2, -y+1/2, z'

loop_
   _atom_site_label
   _atom_site_occupancy
   _atom_site_fract_x
   _atom_site_fract_y
   _atom_site_fract_z
   _atom_site_adp_type
   _atom_site_B_iso_or_equiv
   _atom_site_type_symbol
   Ni1         1.0     0.000000     0.000000     0.000000    Biso  1.000000 Ni
   Ni2         1.0     0.000000     0.000000     0.500000    Biso  1.000000 Ni
   Cl1         1.0     0.665860     0.000000     0.880460    Biso  1.000000 Cl
   Cl2         1.0     0.665860     0.000000     0.380460    Biso  1.000000 Cl

\end{verbatim}
\begin{figure}[h]
    \centering
    \includegraphics[width=\textwidth]{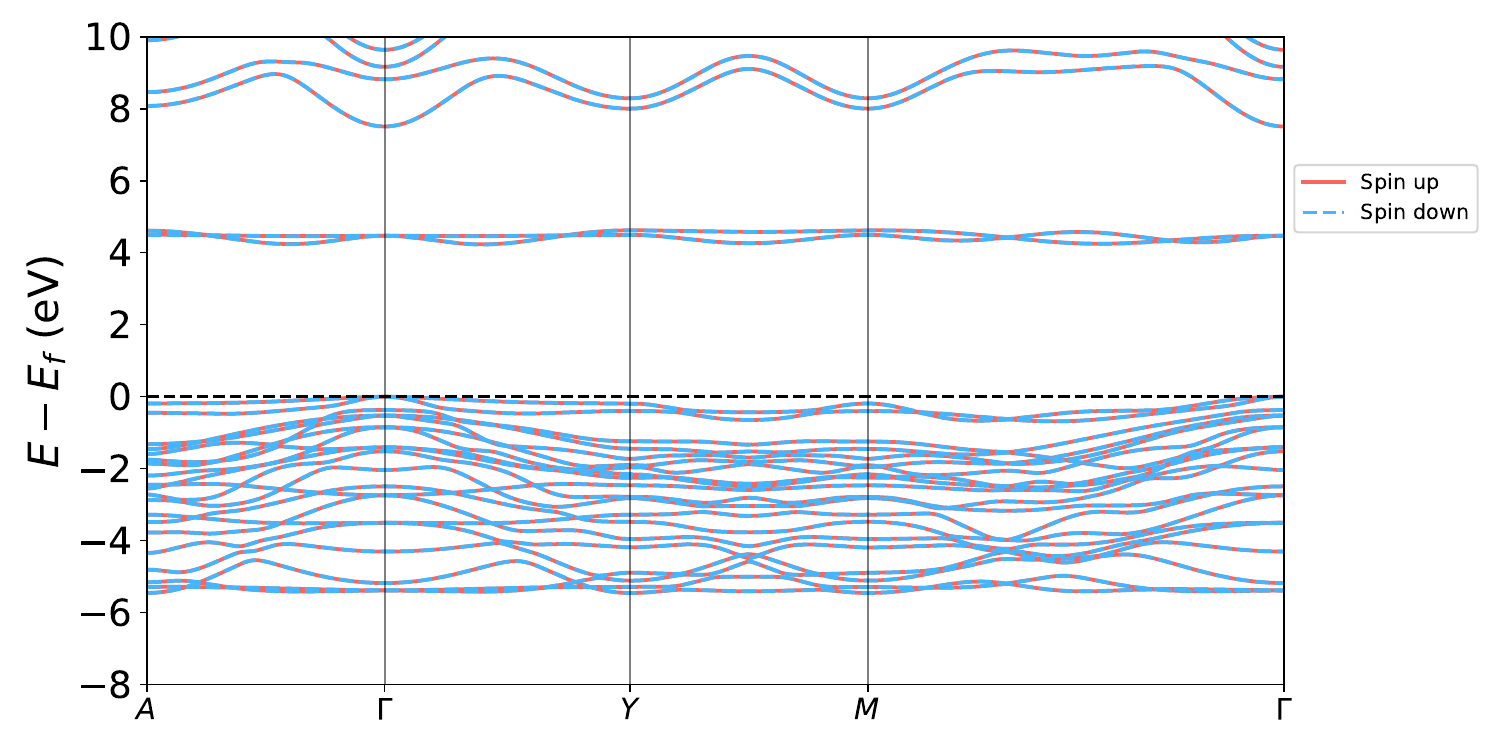}
    \caption{\footnotesize \ce{NiCl2} bulk spin-polarized band structure.}
    \label{fig:NiCl2_band_bulk}
\end{figure}

\clearpage
\newpage

\subsection{\ce{NiCl2} slab}
\begin{verbatim}
_cell_length_a                         3.46112469
_cell_length_b                         3.46112469
_cell_length_c                         40.00000000
_cell_angle_alpha                      90.000000
_cell_angle_beta                       90.000000
_cell_angle_gamma                      119.999441
_cell_volume                           'P 1'
_space_group_name_H-M_alt              'P 1'
_space_group_IT_number                 1

loop_
_space_group_symop_operation_xyz
   'x, y, z'

loop_
   _atom_site_label
   _atom_site_occupancy
   _atom_site_fract_x
   _atom_site_fract_y
   _atom_site_fract_z
   _atom_site_adp_type
   _atom_site_B_ios_or_equiv
   _atom_site_type_symbol
Cl001  1.0  0.333354274179  -0.333354274179  0.533750832292  Biso  1.000000  Cl
Ni002  1.0  -0.000000000000  -0.000000000000  0.5  Biso  1.000000  Ni
Cl003  1.0  -0.333354274179  0.333354274179  0.466249167708  Biso  1.000000  Cl
\end{verbatim}
\begin{figure}[h]
    \centering
    \includegraphics[width=\textwidth]{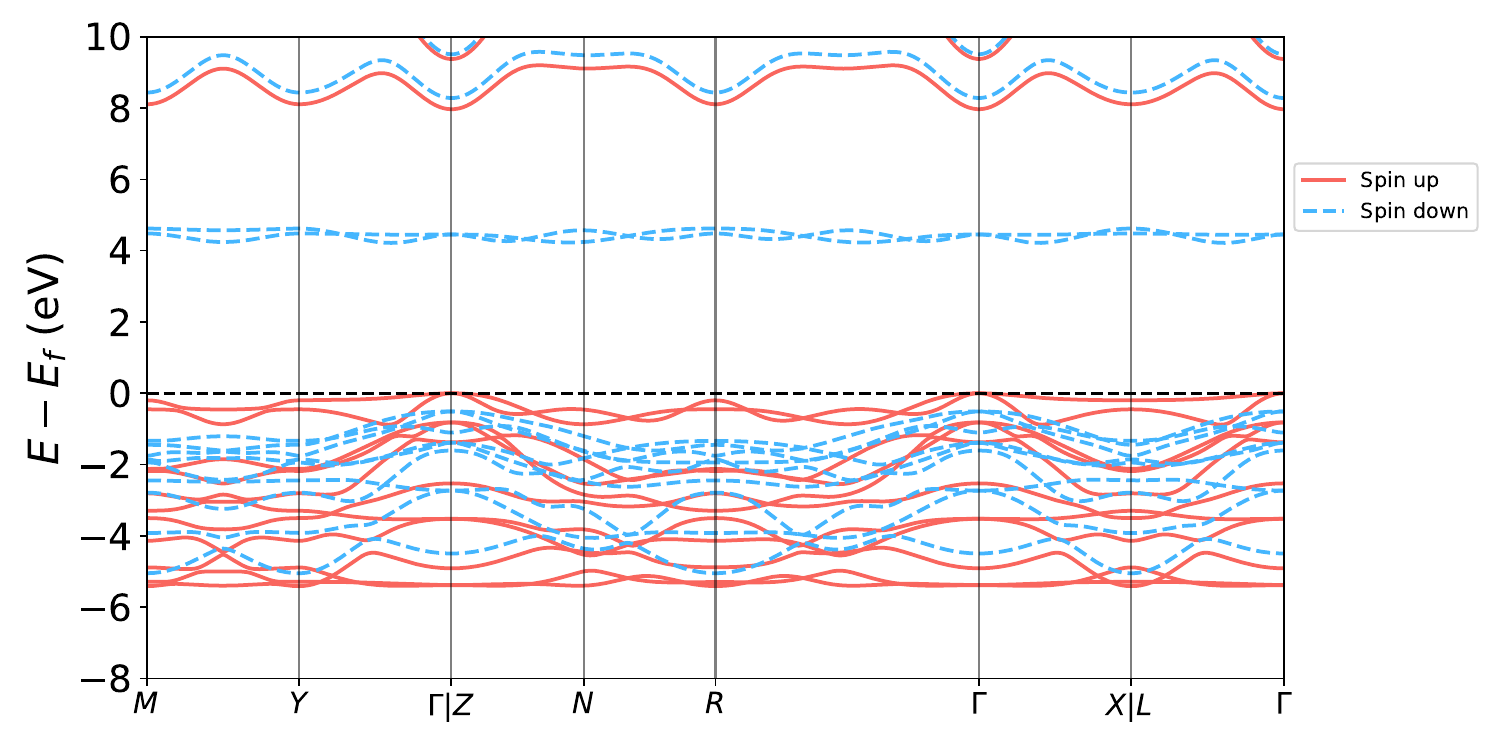}
    \caption{\footnotesize \ce{NiCl2} slab spin-polarized band structure.}
    \label{fig:NiCl2_band_slab}
\end{figure}
\clearpage
\newpage

\subsection{\ce{NiI2} bulk}
\begin{verbatim}
_cell_length_a                         3.86236000
_cell_length_b                         3.86340500
_cell_length_c                         12.39266800
_cell_angle_alpha                      90.000610
_cell_angle_beta                       90.001503
_cell_angle_gamma                      119.988136
_cell_volume                           'P 1'
_space_group_name_H-M_alt              'P 1'
_space_group_IT_number                 1

loop_
_space_group_symop_operation_xyz
   'x, y, z'

loop_
   _atom_site_label
   _atom_site_occupancy
   _atom_site_fract_x
   _atom_site_fract_y
   _atom_site_fract_z
   _atom_site_adp_type
   _atom_site_B_ios_or_equiv
   _atom_site_type_symbol
    I001   1.0   0.166667000000  -0.166667000000  -0.124268000000  Biso  1.000000   I
    I002   1.0   -0.166667000000  0.166667000000  0.124268000000  Biso  1.000000   I
    I003   1.0   0.166664000000  -0.166667000000  0.375729000000  Biso  1.000000   I
    I004   1.0   -0.166664000000  0.166667000000  -0.375729000000  Biso  1.000000   I
   Ni005   1.0   -0.500000000000  -0.500000000000  0.000000000000  Biso  1.000000  Ni
   Ni006   1.0   -0.500000000000  -0.500000000000  -0.500000000000  Biso  1.000000  Ni

\end{verbatim}
\begin{figure}[h]
    \centering
    \includegraphics[width=\textwidth]{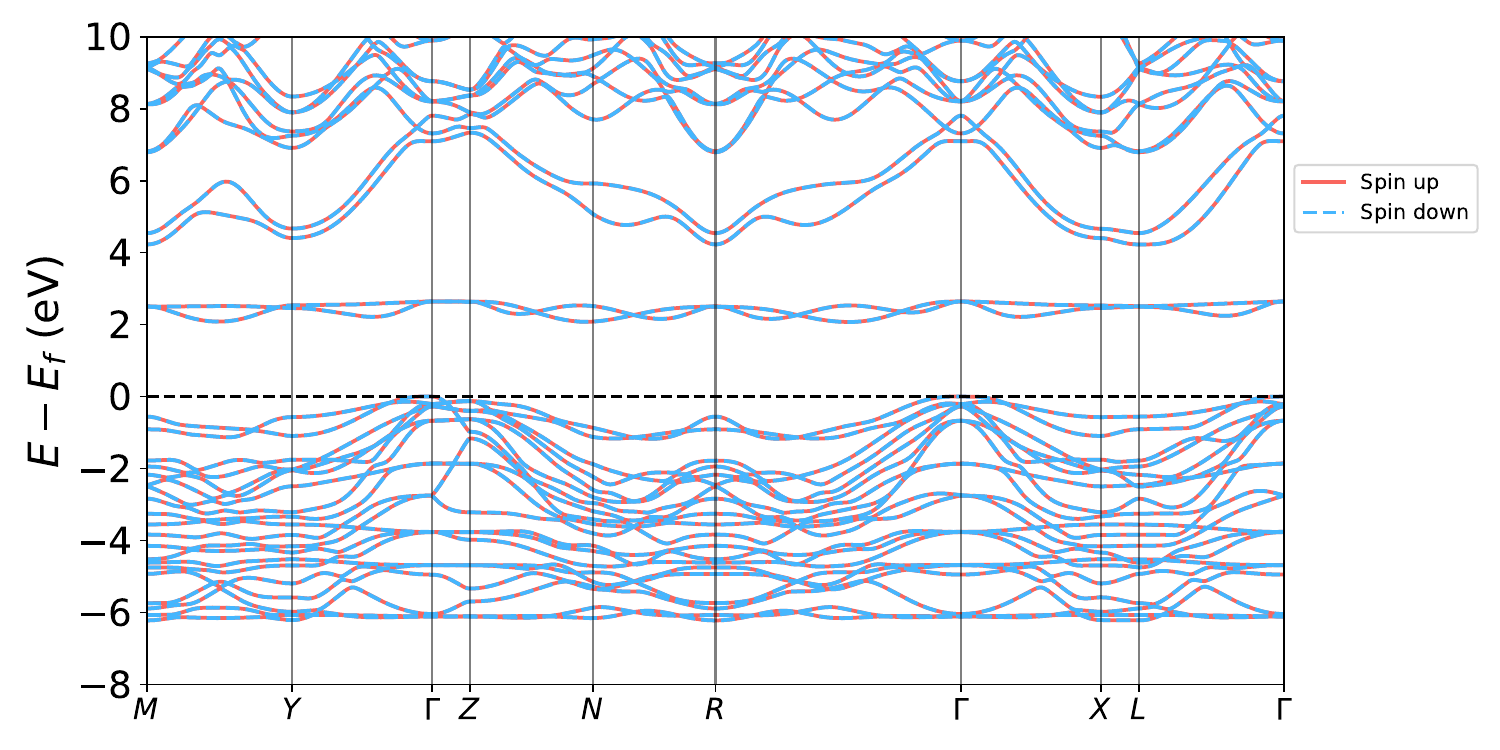}
    \caption{\footnotesize \ce{NiI2} bulk spin-polarized band structure.}
    \label{fig:NiI2_band_bulk}
\end{figure}

\clearpage
\newpage

\subsection{\ce{NiI2} slab}
\begin{verbatim}
_cell_length_a                         3.86849725
_cell_length_b                         3.86849725
_cell_length_c                         40.00000000
_cell_angle_alpha                      90.000000
_cell_angle_beta                       90.000000
_cell_angle_gamma                      119.991928
_cell_volume                           'P 1'
_space_group_name_H-M_alt              'P 1'
_space_group_IT_number                 1

loop_
_space_group_symop_operation_xyz
   'x, y, z'

loop_
   _atom_site_label
   _atom_site_occupancy
   _atom_site_fract_x
   _atom_site_fract_y
   _atom_site_fract_z
   _atom_site_adp_type
   _atom_site_B_ios_or_equiv
   _atom_site_type_symbol
I001  1.0  0.166668392053  -0.166668392053  0.538690905741  Biso  1.000000  I
Ni002  1.0  0.500000000000  0.500000000000  0.5  Biso  1.000000  Ni
I003  1.0  -0.166668392053  0.166668392053  0.461309094259  Biso  1.000000  I

\end{verbatim}
\begin{figure}[h]
    \centering
    \includegraphics[width=\textwidth]{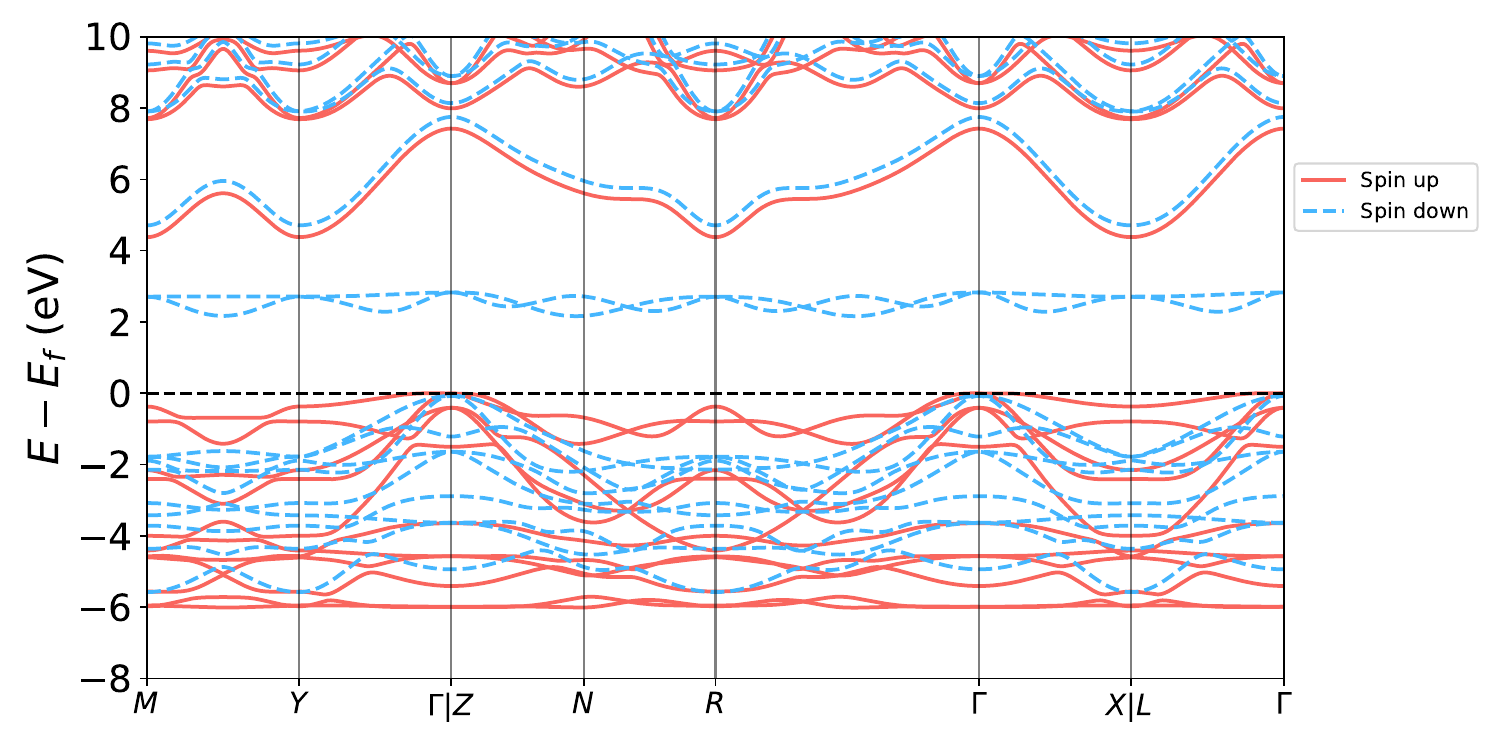}
    \caption{\footnotesize \ce{NiI2} slab spin-polarized band structure.}
    \label{fig:NiI2_band_slab}
\end{figure}
\clearpage
\newpage

\subsection{\ce{PbF2} bulk}
\begin{verbatim}
_cell_length_a 6.35055824
_cell_length_b 3.86218306
_cell_length_c 7.60072821
_cell_angle_alpha 90.000000
_cell_angle_beta 90.000000
_cell_angle_gamma 90.000000
_symmetry_space_group_name_H-M         'P 1'
_symmetry_Int_Tables_number            1

loop_
_symmetry_equiv_pos_as_xyz
   'x, y, z'

loop_
_atom_site_label
_atom_site_type_symbol
_atom_site_fract_x
_atom_site_fract_y
_atom_site_fract_z
F001 F 1.411746814127E-01 2.500000000000E-01 4.366901117232E-01
F002 F 3.588253185873E-01 -2.500000000000E-01 -6.330988827680E-02
F003 F -3.588253185873E-01 2.500000000000E-01 6.330988827680E-02
F004 F -1.411746814127E-01 -2.500000000000E-01 -4.366901117232E-01
F005 F -4.676719250726E-01 2.500000000000E-01 -3.364785429149E-01
F006 F -3.232807492740E-02 -2.500000000000E-01 1.635214570851E-01
F007 F 3.232807492740E-02 2.500000000000E-01 -1.635214570851E-01
F008 F 4.676719250726E-01 -2.500000000000E-01 3.364785429149E-01
Pb009 Pb -2.528324030705E-01 2.500000000000E-01 3.873599154845E-01
Pb010 Pb -2.471675969295E-01 -2.500000000000E-01 -1.126400845155E-01
Pb011 Pb 2.471675969295E-01 2.500000000000E-01 1.126400845155E-01
Pb012 Pb 2.528324030705E-01 -2.500000000000E-01 -3.873599154845E-01

\end{verbatim}
\begin{figure}[h]
    \centering
    \includegraphics[width=\textwidth]{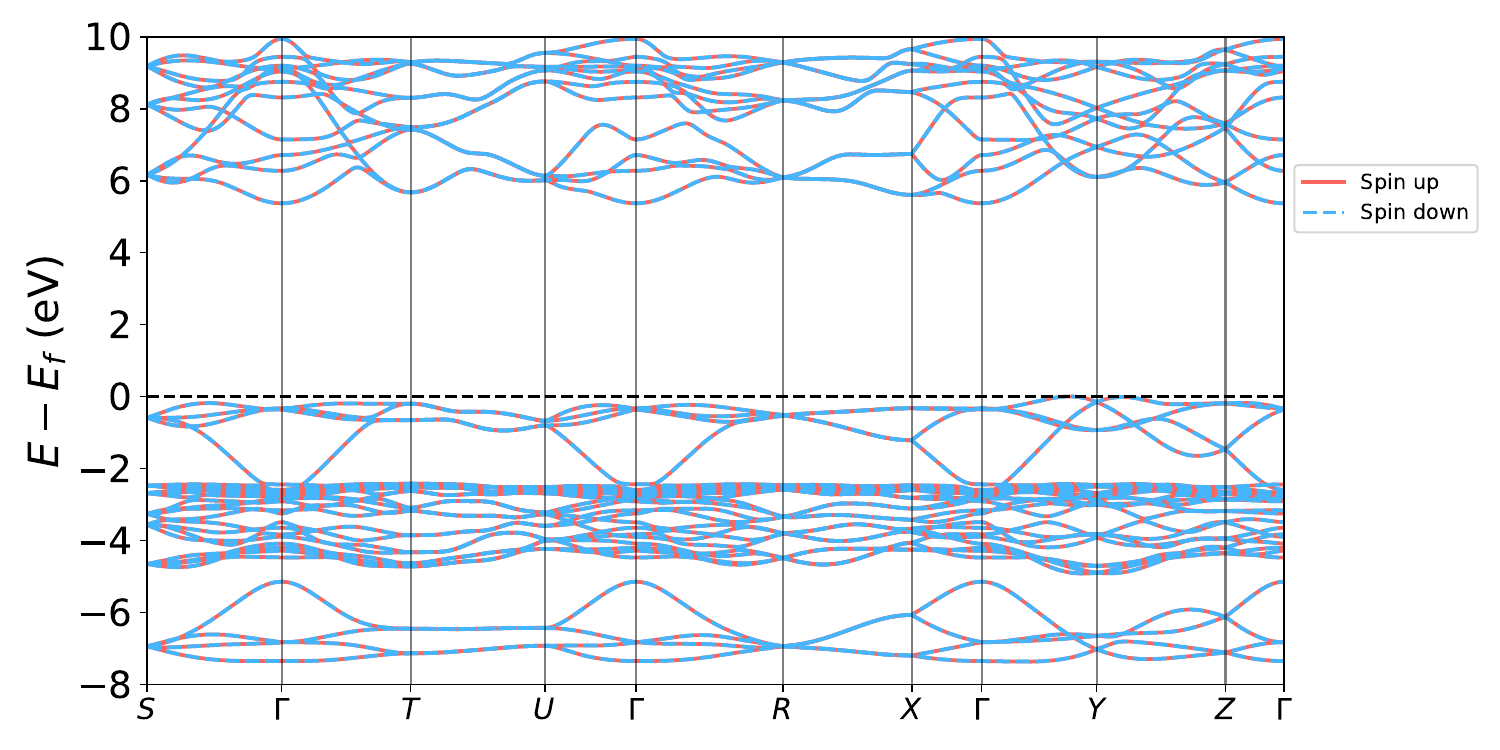}
    \caption{\footnotesize \ce{PbF2} bulk spin-polarized band structure.}
    \label{fig:PbF2_band_bulk}
\end{figure}
\clearpage
\newpage

\subsection{\ce{PbF2} slab}
\begin{verbatim}
_cell_length_a                         5.79612883
_cell_length_b                         8.37258786
_cell_length_c                         40.00000000
_cell_angle_alpha                      90.000000
_cell_angle_beta                       90.000000
_cell_angle_gamma                      90.000000
_cell_volume                           'P 1'
_space_group_name_H-M_alt              'P 1'
_space_group_IT_number                 1

loop_
_space_group_symop_operation_xyz
   'x, y, z'

loop_
   _atom_site_label
   _atom_site_occupancy
   _atom_site_fract_x
   _atom_site_fract_y
   _atom_site_fract_z
   _atom_site_adp_type
   _atom_site_B_ios_or_equiv
   _atom_site_type_symbol
F001  1.0  -0.495526213675  -0.375006241681  0.522152737235  Biso  1.000000  F
F002  1.0  -0.004473786325  0.124993758319  0.522152737235  Biso  1.000000  F
F003  1.0  0.004419048182  0.374999281068  0.477838622911  Biso  1.000000  F
F004  1.0  0.495580951818  -0.125000718932  0.477838622911  Biso  1.000000  F
Pb005  1.0  -0.250002166265  0.374976520309  0.522927105913  Biso  1.000000  Pb
Pb006  1.0  -0.249997833735  -0.125023479691  0.522927105913  Biso  1.000000  Pb
Pb007  1.0  0.249997833735  0.125023479691  0.477072894087  Biso  1.000000  Pb
Pb008  1.0  0.250002166265  -0.374976520309  0.477072894087  Biso  1.000000  Pb
F009  1.0  -0.495580951818  0.125000718932  0.522161377089  Biso  1.000000  F
F010  1.0  -0.004419048182  -0.374999281068  0.522161377089  Biso  1.000000  F
F011  1.0  0.004473786325  -0.124993758319  0.477847262765  Biso  1.000000  F
F012  1.0  0.495526213675  0.375006241681  0.477847262765  Biso  1.000000  F

\end{verbatim}
\begin{figure}[h]
    \centering
    \includegraphics[width=\textwidth]{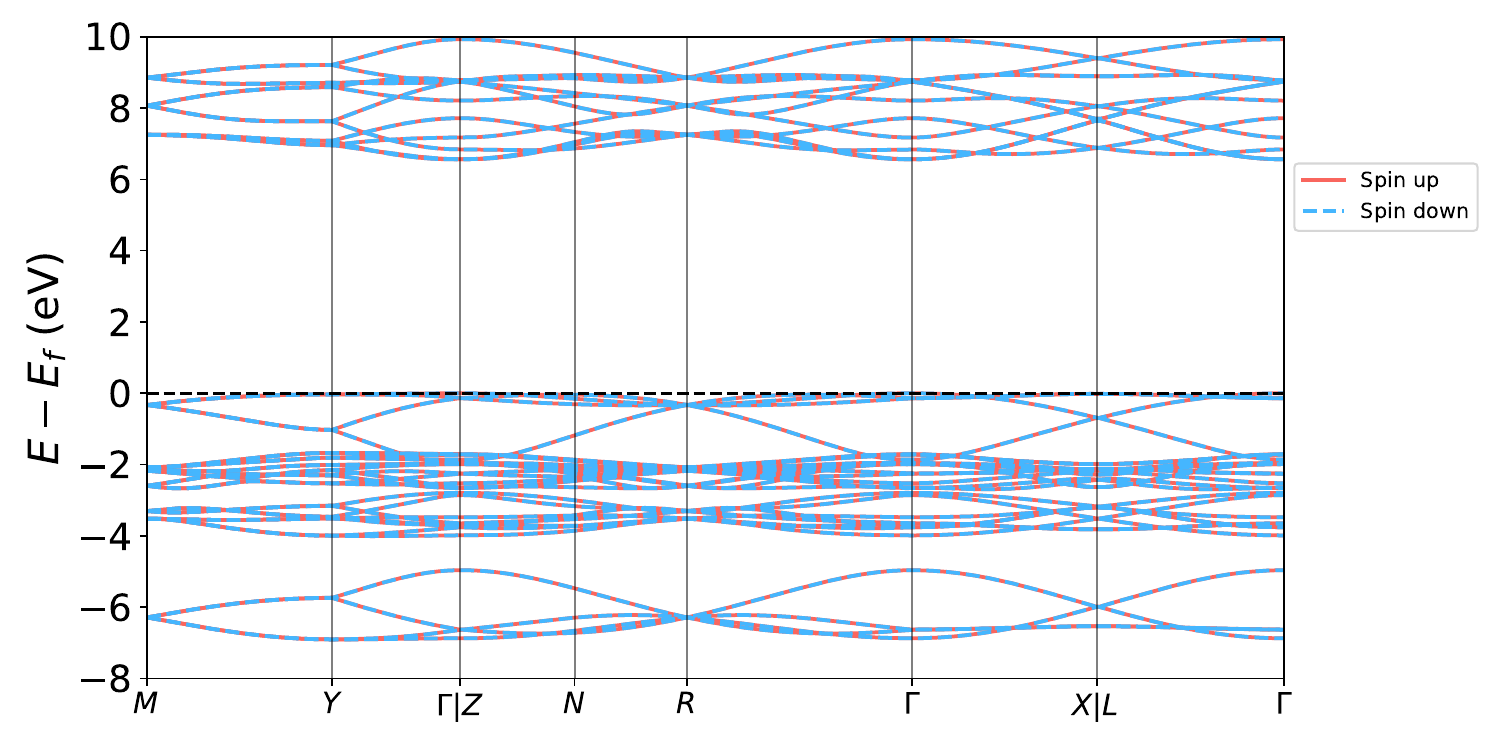}
    \caption{\footnotesize \ce{PbF2} slab spin-polarized band structure.}
    \label{fig:PbF2_band_slab}
\end{figure}
\clearpage
\newpage

\subsection{\ce{PbI2} bulk}
\begin{verbatim}
_cell_length_a 4.49292866
_cell_length_b 4.49292866
_cell_length_c 6.62741647
_cell_angle_alpha 90.000000
_cell_angle_beta 90.000000
_cell_angle_gamma 120.000000
_symmetry_space_group_name_H-M         'P 1'
_symmetry_Int_Tables_number            1

loop_
_symmetry_equiv_pos_as_xyz
   'x, y, z'

loop_
_atom_site_label
_atom_site_type_symbol
_atom_site_fract_x
_atom_site_fract_y
_atom_site_fract_z
I001 I 3.333333333333E-01 -3.333333333333E-01 -2.830675917727E-01
I002 I -3.333333333333E-01 3.333333333333E-01 2.830675917727E-01
Pb003 Pb 0.000000000000E+00 0.000000000000E+00 0.000000000000E+00

\end{verbatim}
\begin{figure}[h]
    \centering
    \includegraphics[width=\textwidth]{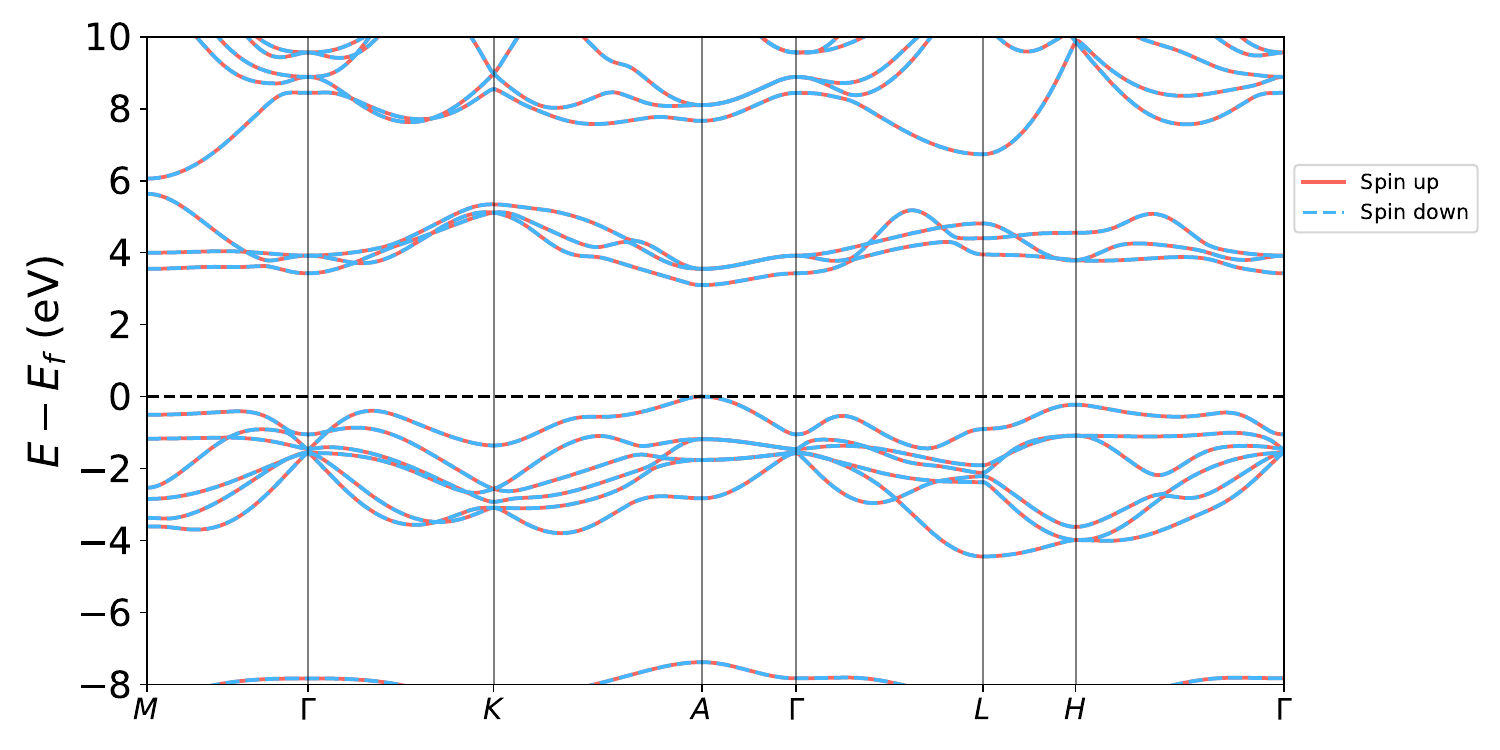}
    \caption{\footnotesize \ce{PbI2} bulk spin-polarized band structure.}
    \label{fig:PbI2_band_bulk}
\end{figure}
\clearpage
\newpage

\subsection{\ce{PbI2} slab}
\begin{verbatim}
_cell_length_a                         4.47035402
_cell_length_b                         4.47035402
_cell_length_c                         40.00000000
_cell_angle_alpha                      90.000000
_cell_angle_beta                       90.000000
_cell_angle_gamma                      120.000000
_cell_volume                           'P 1'
_space_group_name_H-M_alt              'P 1'
_space_group_IT_number                 1

loop_
_space_group_symop_operation_xyz
   'x, y, z'

loop_
   _atom_site_label
   _atom_site_occupancy
   _atom_site_fract_x
   _atom_site_fract_y
   _atom_site_fract_z
   _atom_site_adp_type
   _atom_site_B_ios_or_equiv
   _atom_site_type_symbol
I001  1.0  -0.333333333333  0.333333333333  0.5476653811459999  Biso  1.000000  I
Pb002  1.0  0.000000000000  0.000000000000  0.5  Biso  1.000000  Pb
I003  1.0  0.333333333333  -0.333333333333  0.452334618854  Biso  1.000000  I

\end{verbatim}
\begin{figure}[h]
    \centering
    \includegraphics[width=\textwidth]{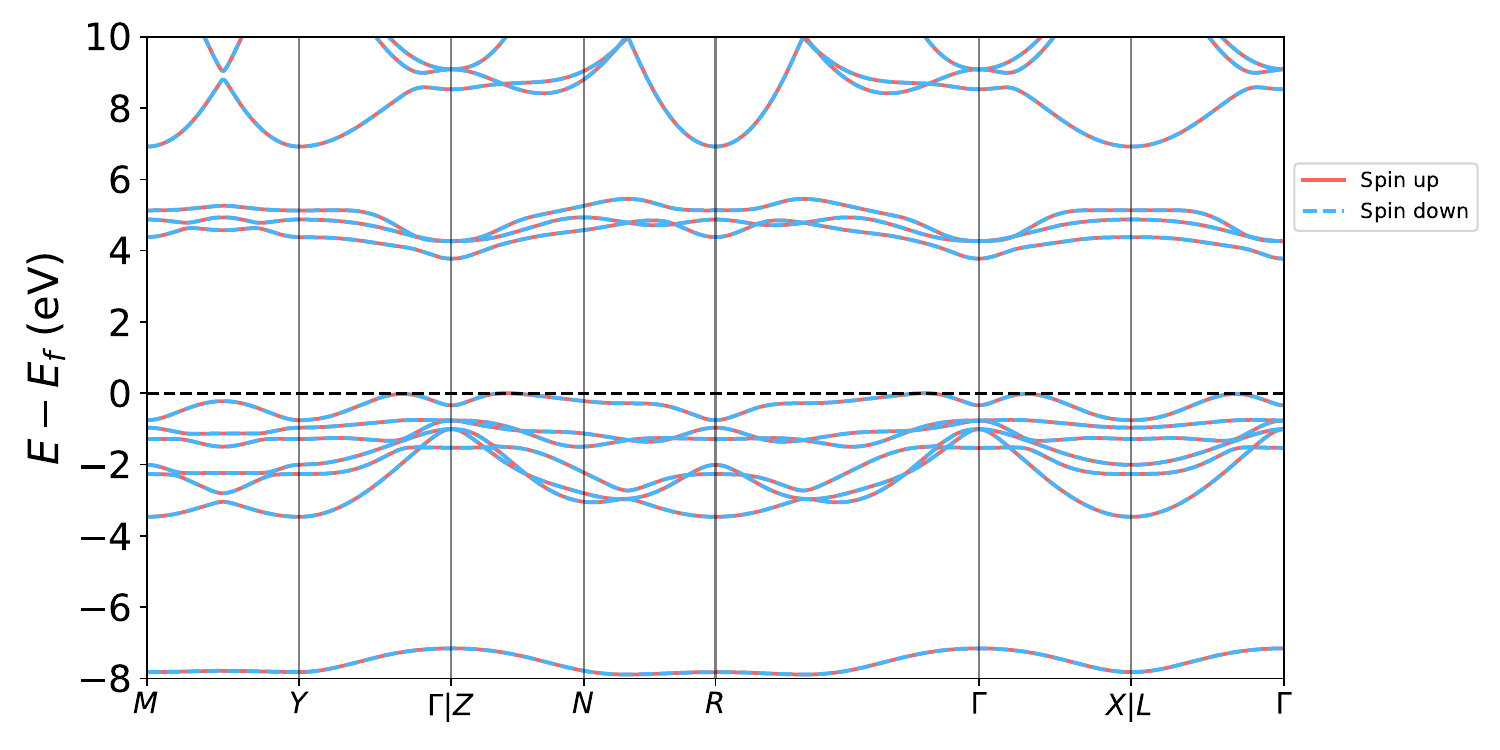}
    \caption{\footnotesize \ce{PbI2} slab spin-polarized band structure.}
    \label{fig:PbI2_band_slab}
\end{figure}
\clearpage
\newpage

\subsection{\ce{PdCl2} bulk}
\begin{verbatim}
_cell_length_a 5.51616147
_cell_length_b 3.60018215
_cell_length_c 6.49381098
_cell_angle_alpha 90.000000
_cell_angle_beta 107.326474
_cell_angle_gamma 90.000000
_symmetry_space_group_name_H-M         'P 1'
_symmetry_Int_Tables_number            1

loop_
_symmetry_equiv_pos_as_xyz
   'x, y, z'

loop_
_atom_site_label
_atom_site_type_symbol
_atom_site_fract_x
_atom_site_fract_y
_atom_site_fract_z
Cl001 Cl -2.652707007453E-01 -2.533018055391E-01 -3.163280104975E-01
Cl002 Cl 2.652707007453E-01 2.466981944609E-01 -1.836719895025E-01
Cl003 Cl 2.652707007453E-01 2.533018055391E-01 3.163280104975E-01
Cl004 Cl -2.652707007453E-01 -2.466981944609E-01 1.836719895025E-01
Pd005 Pd 3.582808947595E-36 6.352783003696E-36 -1.637591001192E-37
Pd006 Pd 2.214503542550E-18 -5.000000000000E-01 -5.000000000000E-01

\end{verbatim}
\begin{figure}[h]
    \centering
    \includegraphics[width=\textwidth]{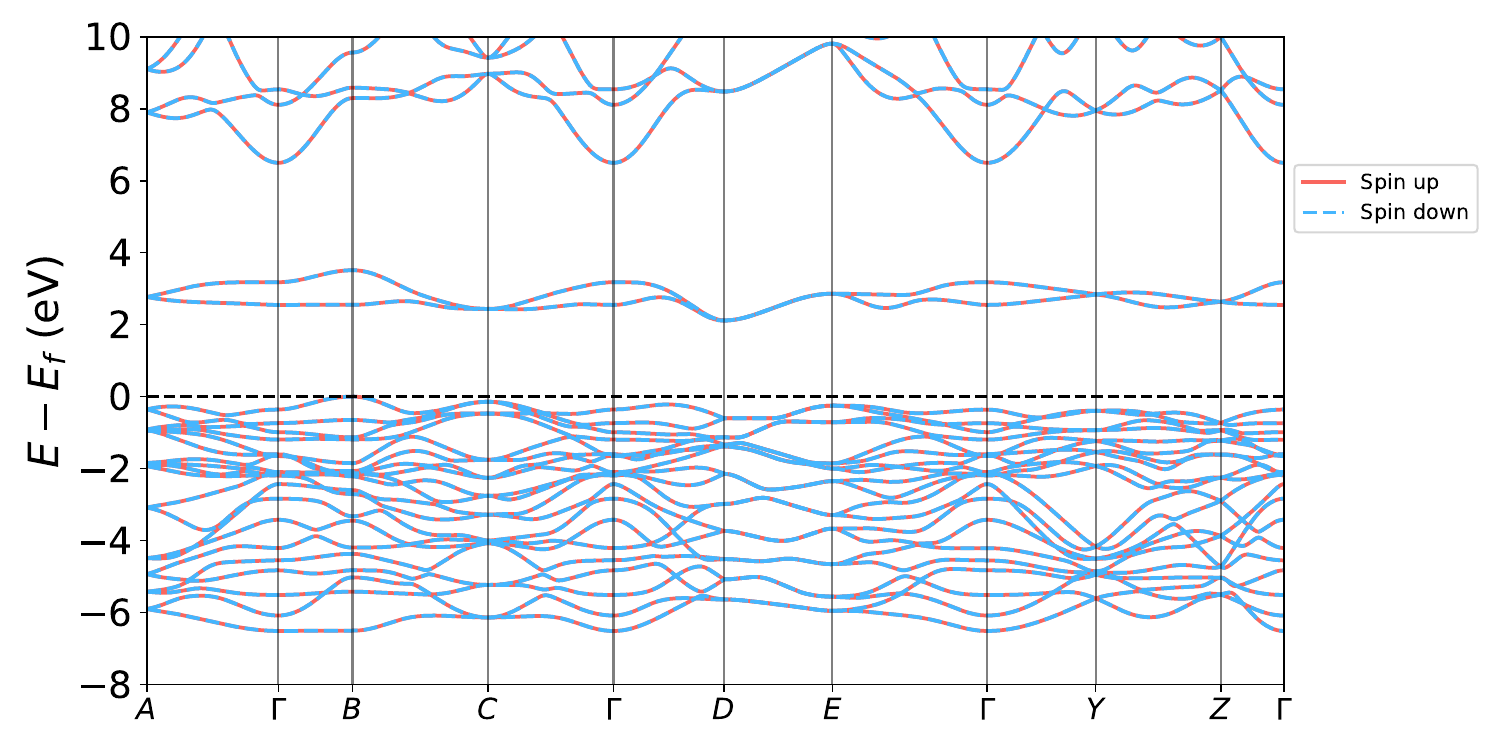}
    \caption{\footnotesize \ce{PdCl2} bulk spin-polarized band structure.}
    \label{fig:PdCl2_band_bulk}
\end{figure}
\clearpage
\newpage

\subsection{\ce{PdCl2} slab}
\begin{verbatim}
_cell_length_a                         3.57155279
_cell_length_b                         6.45138758
_cell_length_c                         40.00000000
_cell_angle_alpha                      90.000000
_cell_angle_beta                       90.000000
_cell_angle_gamma                      90.000000
_cell_volume                           'P 1'
_space_group_name_H-M_alt              'P 1'
_space_group_IT_number                 1

loop_
_space_group_symop_operation_xyz
   'x, y, z'

loop_
   _atom_site_label
   _atom_site_occupancy
   _atom_site_fract_x
   _atom_site_fract_y
   _atom_site_fract_z
   _atom_site_adp_type
   _atom_site_B_ios_or_equiv
   _atom_site_type_symbol
Cl001  1.0  0.249680726574  -0.250098246762  0.53556089477  Biso  1.000000  Cl
Cl002  1.0  0.250319273426  0.249901753238  0.53556089477  Biso  1.000000  Cl
Pd003  1.0  -0.000000000000  0.000000000000  0.5  Biso  1.000000  Pd
Pd004  1.0  0.500000000000  -0.500000000000  0.5  Biso  1.000000  Pd
Cl005  1.0  -0.250319273426  -0.249901753238  0.46443910523  Biso  1.000000  Cl
Cl006  1.0  -0.249680726574  0.250098246762  0.46443910523  Biso  1.000000  Cl

\end{verbatim}
\begin{figure}[h]
    \centering
    \includegraphics[width=\textwidth]{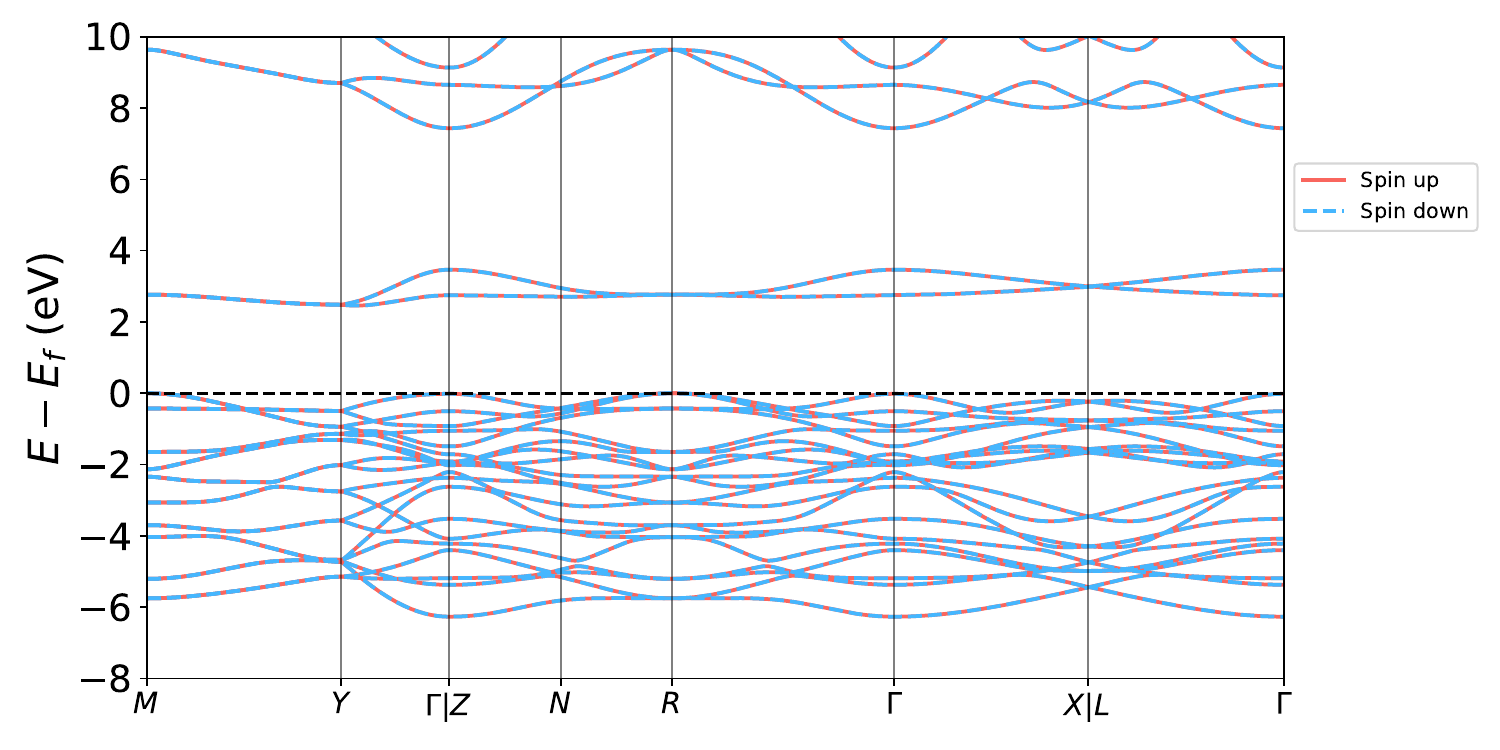}
    \caption{\footnotesize \ce{PdCl2} slab spin-polarized band structure.}
    \label{fig:PdCl2_band_slab}
\end{figure}
\clearpage
\newpage

\subsection{RbCl bulk}
\begin{verbatim}
_cell_length_a 3.84146228
_cell_length_b 3.84146228
_cell_length_c 3.84093796
_cell_angle_alpha 90.000000
_cell_angle_beta 90.000000
_cell_angle_gamma 90.000000
_symmetry_space_group_name_H-M         'P 1'
_symmetry_Int_Tables_number            1

loop_
_symmetry_equiv_pos_as_xyz
   'x, y, z'

loop_
_atom_site_label
_atom_site_type_symbol
_atom_site_fract_x
_atom_site_fract_y
_atom_site_fract_z
Cl001 Cl -5.000000000000E-01 -5.000000000000E-01 -5.000000000000E-01
Rb002 Rb 0.000000000000E+00 0.000000000000E+00 0.000000000000E+00

\end{verbatim}
\begin{figure}[h]
    \centering
    \includegraphics[width=\textwidth]{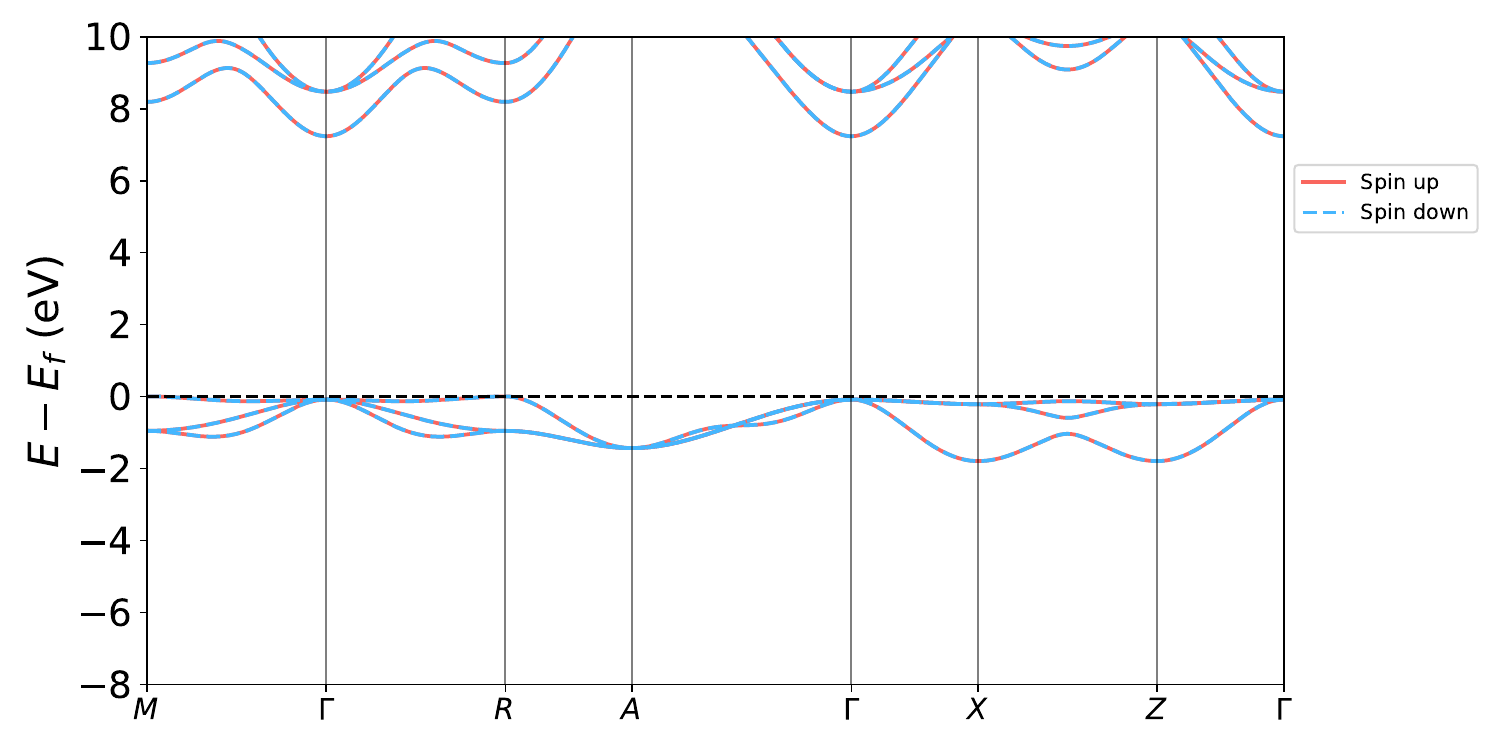}
    \caption{\footnotesize RbCl bulk spin-polarized band structure.}
    \label{fig:RbCl_band_bulk}
\end{figure}
\clearpage
\newpage

\subsection{RbCl slab}
\begin{verbatim}
_cell_length_a                         4.47257143
_cell_length_b                         4.47257143
_cell_length_c                         40.00000000
_cell_angle_alpha                      90.000000
_cell_angle_beta                       90.000000
_cell_angle_gamma                      90.000000
_cell_volume                           'P 1'
_space_group_name_H-M_alt              'P 1'
_space_group_IT_number                 1

loop_
_space_group_symop_operation_xyz
   'x, y, z'

loop_
   _atom_site_label
   _atom_site_occupancy
   _atom_site_fract_x
   _atom_site_fract_y
   _atom_site_fract_z
   _atom_site_adp_type
   _atom_site_B_ios_or_equiv
   _atom_site_type_symbol
Cl001  1.0  -0.500000000000  -0.500000000000  0.50000024889  Biso  1.000000  Cl
Rb002  1.0  0.000000000000  0.000000000000  0.49999975111  Biso  1.000000  Rb

\end{verbatim}
\begin{figure}[h]
    \centering
    \includegraphics[width=\textwidth]{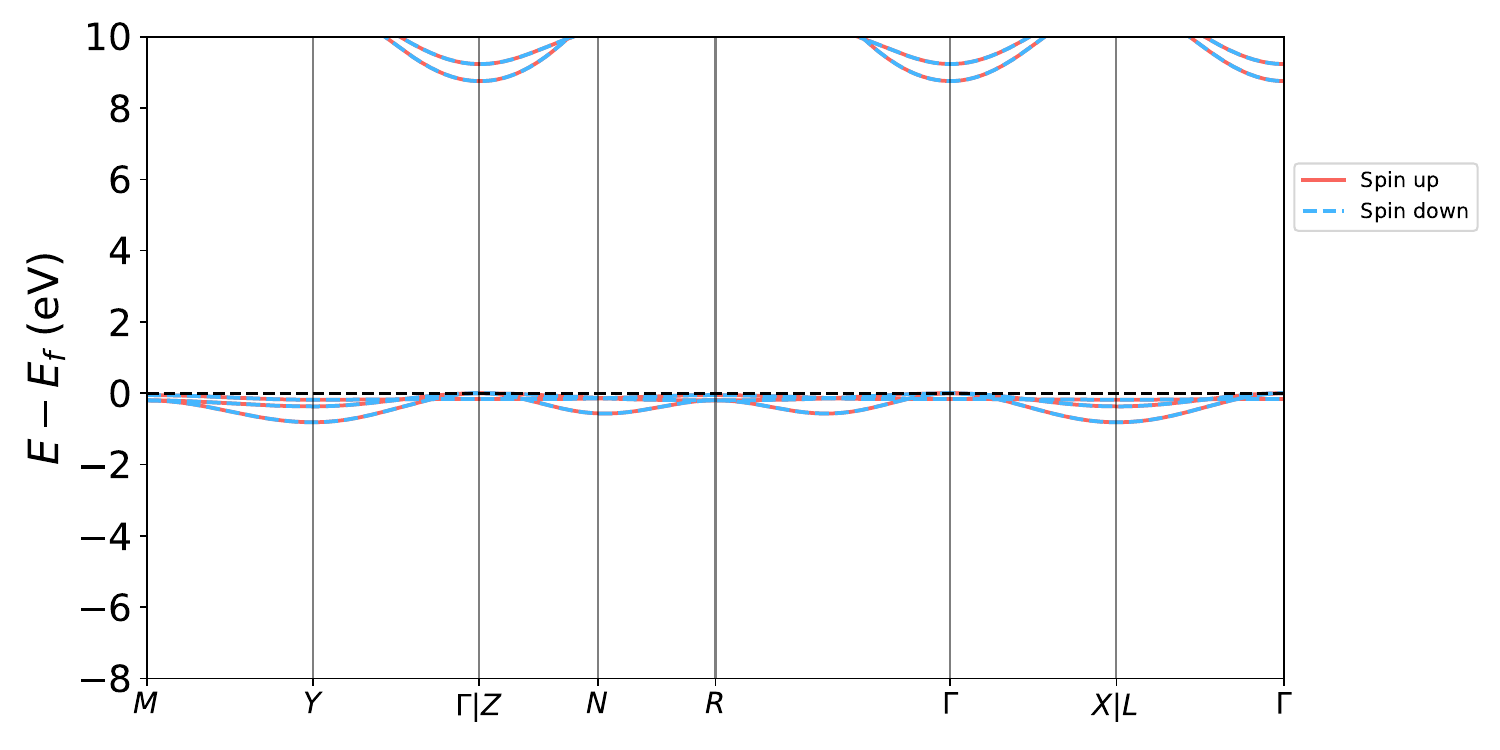}
    \caption{\footnotesize RbCl slab spin-polarized band structure.}
    \label{fig:RbCl_band_slab}
\end{figure}

\clearpage
\newpage

\subsection{ScCl bulk}
\begin{verbatim}
_cell_length_a 3.32630691
_cell_length_b 3.36017393
_cell_length_c 8.73770248
_cell_angle_alpha 100.261608
_cell_angle_beta 90.039281
_cell_angle_gamma 119.638739
_symmetry_space_group_name_H-M         'P 1'
_symmetry_Int_Tables_number            1

loop_
_symmetry_equiv_pos_as_xyz
   'x, y, z'

loop_
_atom_site_label
_atom_site_type_symbol
_atom_site_fract_x
_atom_site_fract_y
_atom_site_fract_z
Sc001 Sc 1.171285335255E-01 2.327923215603E-01 -1.420309982838E-01
Sc002 Sc -1.171285335255E-01 -2.327923215603E-01 1.420309982838E-01
Cl003 Cl 3.875456928210E-01 -2.256545119749E-01 -3.323584458797E-01
Cl004 Cl -3.875456928210E-01 2.256545119749E-01 3.323584458797E-01

\end{verbatim}
\begin{figure}[h]
    \centering
    \includegraphics[width=\textwidth]{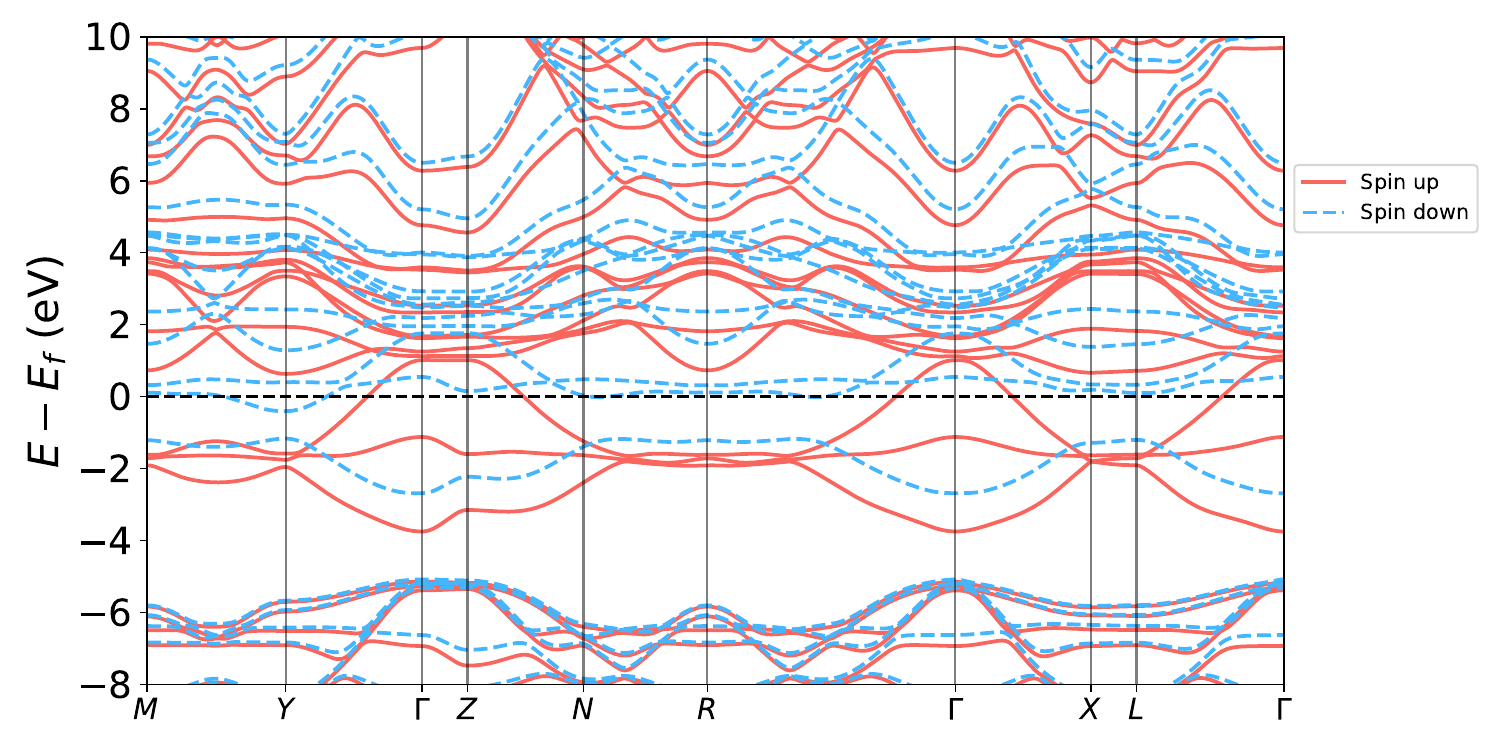}
    \caption{\footnotesize ScCl bulk spin-polarized band structure.}
    \label{fig:ScCl_band_bulk}
\end{figure}
\clearpage
\newpage

\subsection{ScCl slab}
\begin{verbatim}
_cell_length_a                         3.20738122
_cell_length_b                         5.56049461
_cell_length_c                         40.00000000
_cell_angle_alpha                      90.000000
_cell_angle_beta                       90.000000
_cell_angle_gamma                      90.001343
_cell_volume                           'P 1'
_space_group_name_H-M_alt              'P 1'
_space_group_IT_number                 1

loop_
_space_group_symop_operation_xyz
   'x, y, z'

loop_
   _atom_site_label
   _atom_site_occupancy
   _atom_site_fract_x
   _atom_site_fract_y
   _atom_site_fract_z
   _atom_site_adp_type
   _atom_site_B_ios_or_equiv
   _atom_site_type_symbol
Sc001  1.0  -0.118509843423  -0.210055885615  0.521064120185  Biso  1.000000  Sc
Sc002  1.0  0.381568657445  0.289980074548  0.521148733307  Biso  1.000000  Sc
Cl003  1.0  0.381522699170  -0.042606451770  0.478867453294  Biso  1.000000  Cl
Cl004  1.0  -0.118454035808  0.456953513478  0.478910925698  Biso  1.000000  Cl

\end{verbatim}
\begin{figure}[h]
    \centering
    \includegraphics[width=\textwidth]{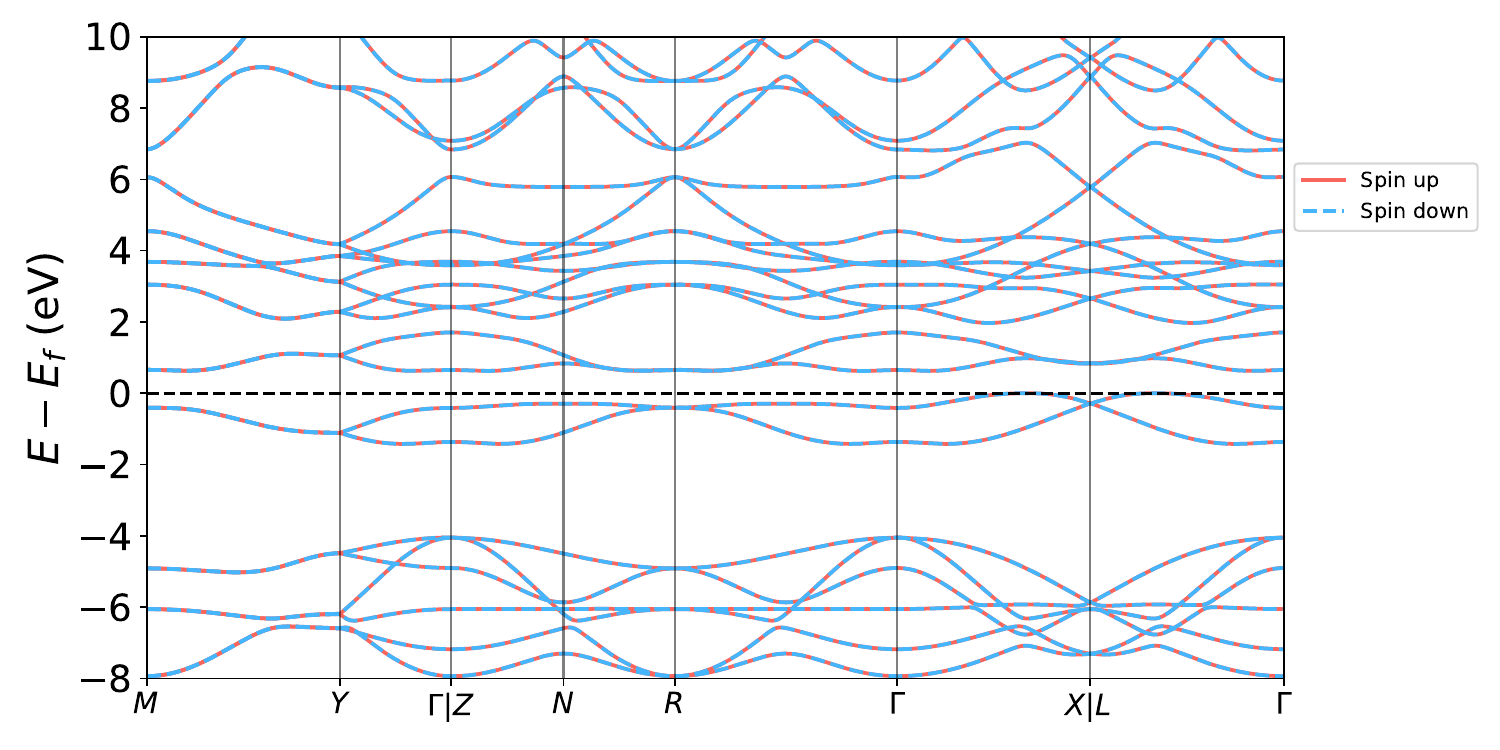}
    \caption{\footnotesize ScCl slab spin-polarized band structure.}
    \label{fig:ScCl_band_slab}
\end{figure}
\clearpage
\newpage

\subsection{\ce{SnCl2} bulk}
\begin{verbatim}
_cell_length_a 9.05454440
_cell_length_b 4.30769982
_cell_length_c 15.11344186
_cell_angle_alpha 90.000000
_cell_angle_beta 89.999932
_cell_angle_gamma 90.000000
_symmetry_space_group_name_H-M         'P 1'
_symmetry_Int_Tables_number            1

loop_
_symmetry_equiv_pos_as_xyz
   'x, y, z'

loop_
_atom_site_label
_atom_site_type_symbol
_atom_site_fract_x
_atom_site_fract_y
_atom_site_fract_z
Sn001 Sn 7.945897025968E-02 -2.500472523954E-01 -3.650992583126E-01
Sn002 Sn -7.945897025968E-02 2.499527476046E-01 -1.349007416874E-01
Sn003 Sn -7.945897025968E-02 2.500472523954E-01 3.650992583126E-01
Sn004 Sn 7.945897025968E-02 -2.499527476046E-01 1.349007416874E-01
Sn005 Sn 4.205389234753E-01 -2.499750903518E-01 3.849024073860E-01
Sn006 Sn -4.205389234753E-01 2.500249096482E-01 1.150975926140E-01
Sn007 Sn -4.205389234753E-01 2.499750903518E-01 -3.849024073860E-01
Sn008 Sn 4.205389234753E-01 -2.500249096482E-01 -1.150975926140E-01
Cl009 Cl 7.615964058208E-02 -2.500023340199E-01 -7.228258023893E-02
Cl010 Cl -7.615964058208E-02 2.499976659801E-01 -4.277174197611E-01
Cl011 Cl -7.615964058208E-02 2.500023340199E-01 7.228258023893E-02
Cl012 Cl 7.615964058208E-02 -2.499976659801E-01 4.277174197611E-01
Cl013 Cl 1.554679558684E-01 2.500199228165E-01 -2.398299719985E-01
Cl014 Cl -1.554679558684E-01 -2.499800771835E-01 -2.601700280015E-01
Cl015 Cl -1.554679558684E-01 -2.500199228165E-01 2.398299719985E-01
Cl016 Cl 1.554679558684E-01 2.499800771835E-01 2.601700280015E-01
Cl017 Cl 3.445241111522E-01 2.500012165825E-01 1.017754701218E-02
Cl018 Cl -3.445241111522E-01 -2.499987834175E-01 4.898224529878E-01
Cl019 Cl -3.445241111522E-01 -2.500012165825E-01 -1.017754701218E-02
Cl020 Cl 3.445241111522E-01 2.499987834175E-01 -4.898224529878E-01
Cl021 Cl 4.238500311370E-01 -2.500126740751E-01 1.777171466092E-01
Cl022 Cl -4.238500311370E-01 2.499873259249E-01 3.222828533908E-01
Cl023 Cl -4.238500311370E-01 2.500126740751E-01 -1.777171466092E-01
Cl024 Cl 4.238500311370E-01 -2.499873259249E-01 -3.222828533908E-01

\end{verbatim}
\begin{figure}[h]
    \centering
    \includegraphics[width=\textwidth]{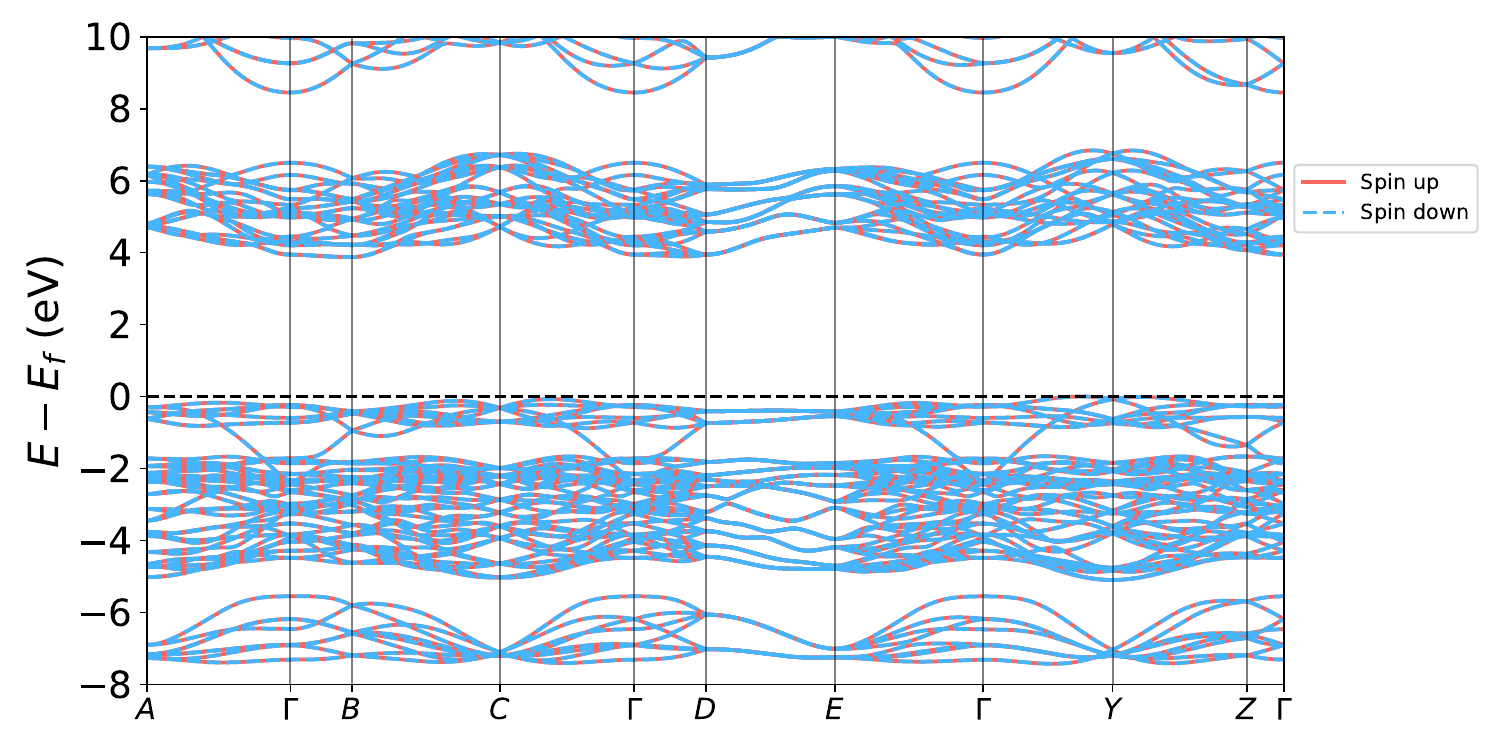}
    \caption{\footnotesize \ce{SnCl2} bulk spin-polarized band structure.}
    \label{fig:SnCl2_band_bulk}
\end{figure}
\clearpage
\newpage

\subsection{\ce{SnCl2} slab}
\begin{verbatim}
_cell_length_a                         4.00856891
_cell_length_b                         15.59771150
_cell_length_c                         40.00000000
_cell_angle_alpha                      90.000000
_cell_angle_beta                       90.000000
_cell_angle_gamma                      90.000000
_cell_volume                           'P 1'
_space_group_name_H-M_alt              'P 1'
_space_group_IT_number                 1

loop_
_space_group_symop_operation_xyz
   'x, y, z'

loop_
   _atom_site_label
   _atom_site_occupancy
   _atom_site_fract_x
   _atom_site_fract_y
   _atom_site_fract_z
   _atom_site_adp_type
   _atom_site_B_ios_or_equiv
   _atom_site_type_symbol
Cl001  1.0  -0.249981183567  0.499192059599  0.536061110046  Biso  1.000000  Cl
Cl002  1.0  -0.250018816433  -0.000807940401  0.536061110046  Biso  1.000000  Cl
Sn003  1.0  0.249942189440  0.117953716287  0.509958371222  Biso  1.000000  Sn
Sn004  1.0  0.250057810561  -0.382046283713  0.509958371222  Biso  1.000000  Sn
Cl005  1.0  0.250048270940  0.319616798982  0.528573444324  Biso  1.000000  Cl
Cl006  1.0  0.249951729060  -0.180383201018  0.528573444324  Biso  1.000000  Cl
Cl007  1.0  -0.249951729060  0.180383201018  0.471426555676  Biso  1.000000  Cl
Cl008  1.0  -0.250048270940  -0.319616798982  0.471426555676  Biso  1.000000  Cl
Sn009  1.0  -0.250057810561  0.382046283713  0.490041628778  Biso  1.000000  Sn
Sn010  1.0  -0.249942189440  -0.117953716287  0.490041628778  Biso  1.000000  Sn
Cl011  1.0  0.250018816433  0.000807940401  0.463938889954  Biso  1.000000  Cl
Cl012  1.0  0.249981183567  -0.499192059599  0.463938889954  Biso  1.000000  Cl

\end{verbatim}
\begin{figure}[h]
    \centering
    \includegraphics[width=\textwidth]{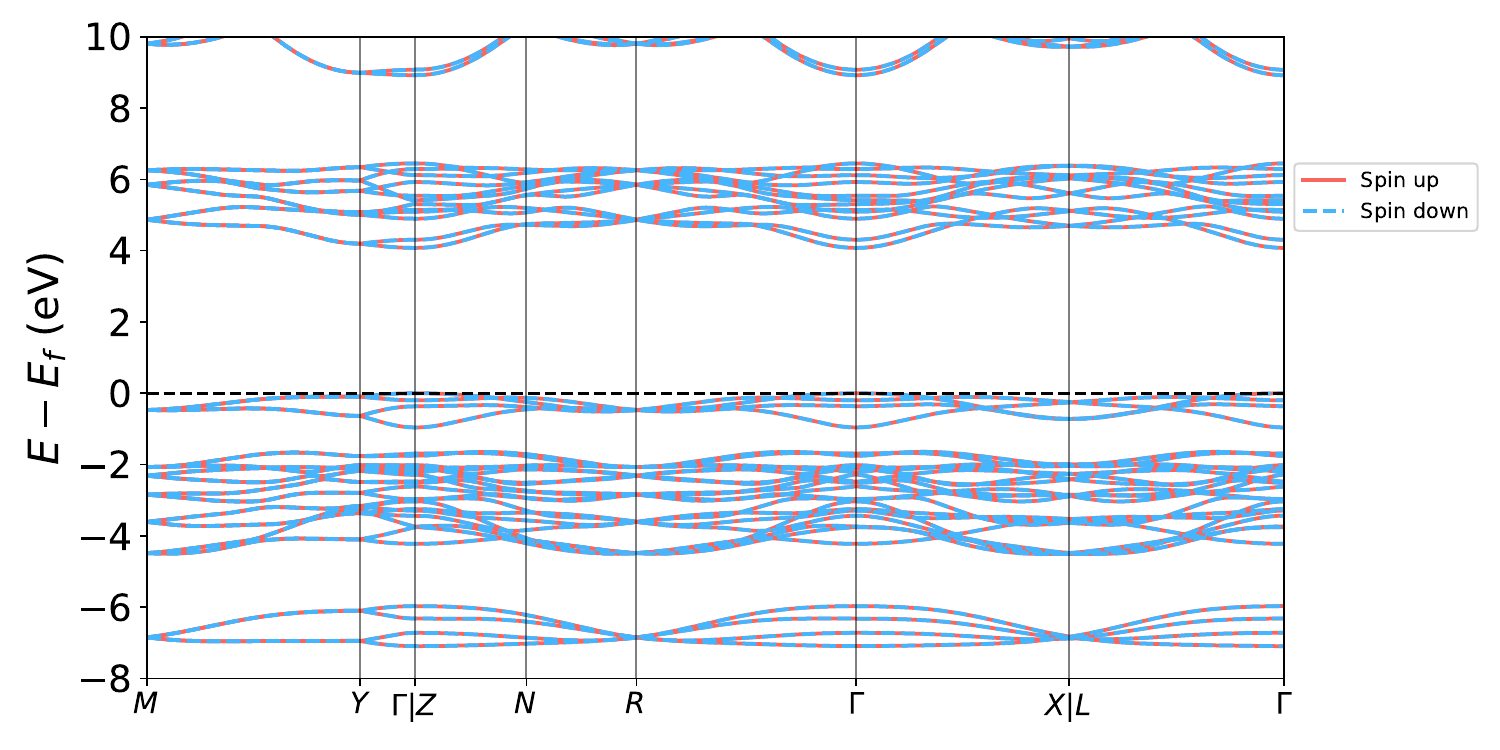}
    \caption{\footnotesize \ce{SnCl2} slab spin-polarized band structure.}
    \label{fig:SnCl2_band_slab}
\end{figure}
\clearpage
\newpage

\subsection{\ce{SrI2} bulk}
\begin{verbatim}
_cell_length_a 5.50381645
_cell_length_b 5.50381645
_cell_length_c 5.50381645
_cell_angle_alpha 120.058316
_cell_angle_beta 119.937667
_cell_angle_gamma 90.003511
_symmetry_space_group_name_H-M         'P 1'
_symmetry_Int_Tables_number            1

loop_
_symmetry_equiv_pos_as_xyz
   'x, y, z'

loop_
_atom_site_label
_atom_site_type_symbol
_atom_site_fract_x
_atom_site_fract_y
_atom_site_fract_z
Sr001 Sr -5.000000000000E-01 3.453357968891E-18 5.000000000000E-01
I002 I -2.497757782728E-01 -2.497757782728E-01 3.269719563730E-17
I003 I 2.497757782728E-01 2.497757782728E-01 -3.317686551286E-17

\end{verbatim}
\begin{figure}[h]
    \centering
    \includegraphics[width=\textwidth]{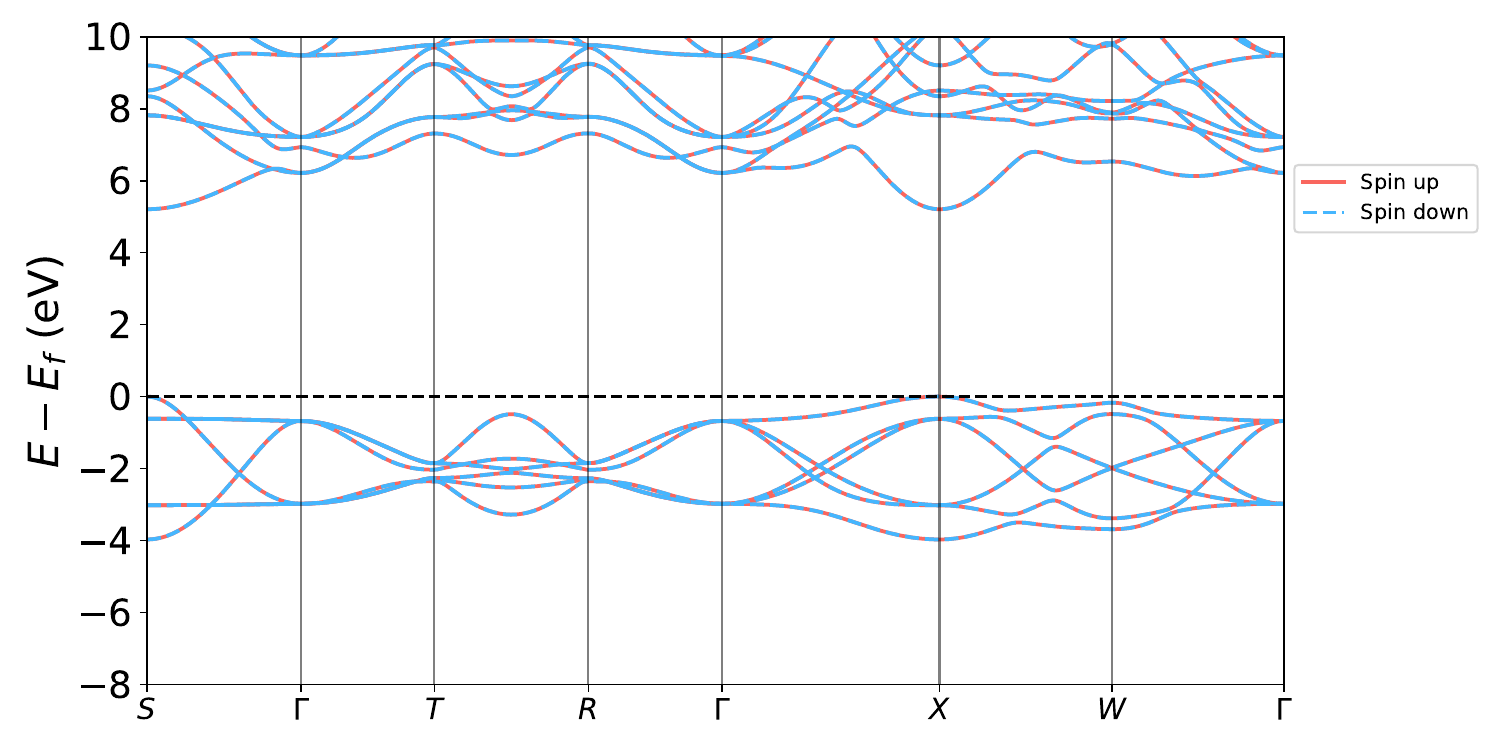}
    \caption{\footnotesize \ce{SrI2} bulk spin-polarized band structure.}
    \label{fig:SrI2_band_bulk}
\end{figure}

\clearpage
\newpage

\subsection{\ce{SrI2} slab}
\begin{verbatim}
_cell_length_a                         5.05213155
_cell_length_b                         4.59010060
_cell_length_c                         40.00000000
_cell_angle_alpha                      90.000000
_cell_angle_beta                       90.000000
_cell_angle_gamma                      90.000000
_cell_volume                           'P 1'
_space_group_name_H-M_alt              'P 1'
_space_group_IT_number                 1

loop_
_space_group_symop_operation_xyz
   'x, y, z'

loop_
   _atom_site_label
   _atom_site_occupancy
   _atom_site_fract_x
   _atom_site_fract_y
   _atom_site_fract_z
   _atom_site_adp_type
   _atom_site_B_ios_or_equiv
   _atom_site_type_symbol
I001  1.0  -0.000000000000  0.000000000000  0.553081480887  Biso  1.000000  I
Sr002  1.0  -0.000000000000  -0.500000000000  0.5  Biso  1.000000  Sr
I003  1.0  0.000000000000  -0.000000000000  0.446918519113  Biso  1.000000  I

\end{verbatim}
\begin{figure}[h]
    \centering
    \includegraphics[width=\textwidth]{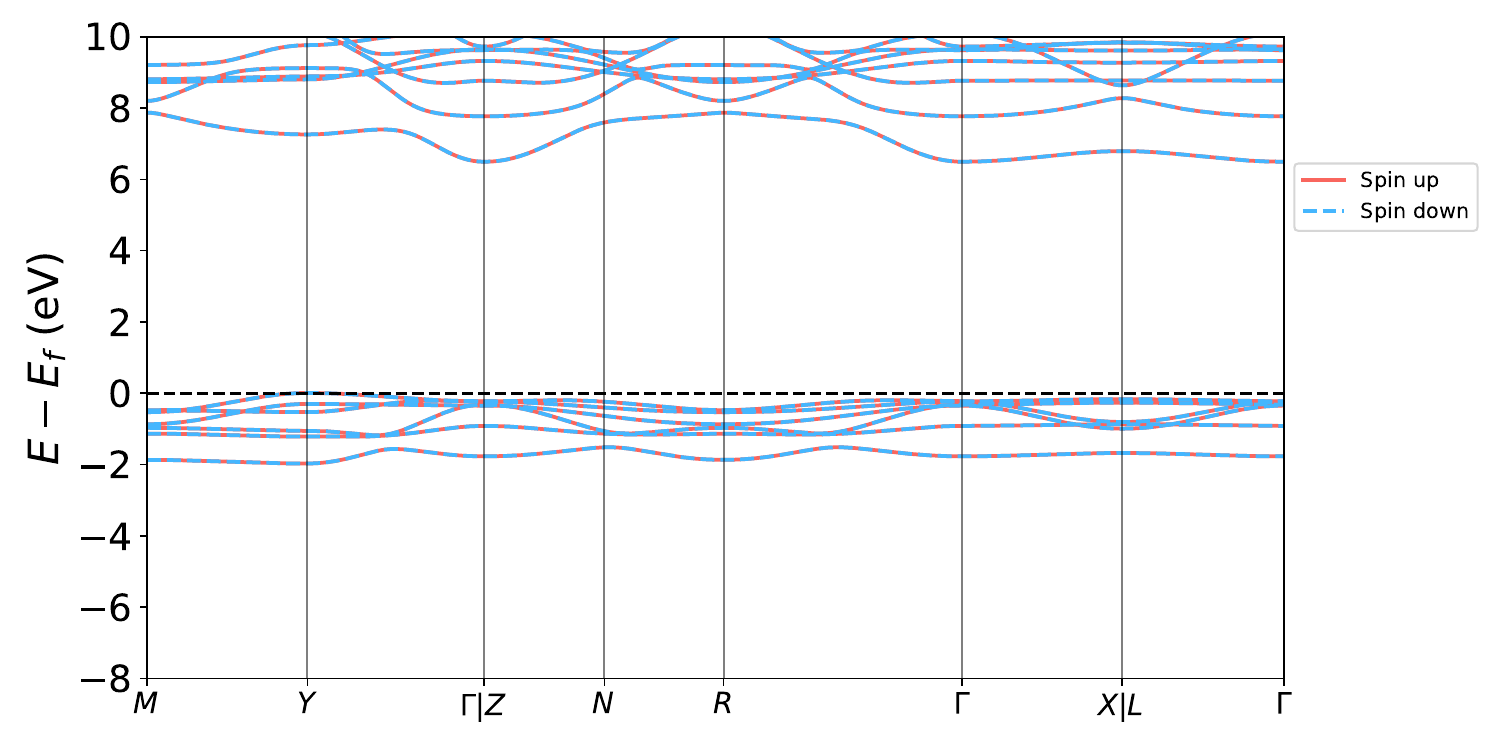}
    \caption{\footnotesize \ce{SrI2} slab spin-polarized band structure.}
    \label{fig:SrI2_band_slab}
\end{figure}
\clearpage
\newpage

\subsection{TeI bulk}
\begin{verbatim}
_cell_length_a 7.68067454
_cell_length_b 7.88783113
_cell_length_c 9.65505273
_cell_angle_alpha 89.675024
_cell_angle_beta 76.589301
_cell_angle_gamma 75.783869
_symmetry_space_group_name_H-M         'P 1'
_symmetry_Int_Tables_number            1

loop_
_symmetry_equiv_pos_as_xyz
   'x, y, z'

loop_
_atom_site_label
_atom_site_type_symbol
_atom_site_fract_x
_atom_site_fract_y
_atom_site_fract_z
Te001 Te -4.887016846764E-01 4.875339604028E-01 -1.935831778131E-01
Te002 Te 4.887016846764E-01 -4.875339604028E-01 1.935831778131E-01
Te003 Te 2.291107642376E-01 3.088880787811E-01 -1.868126459646E-01
Te004 Te -2.291107642376E-01 -3.088880787811E-01 1.868126459646E-01
Te005 Te -2.076451061255E-01 1.777114320686E-01 -2.920855462737E-01
Te006 Te 2.076451061255E-01 -1.777114320686E-01 2.920855462737E-01
Te007 Te -4.897897148073E-01 5.823324976090E-03 -1.961905053850E-01
Te008 Te 4.897897148073E-01 -5.823324976090E-03 1.961905053850E-01
I009 I 1.956184262062E-01 -1.863966971456E-01 -5.563292537369E-02
I010 I -1.956184262062E-01 1.863966971456E-01 5.563292537369E-02
I011 I -1.778855923318E-01 -3.434001876149E-01 -1.660418894701E-01
I012 I 1.778855923318E-01 3.434001876149E-01 1.660418894701E-01
I013 I -2.166941629009E-01 1.811892080299E-01 4.218356705394E-01
I014 I 2.166941629009E-01 -1.811892080299E-01 -4.218356705394E-01
I015 I 2.143562007785E-01 3.171295932474E-01 -4.724285292977E-01
I016 I -2.143562007785E-01 -3.171295932474E-01 4.724285292977E-01

\end{verbatim}
\begin{figure}[h]
    \centering
    \includegraphics[width=\textwidth]{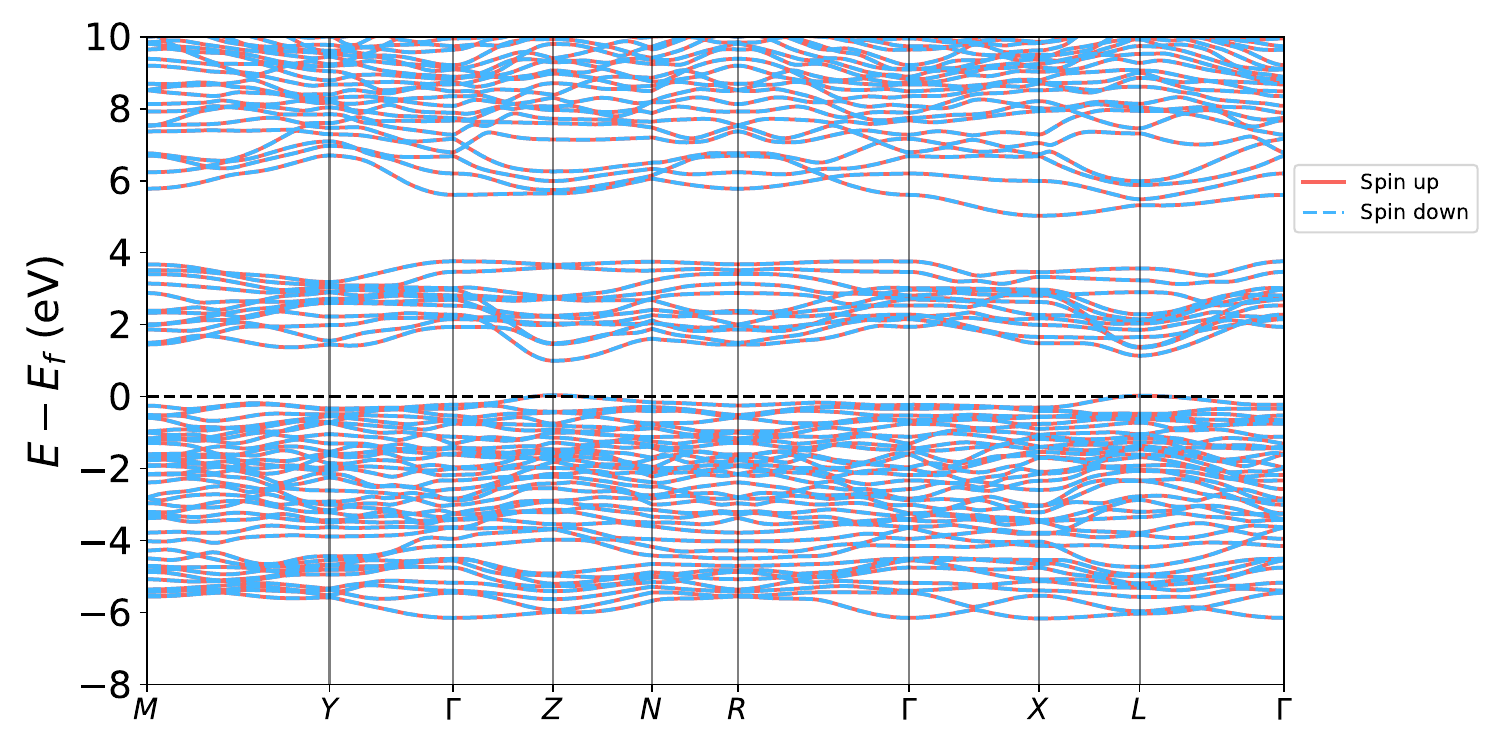}
    \caption{\footnotesize TeI bulk spin-polarized band structure.}
    \label{fig:TeI_band_bulk}
\end{figure}
\clearpage
\newpage

\subsection{TeI slab}
\begin{verbatim}
_cell_length_a                         8.74880490
_cell_length_b                         9.57845542
_cell_length_c                         40.00000000
_cell_angle_alpha                      90.000000
_cell_angle_beta                       90.000000
_cell_angle_gamma                      96.146118
_cell_volume                           'P 1'
_space_group_name_H-M_alt              'P 1'
_space_group_IT_number                 1

loop_
_space_group_symop_operation_xyz
   'x, y, z'

loop_
   _atom_site_label
   _atom_site_occupancy
   _atom_site_fract_x
   _atom_site_fract_y
   _atom_site_fract_z
   _atom_site_adp_type
   _atom_site_B_ios_or_equiv
   _atom_site_type_symbol
Te001  1.0  0.140629863377  -0.205467012637  0.554028361484  Biso  1.000000  Te
I002  1.0  0.499210533051  0.123522226529  0.574184502791  Biso  1.000000  I
I003  1.0  0.435115808156  0.479755402078  0.576716130221  Biso  1.000000  I
Te004  1.0  0.447041436720  -0.229088626342  0.568996648294  Biso  1.000000  Te
I005  1.0  0.207464724949  0.455054862928  0.488756944086  Biso  1.000000  I
Te006  1.0  0.185580713599  -0.255623079004  0.487374649738  Biso  1.000000  Te
I007  1.0  0.175126765797  0.086210905113  0.474650591701  Biso  1.000000  I
Te008  1.0  -0.494686938536  -0.188659960508  0.498474210106  Biso  1.000000  Te
Te009  1.0  0.494686938536  0.188659960508  0.501525789894  Biso  1.000000  Te
I010  1.0  -0.175126765797  -0.086210905113  0.525349408299  Biso  1.000000  I
Te011  1.0  -0.185580713599  0.255623079004  0.512625350262  Biso  1.000000  Te
I012  1.0  -0.207464724949  -0.455054862928  0.511243055914  Biso  1.000000  I
Te013  1.0  -0.447041436720  0.229088626342  0.431003351706  Biso  1.000000  Te
I014  1.0  -0.435115808156  -0.479755402078  0.423283869779  Biso  1.000000  I
I015  1.0  -0.499210533051  -0.123522226529  0.425815497209  Biso  1.000000  I
Te016  1.0  -0.140629863377  0.205467012637  0.445971638516  Biso  1.000000  Te

\end{verbatim}
\begin{figure}[h]
    \centering
    \includegraphics[width=\textwidth]{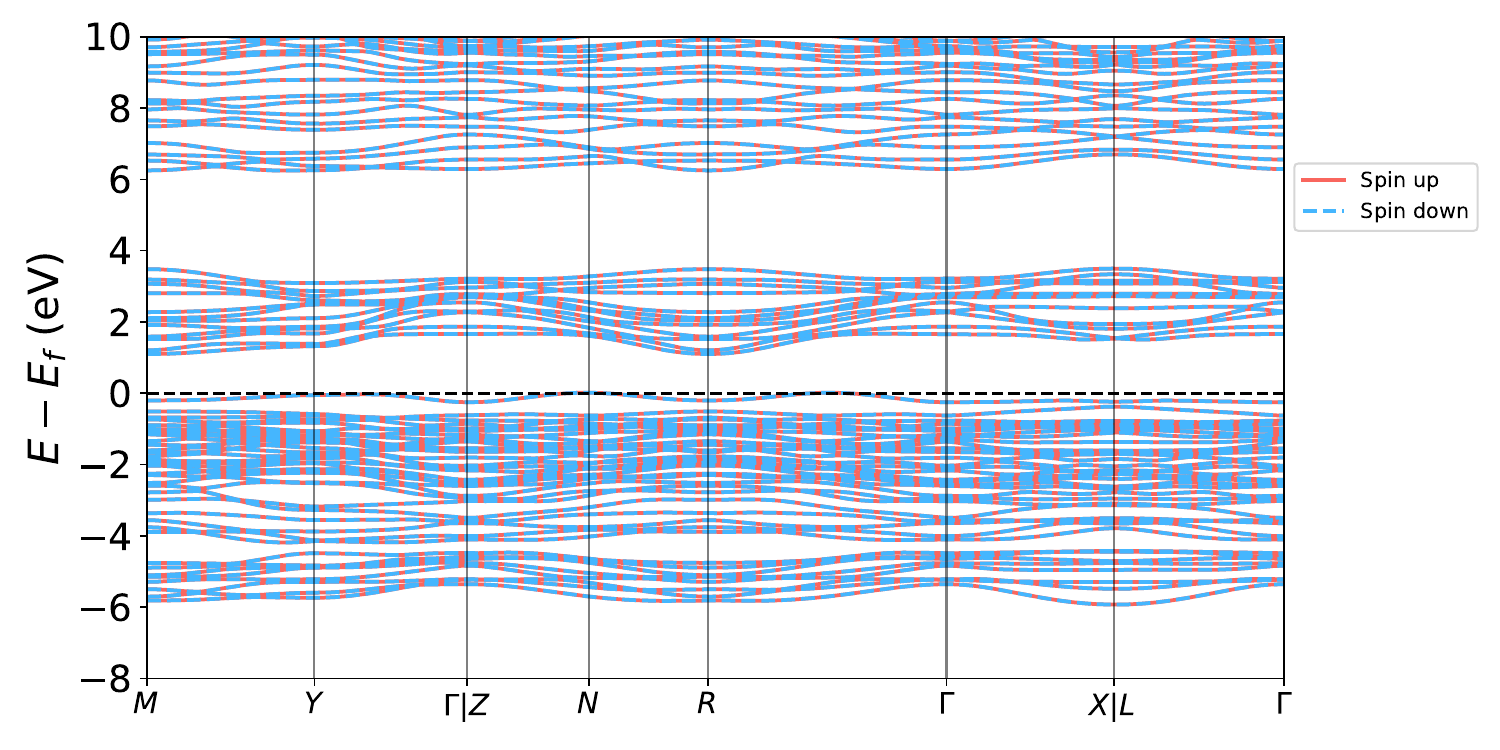}
    \caption{\footnotesize TeI slab spin-polarized band structure.}
    \label{fig:TeI_band_slab}
\end{figure}
\clearpage
\newpage

\subsection{\ce{TiBr2} bulk}
\begin{verbatim}
_cell_length_a                         3.800392
_cell_length_b                         6.553070
_cell_length_c                         6.941540
_cell_angle_alpha                      117.876778
_cell_angle_beta                       105.883438
_cell_angle_gamma                      90.002083
_cell_volume                           145.303042
_space_group_name_H-M_alt              'P -1'
_space_group_IT_number                 2

loop_
_space_group_symop_operation_xyz
   'x, y, z'
   '-x, -y, -z'

loop_
   _atom_site_label
   _atom_site_occupancy
   _atom_site_fract_x
   _atom_site_fract_y
   _atom_site_fract_z
   _atom_site_adp_type
   _atom_site_B_iso_or_equiv
   _atom_site_type_symbol
   Br1         1.0     0.873340     0.540030     0.746600    Biso  1.000000 Br
   Br2         1.0     0.373340     0.039940     0.746610    Biso  1.000000 Br
   Ti1         1.0     0.000000     0.000000     0.000000    Biso  1.000000 Ti
   Ti2         1.0     0.500000     0.500000     0.000000    Biso  1.000000 Ti

\end{verbatim}
\begin{figure}[h]
    \centering
    \includegraphics[width=\textwidth]{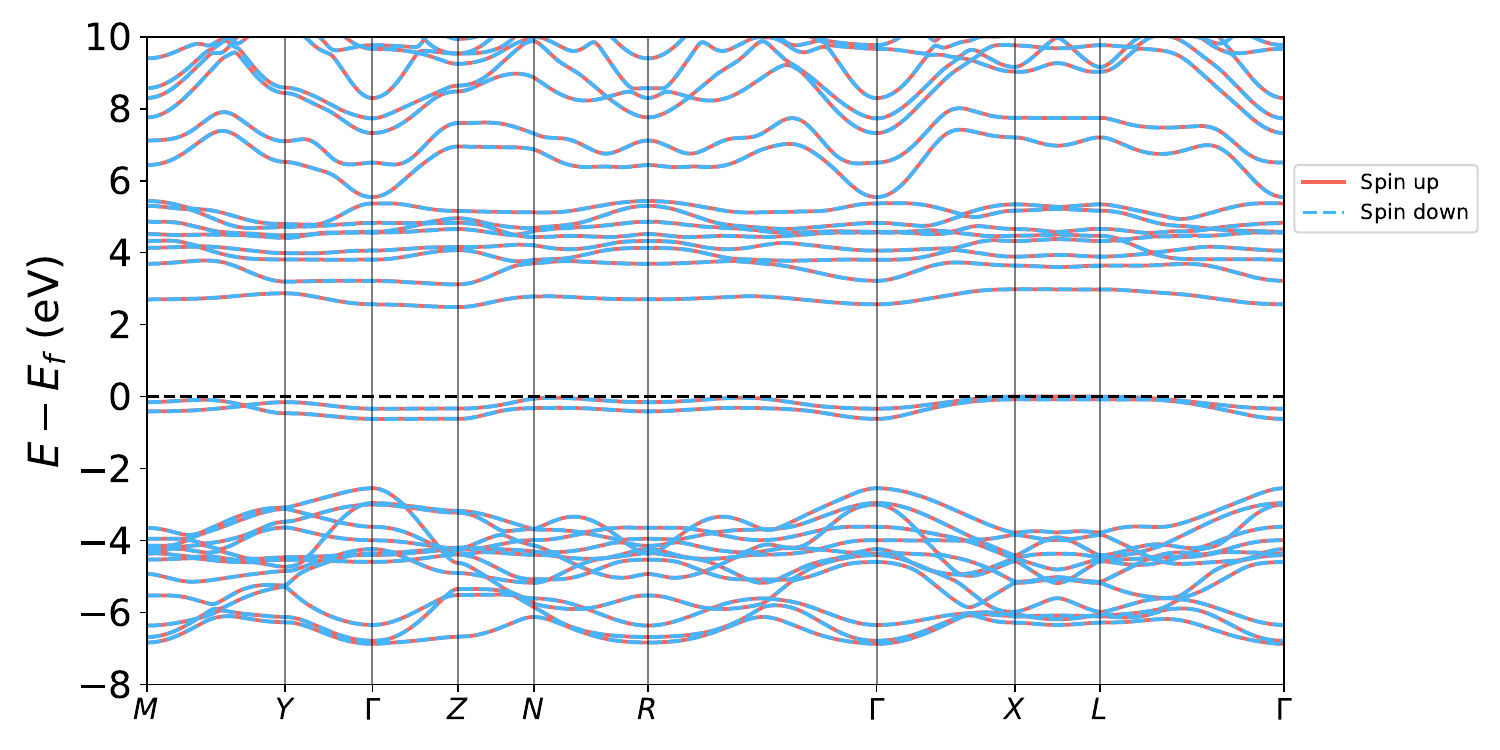}
    \caption{\footnotesize \ce{TiBr2} bulk spin-polarized band structure.}
    \label{fig:TiBr2_band_bulk}
\end{figure}
\clearpage
\newpage

\subsection{\ce{TiBr2} slab}
\begin{verbatim}
_cell_length_a                         3.80644829
_cell_length_b                         6.56304094
_cell_length_c                         40.00000000
_cell_angle_alpha                      90.000000
_cell_angle_beta                       90.000000
_cell_angle_gamma                      90.000339
_cell_volume                           'P 1'
_space_group_name_H-M_alt              'P 1'
_space_group_IT_number                 1

loop_
_space_group_symop_operation_xyz
   'x, y, z'

loop_
   _atom_site_label
   _atom_site_occupancy
   _atom_site_fract_x
   _atom_site_fract_y
   _atom_site_fract_z
   _atom_site_adp_type
   _atom_site_B_ios_or_equiv
   _atom_site_type_symbol
Br001  1.0  0.000001346383  0.334484615116  0.537240791102  Biso  1.000000  Br
Br002  1.0  -0.499997986586  -0.165411139095  0.537236050729  Biso  1.000000  Br
Ti003  1.0  0.000003915926  0.000003478408  0.499998955549  Biso  1.000000  Ti
Ti004  1.0  -0.499997006291  -0.499996553883  0.499998927465  Biso  1.000000  Ti
Br005  1.0  0.000004864933  -0.334477656721  0.462757214883  Biso  1.000000  Br
Br006  1.0  -0.499994137780  0.165418058778  0.46276193407599997  Biso  1.000000  Br

\end{verbatim}
\begin{figure}[h]
    \centering
    \includegraphics[width=\textwidth]{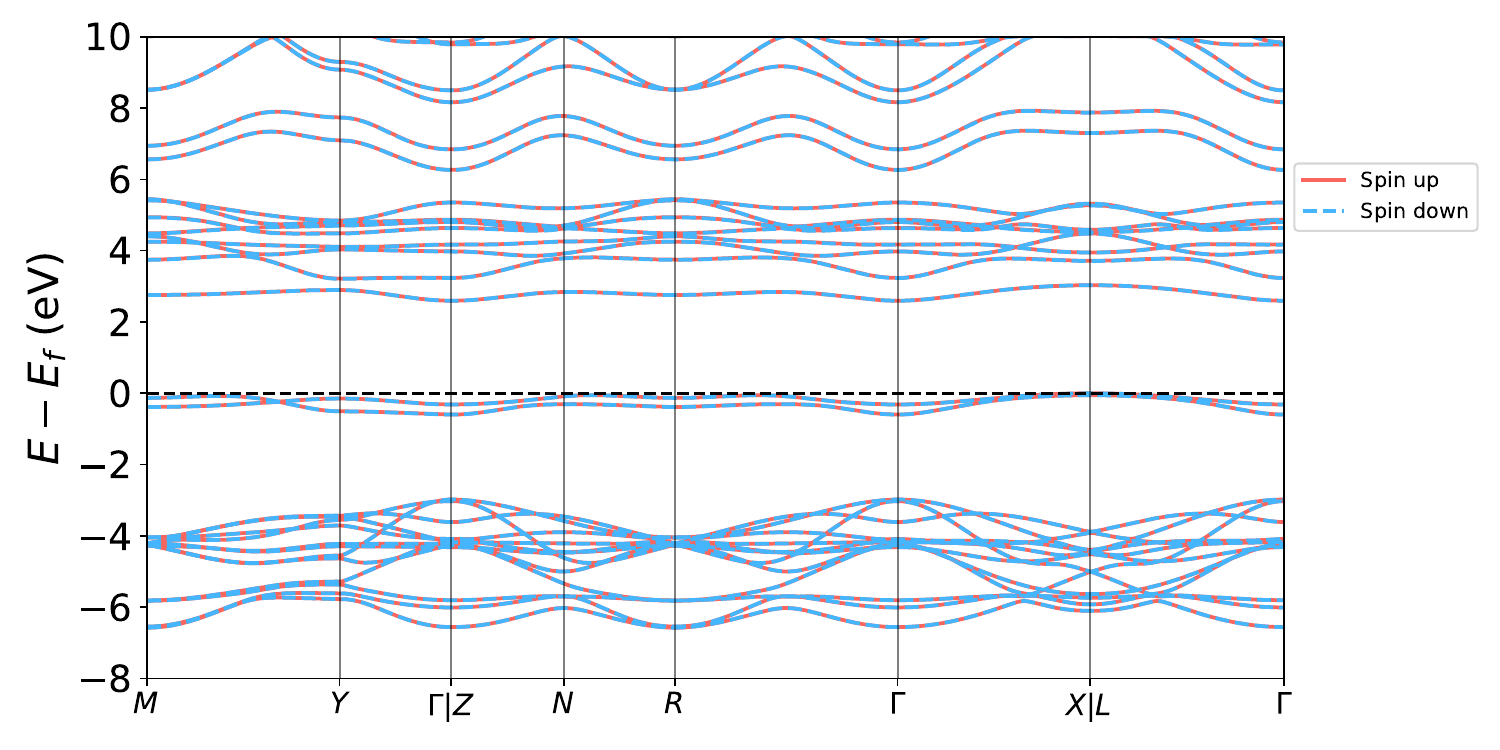}
    \caption{\footnotesize \ce{TiBr2} slab spin-polarized band structure.}
    \label{fig:TiBr2_band_slab}
\end{figure}
\clearpage
\newpage

\subsection{\ce{TiCl2} bulk}
\begin{verbatim}
_cell_length_a                         3.650175
_cell_length_b                         6.272978
_cell_length_c                         6.646604
_cell_angle_alpha                      62.236919
_cell_angle_beta                       74.065254
_cell_angle_gamma                      89.993874
_cell_volume                           128.027206
_space_group_name_H-M_alt              'P -1'
_space_group_IT_number                 2

loop_
_space_group_symop_operation_xyz
   'x, y, z'
   '-x, -y, -z'

loop_
   _atom_site_label
   _atom_site_occupancy
   _atom_site_fract_x
   _atom_site_fract_y
   _atom_site_fract_z
   _atom_site_adp_type
   _atom_site_B_iso_or_equiv
   _atom_site_type_symbol
   Cl1         1.0     0.875540     0.541850     0.248950    Biso  1.000000 Cl
   Cl2         1.0     0.375540     0.041750     0.248940    Biso  1.000000 Cl
   Ti1         1.0     0.000000     0.000000     0.000000    Biso  1.000000 Ti
   Ti2         1.0     0.500000     0.500000     0.000000    Biso  1.000000 Ti

\end{verbatim}
\begin{figure}[h]
    \centering
    \includegraphics[width=\textwidth]{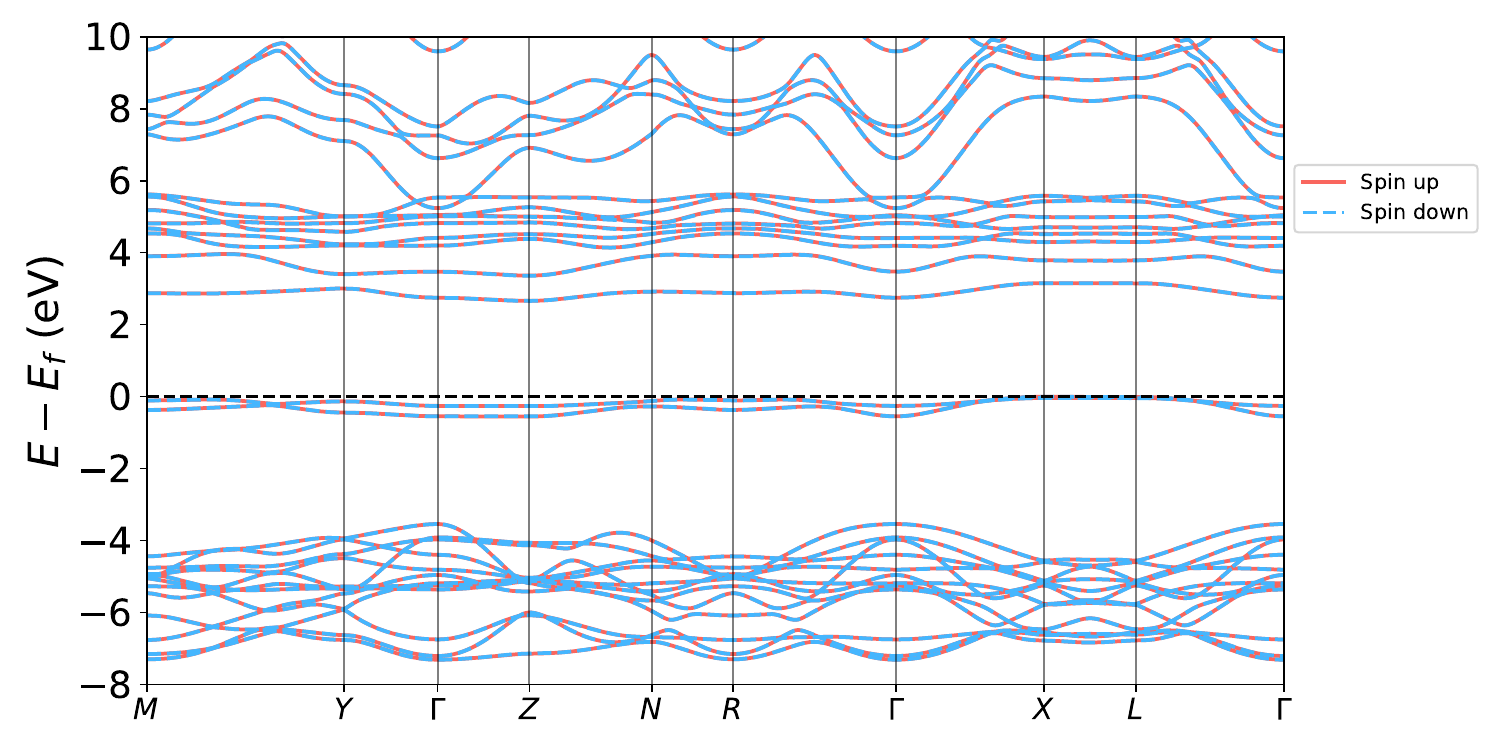}
    \caption{\footnotesize \ce{TiCl2} bulk spin-polarized band structure.}
    \label{fig:TiCl2_band_bulk}
\end{figure}
\clearpage
\newpage

\subsection{\ce{TiCl2} slab}
\begin{verbatim}
_cell_length_a                         3.64126686
_cell_length_b                         6.25827434
_cell_length_c                         40.00000000
_cell_angle_alpha                      90.000000
_cell_angle_beta                       90.000000
_cell_angle_gamma                      89.999729
_cell_volume                           'P 1'
_space_group_name_H-M_alt              'P 1'
_space_group_IT_number                 1

loop_
_space_group_symop_operation_xyz
   'x, y, z'

loop_
   _atom_site_label
   _atom_site_occupancy
   _atom_site_fract_x
   _atom_site_fract_y
   _atom_site_fract_z
   _atom_site_adp_type
   _atom_site_B_ios_or_equiv
   _atom_site_type_symbol
Cl001  1.0  -0.000004403856  -0.335325205845  0.535240022611  Biso  1.000000  Cl
Cl002  1.0  0.499994470473  0.164574304330  0.535236420573  Biso  1.000000  Cl
Ti003  1.0  -0.000004280013  -0.000003117518  0.499999075668  Biso  1.000000  Ti
Ti004  1.0  0.499997193375  0.499996531544  0.499999107255  Biso  1.000000  Ti
Cl005  1.0  -0.000000888885  0.335319058484  0.464758147174  Biso  1.000000  Cl
Cl006  1.0  0.499998415786  -0.164580818667  0.464761775177  Biso  1.000000  Cl

\end{verbatim}
\begin{figure}[h]
    \centering
    \includegraphics[width=\textwidth]{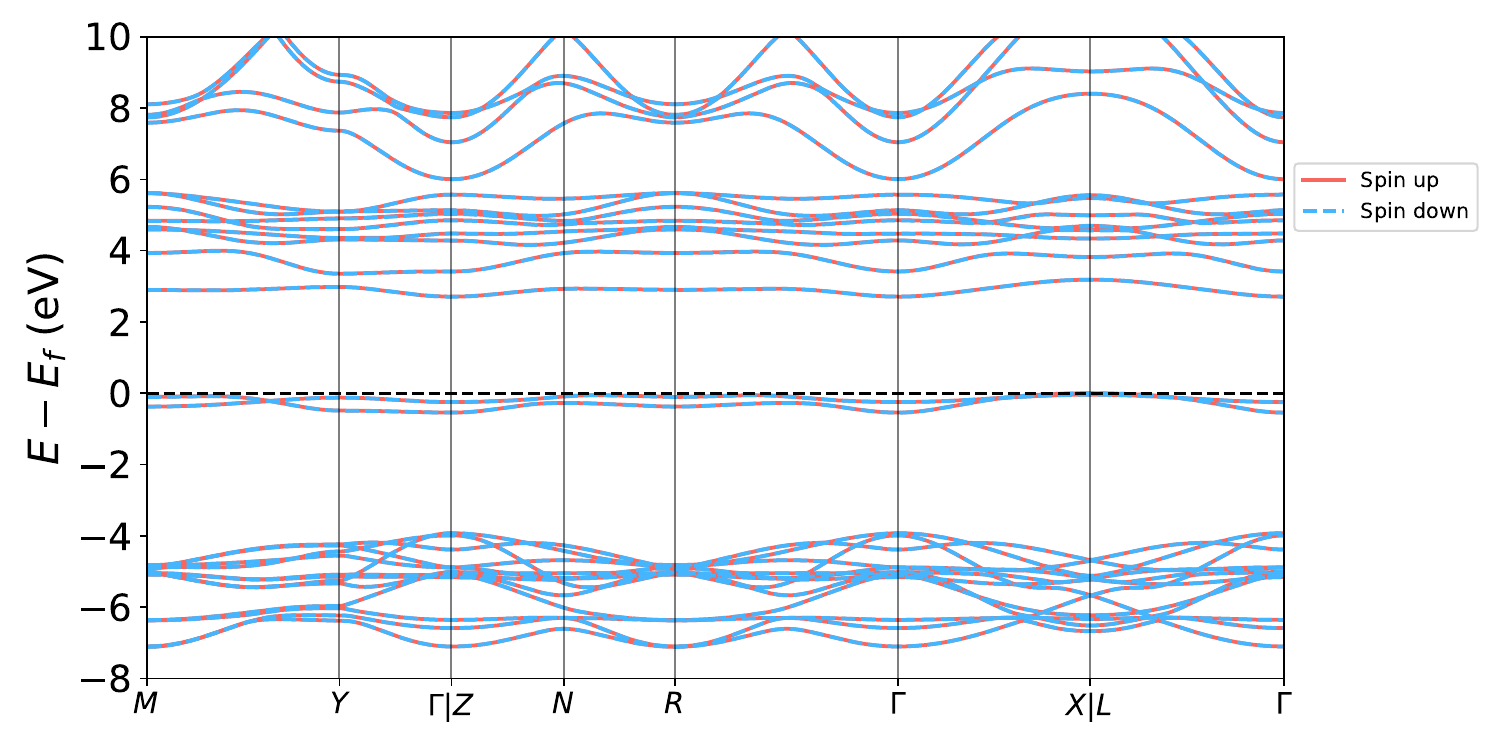}
    \caption{\footnotesize \ce{TiCl2} slab spin-polarized band structure.}
    \label{fig:TiCl2_band_slab}
\end{figure}

\clearpage
\newpage

\subsection{TlF bulk}
\begin{verbatim}
_cell_length_a 3.57209125
_cell_length_b 3.57209125
_cell_length_c 5.81635828
_cell_angle_alpha 90.000000
_cell_angle_beta 90.000000
_cell_angle_gamma 90.000000
_symmetry_space_group_name_H-M         'P 1'
_symmetry_Int_Tables_number            1

loop_
_symmetry_equiv_pos_as_xyz
   'x, y, z'

loop_
_atom_site_label
_atom_site_type_symbol
_atom_site_fract_x
_atom_site_fract_y
_atom_site_fract_z
Tl001 Tl 2.500000000000E-01 2.500000000000E-01 2.806561509055E-01
Tl002 Tl -2.500000000000E-01 -2.500000000000E-01 -2.806561509055E-01
F003 F 2.500000000000E-01 2.500000000000E-01 -1.400922886257E-01
F004 F -2.500000000000E-01 -2.500000000000E-01 1.400922886257E-01

\end{verbatim}
\begin{figure}[h]
    \centering
    \includegraphics[width=\textwidth]{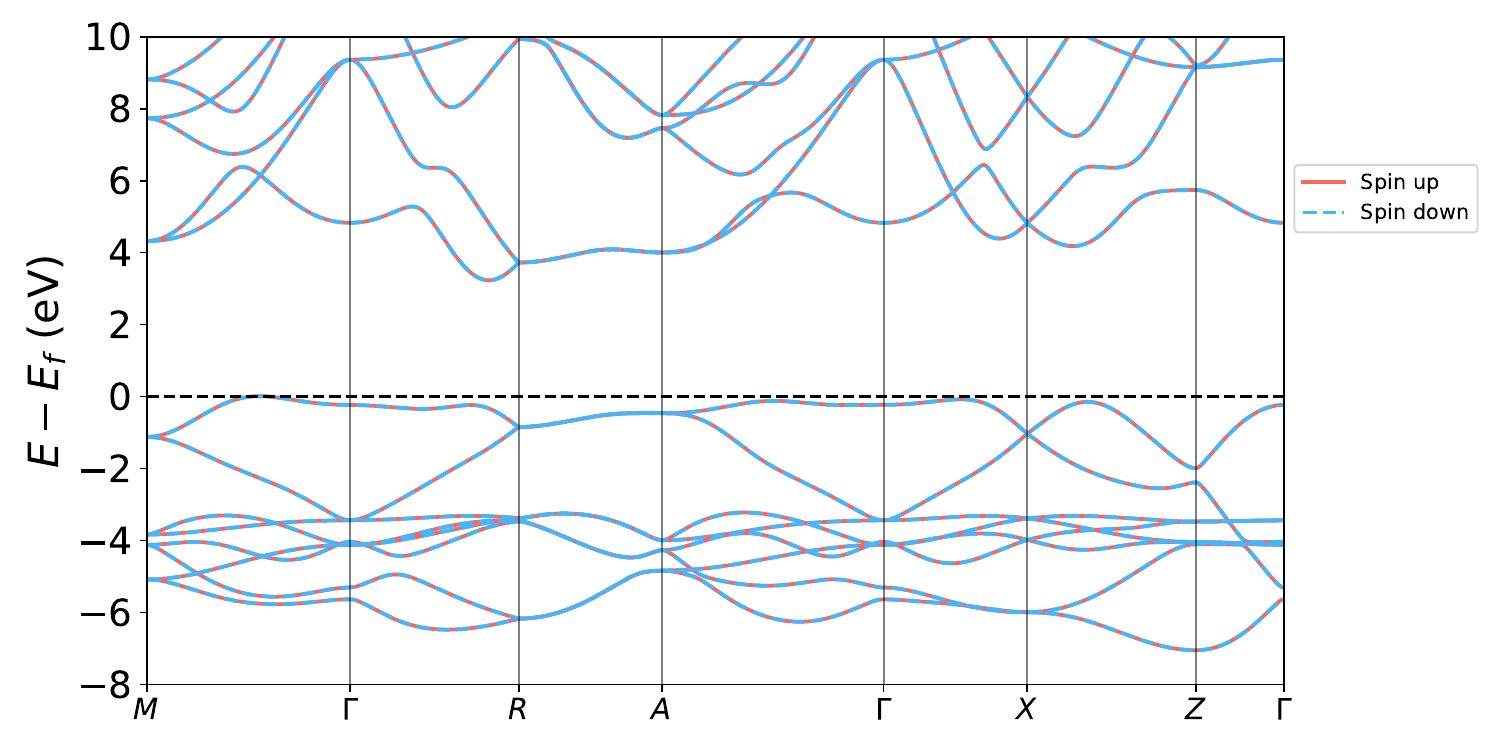}
    \caption{\footnotesize TlF bulk spin-polarized band structure.}
    \label{fig:TlF_band_bulk}
\end{figure}
\clearpage
\newpage

\subsection{TlF slab}
\begin{verbatim}
_cell_length_a                         3.55636781
_cell_length_b                         3.55636781
_cell_length_c                         40.00000000
_cell_angle_alpha                      90.000000
_cell_angle_beta                       90.000000
_cell_angle_gamma                      90.000000
_cell_volume                           'P 1'
_space_group_name_H-M_alt              'P 1'
_space_group_IT_number                 1

loop_
_space_group_symop_operation_xyz
   'x, y, z'

loop_
   _atom_site_label
   _atom_site_occupancy
   _atom_site_fract_x
   _atom_site_fract_y
   _atom_site_fract_z
   _atom_site_adp_type
   _atom_site_B_ios_or_equiv
   _atom_site_type_symbol
Tl001  1.0  0.250000000000  0.250000000000  0.540738638646  Biso  1.000000  Tl
F002  1.0  -0.250000000000  -0.250000000000  0.521441127258  Biso  1.000000  F
F003  1.0  0.250000000000  0.250000000000  0.478558872742  Biso  1.000000  F
Tl004  1.0  -0.250000000000  -0.250000000000  0.459261361354  Biso  1.000000  Tl

\end{verbatim}
\begin{figure}[h]
    \centering
    \includegraphics[width=\textwidth]{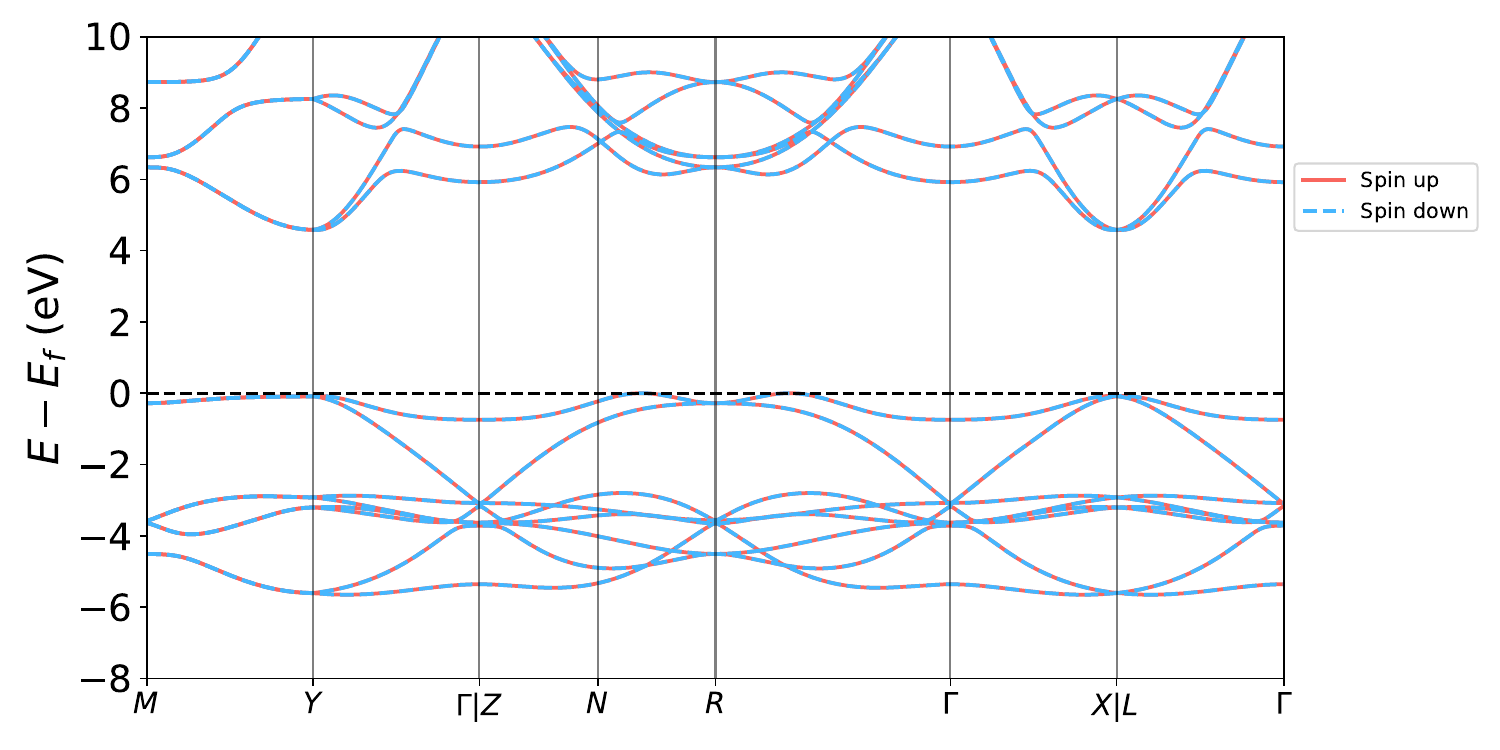}
    \caption{\footnotesize TlF slab spin-polarized band structure.}
    \label{fig:TlF_band_slab}
\end{figure}
\clearpage
\newpage

\subsection{\ce{TmI2} bulk}
\begin{verbatim}
_cell_length_a 4.42126843
_cell_length_b 4.42126843
_cell_length_c 6.35258518
_cell_angle_alpha 90.000000
_cell_angle_beta 90.000000
_cell_angle_gamma 120.000000
_symmetry_space_group_name_H-M         'P 1'
_symmetry_Int_Tables_number            1

loop_
_symmetry_equiv_pos_as_xyz
   'x, y, z'

loop_
_atom_site_label
_atom_site_type_symbol
_atom_site_fract_x
_atom_site_fract_y
_atom_site_fract_z
I001 I 3.333333333333E-01 -3.333333333333E-01 2.714726349742E-01
I002 I -3.333333333333E-01 3.333333333333E-01 -2.714726349742E-01
Tm003 Tm 0.000000000000E+00 0.000000000000E+00 0.000000000000E+00

\end{verbatim}
\begin{figure}[h]
    \centering
    \includegraphics[width=\textwidth]{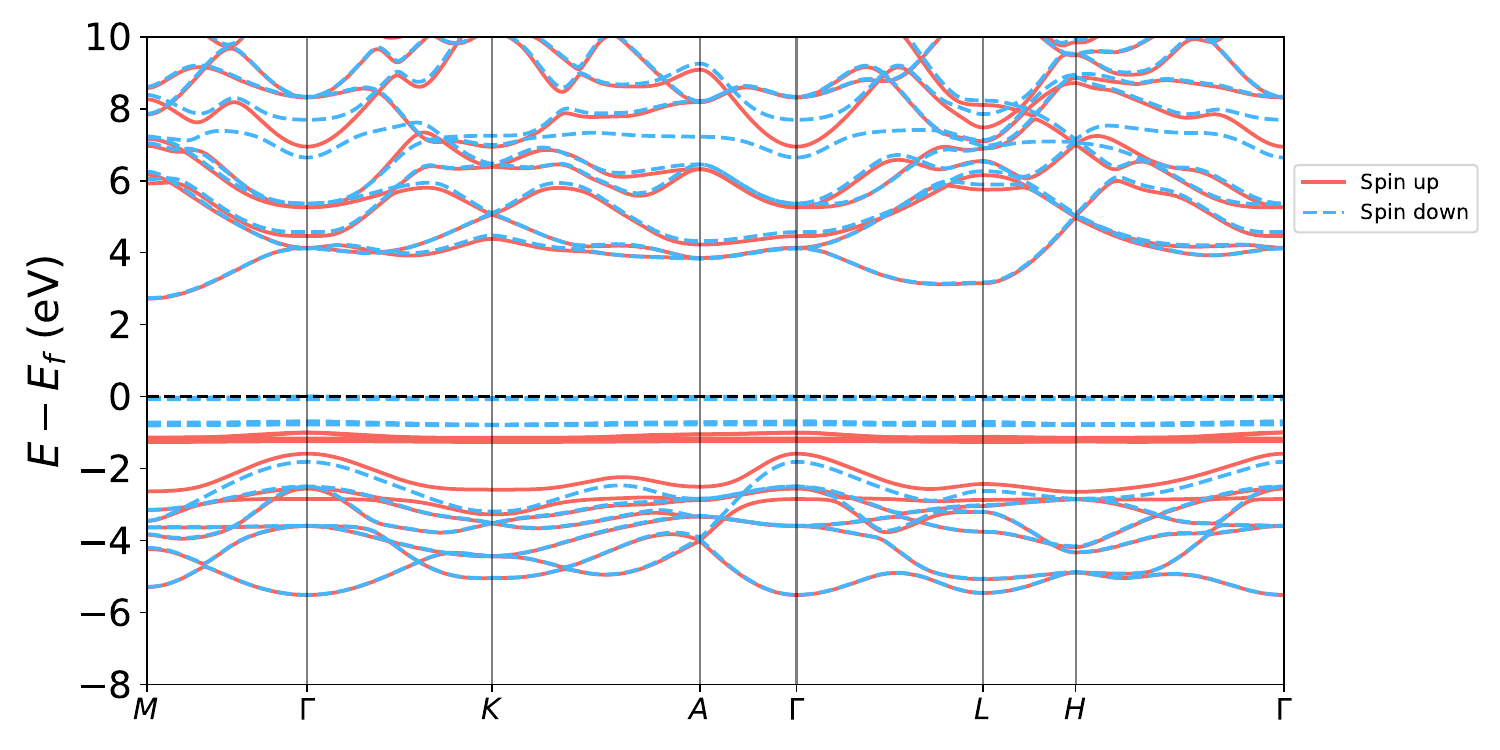}
    \caption{\footnotesize \ce{TmI2} bulk spin-polarized band structure.}
    \label{fig:TmI2_band_bulk}
\end{figure}
\clearpage
\newpage

\subsection{\ce{TmI2} slab}
\begin{verbatim}
_cell_length_a                         4.43833170
_cell_length_b                         4.43833170
_cell_length_c                         40.00000000
_cell_angle_alpha                      90.000000
_cell_angle_beta                       90.000000
_cell_angle_gamma                      120.000000
_cell_volume                           'P 1'
_space_group_name_H-M_alt              'P 1'
_space_group_IT_number                 1

loop_
_space_group_symop_operation_xyz
   'x, y, z'

loop_
   _atom_site_label
   _atom_site_occupancy
   _atom_site_fract_x
   _atom_site_fract_y
   _atom_site_fract_z
   _atom_site_adp_type
   _atom_site_B_ios_or_equiv
   _atom_site_type_symbol
I001  1.0  0.333333333333  -0.333333333333  0.543219572375  Biso  1.000000  I
Tm002  1.0  0.000000000000  0.000000000000  0.5  Biso  1.000000  Tm
I003  1.0  -0.333333333333  0.333333333333  0.456780427625  Biso  1.000000  I

\end{verbatim}
\begin{figure}[h]
    \centering
    \includegraphics[width=\textwidth]{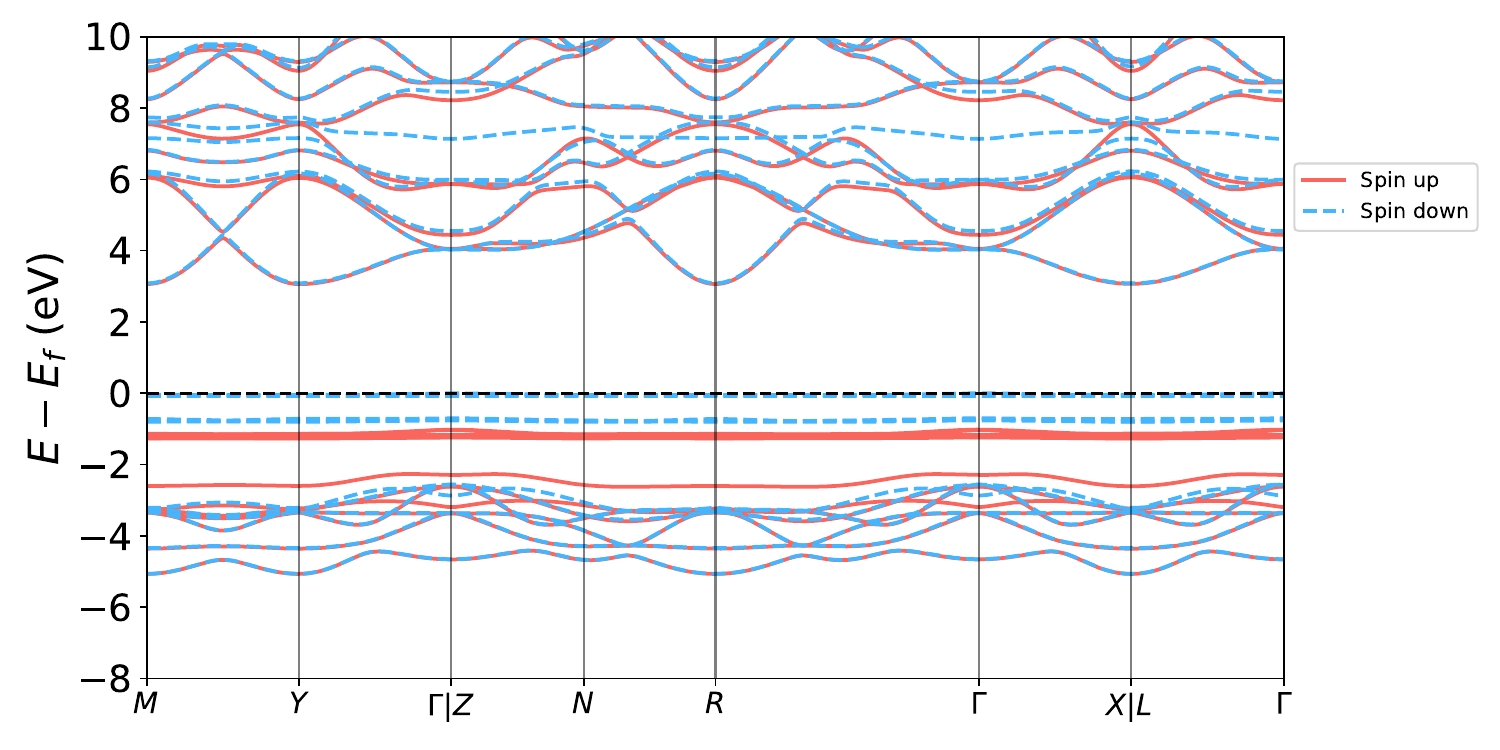}
    \caption{\footnotesize \ce{TmI2} slab spin-polarized band structure.}
    \label{fig:TmI2_band_slab}
\end{figure}
\clearpage
\newpage

\subsection{\ce{VBr2} bulk}
\begin{verbatim}
_cell_length_a                         6.968451
_cell_length_b                         7.048035
_cell_length_c                         7.339302
_cell_angle_alpha                      105.171761
_cell_angle_beta                       104.049614
_cell_angle_gamma                      116.855629
_cell_volume                           281.782742
_space_group_name_H-M_alt              'P -1'
_space_group_IT_number                 2

loop_
_space_group_symop_operation_xyz
   'x, y, z'
   '-x, -y, -z'

loop_
   _atom_site_label
   _atom_site_occupancy
   _atom_site_fract_x
   _atom_site_fract_y
   _atom_site_fract_z
   _atom_site_adp_type
   _atom_site_B_iso_or_equiv
   _atom_site_type_symbol
   V1          1.0     0.749980     0.749970     0.250020    Biso  1.000000 V
   V2         1.0     0.249980     0.249970     0.250020    Biso  1.000000 V
   Br1         1.0     0.791610     0.041600     0.083380    Biso  1.000000 Br
   Br2         1.0     0.291700     0.541660     0.083300    Biso  1.000000 Br
   Br3         1.0     0.791750     0.041730     0.583300    Biso  1.000000 Br
   Br4         1.0     0.291660     0.541680     0.583380    Biso  1.000000 Br

\end{verbatim}
\begin{figure}[h]
    \centering
    \includegraphics[width=\textwidth]{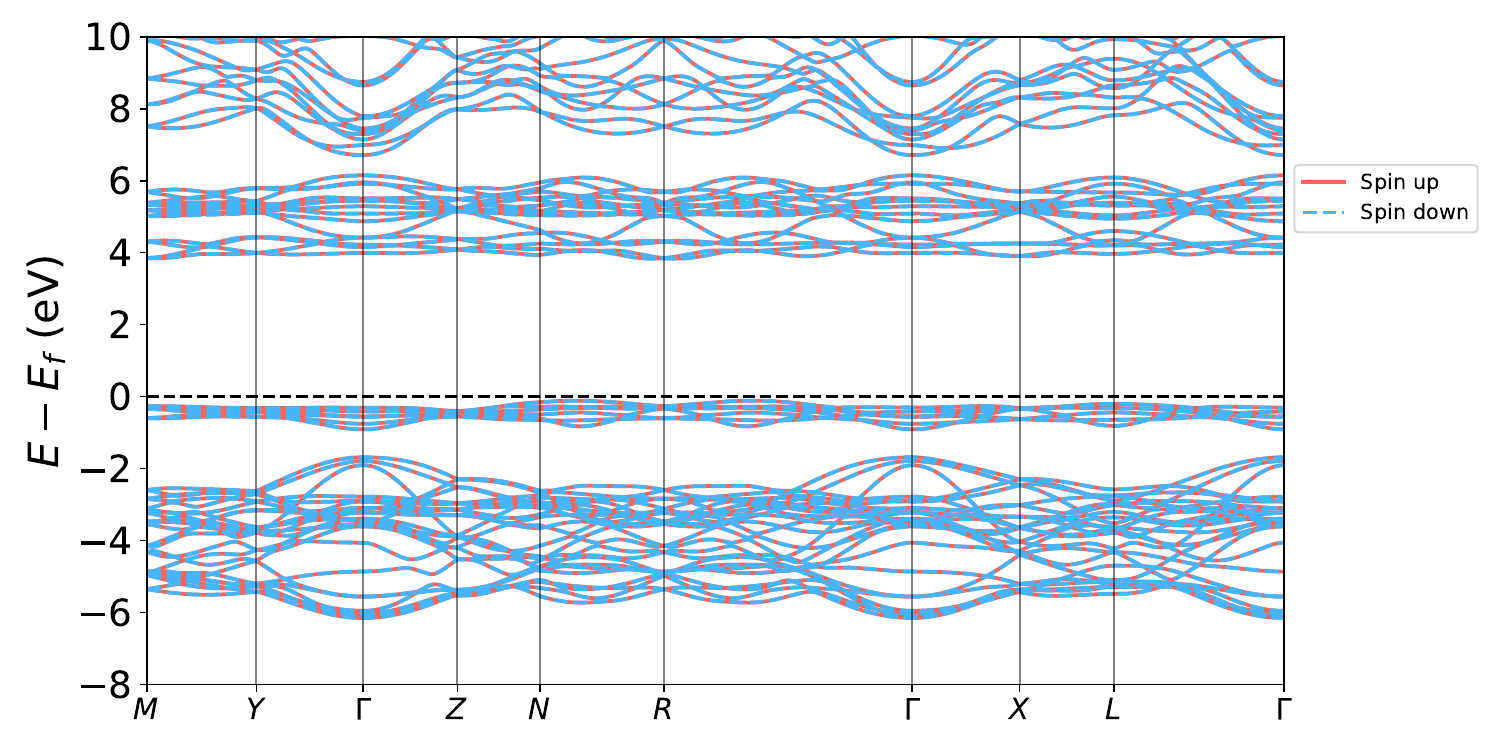}
    \caption{\footnotesize \ce{VBr2} bulk spin-polarized band structure.}
    \label{fig:VBr2_band_bulk}
\end{figure}
\clearpage
\newpage

\subsection{\ce{VBr2} slab}
\begin{verbatim}
_cell_length_a                         7.34548427
_cell_length_b                         7.34579708
_cell_length_c                         40.00000000
_cell_angle_alpha                      90.000000
_cell_angle_beta                       90.000000
_cell_angle_gamma                      118.886484
_cell_volume                           'P 1'
_space_group_name_H-M_alt              'P 1'
_space_group_IT_number                 1

loop_
_space_group_symop_operation_xyz
   'x, y, z'

loop_
   _atom_site_label
   _atom_site_occupancy
   _atom_site_fract_x
   _atom_site_fract_y
   _atom_site_fract_z
   _atom_site_adp_type
   _atom_site_B_ios_or_equiv
   _atom_site_type_symbol
Br001  1.0  0.081755092489  -0.081717224595  0.537682302701  Biso  1.000000  Br
Br002  1.0  -0.418316997560  -0.081659530408  0.5376748627  Biso  1.000000  Br
Br003  1.0  0.081667576369  0.418348849893  0.537677211648  Biso  1.000000  Br
Br004  1.0  -0.418269468028  0.418271039425  0.537684468267  Biso  1.000000  Br
V005  1.0  -0.250000512201  0.250003514613  0.500002573333  Biso  1.000000  V
V006  1.0  0.250013856565  -0.250004172393  0.500001466882  Biso  1.000000  V
V007  1.0  0.249995671483  0.250014157961  0.500001616725  Biso  1.000000  V
V008  1.0  -0.249981898469  -0.250014372386  0.500002466212  Biso  1.000000  V
Br009  1.0  -0.081739803720  0.081717689765  0.462322353803  Biso  1.000000  Br
Br010  1.0  0.418332395535  0.081660111941  0.462329844915  Biso  1.000000  Br
Br011  1.0  -0.081652059758  -0.418348127866  0.46232751505  Biso  1.000000  Br
Br012  1.0  0.418284896315  -0.418270663838  0.462320183983  Biso  1.000000  Br

\end{verbatim}
\begin{figure}[h]
    \centering
    \includegraphics[width=\textwidth]{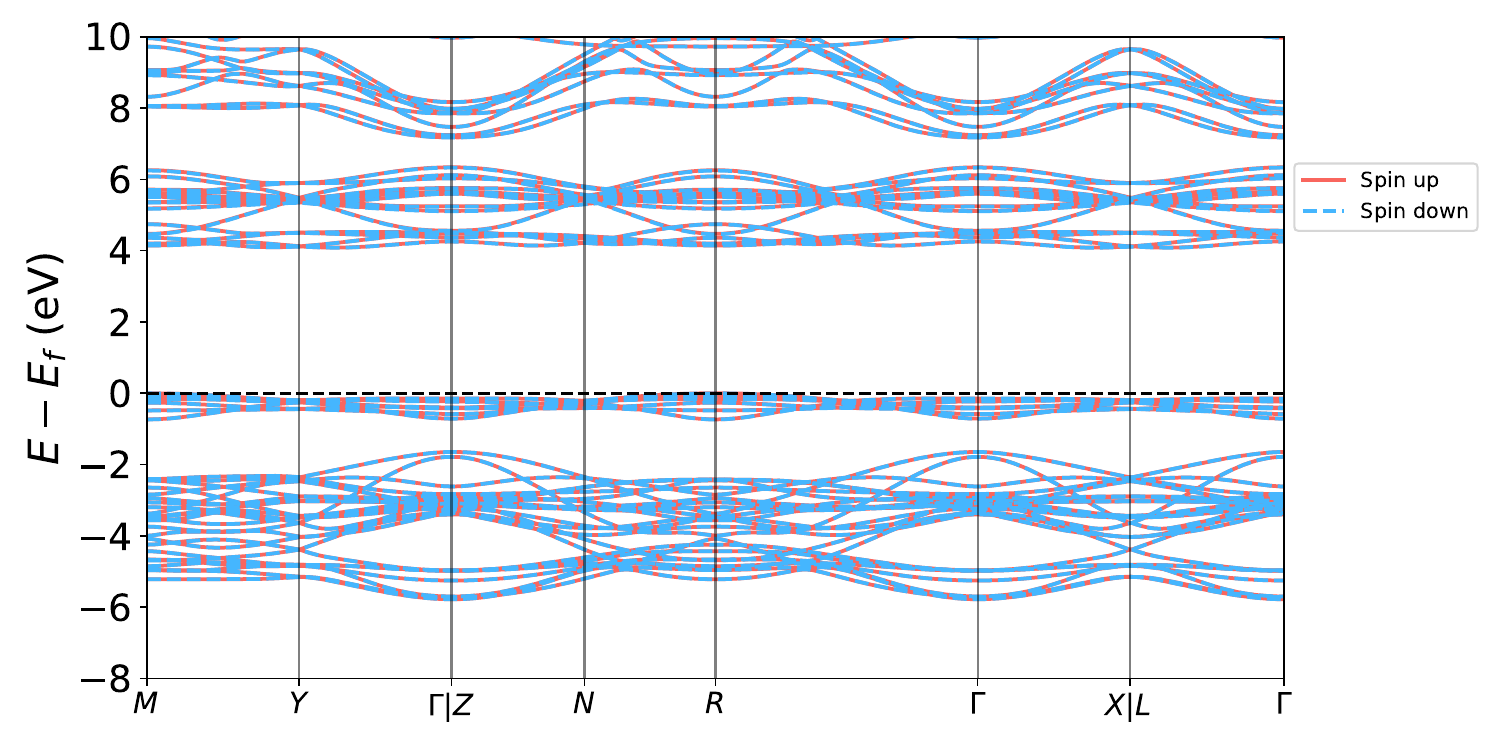}
    \caption{\footnotesize \ce{VBr2} slab spin-polarized band structure.}
    \label{fig:VBr_band_slab}
\end{figure}
\clearpage
\newpage

\subsection{\ce{VCl2} bulk}
\begin{verbatim}
_cell_length_a                         7.051878
_cell_length_b                         7.051999
_cell_length_c                         11.313018
_cell_angle_alpha                      89.312943
_cell_angle_beta                       89.315300
_cell_angle_gamma                      61.206039
_cell_volume                           492.986230
_space_group_name_H-M_alt              'P -1'
_space_group_IT_number                 2

loop_
_space_group_symop_operation_xyz
   'x, y, z'
   '-x, -y, -z'

loop_
   _atom_site_label
   _atom_site_occupancy
   _atom_site_fract_x
   _atom_site_fract_y
   _atom_site_fract_z
   _atom_site_adp_type
   _atom_site_B_iso_or_equiv
   _atom_site_type_symbol
   V1          1.0     0.749990     0.250040     0.000000    Biso  1.000000 V
   V2         1.0     0.749990     0.250040     0.499990    Biso  1.000000 V
   V3         1.0     0.249950     0.249940    -0.000030    Biso  1.000000 V
   V4          1.0     0.249950     0.249930     0.499970    Biso  1.000000 V
   Cl1         1.0     0.416900     0.416870     0.123680    Biso  1.000000 Cl
   Cl2         1.0     0.416840     0.416820     0.623690    Biso  1.000000 Cl
   Cl3         1.0    -0.083150     0.416900     0.123710    Biso  1.000000 Cl
   Cl4         1.0    -0.083080     0.416960     0.623690    Biso  1.000000 Cl
   Cl5         1.0     0.416880    -0.083170     0.123700    Biso  1.000000 Cl
   Cl6         1.0     0.416940    -0.083100     0.623690    Biso  1.000000 Cl
   Cl7         1.0    -0.083040    -0.083010     0.123710    Biso  1.000000 Cl
   Cl8         1.0    -0.083110    -0.083090     0.623720    Biso  1.000000 Cl

\end{verbatim}
\begin{figure}[h]
    \centering
    \includegraphics[width=\textwidth]{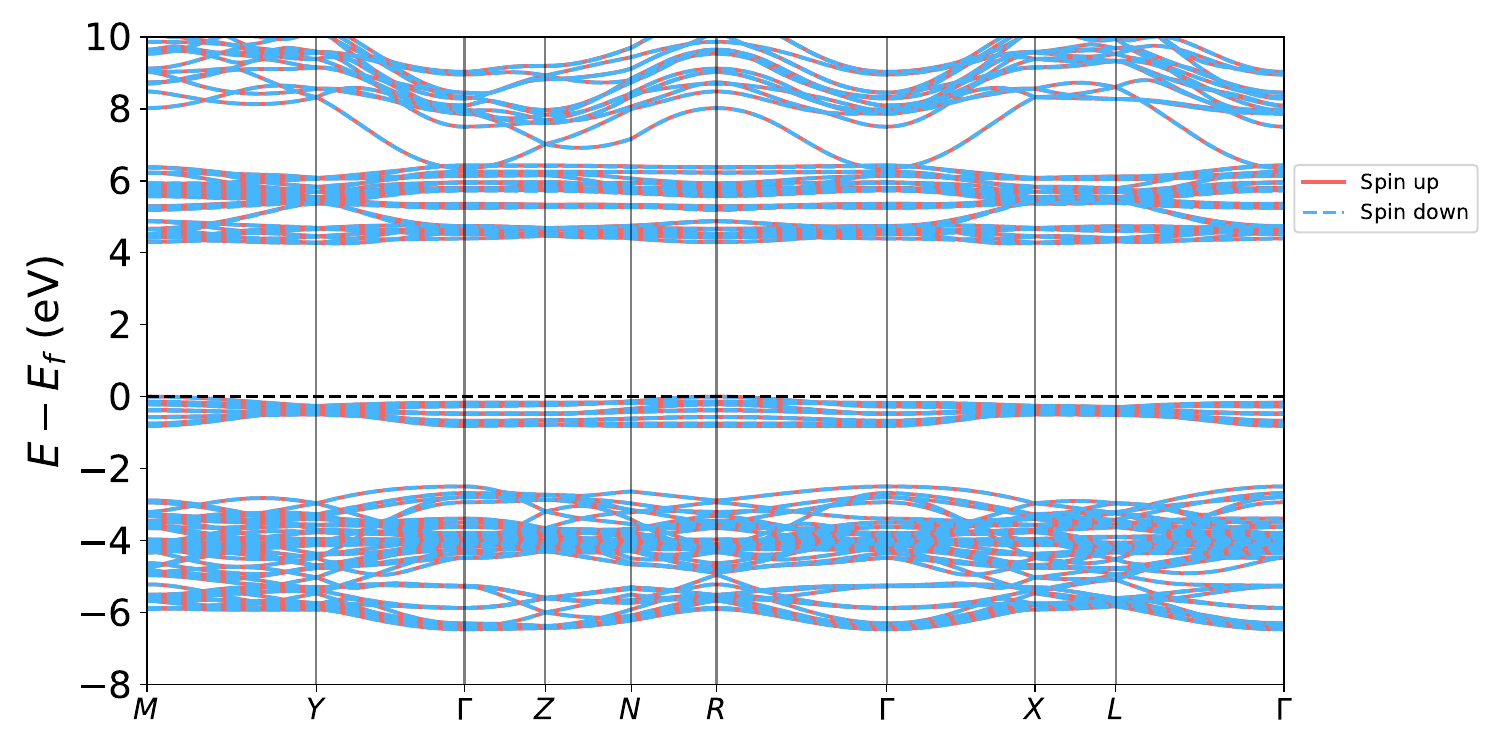}
    \caption{\footnotesize \ce{VCl2} bulk spin-polarized band structure.}
    \label{fig:VCl2_band_bulk}
\end{figure}
\clearpage
\newpage

\subsection{\ce{VCl2} slab}
\begin{verbatim}
_cell_length_a                         7.04348225
_cell_length_b                         7.04374375
_cell_length_c                         40.00000000
_cell_angle_alpha                      90.000000
_cell_angle_beta                       90.000000
_cell_angle_gamma                      118.835795
_cell_volume                           'P 1'
_space_group_name_H-M_alt              'P 1'
_space_group_IT_number                 1

loop_
_space_group_symop_operation_xyz
   'x, y, z'

loop_
   _atom_site_label
   _atom_site_occupancy
   _atom_site_fract_x
   _atom_site_fract_y
   _atom_site_fract_z
   _atom_site_adp_type
   _atom_site_B_ios_or_equiv
   _atom_site_type_symbol
Cl001  1.0  0.418489561387  -0.418499633849  0.535391425805  Biso  1.000000  Cl
Cl002  1.0  0.418404445107  0.081583515418  0.535399371129  Biso  1.000000  Cl
Cl003  1.0  -0.081601077295  -0.418432445950  0.535399657709  Biso  1.000000  Cl
Cl004  1.0  -0.081491924161  0.081488626508  0.535395933335  Biso  1.000000  Cl
V005  1.0  0.250009823130  0.250004420839  0.500001829199  Biso  1.000000  V
V006  1.0  -0.250010493367  -0.250004814719  0.500006362031  Biso  1.000000  V
V007  1.0  0.249991037996  -0.249986720179  0.500003493436  Biso  1.000000  V
V008  1.0  -0.249991691649  0.249986402005  0.500004742171  Biso  1.000000  V
Cl009  1.0  -0.418488967280  0.418499394776  0.464616772435  Biso  1.000000  Cl
Cl010  1.0  -0.418403128416  -0.081583683372  0.46460887628  Biso  1.000000  Cl
Cl011  1.0  0.081601885031  0.418432114339  0.46460845111  Biso  1.000000  Cl
Cl012  1.0  0.081492795771  -0.081489419957  0.464612292229  Biso  1.000000  Cl

\end{verbatim}
\begin{figure}[h]
    \centering
    \includegraphics[width=\textwidth]{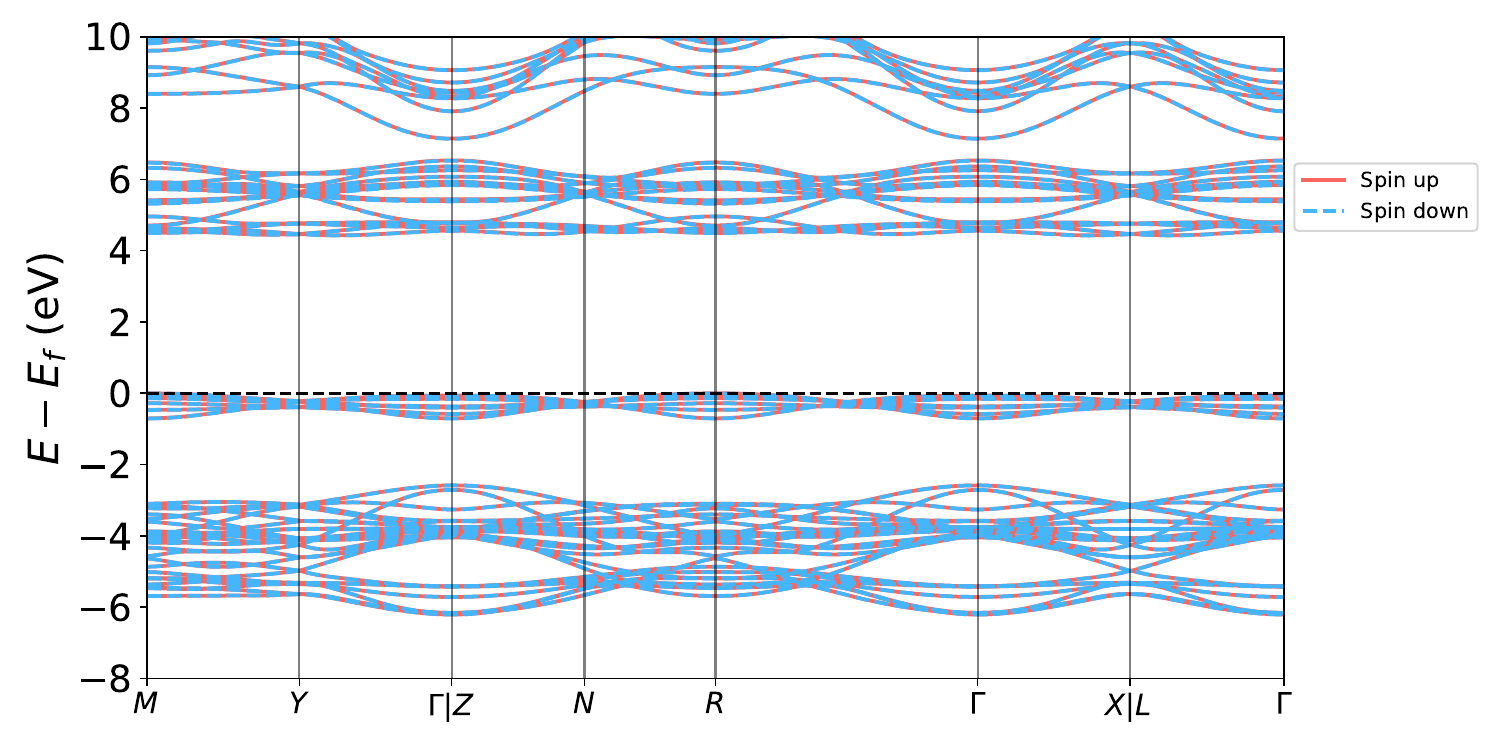}
    \caption{\footnotesize \ce{VCl2} slab spin-polarized band structure.}
    \label{fig:VCl2_band_slab}
\end{figure}
\clearpage
\newpage

\subsection{\ce{VI2} bulk}
\begin{verbatim}
_cell_length_a 4.00393702
_cell_length_b 4.00393702
_cell_length_c 6.28534848
_cell_angle_alpha 90.131708
_cell_angle_beta 89.868292
_cell_angle_gamma 120.043425
_symmetry_space_group_name_H-M         'P 1'
_symmetry_Int_Tables_number            1

loop_
_symmetry_equiv_pos_as_xyz
   'x, y, z'

loop_
_atom_site_label
_atom_site_type_symbol
_atom_site_fract_x
_atom_site_fract_y
_atom_site_fract_z
V001 V -1.552623493132E-21 1.552623493132E-21 -6.565763895986E-39
I002 I -3.337229697205E-01 3.337229697205E-01 2.527611037790E-01
I003 I 3.337229697205E-01 -3.337229697205E-01 -2.527611037790E-01

\end{verbatim}
\begin{figure}[h]
    \centering
    \includegraphics[width=\textwidth]{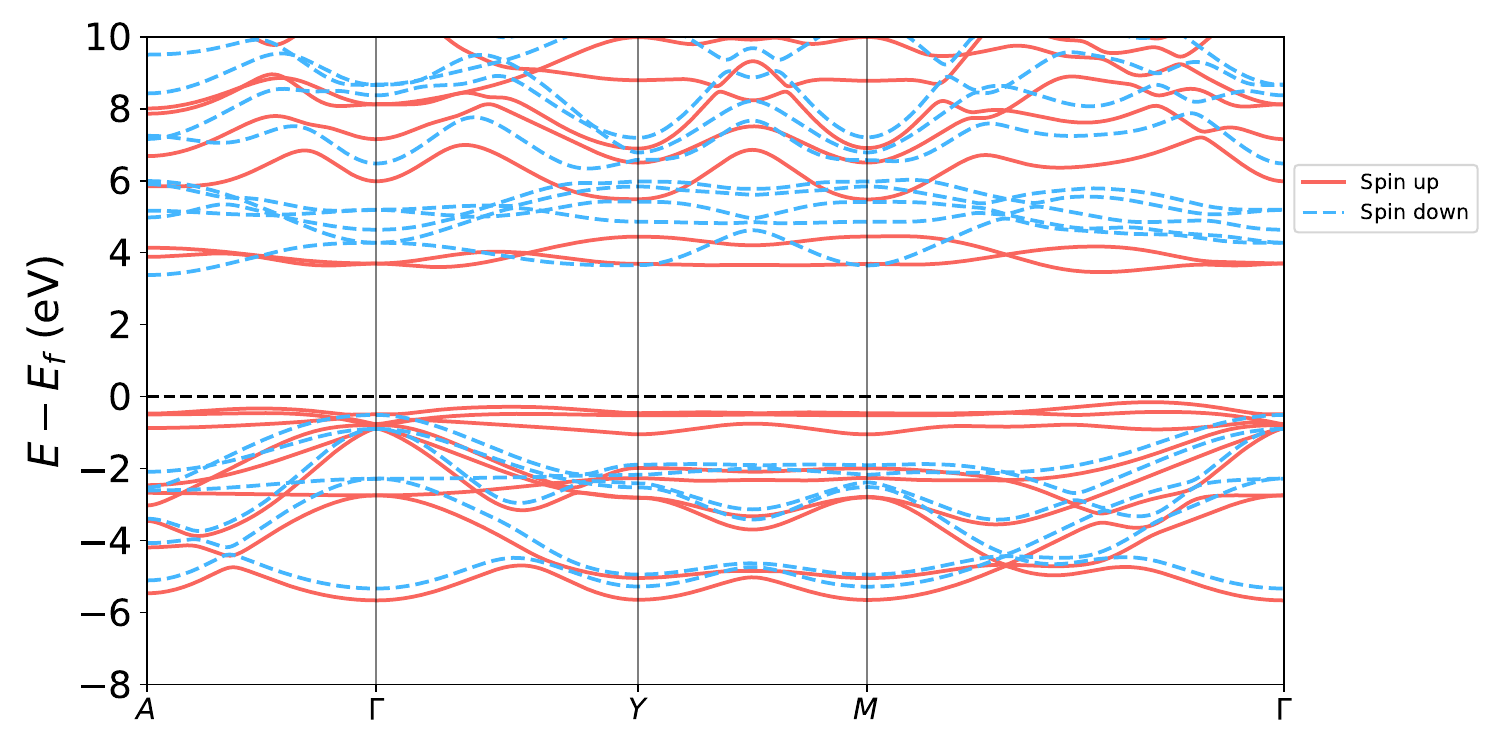}
    \caption{\footnotesize \ce{VI2} bulk spin-polarized band structure.}
    \label{fig:VI2_band_bulk}
\end{figure}
\clearpage
\newpage

\subsection{\ce{VI2} slab}
\begin{verbatim}
_cell_length_a                         7.92308209
_cell_length_b                         7.92323282
_cell_length_c                         40.00000000
_cell_angle_alpha                      90.000000
_cell_angle_beta                       90.000000
_cell_angle_gamma                      119.302105
_cell_volume                           'P 1'
_space_group_name_H-M_alt              'P 1'
_space_group_IT_number                 1

loop_
_space_group_symop_operation_xyz
   'x, y, z'

loop_
   _atom_site_label
   _atom_site_occupancy
   _atom_site_fract_x
   _atom_site_fract_y
   _atom_site_fract_z
   _atom_site_adp_type
   _atom_site_B_ios_or_equiv
   _atom_site_type_symbol
I001  1.0  0.167820364841  -0.167891876454  0.54072504747  Biso  1.000000  I
I002  1.0  0.167814318810  0.332187730383  0.540725850787  Biso  1.000000  I
I003  1.0  -0.332213322082  -0.167813863280  0.540728283919  Biso  1.000000  I
I004  1.0  -0.332164107901  0.332128257462  0.540724200028  Biso  1.000000  I
V005  1.0  0.000026516509  0.499926144861  0.499992815037  Biso  1.000000  V
V006  1.0  -0.499997652751  -0.000076404615  0.49999520797  Biso  1.000000  V
V007  1.0  0.000039851283  -0.000064052965  0.499983391077  Biso  1.000000  V
V008  1.0  -0.499933329924  0.499948059300  0.499980788633  Biso  1.000000  V
I009  1.0  -0.167801836608  0.167753916503  0.459261413411  Biso  1.000000  I
I010  1.0  -0.167683264970  -0.332318100375  0.4592417213  Biso  1.000000  I
I011  1.0  0.332295312214  0.167679658575  0.459243852991  Biso  1.000000  I
I012  1.0  0.332217115088  -0.332236619971  0.459256172653  Biso  1.000000  I

\end{verbatim}
\begin{figure}[h]
    \centering
    \includegraphics[width=\textwidth]{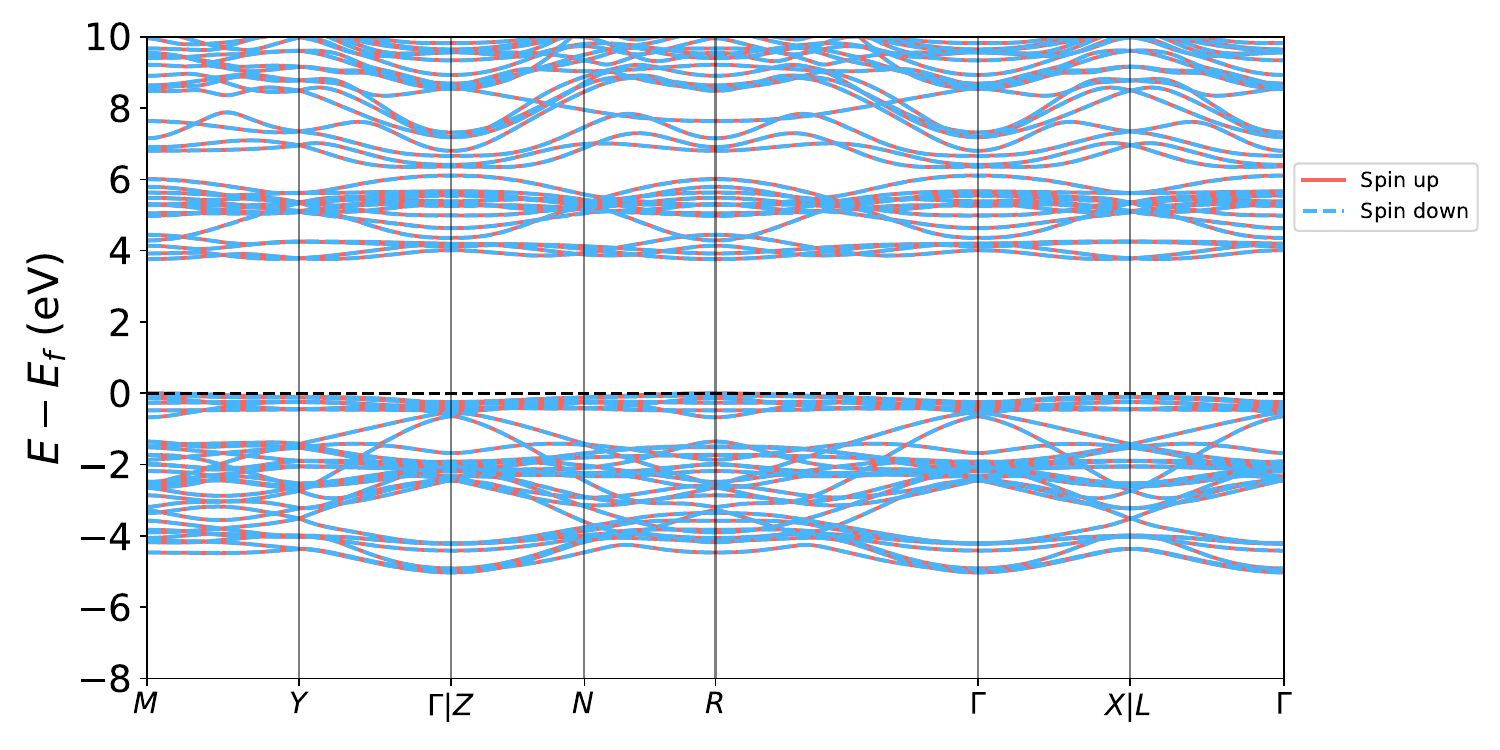}
    \caption{\footnotesize \ce{VI2} slab spin-polarized band structure.}
    \label{fig:VI2_band_slab}
\end{figure}

\clearpage
\newpage

\subsection{YCl bulk}
\begin{verbatim}
_cell_length_a                         3.523126
_cell_length_b                         6.102166
_cell_length_c                         9.189466
_cell_angle_alpha                      96.357010
_cell_angle_beta                       101.050117
_cell_angle_gamma                      90.009399
_cell_volume                           192.660203
_space_group_name_H-M_alt              'P 1'
_space_group_IT_number                 1

loop_
_space_group_symop_operation_xyz
   'x, y, z'

loop_
   _atom_site_label
   _atom_site_occupancy
   _atom_site_fract_x
   _atom_site_fract_y
   _atom_site_fract_z
   _atom_site_adp_type
   _atom_site_B_iso_or_equiv
   _atom_site_type_symbol
   Y1          1.0     0.574160     0.191410     0.148540    Biso  1.000000 Y
   Y2          1.0     0.074160     0.691360     0.148580    Biso  1.000000 Y
   Y3         1.0     0.425840     0.808590     0.851470    Biso  1.000000 Y
   Y4         1.0    -0.074150     0.308640     0.851420    Biso  1.000000 Y
   Cl1         1.0     0.171690     0.057380     0.343750    Biso  1.000000 Cl
   Cl2         1.0     0.671770     0.557330     0.343810    Biso  1.000000 Cl
   Cl3         1.0     0.828300    -0.057380     0.656250    Biso  1.000000 Cl
   Cl4         1.0     0.328240     0.442670     0.656190    Biso  1.000000 Cl

\end{verbatim}
\begin{figure}[h]
    \centering
    \includegraphics[width=\textwidth]{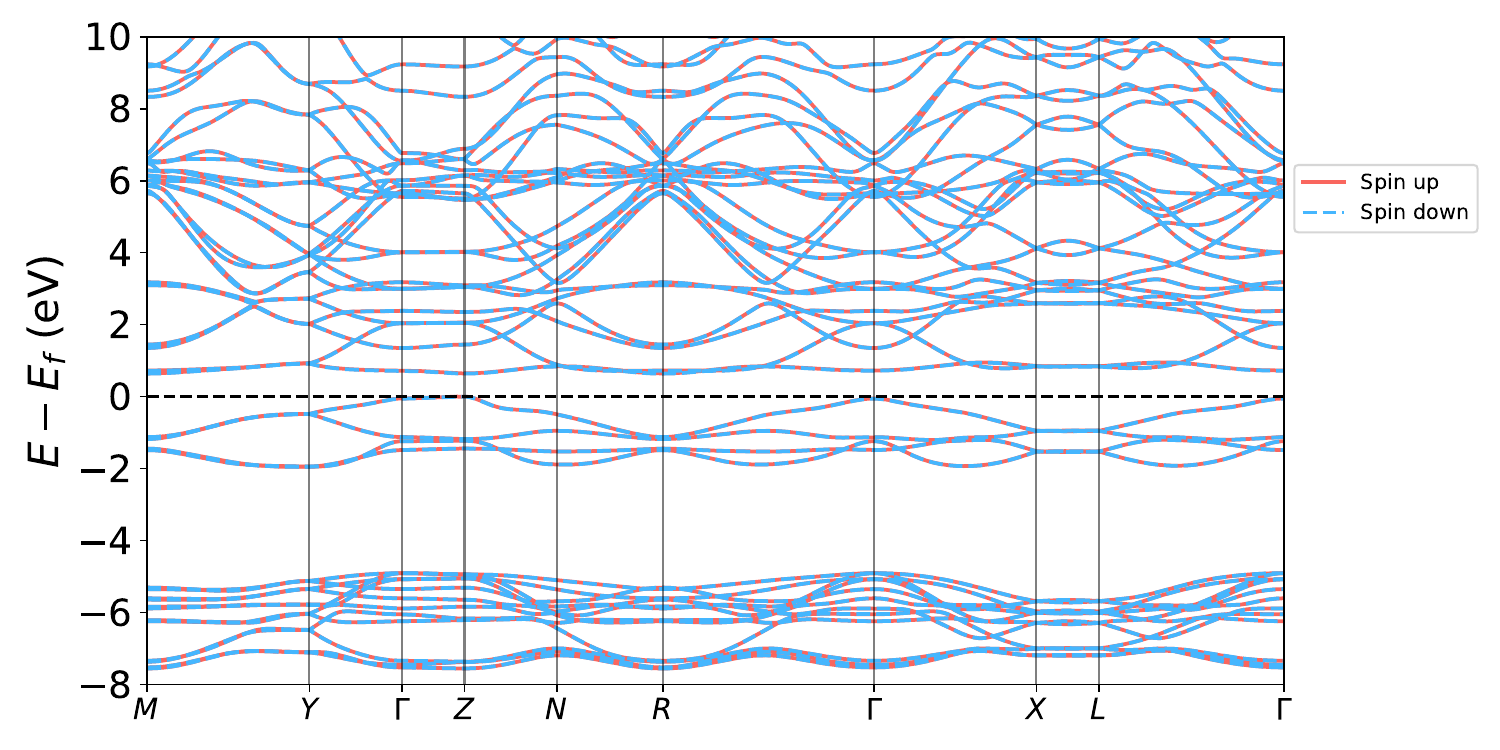}
    \caption{\footnotesize YCl bulk spin-polarized band structure.}
    \label{fig:YCl_band_bulk}
\end{figure}
\clearpage
\newpage

\subsection{YCl slab}
\begin{verbatim}
_cell_length_a                         3.41821501
_cell_length_b                         5.92271038
_cell_length_c                         40.00000000
_cell_angle_alpha                      90.000000
_cell_angle_beta                       90.000000
_cell_angle_gamma                      90.000098
_cell_volume                           'P 1'
_space_group_name_H-M_alt              'P 1'
_space_group_IT_number                 1

loop_
_space_group_symop_operation_xyz
   'x, y, z'

loop_
   _atom_site_label
   _atom_site_occupancy
   _atom_site_fract_x
   _atom_site_fract_y
   _atom_site_fract_z
   _atom_site_adp_type
   _atom_site_B_ios_or_equiv
   _atom_site_type_symbol
Y001  1.0  0.377235191368  0.210977372282  0.52208857094  Biso  1.000000  Y
Y002  1.0  -0.122797773614  -0.289020599803  0.522087886378  Biso  1.000000  Y
Cl003  1.0  -0.122757510629  0.044291869585  0.477920523229  Biso  1.000000  Cl
Cl004  1.0  0.377193659507  -0.455713087738  0.477920626426  Biso  1.000000  Cl

\end{verbatim}

\begin{figure}[h]
    \centering
    \includegraphics[width=\textwidth]{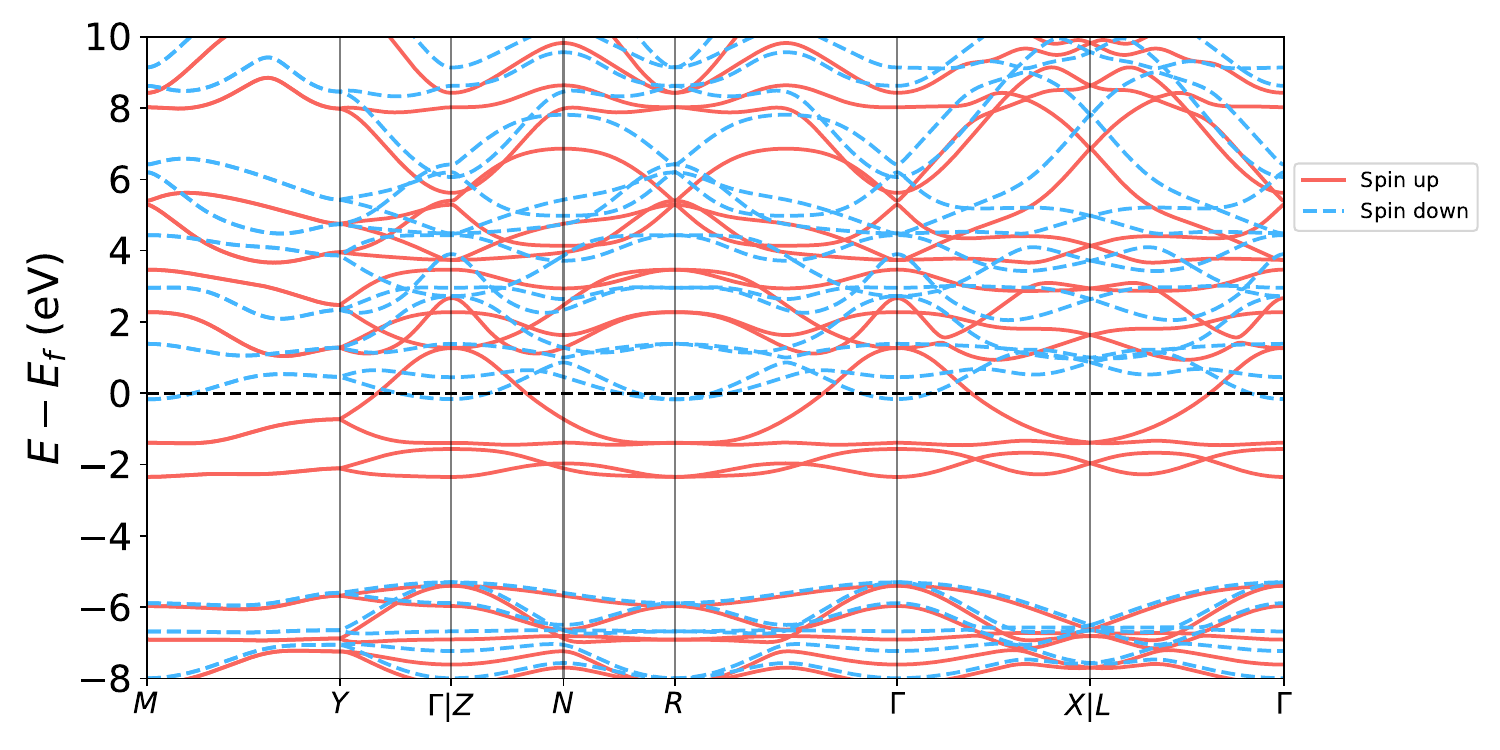}
    \caption{\footnotesize YCl slab spin-polarized band structure.}
    \label{fig:YCl_band_slab}
\end{figure}
\clearpage
\newpage

\subsection{\ce{ZnBr2} bulk}
\begin{verbatim}
_cell_length_a 3.68055906
_cell_length_b 3.68055906
_cell_length_c 5.95504394
_cell_angle_alpha 90.000000
_cell_angle_beta 90.000000
_cell_angle_gamma 120.000000
_symmetry_space_group_name_H-M         'P 1'
_symmetry_Int_Tables_number            1

loop_
_symmetry_equiv_pos_as_xyz
   'x, y, z'

loop_
_atom_site_label
_atom_site_type_symbol
_atom_site_fract_x
_atom_site_fract_y
_atom_site_fract_z
Zn001 Zn 0.000000000000E+00 0.000000000000E+00 0.000000000000E+00
Br002 Br 3.333333333333E-01 -3.333333333333E-01 -2.470768292339E-01
Br003 Br -3.333333333333E-01 3.333333333333E-01 2.470768292339E-01

\end{verbatim}
\begin{figure}[h]
    \centering
    \includegraphics[width=\textwidth]{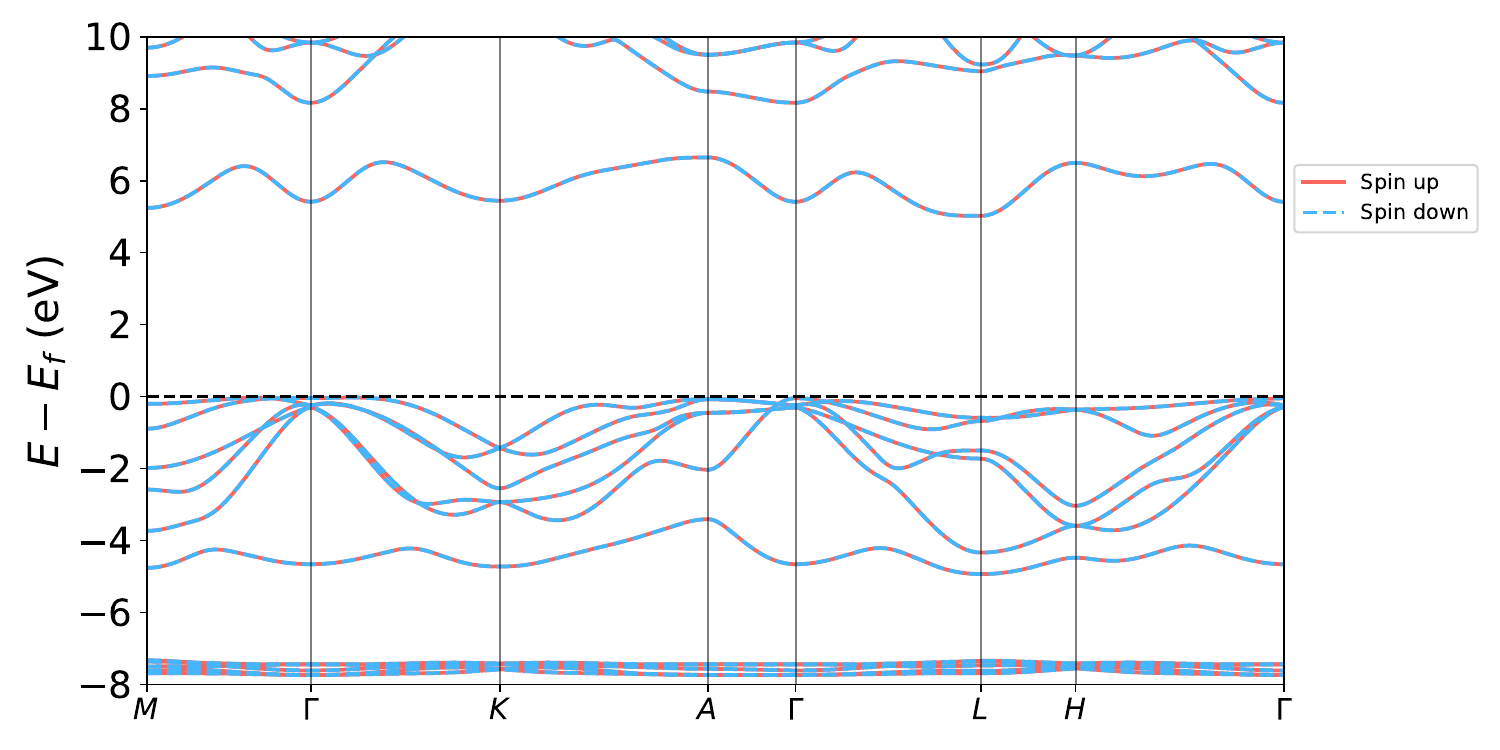}
    \caption{\footnotesize \ce{ZnBr2} bulk spin-polarized band structure.}
    \label{fig:ZnBr2_band_bulk}
\end{figure}

\clearpage
\newpage

\subsection{\ce{ZnBr2} slab}
\begin{verbatim}
_cell_length_a                         3.68559637
_cell_length_b                         3.68559637
_cell_length_c                         40.00000000
_cell_angle_alpha                      90.000000
_cell_angle_beta                       90.000000
_cell_angle_gamma                      120.000000
_cell_volume                           'P 1'
_space_group_name_H-M_alt              'P 1'
_space_group_IT_number                 1

loop_
_space_group_symop_operation_xyz
   'x, y, z'

loop_
   _atom_site_label
   _atom_site_occupancy
   _atom_site_fract_x
   _atom_site_fract_y
   _atom_site_fract_z
   _atom_site_adp_type
   _atom_site_B_ios_or_equiv
   _atom_site_type_symbol
Br001  1.0  -0.333333333333  0.333333333333  0.536893618086  Biso  1.000000  Br
Zn002  1.0  0.000000000000  0.000000000000  0.5  Biso  1.000000  Zn
Br003  1.0  0.333333333333  -0.333333333333  0.463106381914  Biso  1.000000  Br

\end{verbatim}
\begin{figure}[h]
    \centering
    \includegraphics[width=\textwidth]{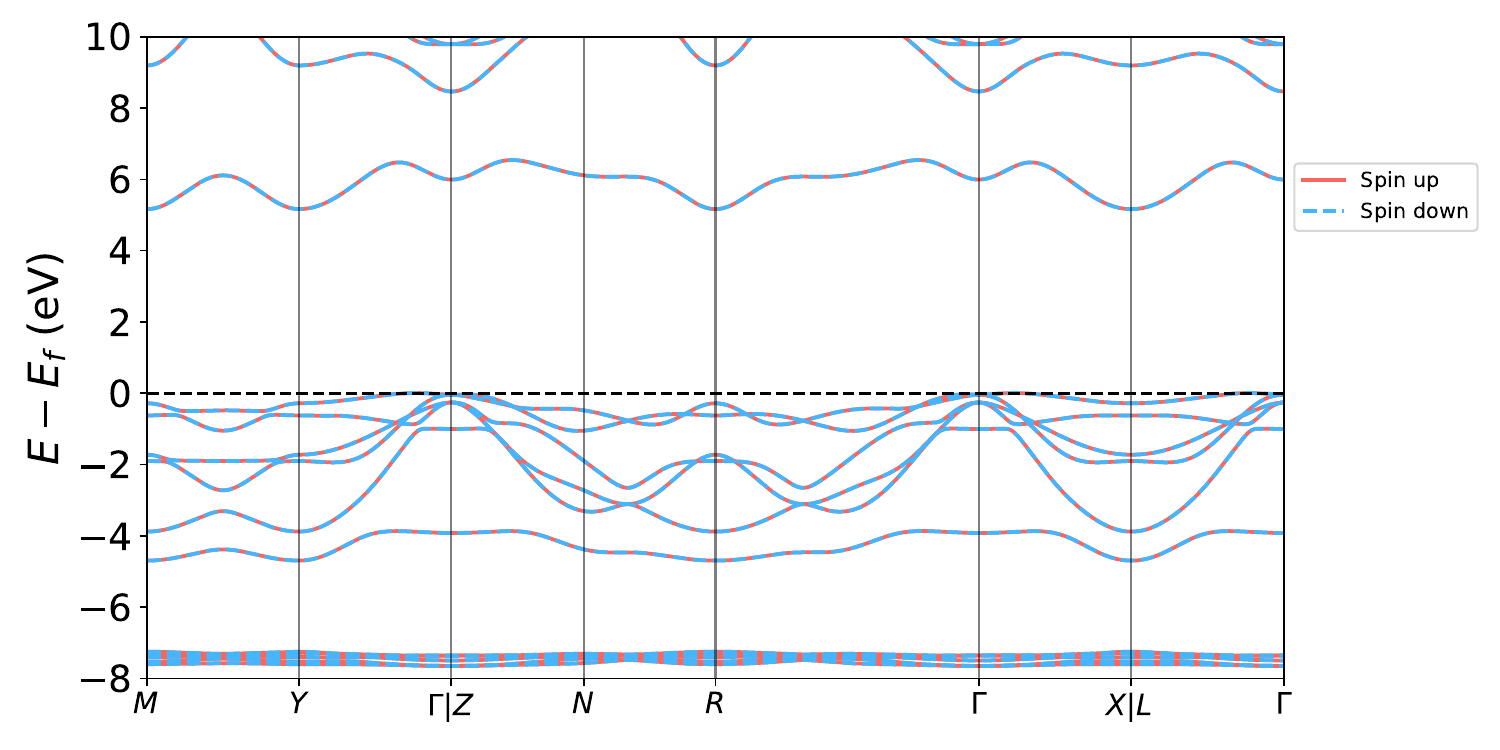}
    \caption{\footnotesize \ce{ZnBr2} slab spin-polarized band structure.}
    \label{fig:ZnBr2_band_slab}
\end{figure}
\clearpage
\newpage

\subsection{\ce{ZnCl2} bulk}
\begin{verbatim}
_cell_length_a 3.51273516
_cell_length_b 3.51273516
_cell_length_c 5.65207304
_cell_angle_alpha 89.955040
_cell_angle_beta 90.044960
_cell_angle_gamma 119.999222
_symmetry_space_group_name_H-M         'P 1'
_symmetry_Int_Tables_number            1

loop_
_symmetry_equiv_pos_as_xyz
   'x, y, z'

loop_
_atom_site_label
_atom_site_type_symbol
_atom_site_fract_x
_atom_site_fract_y
_atom_site_fract_z
Cl001 Cl 3.330984179554E-01 -3.330984179554E-01 -2.424218082597E-01
Cl002 Cl -3.330984179554E-01 3.330984179554E-01 2.424218082597E-01
Zn003 Zn -4.063879496324E-37 -2.369320658322E-37 -1.665162657906E-36

\end{verbatim}
\begin{figure}[h]
    \centering
    \includegraphics[width=\textwidth]{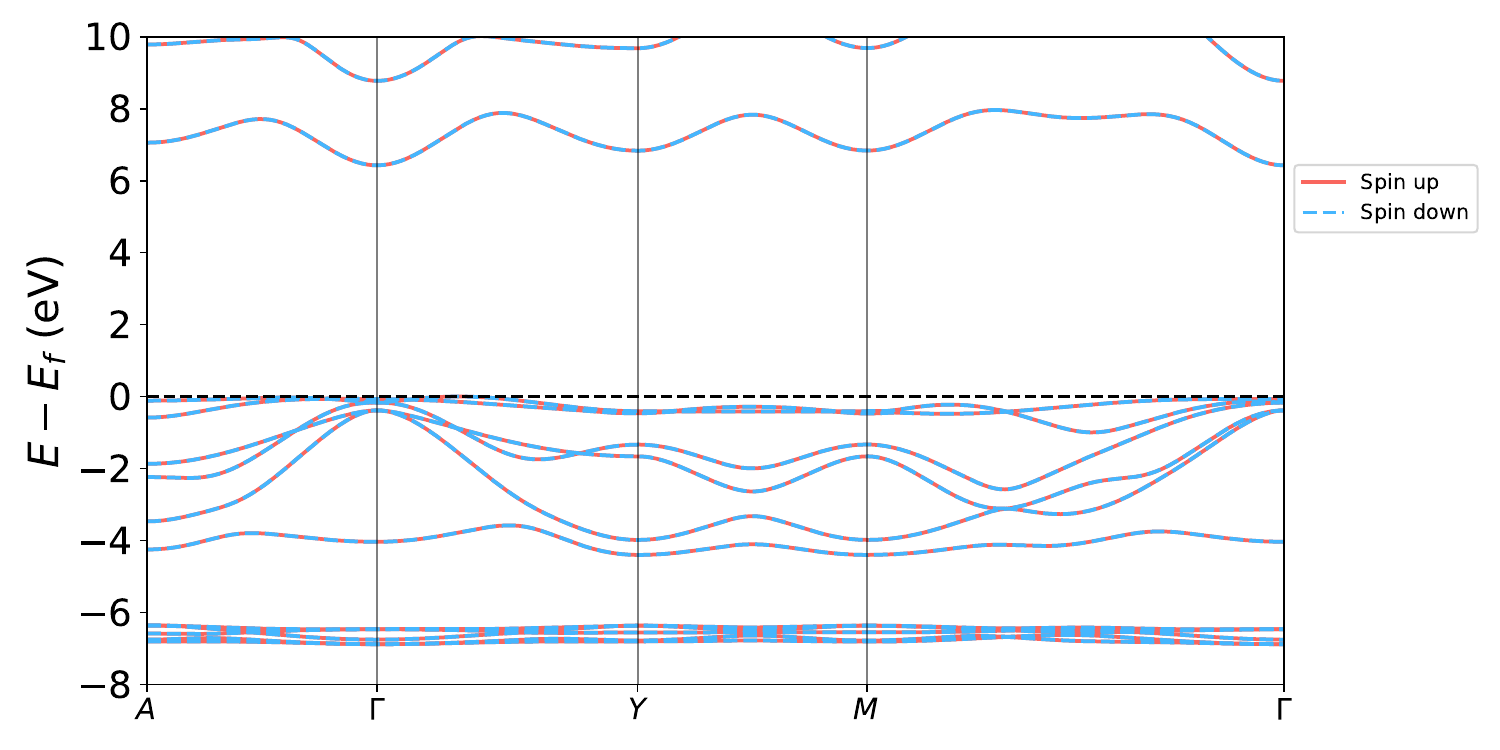}
    \caption{\footnotesize \ce{ZnCl2} bulk spin-polarized band structure.}
    \label{fig:ZnCl2_band_bulk}
\end{figure}
\clearpage
\newpage

\subsection{\ce{ZnCl2} slab}
\begin{verbatim}
_cell_length_a                         3.51308511
_cell_length_b                         3.51308511
_cell_length_c                         40.00000000
_cell_angle_alpha                      90.000000
_cell_angle_beta                       90.000000
_cell_angle_gamma                      119.998893
_cell_volume                           'P 1'
_space_group_name_H-M_alt              'P 1'
_space_group_IT_number                 1

loop_
_space_group_symop_operation_xyz
   'x, y, z'

loop_
   _atom_site_label
   _atom_site_occupancy
   _atom_site_fract_x
   _atom_site_fract_y
   _atom_site_fract_z
   _atom_site_adp_type
   _atom_site_B_ios_or_equiv
   _atom_site_type_symbol
Cl001  1.0  -0.333319718788  0.333319718788  0.534463026039  Biso  1.000000  Cl
Zn002  1.0  0.000000000000  0.000000000000  0.5  Biso  1.000000  Zn
Cl003  1.0  0.333319718788  -0.333319718788  0.465536973961  Biso  1.000000  Cl

\end{verbatim}
\begin{figure}[h]
    \centering
    \includegraphics[width=\textwidth]{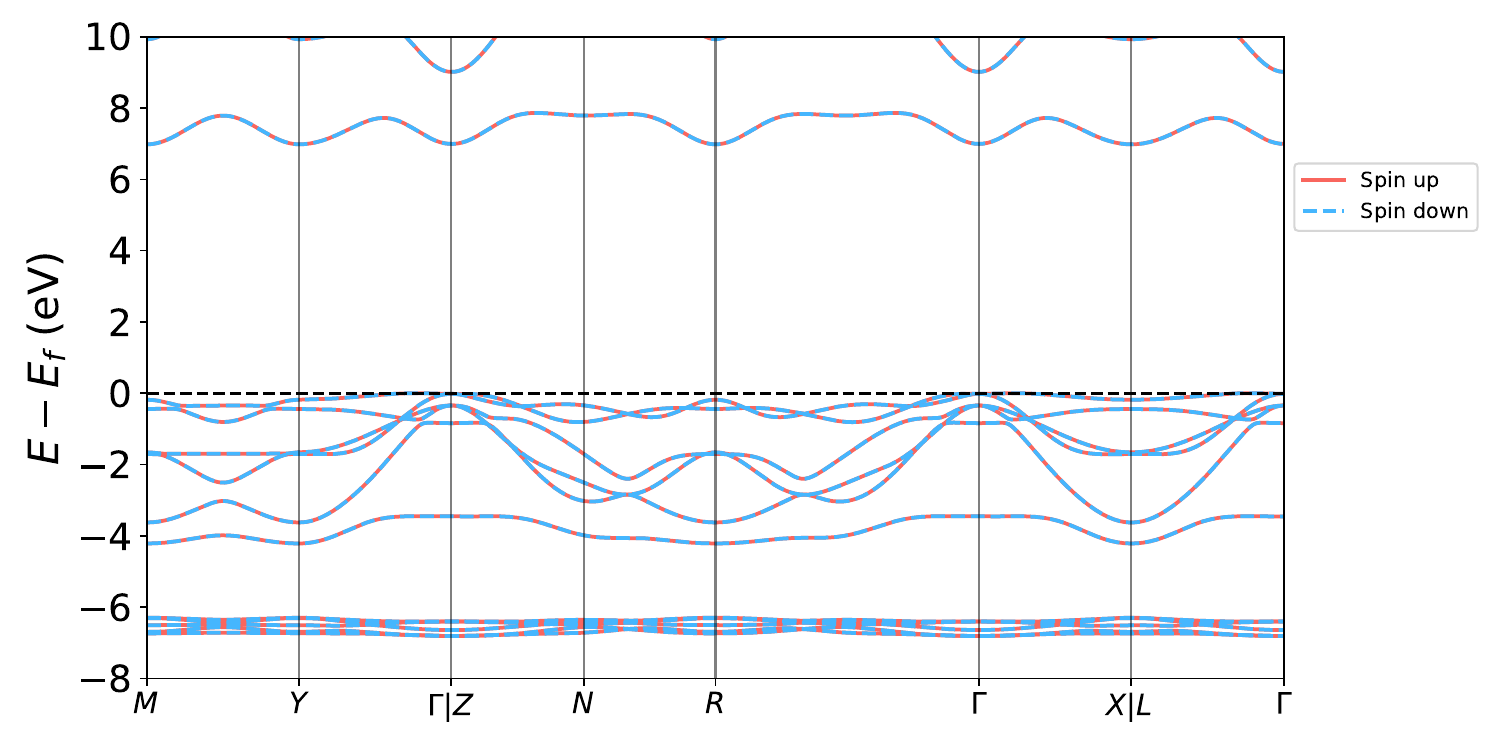}
    \caption{\footnotesize \ce{ZnCl2} slab spin-polarized band structure.}
    \label{fig:ZnCl2_band_slab}
\end{figure}

\clearpage
\newpage

\subsection{\ce{ZnI2} bulk}
\begin{verbatim}
_cell_length_a 3.95534587
_cell_length_b 3.95534587
_cell_length_c 6.36661350
_cell_angle_alpha 90.000000
_cell_angle_beta 90.000000
_cell_angle_gamma 120.000000
_symmetry_space_group_name_H-M         'P 1'
_symmetry_Int_Tables_number            1

loop_
_symmetry_equiv_pos_as_xyz
   'x, y, z'

loop_
_atom_site_label
_atom_site_type_symbol
_atom_site_fract_x
_atom_site_fract_y
_atom_site_fract_z
I001 I 3.333333333333E-01 -3.333333333333E-01 2.527841191049E-01
I002 I -3.333333333333E-01 3.333333333333E-01 -2.527841191049E-01
Zn003 Zn 0.000000000000E+00 0.000000000000E+00 0.000000000000E+00

\end{verbatim}
\begin{figure}[h]
    \centering
    \includegraphics[width=\textwidth]{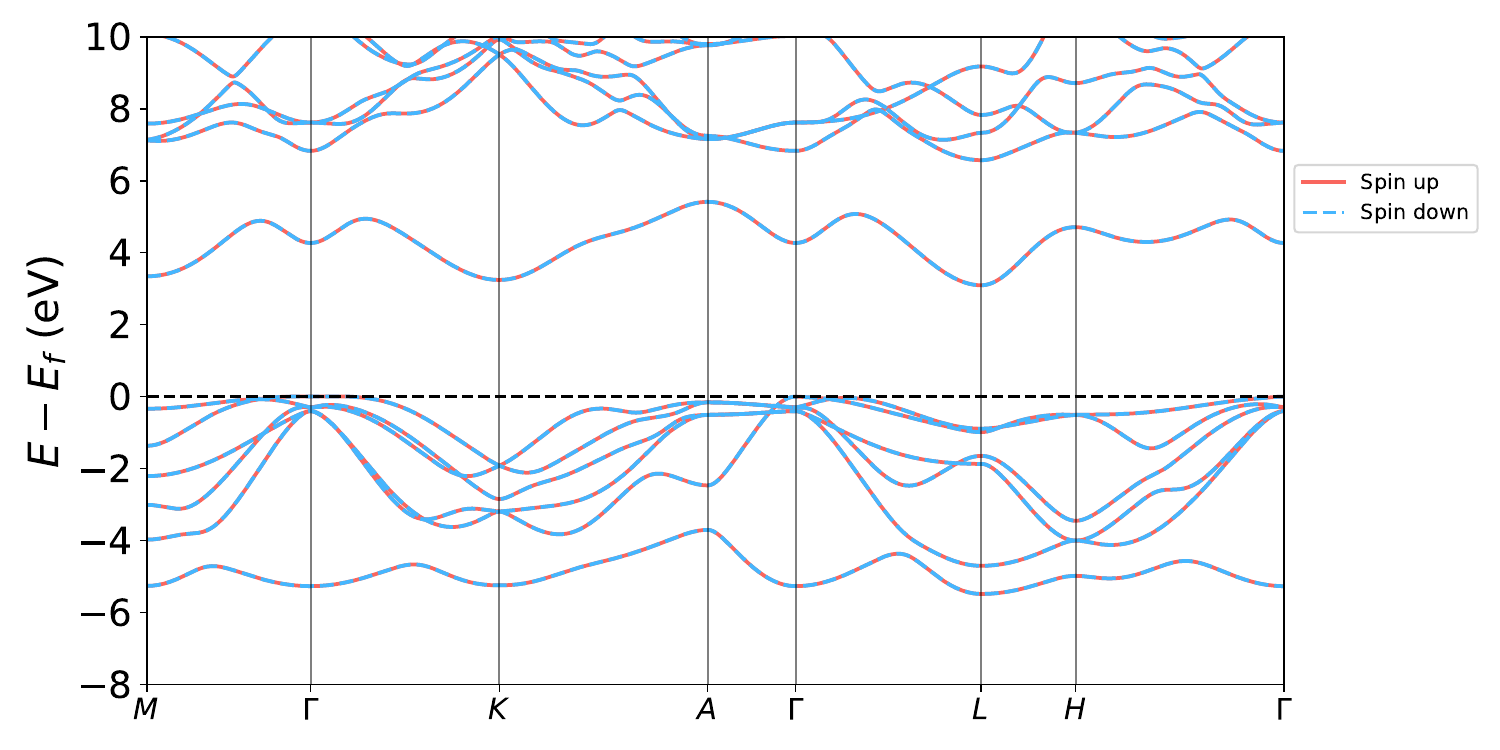}
    \caption{\footnotesize \ce{ZnI2} bulk spin-polarized band structure.}
    \label{fig:ZnI2_band_bulk}
\end{figure}
\clearpage
\newpage

\subsection{\ce{ZnI2} slab}
\begin{verbatim}
_cell_length_a                         3.96327246
_cell_length_b                         3.96327246
_cell_length_c                         40.00000000
_cell_angle_alpha                      90.000000
_cell_angle_beta                       90.000000
_cell_angle_gamma                      120.000000
_cell_volume                           'P 1'
_space_group_name_H-M_alt              'P 1'
_space_group_IT_number                 1

loop_
_space_group_symop_operation_xyz
   'x, y, z'

loop_
   _atom_site_label
   _atom_site_occupancy
   _atom_site_fract_x
   _atom_site_fract_y
   _atom_site_fract_z
   _atom_site_adp_type
   _atom_site_B_ios_or_equiv
   _atom_site_type_symbol
I001  1.0  0.333333333333  -0.333333333333  0.540361893847  Biso  1.000000  I
Zn002  1.0  0.000000000000  0.000000000000  0.5  Biso  1.000000  Zn
I003  1.0  -0.333333333333  0.333333333333  0.459638106153  Biso  1.000000  I

\end{verbatim}
\begin{figure}[h]
    \centering
    \includegraphics[width=\textwidth]{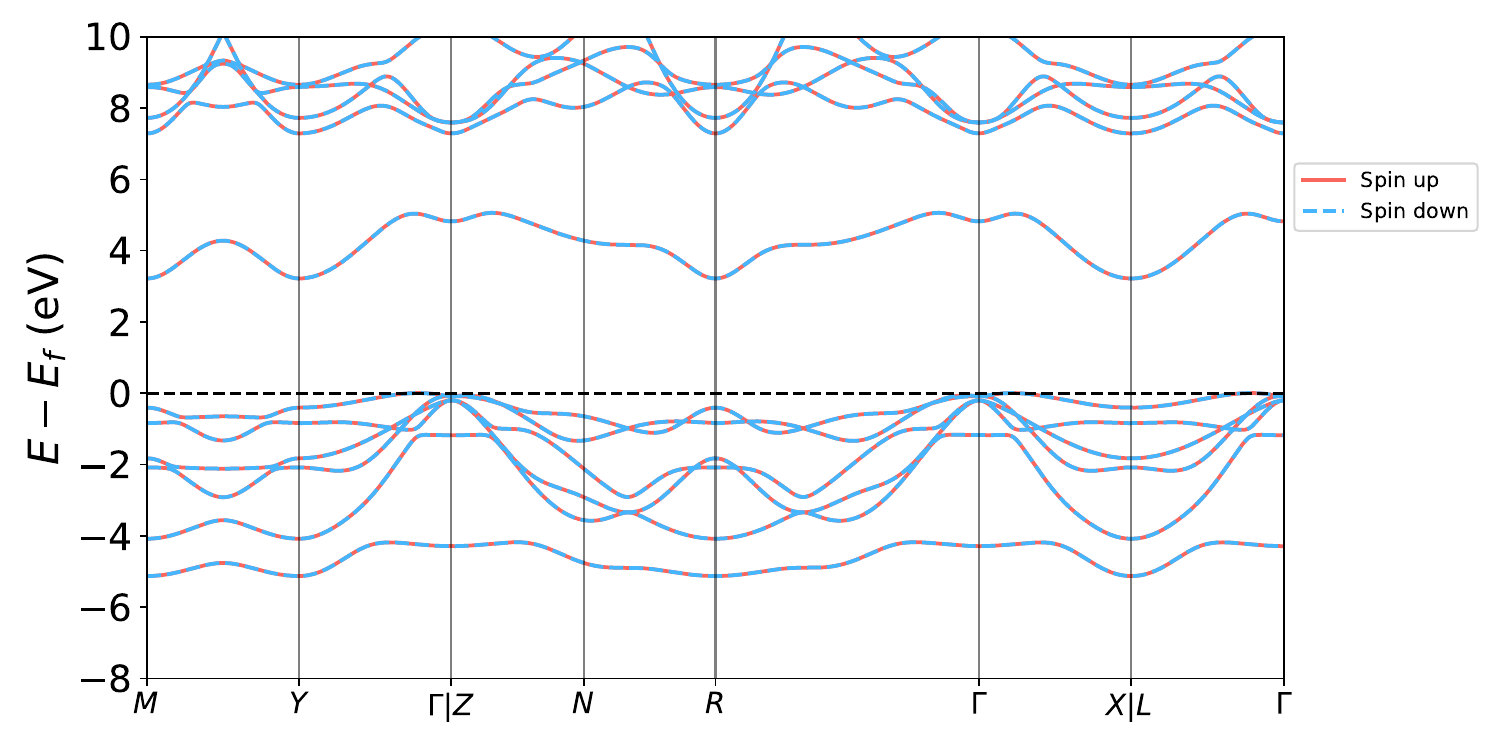}
    \caption{\footnotesize \ce{ZnI2} slab spin-polarized band structure.}
    \label{fig:ZnI2_band_slab}
\end{figure}

\clearpage
\newpage

\subsection{\ce{ZrCl2} bulk}
\begin{verbatim}
_cell_length_a 3.33795740
_cell_length_b 3.33795740
_cell_length_c 6.68459329
_cell_angle_alpha 90.000000
_cell_angle_beta 90.000000
_cell_angle_gamma 120.000000
_symmetry_space_group_name_H-M         'P 1'
_symmetry_Int_Tables_number            1

loop_
_symmetry_equiv_pos_as_xyz
   'x, y, z'

loop_
_atom_site_label
_atom_site_type_symbol
_atom_site_fract_x
_atom_site_fract_y
_atom_site_fract_z
Cl001 Cl -3.333333333333E-01 3.333333333333E-01 -2.543952580837E-01
Cl002 Cl -3.333333333333E-01 3.333333333333E-01 2.543952580837E-01
Zr003 Zr 3.333333333333E-01 -3.333333333333E-01 0.000000000000E+00

\end{verbatim}
\begin{figure}[h]
    \centering
    \includegraphics[width=\textwidth]{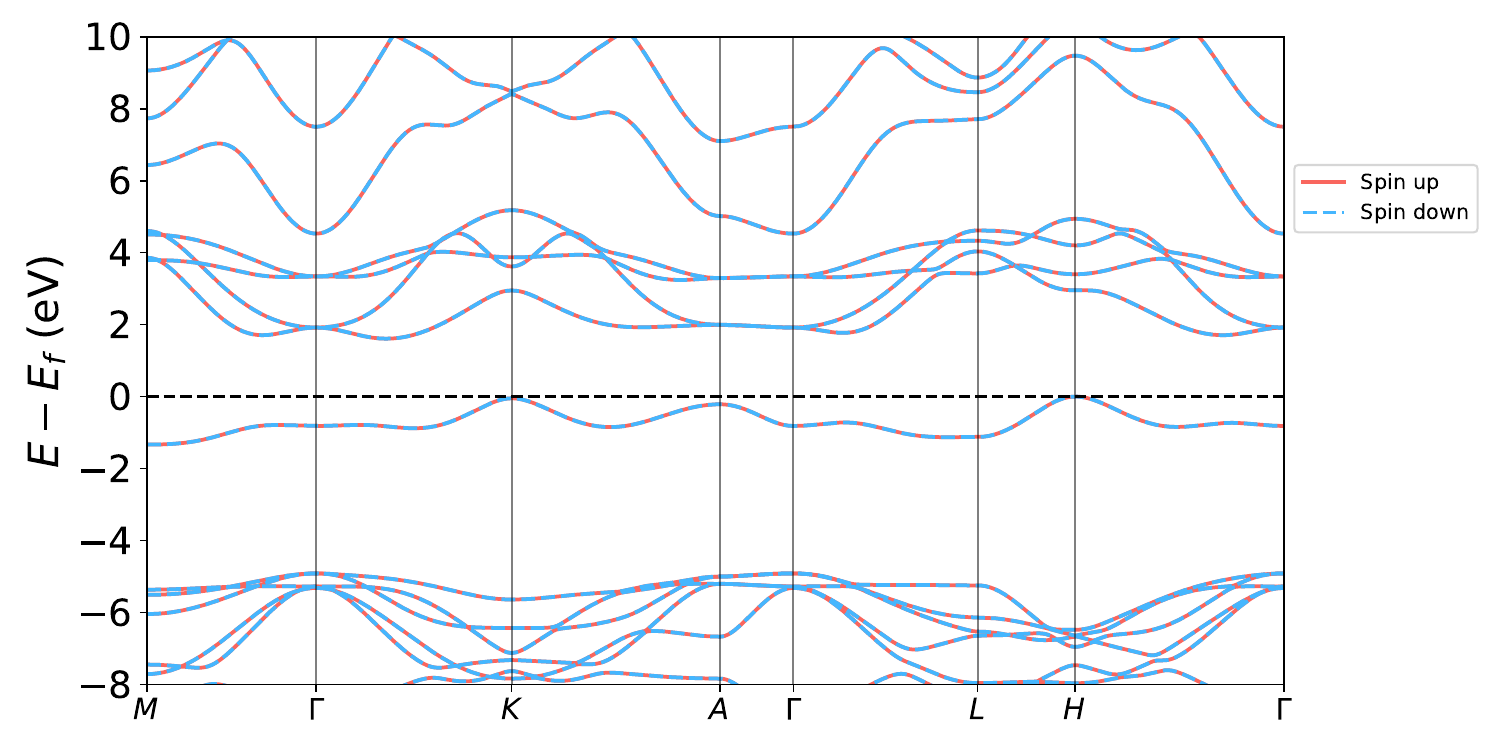}
    \caption{\footnotesize \ce{ZrCl2} bulk spin-polarized band structure.}
    \label{fig:ZrCl2_band_bulk}
\end{figure}

\clearpage
\newpage

\subsection{\ce{ZrCl2} slab}
\begin{verbatim}
_cell_length_a                         3.34506017
_cell_length_b                         3.34506017
_cell_length_c                         40.00000000
_cell_angle_alpha                      90.000000
_cell_angle_beta                       90.000000
_cell_angle_gamma                      120.000000
_cell_volume                           'P 1'
_space_group_name_H-M_alt              'P 1'
_space_group_IT_number                 1

loop_
_space_group_symop_operation_xyz
   'x, y, z'

loop_
   _atom_site_label
   _atom_site_occupancy
   _atom_site_fract_x
   _atom_site_fract_y
   _atom_site_fract_z
   _atom_site_adp_type
   _atom_site_B_ios_or_equiv
   _atom_site_type_symbol
Cl001  1.0  -0.333333333333  0.333333333333  0.5426418855490001  Biso  1.000000  Cl
Zr002  1.0  0.333333333333  -0.333333333333  0.5  Biso  1.000000  Zr
Cl003  1.0  -0.333333333333  0.333333333333  0.457358114451  Biso  1.000000  Cl

\end{verbatim}
\begin{figure}[h]
    \centering
    \includegraphics[width=\textwidth]{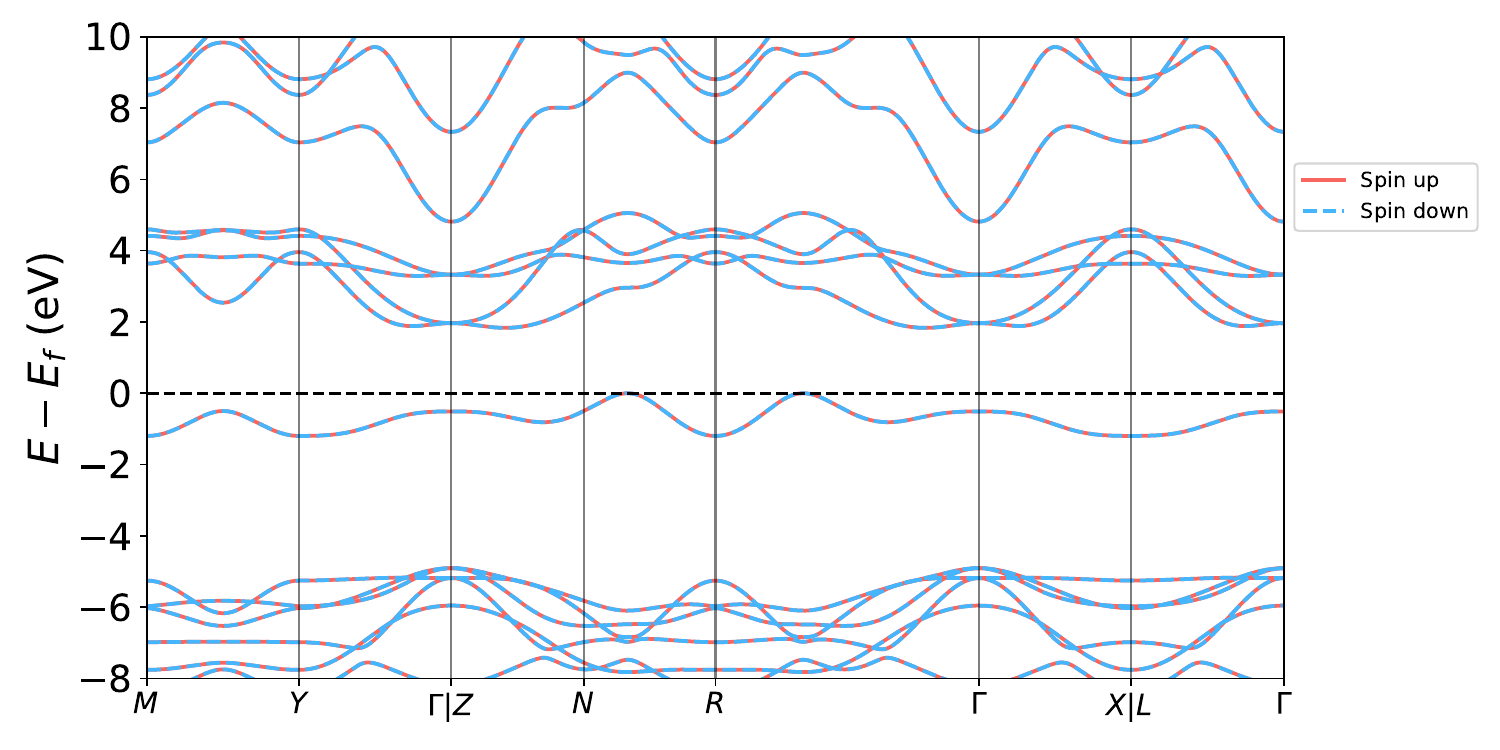}
    \caption{\footnotesize \ce{ZrCl2} slab spin-polarized band structure.}
    \label{fig:ZrCl2_band_slab}
\end{figure}

\clearpage
\newpage

\subsection{ZrCl bulk}
\begin{verbatim}
_cell_length_a 3.27483162
_cell_length_b 3.31257447
_cell_length_c 8.93288254
_cell_angle_alpha 89.866558
_cell_angle_beta 89.886613
_cell_angle_gamma 60.602468
_symmetry_space_group_name_H-M         'P 1'
_symmetry_Int_Tables_number            1

loop_
_symmetry_equiv_pos_as_xyz
   'x, y, z'

loop_
_atom_site_label
_atom_site_type_symbol
_atom_site_fract_x
_atom_site_fract_y
_atom_site_fract_z
Zr001 Zr 3.346495671453E-01 3.300935604069E-01 -1.263728513776E-01
Zr002 Zr -3.346495671453E-01 -3.300935604069E-01 1.263728513776E-01
Cl003 Cl -3.211125716628E-03 6.564293017944E-03 3.183359143719E-01
Cl004 Cl 3.211125716628E-03 -6.564293017944E-03 -3.183359143719E-01

\end{verbatim}
\begin{figure}[h]
    \centering
    \includegraphics[width=\textwidth]{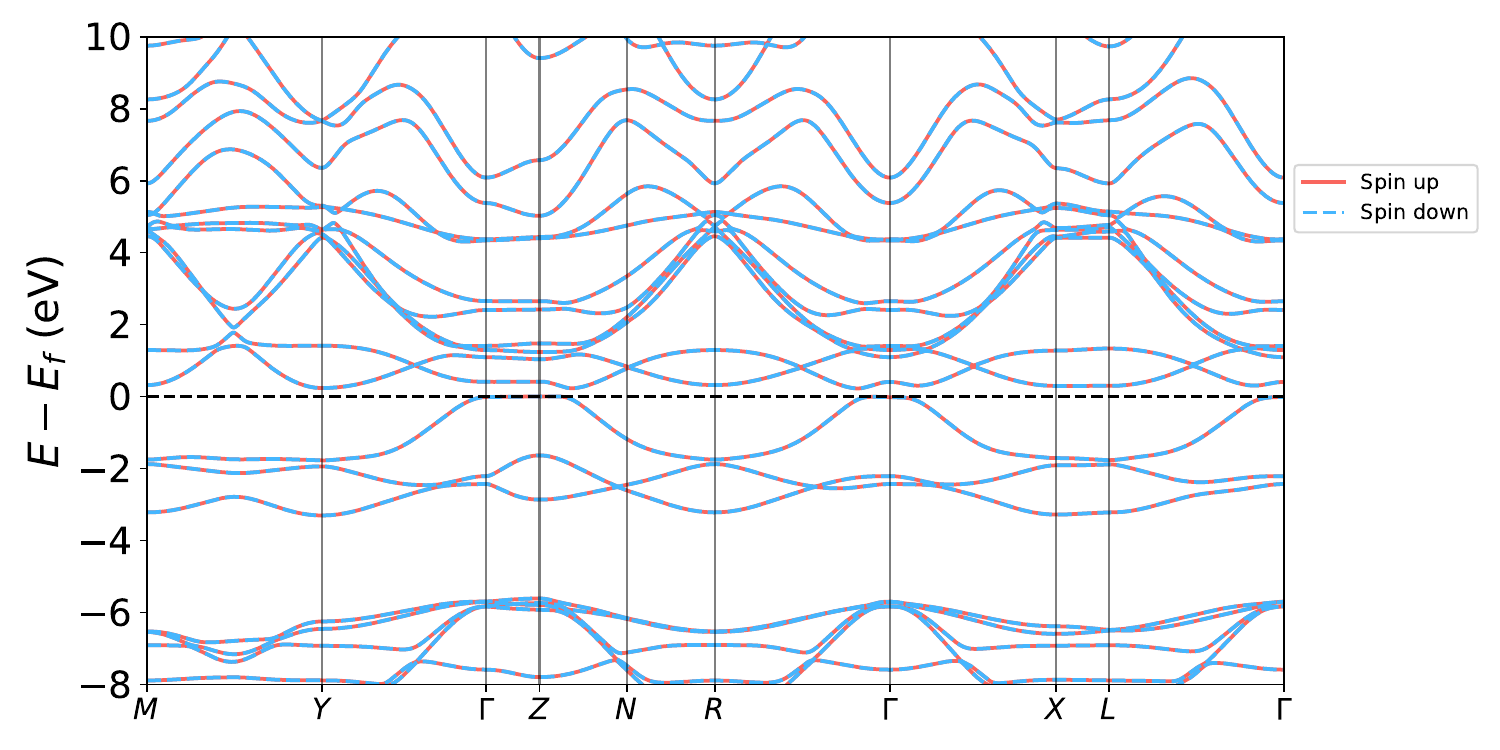}
    \caption{\footnotesize ZrCl bulk spin-polarized band structure.}
    \label{fig:ZrCl_band_bulk}
\end{figure}
\clearpage
\newpage

\subsection{ZrCl slab}
\begin{verbatim}
_cell_length_a                         3.27892198
_cell_length_b                         3.31575827
_cell_length_c                         40.00000000
_cell_angle_alpha                      90.000000
_cell_angle_beta                       90.000000
_cell_angle_gamma                      119.365823
_cell_volume                           'P 1'
_space_group_name_H-M_alt              'P 1'
_space_group_IT_number                 1

loop_
_space_group_symop_operation_xyz
   'x, y, z'

loop_
   _atom_site_label
   _atom_site_occupancy
   _atom_site_fract_x
   _atom_site_fract_y
   _atom_site_fract_z
   _atom_site_adp_type
   _atom_site_B_ios_or_equiv
   _atom_site_type_symbol
Cl001  1.0  0.002077355206  0.008441250595  0.571269992688  Biso  1.000000  Cl
Zr002  1.0  0.334111135550  -0.329197663741  0.528244306205  Biso  1.000000  Zr
Zr003  1.0  -0.334111135550  0.329197663741  0.471755693795  Biso  1.000000  Zr
Cl004  1.0  -0.002077355206  -0.008441250595  0.428730007312  Biso  1.000000  Cl

\end{verbatim}
\begin{figure}[h]
    \centering
    \includegraphics[width=\textwidth]{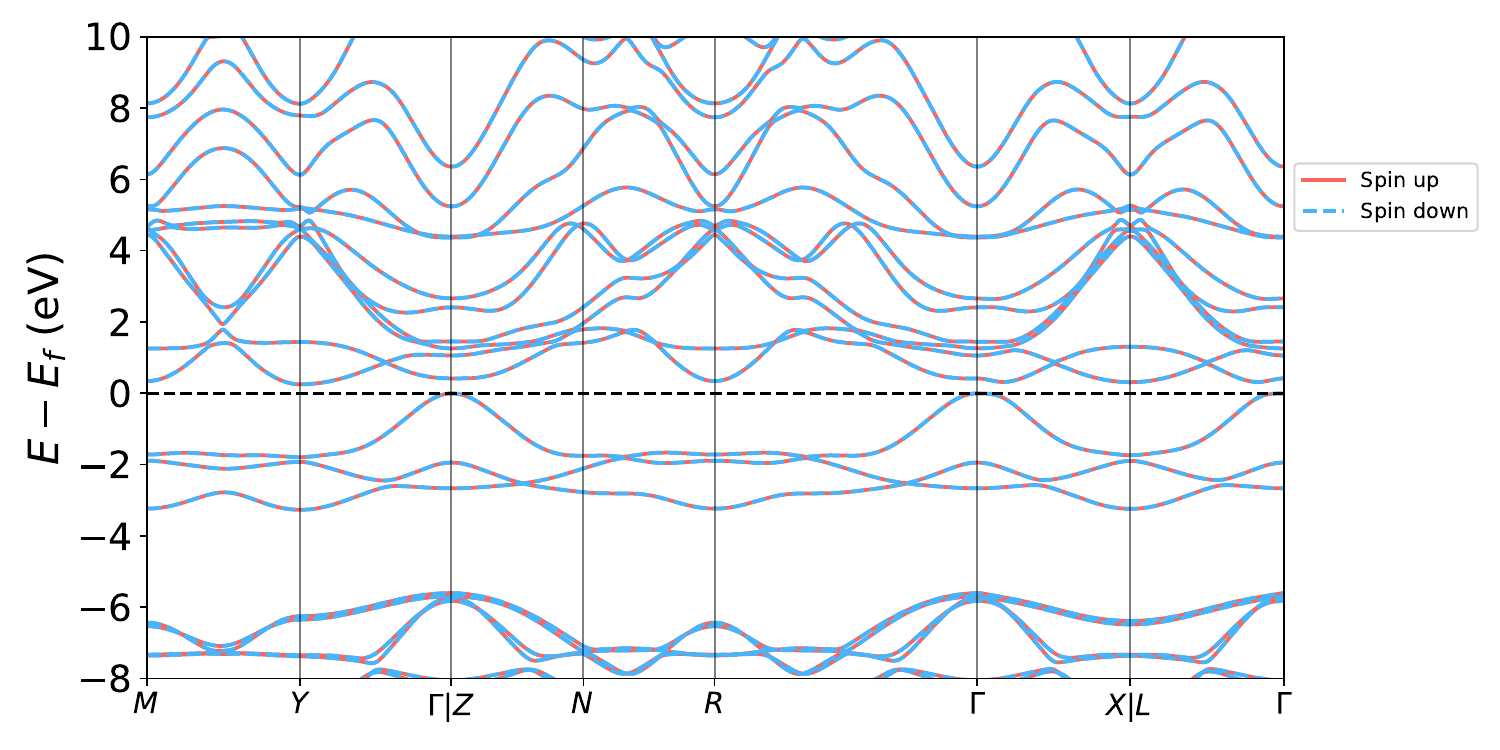}
    \caption{\footnotesize ZrCl slab spin-polarized band structure.}
    \label{fig:ZrCl_band_slab}
\end{figure}
\clearpage
\newpage

\subsection{\ce{ZrI2} bulk}
\begin{verbatim}
_cell_length_a 6.65626353
_cell_length_b 3.64187829
_cell_length_c 7.08656638
_cell_angle_alpha 90.000000
_cell_angle_beta 93.484326
_cell_angle_gamma 90.000000
_symmetry_space_group_name_H-M         'P 1'
_symmetry_Int_Tables_number            1

loop_
_symmetry_equiv_pos_as_xyz
   'x, y, z'

loop_
_atom_site_label
_atom_site_type_symbol
_atom_site_fract_x
_atom_site_fract_y
_atom_site_fract_z
Zr001 Zr -1.921543928987E-01 2.500000000000E-01 -4.179658456218E-03
Zr002 Zr 1.921543928987E-01 -2.500000000000E-01 4.179658456218E-03
I003 I 4.100144765109E-01 2.500000000000E-01 -2.332049898198E-01
I004 I -4.100144765109E-01 -2.500000000000E-01 2.332049898198E-01
I005 I 1.044685313115E-01 2.500000000000E-01 3.000276012387E-01
I006 I -1.044685313115E-01 -2.500000000000E-01 -3.000276012387E-01

\end{verbatim}
\begin{figure}[h]
    \centering
    \includegraphics[width=\textwidth]{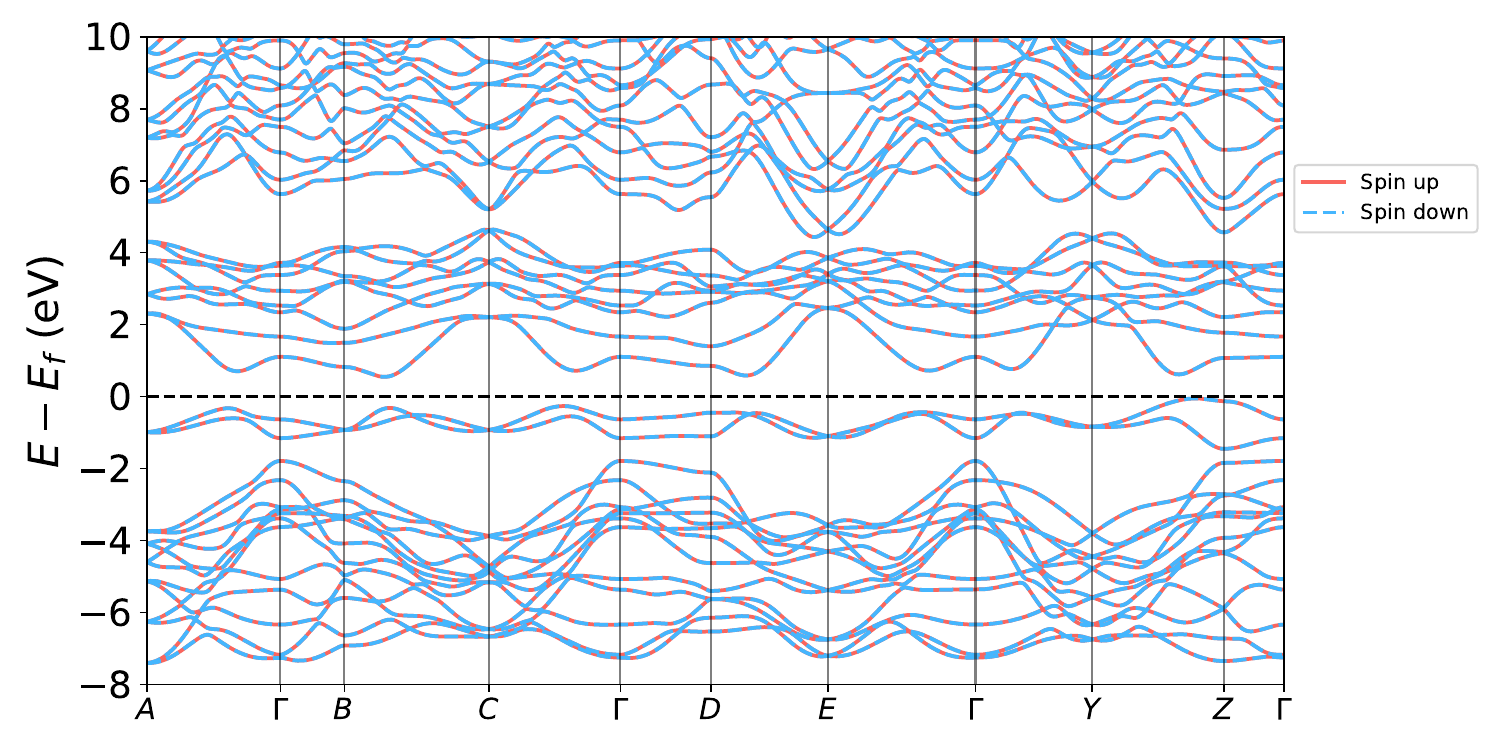}
    \caption{\footnotesize \ce{ZrI2} bulk spin-polarized band structure.}
    \label{fig:ZrI2_band_bulk}
\end{figure}

\clearpage
\newpage

\subsection{\ce{ZrI2} slab}
\begin{verbatim}
_cell_length_a                         3.64736121
_cell_length_b                         6.65245161
_cell_length_c                         40.00000000
_cell_angle_alpha                      90.000000
_cell_angle_beta                       90.000000
_cell_angle_gamma                      90.000000
_cell_volume                           'P 1'
_space_group_name_H-M_alt              'P 1'
_space_group_IT_number                 1

loop_
_space_group_symop_operation_xyz
   'x, y, z'

loop_
   _atom_site_label
   _atom_site_occupancy
   _atom_site_fract_x
   _atom_site_fract_y
   _atom_site_fract_z
   _atom_site_adp_type
   _atom_site_B_ios_or_equiv
   _atom_site_type_symbol
I001  1.0  0.250000000000  -0.079960418308  0.553855565416  Biso  1.000000  I
I002  1.0  -0.250000000000  0.423570969118  0.540832826405  Biso  1.000000  I
Zr003  1.0  -0.250000000000  -0.191024365578  0.501212791177  Biso  1.000000  Zr
Zr004  1.0  0.250000000000  0.191024365578  0.498787208823  Biso  1.000000  Zr
I005  1.0  0.250000000000  -0.423570969118  0.459167173595  Biso  1.000000  I
I006  1.0  -0.250000000000  0.079960418308  0.446144434584  Biso  1.000000  I

\end{verbatim}
\begin{figure}[h]
    \centering
    \includegraphics[width=\textwidth]{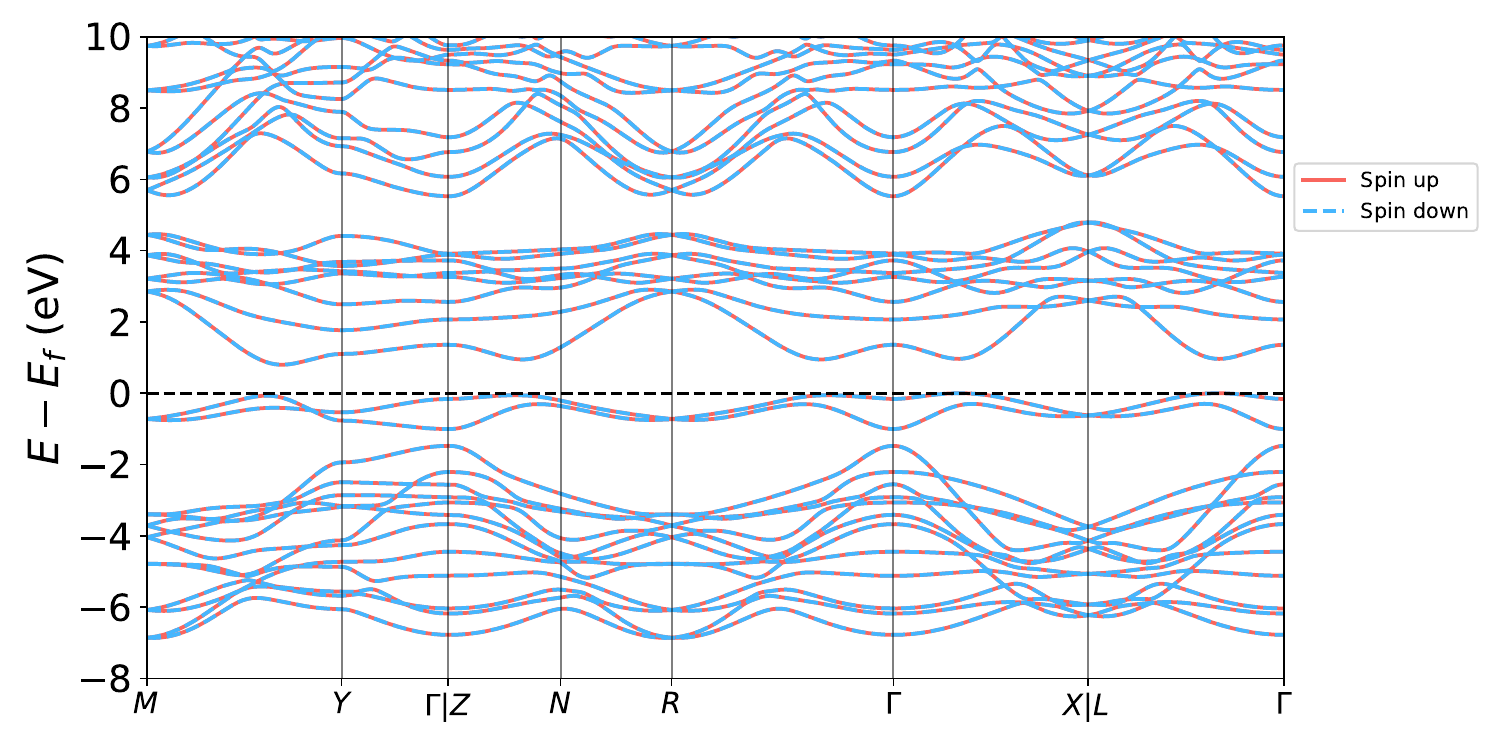}
    \caption{\footnotesize \ce{ZrI2} slab spin-polarized band structure.}
    \label{fig:ZrI2_band_slab}
\end{figure}
\clearpage

\end{document}